\documentclass[structabstract]{aa}
\usepackage{graphicx}
\usepackage{txfonts}
\newcommand{\Msol}{M_\odot}

\newcommand{\degree}{^\circ}
\newcommand{\gsct}{\gamma \ {\rm Sct} }
\newcommand{\gnor}{\gamma \ {\rm Nor} }
\newcommand{\bsct}{\beta \ {\rm Sct} }
\newcommand{\tmus}{\theta \ {\rm Mus} }
\newcommand{\ud}{{\mathrm d}}

\begin{document}

\title{The EROS2 search for microlensing events towards the spiral arms:
the complete seven season results}
\author{
Y.R.~Rahal\inst{1}\thanks{Now at Electronics Arts Canada, Vancouver, Canada},
C.~Afonso\inst{2}\thanks{Now at  Max-Planck-Institut f\"ur Astronomie,
Koenigstuhl 17, D-69117 Heidelberg, Germany},
J.-N.~Albert\inst{1},
J.~Andersen\inst{3},
R.~Ansari\inst{1}, 
\'E.~Aubourg\inst{2}\thanks{Also at APC, 10 rue Alice Domon et Lonie Duquet, F-75205 Paris Cedex 13, France}, 
P.~Bareyre\inst{2}, 
J.-P.~Beaulieu\inst{4},
X.~Charlot\inst{2},
F.~Couchot\inst{1}, 
C.~Coutures\inst{2,4}, 
F.~Derue\inst{1}\thanks{Now at LPNHE, 4 place Jussieu, F-75252 Paris Cedex 5, France}, 
R.~Ferlet\inst{4},
P.~Fouqu\'e\inst{7,8},
J.-F.~Glicenstein\inst{2},
B.~Goldman\inst{2}\thanks{Now at  Max-Planck-Institut f\"ur Astronomie,
Koenigstuhl 17, D-69117 Heidelberg, Germany}, 
A.~Gould\inst{5},
D.~Graff\inst{5}\thanks{Now at Division of Medical Imaging Physics,
Johns Hopkins University
Baltimore, MD 21287-0859, USA}, 
M.~Gros\inst{2}, 
J.~Ha\"{\i}ssinski\inst{1}, 
C. Hamadache\inst{2}\thanks{Also at CSNSM, UniversitŽ Paris Sud 11, IN2P3-CNRS, 91405 Orsay Campus, France},
J.~de Kat\inst{2},
\'E.~Lesquoy\inst{2,4},
C.~Loup\inst{4},
L.~Le Guillou\inst{2}\thanks{Now at LPNHE, 4 place Jussieu, F-75252 Paris Cedex 5, France},
C.~Magneville \inst{2}, 
B.~Mansoux \inst{1},
J.-B.~Marquette\inst{4},
\'E.~Maurice\inst{6},
A.~Maury\inst{8}\thanks{Now at San Pedro de Atacama Celestial Exploration,
Casilla 21, San Pedro de Atacama, Chile},
A.~Milsztajn \inst{2}\thanks{Deceased},  
M.~Moniez\inst{1},
N.~Palanque-Delabrouille\inst{2},
O.~Perdereau\inst{1},
S.~Rahvar\inst{9},
J.~Rich\inst{2}, 
M.~Spiro\inst{2},
P.~Tisserand\inst{2}\thanks{Now at Mount Stromlo Observatory, Weston P.O., ACT, 2611, Australia},
A.~Vidal-Madjar\inst{4},
\\   \indent   \indent
The {\sc EROS-2} collaboration\\
}
\institute{
Laboratoire de l'Acc\'{e}l\'{e}rateur Lin\'{e}aire,
{\sc IN2P3-CNRS}, Universit\'e de Paris-Sud, B.P. 34, 91898 Orsay Cedex, France
\and
{\sc CEA}, {\sc DSM}, {\sc DAPNIA},
Centre d'\'Etudes de Saclay, 91191 Gif-sur-Yvette Cedex, France
\and
The Niels Bohr Institute, Astronomy Group, Juliane Maries vej 30,
DK - 2100 Copenhagen, Denmark
\and
Institut d'Astrophysique de Paris, UMR 7095 CNRS, Universit\'e Pierre \&
Marie Curie, 98~bis Boulevard Arago, 75014 Paris, France
\and
Department of Astronomy, Ohio State University, Columbus, Ohio 43210, U.S.A.
\and
Observatoire de Marseille, {\sc INSU-CNRS},
2 place Le Verrier, 13248 Marseille Cedex 04, France
\and
Observatoire Midi-Pyr\'en\'ees, LATT, Universit\'e de Toulouse, CNRS,
14 av. E. Belin, F-31400 Toulouse, France
\and
European Southern Observatory (ESO), Casilla 19001, Santiago 19, Chile
\and
Dept. of Physics, Sharif University of Technology, Tehran, Iran
}

\offprints{M. Moniez, \email{ moniez@lal.in2p3.fr} \\
{\it see also our WWW server at  URL :} \\
{\tt http://www.lal.in2p3.fr/recherche/eros}}

\date{Received ??/??/2008, accepted }
%
%
%

\abstract
{}{
The EROS-2 project has been designed to search for microlensing events
towards any dense stellar field. The densest parts of the
Galactic spiral arms have been monitored to maximize the microlensing
signal expected from the stars of the Galactic disk and bulge.\\
}
{
12.9 million stars have been monitored during 7 seasons towards
4 directions in the Galactic plane, away from the Galactic center.\\
}
{
A total of 27 microlensing event candidates have been found.
Estimates of the optical depths from the 22 best events are
provided. A first order interpretation shows that
simple Galactic models with a standard disk and an elongated bulge
are in agreement with our observations.
We find that the average microlensing optical depth towards the
complete EROS-cataloged stars of the spiral arms is $\bar{\tau} =0.51\pm .13\times 10^{-6}$,
a number that is stable when the selection criteria are moderately varied.
As the EROS catalog is almost complete up to $I_C=18.5$,
the optical depth estimated for the sub-sample of bright target stars
with $I_C<18.5$ ($\bar{\tau}=0.39\pm .11\times 10^{-6}$) is easier to interpret.\\
}
{
The set of microlensing events that we have observed
is consistent with a simple Galactic model.
A more precise interpretation would
require either a better knowledge of the
distance distribution of the target stars, or a simulation based on a
Galactic model. For this purpose, we define and discuss the concept of
optical depth for a given catalog or for a limiting magnitude.
}

\titlerunning{Microlensing towards the spiral arms}
\authorrunning{EROS collaboration}
\maketitle

\section{Introduction}
After the first reports of microlensing candidates
(\cite{eroslmc}, \cite{machlmc}, \cite{oglpr}),
the EROS team has performed extensive microlensing surveys
from 1996 to 2003,
that monitored the Magellanic clouds and large regions in the Galactic plane.
The EROS-2 search for lensing towards the Magellanic clouds
(\cite{ErosLMCfinal}) yielded significant
upper limits on the fraction of the Milky Way halo that
can be comprised of dark objects with masses between $10^{-7}M_\odot$
and $10M_\odot$.  For objects of mass $0.4M_\odot$ the 95\% CL limit is
8\%, in conflict with the suggestion by the MACHO collaboration
(\cite{macho2000LMC}) that between 7\% and 50\% of the halo is
made up of such objects.  The EROS-2 search for microlensing of Galactic
Bulge clump giants yielded 120 events (\cite{Hamadache}) giving
a Galactic-latitude dependent optical depth of
\begin{equation}
\tau/10^{-6}=(1.62\,\pm 0.23)\exp[-a(|b|-3\degr)] \;,
\label{profoptfit}
\end{equation}
with
\begin{equation}
a\;=\;(0.43\,\pm0.16)\,\deg^{-1} \;.
\end{equation}
This optical depth agrees with Galactic models
(\cite{evans} ; \cite{Bissantz}) and with the
results of the MACHO (\cite{machobul2005}) and Ogle-II (\cite{Sumi2006}) collaborations.
The duration distribution of the events discovered by the three collaborations
have been recently analyzed by \cite{Calchi} to constrain the Galactic
Bulge Initial Mass Function.

Our team has devoted about $15\%$ of the observing time during
7 seasons to the search for microlensing events towards
the Galactic Spiral Arms (GSA),
as far as 55 degrees in longitude away from the Galactic center.
In our previous publications (\cite{GSA2y}, \cite{GSA3y}, hereafter referred as
papers I and II) describing the detection
of respectively 3 and 7 events, our attention was called on
a possible optical depth asymmetry, accompanied by an asymmetric event
dynamics with respect to the Galactic center. This marginal
effect (a $9\%$ probability to be accidental) could
be interpreted as an indication of a long Galactic bar within the bulge.
Its investigation required a
significant increase in the number of events.

In addition to the observing time increase (by more than a factor 2),
we improved our catalog of monitored stars by
increasing the limiting magnitude as well as by
recovering some fields and sub-fields that were not analyzed previously.
These improvements allowed us to recover another factor $\sim 1.5$ in
sensitivity.
Moreover the discrimination power for microlensing event identification
has been significantly improved, partly because
the light curves are longer and thus provide
a better rejection of recurrent variable objects.

A specific difficulty in the analysis of the spiral arms survey
comes from the poor knownledge of the source distance distribution;
in contrast with the LMC, the SMC and the Galactic center red giant clump,
the monitored sources in the Galactic disk
span a wide range of distances ($\pm 5$ kpc according to preliminary
studies, see Sect. \ref{sec:population}).
Their mean distance is also uncertain and has been estimated to be
$7\pm 1\,kpc$ (\cite{derueb}).
We define in this paper
the notion of ``catalog optical depth'' (Sect. \ref{sec:guidelines})
and provide all the necessary data to test Galactic models.
\section{Microlensing basics}
\label{sec:basics}
Gravitational microlensing (\cite{pacz1986}) occurs when a massive
compact object passes close enough to the line of sight of a star,
temporarily magnifying the received light.
In the approximation of a single point-like object acting as a
deflector on a single point-like source,
the total magnification of the source luminosity
at a given time $t$ is the sum of the contributions of
two images, given by
\begin{equation}
\label{magnification}
A(t)=\frac{u(t)^2+2}{u(t)\sqrt{u(t)^2+4}}\ ,
\end{equation}
where $u(t)$ is the distance of the deflecting object
to the undeflected line of sight, expressed
in units of the ``Einstein Radius" $R_E$:
\begin{eqnarray}
R_E\ &=& \sqrt{\frac{4GM}{c^2}\ Lx(1-x)}\ ,
\\
&\simeq&\ 4.54\ A.U. \times\left[\frac{M}{\Msol}\right]^{\frac{1}{2}}
\times\left[\frac{L}{10\ kpc}\right]^{\frac{1}{2}}
\times\frac{\left[x(1-x)\right]^{\frac{1}{2}}}{0.5}. \nonumber
\end{eqnarray}
Here $G$ is the Newtonian gravitational constant,
$L$ is the distance of the observer to the source and
$xL$ is its distance to the deflector of mass $M$.
The motion of the deflector relative to the line of sight
makes the magnification vary with time.
Assuming a deflector moving at a constant relative transverse
speed $V_T$, reaching its minimum
distance $u_0$ (impact parameter) to the undeflected line of sight
at time $t_0$, $u(t)$ is given by
\begin{equation}
\label{impact}
u(t)=\sqrt{u_0^2+((t-t_0)/t_E)^2},
\end{equation}
where $t_E=\frac{R_E}{V_T}$, the ``lensing time scale", is
the only measurable parameter
bringing useful information on the lens configuration in the
approximation of simple microlensing:
\begin{eqnarray}
t_E (days)=
79.\left[\frac{V_T}{100\, km/s}\right]^{-1}
\left[\frac{M}{\Msol}\right]^{\frac{1}{2}}
\left[\frac{L}{10\, kpc}\right]^{\frac{1}{2}}
\frac{[x(1-x)]^{\frac{1}{2}}}{0.5}\; . 
\end{eqnarray}

This simple microlensing description can be broken in
many different ways : double lens (\cite{mao1995}), extended source, deviations
from a uniform motion due either to the rotation of the Earth around
the Sun (parallax effect)(\cite{Gould92}, \cite{Hardy95}),
or to the orbital motion of the source around
the center-of-mass of a multiple system, or to a similar motion of
the deflector (see for example \cite{Mollerach}).

The optical depth $\tau$ towards a particular set of target stars
is defined as the
average probability for the line of sight to intercept
the Einstein disk of a deflector (magnification $A > 1.34$).
This probability is independent of the deflector mass function,
since the surface of the Einstein disk is proportional to
the deflector's mass.
When the target consists of a population of stars,
the {\it measured} optical depth is obtained from
\begin{equation}
\tau =\frac{1}{N_{obs}\Delta T_{obs}}\frac{\pi}{2}\sum_{events}
\frac{t_E}{\epsilon (t_E)},
\end{equation}
where $N_{obs}$ is the number of monitored stars;
$\Delta T_{obs}$ is the duration of the observing period;
$\epsilon (t_E)$ is the average detection efficiency
of microlensing events with a time scale $t_E$,
defined as the ratio of detected events
to the number of events with $u_0<1$
whose magnification reaches its maximum during the observing period.
Similarly, the event rate corrected for the detection efficiency is
\begin{equation}
\Gamma = \frac{1}{N_{obs}\Delta T_{obs}}\times\sum_{events}\frac{1}{\epsilon(t_E)}.
\end{equation}

\section{Experimental setup and observations}
\begin{figure*}[htbp]
\begin{center}
\includegraphics[width=18cm]{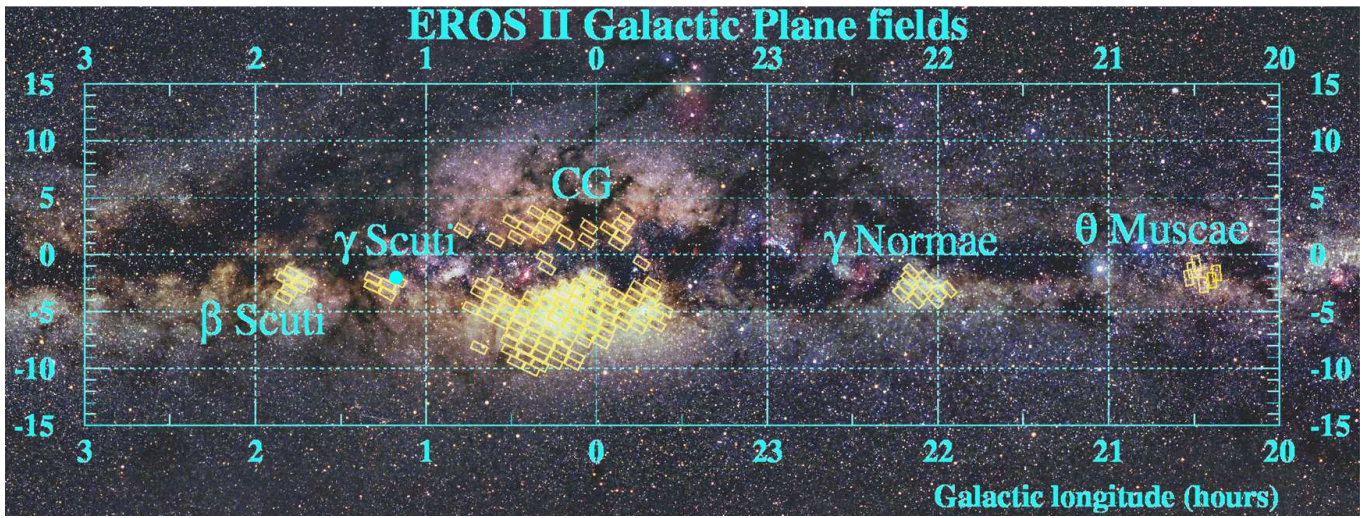}
\caption[]{The Galactic plane fields (Galactic coordinates)
monitored by {\sc EROS} superimposed on the image of the Milky-way.
The locations of our fields towards the spiral arms, as
well as our Galactic bulge fields (not discussed in this paper) 
are shown.
The large blue dot towards $\gsct$ indicates the position of the HST field
used to estimate our star detection efficiency (see text).
North is up, East is left.
}
\label{figfields}
\end{center}
\end{figure*}

\begin{figure}[htbp]
\begin{center}
\includegraphics[width=9.5cm]{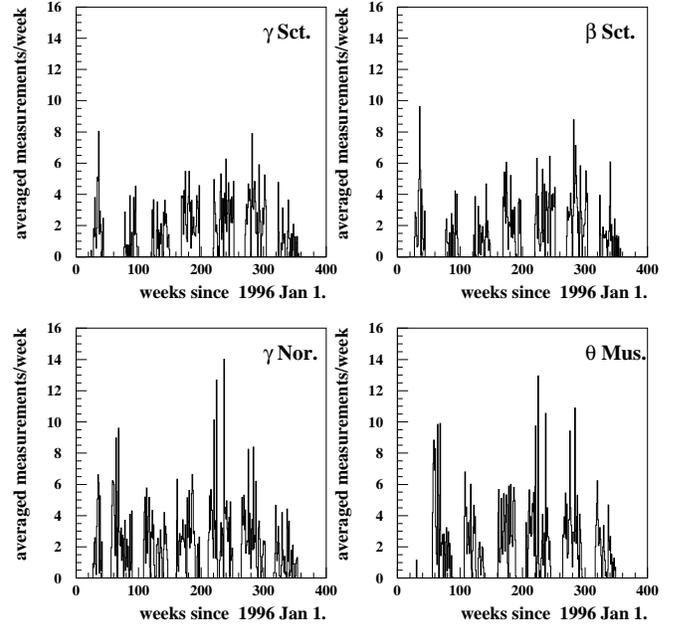}
\caption[]{Time sampling for each monitored direction: weekly average
number of measurements per star since January 1rst, 1996.}
\label{figsampling}
\end{center}
\end{figure}

The telescope, the camera and the observations, as well as the operations and 
data reduction are described in paper I and references therein.
The star population locations and the amount of data collected towards the 29 fields
that have been monitored in four different regions ($\bsct$, $\gsct$, $\gnor$
and $\tmus$) are given in Fig. \ref{figfields} and table 1.
Taking into account the dead zones, the lower efficiency sectors
of our CCDs and the blind zones around the brightest stars,
we estimate that $75\pm 4\%$ of the total CCD area ($0.95\ deg^2$)
was effectively sensitive.
This number was obtained by estimating the excess of
$10\times 10$ pixel domains ($6''\times 6''$) containing zero star,
with respect to the number of void domains expected from the Poissonian
distribution of the stellar number density.
It is in agreement with the ratio
between the total number of detected stars (summed over all
fields) and the number extrapolated from the stellar density
observed in the CCD best zones.
We took exposures of $120\ s$ towards $\bsct$, $\gsct$ and $\gnor$
and $180\ s$ towards $\tmus$.
The observations span a period of $\Delta T_{obs}=2325\ days$, starting July 1996
and ending October 2002;
369 measurements per field were obtained on average
in each of the $R_{EROS}$ and $B_{EROS}$ bands.
Our fields were calibrated using the DENIS catalog (\cite{DENIS})
and the calibration was checked with the OGLE-II catalog (\cite{ogle2000b}).
We found that $R_{EROS}$ and $B_{EROS}$ bands
are related to the Cousins I and Johnson V magnitudes through
the following color equations, to a precision of $\sim 0.1$ mag:
\begin{equation}
R_{EROS}=I_C\ ,\ \ \ \  B_{EROS}=V_J-0.4(V_J-I_C).
\label{eqcolour}
\end{equation}
Figure \ref{figsampling} shows 
the observation time span and the average sampling for the four
different directions.
\begin{table}[h!] 
\begin{center}
\caption{Characteristics of the 29 fields which were monitored 
in the EROS spiral arm program:
Locations of the field centers, average sampling 
(number of photometric measurements per light curve and per color)
%
%
and number of stars monitored
for each field. The observing time was $\Delta T_{obs}=2325\ days$.
The total numbers of observed stars towards $\gnor$ and $\tmus$ are smaller
than the sum of the numbers given for each field because of
some overlap between contiguous fields.
The total fields of view (f.o.v.) are the areas effectively monitored
($0.71\ deg^2$ per field, see text),
corrected for the overlap between fields.
}
\label{tabfields}
{\scriptsize
\begin{tabular}{|c|c c|c c|c|c|}\hline
\multicolumn{1}{|c}{\bf Field}&
\multicolumn{1}{|c}{$\alpha ^{\circ}\,(J2000)$} &
\multicolumn{1}{c|}{$\delta ^{\circ}\,(J2000)$} &
$b^{\circ}$ & $l^{\circ}$ & $N_{meas}$  & $N_{obs}\,(10^6)$ \\
\hline
\multicolumn{5}{|c|} {$\bf \bsct\  Exposure=120s.\  f.o.v.=4.3\ deg^2.$} & 268 & 3.00 \\ \hline   
bs300 & 280.8417	& -7.6814 & -1.75 & 25.20 & 269 & 0.48 \\        
bs301 & 280.8625	& -6.2283  & -1.11 & 26.51 & 266 & 0.47 \\        
bs302 & 281.5667	& -7.3792  & -2.25 & 26.80 & 272 & 0.50 \\         
bs303 & 281.5833	& -5.9264  & -1.60 & 27.09 & 261 & 0.47 \\         
bs304 & 282.3375	& -6.7642  & -2.70 & 26.71 & 269 & 0.52 \\        
bs305 & 283.1083	& -6.5956  & -3.26 & 27.19 & 271 & 0.54 \\ \hline  

\multicolumn{5}{|c|}{$\bf \gsct\  Exposure=120s.\  f.o.v.=3.6\ deg^2.$} & 277 & 2.38 \\ \hline
gs200 & 277.0125	& -14.8517 & -1.64 & 17.72  & 282 & 0.47\\        
gs201 & 277.8125	& -14.2439 & -2.12 & 18.00  & 266 & 0.49\\        
gs202 & 277.8875	& -12.8147 & -1.52 & 19.30  & 291 & 0.49\\        
gs203 & 278.5917	& -14.5275 & -2.92 & 18.09  & 281 & 0.46\\         
gs204 & 278.6167	& -13.0753 & -2.28 & 19.40  & 267 & 0.47\\ \hline 
                                                                           
\multicolumn{5}{|c|} {$\bf \gnor\  Exposure=120s.\  f.o.v.=8.4\ deg^2.$}  & 454 & 5.24 \\ \hline 
gn400 & 242.4375	& -53.1175  & -1.17 & 330.49 & 496 & 0.42 \\         
gn401 & 244.5917	& -51.7453  & -0.99 & 332.04 & 475 & 0.41 \\         
gn402 & 243.7375	& -53.0764  & -1.59 & 330.74 & 463 &0.45\\
gn403 & 245.6167	& -52.1056  & -1.69 & 332.24 & 420 &0.42\\         
gn404 & 244.7875	& -53.4439  & -2.29 & 330.94 & 435 &0.43\\         
gn405 & 246.7167	& -52.3506  & -2.35 & 332.54 & 445 &0.44\\         
gn406 & 245.9750	& -53.7314  & -2.99 & 331.23 & 443 &0.46\\         
gn407 & 247.8792	& -52.4789  & -2.95 & 332.93 & 443 &0.47\\         
gn408 & 247.1750	& -53.8661  & -3.60 & 331.63 & 453 &0.47\\         
gn409 & 243.9625	& -54.8125  & -2.86 & 329.82 & 482 &0.47\\         
gn410 & 245.1250	& -55.0717  & -3.59 & 329.93 & 443 &0.46\\         
gn411 & 242.4042	& -55.1686  & -2.54 & 328.78 & 449 &0.48\\ \hline  
                                                                           
\multicolumn{5}{|c|} {$\bf \tmus\  Exposure=180s.\  f.o.v.=3.8\ deg^2.$} & 375 & 2.28 \\ \hline  
tm500 & 201.7667	& -63.0383 & -0.47 & 306.98 & 391 &0.44 \\          
tm501 & 202.8250	& -63.5781 & -1.07 & 307.37 & 355 &0.44 \\      
tm502 & 203.7167	& -64.1750 & -1.72 & 307.66 & 376 &0.47 \\      
tm503 & 200.9917	& -64.9978 & -2.36 & 306.38 & 375 &0.36 \\          
tm504 & 198.0500	& -64.1136 & -1.35 & 305.22 & 392 &0.43\\          
tm505 & 199.0625	& -64.6806 & -1.96 & 305.60 & 360 &0.43 \\ \hline
\multicolumn{6}{r|} {\bf Total} & 12.9 \\ \cline{7-7}  
  
\end{tabular}
}
\end{center}
\end{table}


\section{The catalogs}
The catalogs of monitored stars have been produced following the procedure
described in papers I and II, based on the PEIDA photometric software
(\cite{PEIDA}).
All objects are well identified in both colors
and unambiguously associated between these two colors.
We have removed objects that suffer from a strong contamination
by a nearby bright star; the contribution to the background flux
from such a nearby star
at the position of the object should not exceed $150\%$ of its peak flux.

The seven season data set contains 12.9 million objects measured
in the two colors:
3.0 towards $\bsct$, 2.4 towards $\gsct$, 5.2 towards $\gnor$  
and 2.3 towards $\tmus$.
The number of monitored stars was increased by $\sim 50\%$ since the
analysis of
papers I and II, by producing a richer catalog from a wider choice of
good quality images than available before.
We were also able to solve some
technical problems that prevented us from producing the catalog for
some fields (\cite{thesetisserand}, \cite{TheseRahal}).
The recovered stars are mainly faint stars
with a comparatively low
microlensing sensitivity.
\subsection{Completeness, blending}
\label{sec:catalog}
\begin{figure}[htbp]
\begin{center}
\mbox{\includegraphics[width=4.5cm]{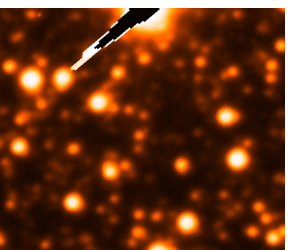}
\includegraphics[width=4.5cm]{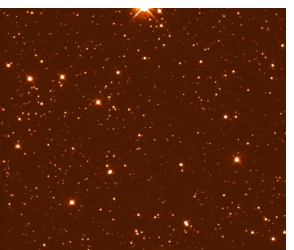}
}
\caption[]{
(a) The $R_{EROS}$ composite image (used to detect the cataloged stars) and
(b) the U6FQ1104B-HST image of the same sub-field towards gs201.
}
\label{Compare-images}
\end{center}
\end{figure}

We have compared a subset of the gs201 EROS field catalog
(Fig. \ref{Compare-images}a)
with the catalog extracted from the deeper HST-WFPC2
(Wide Field Planetary Camera 2) images
(Fig. \ref{Compare-images}b)
named U6FQ1102B (exposure 210s with filter F606W)
and U6FQ1104B (exposure 126s with filter F814W),
centered at
$(\alpha =277.6281\degree,\ \delta =-14.4823\degree)$
or $(b=17.689814\degree, l=-2.039549\degree)$, obtained from the HST archive
(\cite{HSTarchive}).
We detected 3518 stars in both colors in the HST images
corresponding to the EROS monitored field;
we systematically tried to associate these stars with an EROS
object within 1 arcsec.
All of the 869 EROS-objects properly identified as stars
in the field were associated with one or more HST star,
allowing a study of the blending and of the
detection efficiency as a function of the magnitude.
We found that $56\%$ of the EROS objects are blends with
more than one HST-star within 1 arcsec distance.
In this case the brightest HST-star accounts for an average
of $72\%$ of the $B_{EROS}$ flux of the EROS object.
These numbers vary with the EROS object magnitude as follows:
\begin{table}[h!] 
\begin{center}
{
\begin{tabular}{r|c|c|c|}
$B_{EROS}$		& {\bf 15-17}	& {\bf 17-19}	& {\bf 19-21}\\ 
\hline       
fraction of blended EROS-objects	& $70\%$	& $59\%$ 	& $54\%$  \\        
contribution of main HST star	& $88\%$	& $77\%$	& $68\%$
\end{tabular}
}
\end{center}
\end{table}

The comparison of these numbers with the ones found from
a similar study of a LMC dense field (\cite{ErosLMCfinal})
indicates that EROS cataloged objects towards gs201
are on average less blended than the objects
found towards dense regions of LMC.
Such blending could affect the microlensing optical depth determination, as
discussed in detail for SMC fields in (\cite{SMC5ans}),
and the distribution of the lensing time scale $t_E$
(\cite{Rahvar}; \cite{Bennett}).
The average densities of the EROS catalogs are similar
towards the SMC and the spiral arms; therefore
one should expect differences in blending only
if there is a difference between the spatial repartitions
of the 2 stellar populations.
Since our studies of HST images have shown
that the spiral arm stars are less blended than the
LMC stars, and considering the similarity between
the SMC and the LMC populations, we conclude that
blending should have less impact towards the spiral arms than
towards the SMC.
Therefore, to be conservative, we will use the estimates
of (\cite{SMC5ans}) as upper limits on the
optical depth systematic uncertainties
in Section \ref{sec:determination}.

\begin{figure}[htbp]
\begin{center}
\includegraphics[width=9.5cm]{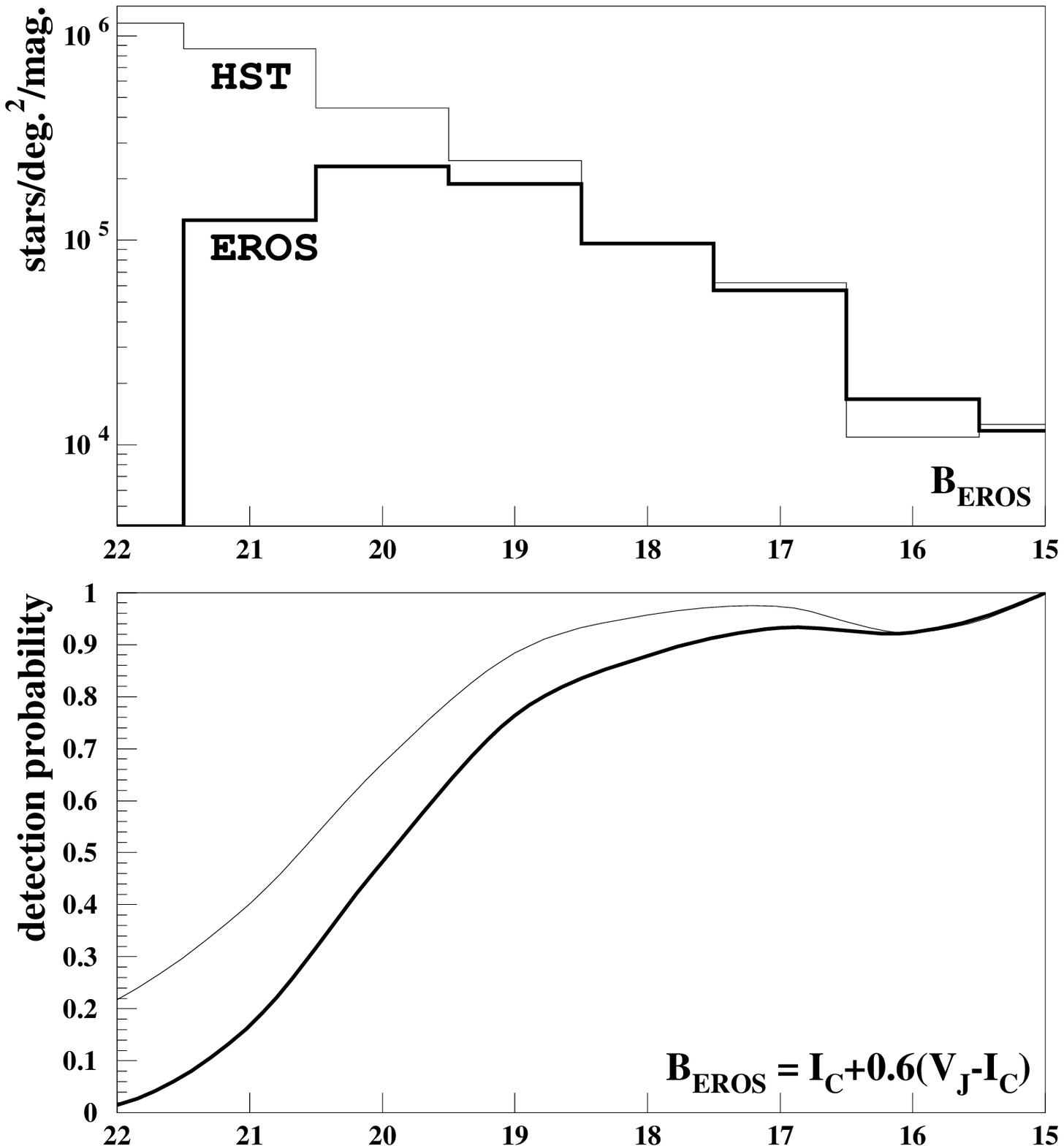}
\caption[]{
- Top panel: The EROS (thick line) and HST (thin line)
$B_{EROS}=I_C+0.6(V_J-I_C)$ magnitude distributions
of the identified objects in a sub-field of gs201.
Objects brighter than $B_{EROS}=16$ are all identified
in both images, but their magnitudes are systematically
overestimated by our photometry in the HST image, explaining the apparent
deficiency of bright HST objects.
\newline
- Lower panel:
The thin line shows the probability for an HST star to contribute
to an EROS object, i.e. to be closer than 1 arcsec from such an object
versus $B_{EROS}=I_C+0.6(V_J-I_C)$.

The thick line gives the probability for an HST star to be the
main contributor to the flux of an EROS object found within 1 arcsec.
}
\label{effdet}
\end{center}
\end{figure}

From the HST-EROS star association, we have extracted our
detection efficiencies as a function of the $B_{EROS}$
stellar magnitudes
(see figure \ref{effdet}).
As F814W and F606W HST-WFPC2 filters
are respectively very close to our $R_{EROS}$ and $B_{EROS}$ bands, we
could directly measure detection efficiencies for HST objects.
We found that every HST star that is detected in the EROS images
(i.e. that is located within 1 arcsec of an EROS object)
in the $B_{EROS}$ band is automatically detected in the $R_{EROS}$
band (the reverse is false). This is due to the different limit magnitudes
of the $B_{EROS}$ and $R_{EROS}$ templates.
Therefore the efficiency to detect a HST star in EROS
is the probability for that star to be found in the $B_{EROS}$ band.
The color-magnitude diagrams of Fig. \ref{HRdiagrams} show
that the diagonal delimitations of the populations in the bottom right
sector follow a $B_{EROS}=constant$ line, thus confirming that the detection
threshold is set by $B_{EROS}$.
We estimate the efficiency within the active region of the CCD-array,
corresponding to the effective field of $0.71\ deg^2$ for the full mosaic.
We provide in Fig. \ref{effdet} the probability for a HST star to be
the main contributor of an EROS object.
A star can also have a minor contribution to the flux of an EROS object, as
a result of blending; we show also the probability
for HST stars to contribute to an EROS object
(even if not as the main contributor).
\subsection{The color-magnitude diagram}
\label{sec:colmag}
\begin{figure*}[htbp]
\begin{center}
\includegraphics[width=18cm]{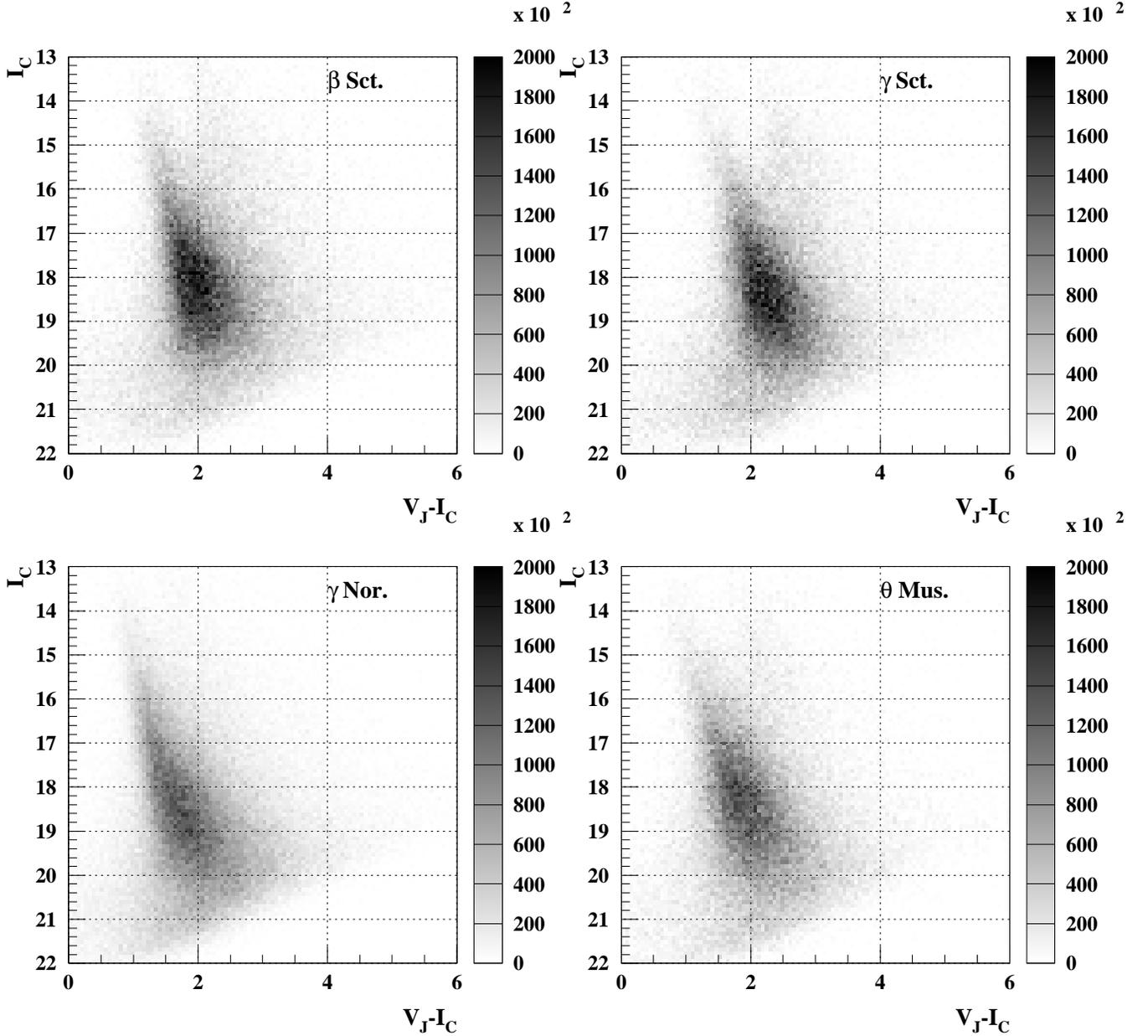}
\caption[]{
Color-magnitude diagrams $n_{eros}(I,V-I)$
of our catalogs towards the 4 monitored directions.
The grey scale gives the number density of stars per square degree,
per unit of magnitude and per unit of color index.
}
\label{HRdiagrams}
\end{center}
\end{figure*}

Figure \ref{HRdiagrams} gives the color-magnitude diagrams
$n_{eros}(I,V-I)$
of our catalogs\footnote{2D-tables of these diagrams can be found
on the Web-site: http://users.lal.in2p3.fr/moniez/
}.
The global pattern of these diagrams follows
the expected magnitude versus color lines resulting from
the light absorption of a distance-distributed stellar population.
Two parallel features are visible, with very different densities.

We were able to qualitatively reproduce these features
with a (simple) simulated catalog (Fig. \ref{diaghrsimu}).
The color-magnitude diagram of this synthesized catalog
shows two parallel features due to the main sequence and the red giant clump,
that are similar to the ones observed in the data.
Without spectroscopic data or a more detailed simulation,
it is not possible to go further
than this qualitative comparison for the
interpretation of the observed color-magnitude diagrams.
\subsection{Photometric precision}
\begin{figure}[htbp]
\begin{center}
\includegraphics[width=9.5cm]{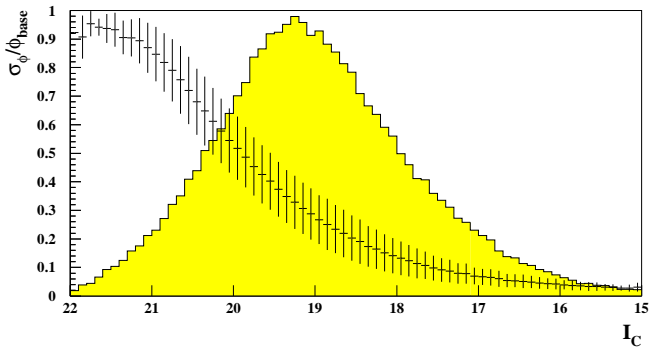}
\caption[]{
Average photometric point-to-point precision along the light-curves
versus $I_C$. The vertical bars show the dispersion of this
precision in our source sample. The histogram shows the magnitude
distribution of the full catalog (average over 4 directions).
}
\label{resolution}
\end{center}
\end{figure}

To complete the description of our observations,
Fig. \ref{resolution} gives the average point-to-point photometric dispersion
along the light-curves as a function of the magnitude $I_C$.

\section{The search for lensed stars}
Our microlensing event detection scheme
is the same as the one described in papers I and II.
In the following, we will outline
the few specificities that arise because of analysis improvements,
specific seasonal conditions or particular problems, and because of
the fact that the time baseline is twice to three times longer
than in our previous publications.

\subsection{Prefiltering}

We used the same non specific prefiltering described in paper II,
and preselected the 
most variable light curves
satisfying at least one of the following criteria:
\begin{itemize}
\item
The strongest fluctuation along the light curve (a series of
consecutive flux measurements that lie below or above the ``base flux'',
i.e. the average flux calculated in time regions devoid of significant
fluctuations)
has a small probability (typically smaller than $10^{-10}$)
to happen for a stable star, assuming Gaussian errors;
\item
The dispersion of the flux measurements is significantly larger
than expected from the photometric precision;
\item
The distribution of the deviations with respect to the base flux
is incompatible with the distribution expected from the measurements of a
stable source with Gaussian errors (using the Kolmogorov-Smirnov test).
\end{itemize}
The thresholds of these three criteria have been tuned
to select a total of $\sim 20\%$ of the light curves.
After this prefiltering, 2446843 light curves are entering the more
discriminating analysis described below.
We also included a randomly selected set of light curves ($\sim 2\%$)
to produce unbiased color-magnitude diagrams and for our
efficiency calculation (Sect. \ref{sec:simulation}).
Furthermore, we have corrected the photometric measurements
presenting a significant
correlation between the flux and the seeing
in a way that is described in (\cite{thesetisserand}).
\subsection{Filtering}
\begin{itemize}
\item
As in paper II, we first searched
for bumps in each light curve.
A bump is defined as a series of consecutive flux measurements that starts 
with a positive fluctuation of more than one
standard deviation ($+1\sigma$) from the base flux,
ends when 3 consecutive measurements lie below $1\sigma$ from the
base flux and contains at least four measurements deviating 
by more than $+1\sigma$.
We characterize such a bump by
the parameter $Q=-log_{10}(P)$ where $P$ is the probability
that the bump be due to an accidental occurrence
in a stable star light curve, assuming Gaussian errors.
We select the light curves whose most significant
fluctuation (bump 1) is positive in both colors.
\item
Then we require the time overlap between the main bumps
in each color to be at least $10\%$ of the combined time
intervals of the two bumps.
\item
To reject most of the periodic or irregular variable stars, we
remove those light curves that have a second bump
(bump 2, {\it positive or negative})
with $Q_2>Q_1/2$ in one color.
\end{itemize}

After this filtering, the 1097 remaining light curves can be fitted
assuming the simplest microlensing hypothesis, {\it i.e.} a
point-like source and a point-like deflector with a constant
speed.
\subsection{Candidate selection}
\label{sec:finalselect}
The observed flux versus time data $\Phi_{obs}(t_i)$ is fitted with
the expression $\Phi(t)=\Phi_{base}\times A(t)$,
where $\Phi_{base}$ is the unmagnified flux and $A(t)$
is given by expression (\ref{magnification}).
The candidate selection is based on the fit quality ($\chi^2$) and on
variables obtained from the $\Phi_{base}$, $t_0$, $t_E$ and $u_0$
fitted parameters.
We apply the following criteria,
tuned to select not only the ``simple'' microlensing events,
but also events that are affected by small deviations due to
parallax, source extension, binary lens effects... mentioned
in Sect. \ref{sec:basics}. The efficiency to detect caustics should be very
limited with this set of cuts, but none was found from a systematic
visual inspection of the 1097 light curves.

\begin{figure}
\begin{center}
\includegraphics[width=9.7cm]{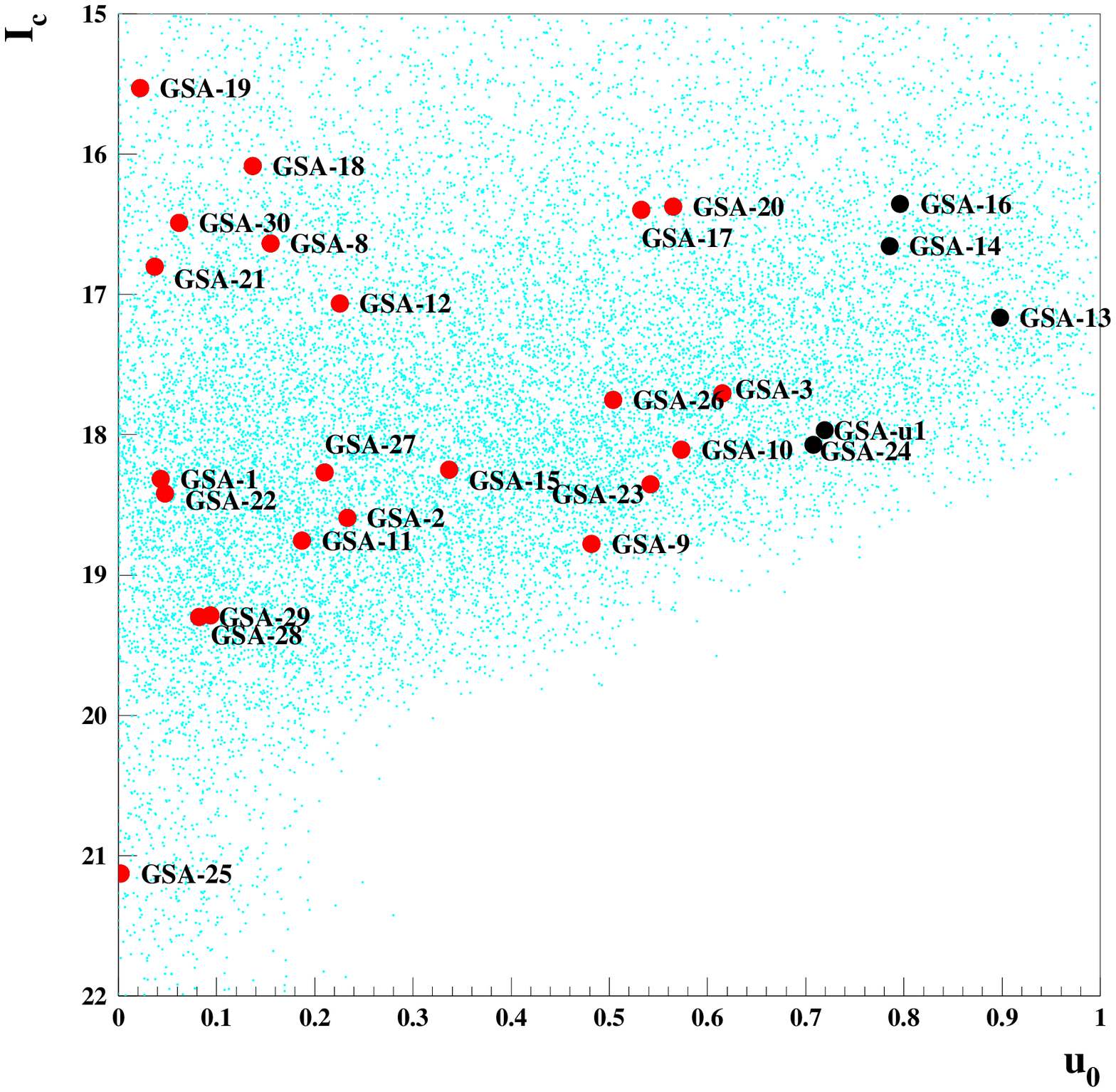}
\includegraphics[width=9.7cm]{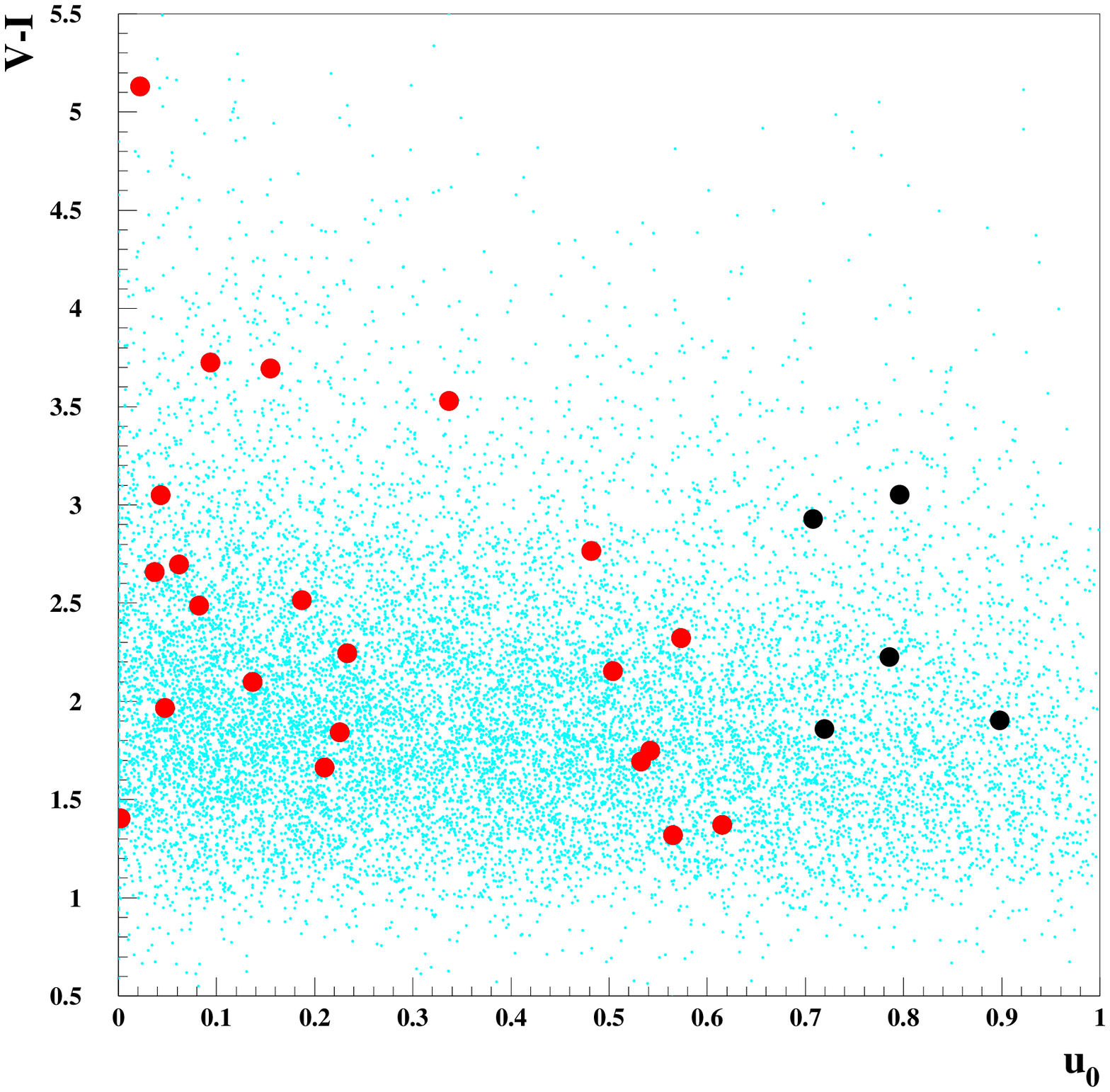}
\caption[]{
Top panel: $I_C$ versus fitted $u_0$ for the 27 microlensing candidates.
\newline
Bottom panel: $V_J-I_C$ versus fitted $u_0$.
$u_0$ is the fitted value assuming a point-like source and
a point-like deflector with a constant speed.
The red dots (respectively black dots) correspond to events with
$u_0<0.7$ used for optical depth studies (resp. $0.7<u_0<1.$).
The small dots
are the simulated events that satisfy the microlensing selection criteria.
}
\label{Ivsu0}
\end{center}
\end{figure}

\begin{itemize}
\item {\bf C1. Minimum observation of the unmagnified epoch :}
We first reduce the background due to instrumental effects
and to field crowding problems by selecting light curves that are sufficiently
sampled both during the unmagnified and the magnified stages.
For this purpose we define the ``high'' magnification epoch (called $peak$, labeled ``$u<2$'')
as the period of time during which the
fitted magnification $A$ is above $1.06$, associated to
an impact parameter $u<2$. The complementary ``low'' magnification epochs,
during which $A<1.06$, are labeled $''base''$.
We require that
\begin{equation}
\Delta T_{obs}-\Delta T_{u<2}>600\ days,
\end{equation}
where $\Delta T_{obs}=2325.\ days$ is the
observation duration, and
$\Delta T_{u<2}$ is the duration of the ``high'' magnification epoch. \\
\item {\bf C2. Sampling during the magnified epoch :} 
We also require that the interval between the peak magnification time $t_0$ and
the nearest measurement is smaller than $0.4\times \Delta T_{u<2}$.
\item {\bf C3. Goodness of a simple microlensing fit :}
To ensure the fit quality, 
we require $\chi_{ml}^2/N_{dof} <1.8$ separately for both colors,
where $\chi_{ml}^2$ and the number of degrees of freedom $N_{dof}$
are obtained from the full light-curve.
\item {\bf C4. Impact parameter :}
We also require that the fitted impact parameter $u_0$ be less than 1
for both colors.
\item {\bf C5. Stability of the unmagnified object :}
One important feature of a microlensing light curve is its stability during the
low magnification epochs, except for the rare configurations of microlensed variable stars.
We reject light curves with
\begin{equation}
\frac{\chi^2_{base}(R)+\chi^2_{base}(B)}{N_{dof}(R)+N_{dof}(B)}>8 ,
\end{equation}
where the $\chi^2_{base}$ and $N_{dof}$ values correspond to the measurements
obtained during the low magnification epochs.
\item {\bf C6. Improvement brought by the microlensing fit compared to a constant fit :}
We use the same $\Delta \chi^2$ variables as in paper II
to select light curves for which a simple microlensing fit is significantly 
better than a constant value fit:
\begin{equation}
\Delta \chi^2 _{B,R} = \frac{\chi^{2}_{cst} - 
\chi^{2}_{ml}}{\chi^{2}_{ml} / N_{dof}} \frac{1}{\sqrt{2N_{dof}}}\Big\vert_{B,R}. \\
\end{equation}
We select light curves with 
$\Delta \chi^2_B+\Delta \chi^2_R>60$.
\item {\bf C7. Overlap in the two colors :}
Defining $\Delta T_{u<1}$ as the time interval during which the fitted magnification is
larger than 1.34 ($u<1$), we require a minimum overlap between the time
intervals found in the two colors:
\begin{equation}
\frac{\Delta T_{u<1}(R)\cap \Delta T_{u<1}(B)}{\Delta T_{u<1}(R)\cup \Delta T_{u<1}(B)}>0.4\ .
\end{equation}
This loose requirement on the simultaneity of the magnifications in the two colors
allows one to keep a good sensitivity to ``complex'' microlensing events;
for example,
this cut tolerates some difference between the fitted impact parameters obtained
in the two colors (which may occur in the case of strong blending).
\end{itemize}
The number of microlensing candidates so far is 27 including an
uncertain one, labeled GSA-u1 (see below). 
The $I_C$ magnitudes and ($V_J-I_C$) colors versus $u_0$
of these candidates are shown in Fig. \ref{Ivsu0}
together with a sample of points representing the population obtained after
selection of simulated events as explained in Sect. \ref{sec:simulation}.
One clearly sees how
the maximal source magnitude required for detection decreases when the impact
parameter increases.
Annex A shows the light-curves and the finding charts of the 27
candidates, and table \ref{candidates} gives their characteristics.
The finding charts are obtained from the reference images used for
the production of the catalogs.

\subsection{Non standard microlensing events}
\label{sec:complications}
Some of our candidates are significantly better fitted with microlensing
curves resulting from complex configurations than with the basic
point-like source, point-like deflector with a
constant-speed microlensing curve.
The refinements that have been introduced in these cases are:
\begin{itemize}
\item
The blending of the lensed source with a nearby, unresolved object.
In that case, the light-curve $\Phi_{obs}(t_i)$ has to be fitted by
the following expression
\begin{equation}
\Phi(t)=C\times \Phi_{base}\times A(t)+(1-C)\times \Phi_{base}\ ,
\end{equation}
where $C$ depends on the color.
In supplement to the standard fit, this fit provides the
$C_R$ and $C_B$ parameters, where
C=(base flux of magnified component)/(total base flux).
In the notes of table \ref{candidates},
we give the magnitudes and colors of the microlensed components
that take into account the color equations (\ref{eqcolour}).
\item
Parallax.
Due to the rotation of the Earth around the Sun,
the apparent trajectory of the deflector with
respect to the line of sight is a cycloid instead of a straight line.
For some configurations (a nearby deflector and an event
that lasts a few months),
the resulting magnification versus time curve may be
affected by this parallax effect (\cite{Gould92}, \cite{Hardy95}).
The specific parameters that can
be fitted in this case are the Einstein radius
$\tilde{r}_E$ and an orientation angle,
both projected on the observer's plane which is orthogonal
to the line of sight.
\item
``Xallarap''. 
This effect is due to the rotation of the source around the center-of-mass
of a multiple system. In this case, the light-curve exhibits modulations with
a characteristic time given by the period of the source rotation
(\cite{GSA2y}, \cite{Mollerach}).
Assuming a circular orbit,
the extra-parameters to be fitted or estimated are the orbital period $P$,
the luminosity ratio of the lensed object to the multiple system,
and the projected orbit radius in the deflector's plane $\rho=ax/R_E$,
where $a$ is the orbit radius and $x=D_{lens}/D_{source}$.
\end{itemize}
In some cases, the values obtained for $t_E$
with the basic fit and the refined one may differ considerably.
For time duration studies and
for optical depth calculations, we use the $t_E$ values given by the best fit.
But as far as efficiency values are concerned, we must use those obtained
with the standard fit since they are the ones that enter the selection
procedure.
\begin{table*}
\caption{
Characteristics of the 27 microlensing candidates.
For those events that have a better fit than the point-like point-source constant speed
microlensing fit (the so-called standard fit), we also provide the standard
fit parameters.
\newline
- Names in bold type correspond to events selected for the optical depth
and duration analysis (with $u_0<0.7$).
\newline
- $I_C$ ($V_J-I_C$) are the fitted unmagnified magnitudes of the lensed object
(including the contribution of a possible blend).
\newline
- $t_0$ is the time of maximum magnification, given in HJD-2,450,000.
\newline
- $t_E$ is the Einstein disk crossing time, in days.
\newline
- $u_0$ is the dimensionless impact parameter.
\newline
- $\chi^2/dof$ corresponds to the best microlensing fit.
\newline
- $\tau$ is the individual contribution of each event to the optical depth
towards the corresponding target.
In the case of ``non standard'' events, we use the $t_E$ value
obtained from the best (non standard) fit
and the efficiency evaluated at $t_E$ of the standard fit (see text)
for the calculation of $\tau$.
}
\label{candidates}
{ 
\begin{tabular}{|c|c|c c|c|c|c|c|c|c|c|}
\hline    
candidate & field & $\alpha ^{\circ}$ & $\delta ^{\circ}$ (J2000) & $I_C$ ($V_J-I_C$) & $t_0$(days) & $t_E$(days) & $u_0$ & $\chi^2/d.o.f$ & $\tau (10^{-6})$ & note \\  
\hline
\multicolumn{10}{|c}{\large $\gsct$} & \\
\hline
 {\bf GSA1}    & 200 & 277.2888	& -14.2528   & 18.3 (3.1) & 301.2$\pm$ 0.1 &  64.0$\pm$ 1.2 & .043$\pm$.0010& 299.7/435& 0.146 &  \\
 {\bf GSA8}    & 200 & 276.8042	& -15.0311  & 16.6 (3.7) & 996.9$\pm$ 0.1 &  40.6$\pm$ 1.0 & .145$\pm$.003& 981./533& 0.114 & (1) \\
	 & & & \multicolumn{2}{r|}{Standard fit parameters:}  & 993.0$\pm$ 0.1 &  35.2$\pm$ 0.7 & .155$\pm$.002 & 1400./535& & \\
 {\bf GSA9}    & 200 & 277.1750	& -15.1644 & 18.8 (2.8) & 1760.3$\pm$ 1.7 &  57.9$\pm$ 3.6 & .482$\pm$.0158& 262.9/565& 0.142 & \\
 {\bf GSA10}   & 200 & 277.2813	& -14.8931 & 18.1 (2.3) & 1806.1$\pm$ 0.8 &  24.6$\pm$ 1.3 & .574$\pm$.0179& 237.9/596& 0.081 & \\
 {\bf GSA11}   & 201 & 278.1650	& -14.1094 & 18.8 (2.5) & 1725.3$\pm$ 0.4 &  44.3$\pm$ 1.4 & .187$\pm$.0049& 367.2/558& 0.116 & \\
 {\bf GSA12}   & 203 & 278.5875	& -13.9794 & 17.1 (1.8) & 1378.6$\pm$ 0.2 &  50.1$\pm$ 0.7 & .225$\pm$.0030&  89.3/359& 0.123 & \\
 GSA13   & 203 & 278.9404	& -14.5803 & 17.2 (1.9) & 313.9$\pm$ 1.2 &  37.2$\pm$ 2.1 & .898$\pm$.0153& 244.9/617& - & \\
 GSA14   & 204 & 278.4388	& -12.8678 & 16.7 (2.2) & 1637.7$\pm$ 3.4 &  68.4$\pm$ 3.7 & .785$\pm$.0097& 421.2/392& - & (2) \\
\hline
\multicolumn{10}{|c}{\large $\bsct$} & \\
\hline
 {\bf GSA15} & 301 & 281.0654	& -6.0339 & 18.3 (3.5) & 1399.8$\pm$ 1.4 &  72.2$\pm$ 2.8 & .337$\pm$.0126& 212.2/411& 0.110 & edge\\
 GSA16   & 301 & 280.7646	& -6.7583 & 16.4 (3.1) & 1997.0$\pm$ 3.2 &  60.6$\pm$ 4.0 & .796$\pm$.0141& 341.8/361& - & \\
 {\bf GSA17}   & 302 & 281.3950	& -7.8867 & 16.4 (1.7) & 1947.2$\pm$ 3.8 &  50.0$\pm$ 2.4 & .532$\pm$.1079& 156.9/400& 0.096 & \\
 {\bf GSA18}   & 304 & 282.2879	& -7.2500 & 16.1 (2.1) & 1718.7$\pm$ 0.1 &  55.0$\pm$ 2.0 & .137$\pm$.0009& 133./514& 0.098 & (3) \\
	 & & & \multicolumn{2}{r|}{Standard fit parameters:} & 1718.4$\pm$ 0.1 &  58.0$\pm$ 0.3 & .137$\pm$.0009& 155.6/516& & \\
\hline
\multicolumn{10}{|c}{\large $\gnor$} & \\
\hline
 {\bf GSA2}    & 400 & 242.9592	& -52.9464 & 18.6 (2.3) & 534.4$\pm$ 0.2 & 98.3$\pm$ 0.9 & .342$\pm$.002& 973.4/934 & 0.059 & (4) \\
         & & & \multicolumn{2}{r|}{Standard fit parameters:} & 533.6$\pm$ 0.5 & 137.8$\pm$ 2.6 & .233$\pm$.0029&1196.5/937& & \\
{\bf  GSA19}   & 401 & 244.1379	& -52.0272 & 15.5 (5.1) & 2367.7$\pm$ 1.3 &  90.4$\pm$ 3.0 & .043$\pm$.025 &893./880& 0.063 & (5) \\
	 & & & \multicolumn{2}{r|}{Standard fit parameters:}& 2373.5$\pm$ 0.1 &  93.1$\pm$ 0.7 & .022$\pm$.007&1388.5/888& & \\
 {\bf GSA20}   & 402 & 243.7758	& -52.9700 & 16.4 (1.3) & 2465.5$\pm$ 1.0 &  40.$\pm$ 5.0 & .72$\pm$.02& 414./696& 0.039 & (6) \\
	 & & & \multicolumn{2}{r|}{Standard fit parameters:} & 2487.1$\pm$ 0.3 &  46.3$\pm$ 0.7 & .565$\pm$.0050& 712.4/698& & \\
 {\bf GSA21}   & 404 & 244.3063	& -53.1100 & 16.8 (2.7) & 1587.3$\pm$ .03 &  74.$\pm$ 3.0 & .0142$\pm$.0008&259./565& 0.077 & (7) \\
	 & & & \multicolumn{2}{r|}{Standard fit parameters:} & 1587.2$\pm$ .03 &  39.1$\pm$ 0.2 & .037$\pm$.0007&1884.4/567& & \\
 GSA22   & 404 & 244.4263	& -54.0508 & 18.4 (2.0) & 2182.4$\pm$ 0.2 &  26.6$\pm$ 1.1 & .048$\pm$.0180& 429.5/742& 0.031 & \\
 {\bf GSA23}   & 404 & 245.1208	& -53.9825 & 18.3 (1.7) & 1573.8$\pm$ 3.8 &  78.5$\pm$ 5.7 & .542$\pm$.0152& 522.3/784& 0.062 & corner\\
 GSA24   & 406 & 246.5442	& -54.0394 & 18.0 (1.9) & 2002.5$\pm$ 1.4 &  55.5$\pm$ 2.6 & .720$\pm$.0158& 579.4/786& - & \\
 {\bf GSA25}   & 408 & 247.6917	& -53.9281 & 21.1 (1.4) & 850.9$\pm$ .03 &  67.6$\pm$ 2.9 & .003$\pm$.0001& 876.2/771& 0.057 & (8) \\
 {\bf GSA3}    & 409 & 244.1129	& -54.6303 & 17.7 (1.4) & 696.0$\pm$ 2.0 &  60.4$\pm$ 3.0 & .615$\pm$.0102& 606.7/1090& 0.051 & \\
 {\bf GSA26}   & 411 & 241.8729	& -55.3814 & 17.8 (2.2) & 1642.1$\pm$ 0.3 &  23.2$\pm$ 0.8 & .504$\pm$.0138& 441.5/759& 0.030 & \\
 {\bf GSA27}   & 411 & 242.4846	& -55.2292 & 18.3 (1.7) & 2193.8$\pm$ 0.1 &   6.8$\pm$ 0.4 & .210$\pm$.0068& 433.6/831& 0.022 & \\
\hline
\multicolumn{10}{|c}{\large $\tmus$} & \\
\hline
 {\bf GSA28}   & 501 & 202.2838	& -64.2750 & 19.3 (3.7) & 1992.2$\pm$ 0.4 &  205.$\pm$ 20.0 & .029$\pm$.004& 717/499& 0.431 & (9) \\
	 & & & \multicolumn{2}{r|}{Standard fit parameters:} & 1992.0$\pm$ 0.4 &  87.3$\pm$ 3.0 & .094$\pm$.0046& 868.5/500& & \\
 {\bf GSA29}   & 502 & 204.0683	& -63.7117 & 19.3 (2.5) & 1229.7$\pm$ 0.3 &  74.2$\pm$ 2.7 & .082$\pm$.0042& 161.7/354& 0.166 & \\
 {\bf GSA30}   & 505 & 199.2942	& -64.2592 & 16.5 (2.7) & 2396.9$\pm$ 0.1 &  12.4$\pm$ 0.2 & .062$\pm$.0023& 792.0/856& 0.073 & \\
\hline
\hline
\multicolumn{10}{|c}{\large Uncertain candidate} & \\
\hline
{\it GSAu1}   & 202 & 278.0371	& -13.2851 & 18.1 (2.9) & 1695.8$\pm$ 6.9 & 409.3$\pm$20.9 & .708$\pm$.0155& 426.9/613& & \\
\hline
\end{tabular}
{
{\bf Notes:}
{\bf (1) GSA8}: Blended; $C_R=1.00\pm 0.03,\ C_B=0.68 \pm 0.02$; lensed star has $I_C^*\ (V_J^*-I_C^*)=16.6\ (4.4)$.
{\bf (2) GSA14}: Light-curve exhibits typical features of a binary lens system.
Given the small number of measurements with significant magnification,
no reliable analysis of the shape can be performed.
{\bf (3) GSA18}: Parallax; projected Einstein radius in the solar plane $\tilde{r}_E=12.5\pm 7.0 AU$.
{\bf (4) GSA2}: Described in paper I. Found
at that time as the first candidate for a binary lensed source (Xallarap).
{\bf (5) GSA19}: Xallarap and blend; the best fit is performed ignoring
the 3 most magnified measurements, that are affected by the non-linearity of the CCD.
$\ C_R=1.,\ C_B=0.160\pm 0.013$; lensed star has $I_C^*\ (V_J^*-I_C^*)=15.5\ (8.4)$.
The light-curve distortion could be due to the face-on circular orbiting of
the source around the center-of-mass of a system including
a non luminous object, with period $P_0=294.\pm 47.\ days$,
and with a projected orbit radius of $\rho=ax/R_E=0.081\pm 0.023$,
where $a$ is the orbit radius and $x=D_{lens}/D_{source}$.
See also text.
{\bf (6) GSA20}: Parallax; $\tilde{r}_E=0.94\pm 0.07 AU$.
{\bf (7) GSA21}: Blended; $C_R=0.53\pm 0.02,\ C_B=0.34\pm 0.02$; lensed star has $I_C^*\ (V_J^*-I_C^*)=17.5\ (3.5)$.
{\bf (8) GSA25}: An improbable configuration, but a genuine one (very small $u_0$ on a very faint star).
{\bf (9) GSA28}: Blended; $C_R=0.30\pm 0.03,\ C_B=1$; lensed star has $I_C^*\ (V_J^*-I_C^*)=20.6\ (1.5)$.
The $\chi^2/dof$ of the fit is affected by
an underestimate of the errors due to bright neighboring stars.
}
}
\end{table*} 
\section{The microlensing candidates}
\subsection{General features}
\begin{figure}
\begin{center}
\includegraphics[width=9.7cm]{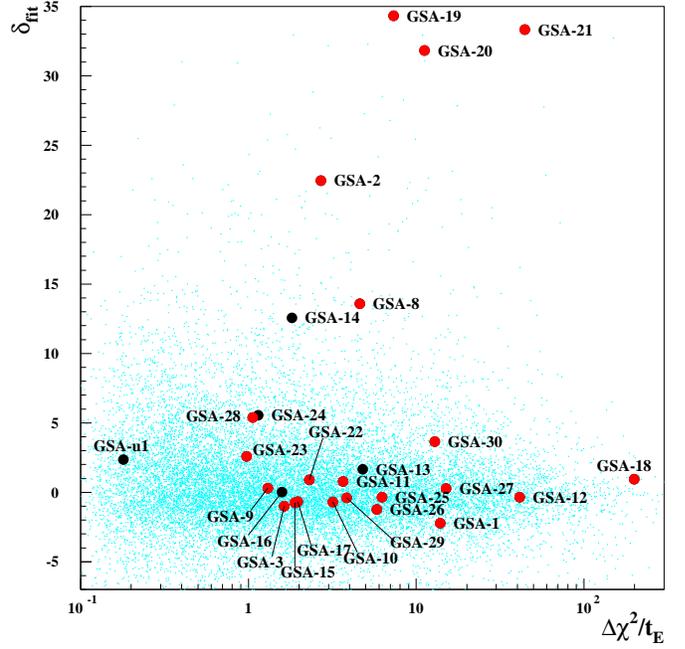}
\caption[]{$\delta_{fit}$ versus $(\Delta \chi^2_B+\Delta \chi^2_R)/t_E$
for the microlensing candidates. Red dots correspond to events with fitted $u_0<0.7$.
The small dots represent the simulated events that satisfy the microlensing selection.
}
\label{qualite}
\end{center}
\end{figure}

In order to quantify the relevance
of the interpretation of the 27 selected objects
as microlensing events, we define two variables as follows: 
\begin{itemize}
\item 
Ideally, the goodness of the microlensing fit should be uniform
throughout the observation duration.
Here we use fits made separately in the two colors.
Let $\chi^2_{u<2}$ and $n_{u<2}$ be the microlensing fit $\chi^2$
and the number of degrees of freedom, restricted to the high magnification epoch
($u<2,\ A>1.06$, see Sect. \ref{sec:finalselect}).
Let $\chi^2_{base}$ and $n_{base}$ be the complementary
variables outside the peak period.
The variable
\begin{eqnarray}
\delta_{fit}=
\left[\frac{\chi^2_{u<2}(R)+\chi^2_{u<2}(B)}{n_{u<2}(R)+n_{u<2}(B)}-
	\frac{\chi^2_{base}(R)+\chi^2_{base}(B)}{n_{base}(R)+n_{base}(B)}\right]\nonumber \\
\times\left[\frac{1}{n_{u<2}(R)+n_{u<2}(B)}+\frac{1}{n_{base}(R)+n_{base}(B)}\right]^{-\frac{1}{2}}
\end{eqnarray}
quantifies the difference of the standard fit quality during and outside the microlensing
peak, expressed in standard deviations (thanks to the second factor).
A negative value of $\delta_{fit}$ ($<-5$) is an indication of a non constant
base, and points to a variable star instead of a microlensing event.
For non-standard microlensing (parallax, blending...) $\delta_{fit}$ will be positive
and may be large ($>10$),
because the fit is expected to be less good in the peak than in the base.
\item
Many of the EROS instrumental defects ---such as bad pixels or
diffraction features--- have a long lifetime, and last for entire
observing seasons. This produces long time scale false candidates.
A signal to noise indicator is provided by the ratio
$(\Delta \chi^2_B+\Delta \chi^2_R)/(t_E/1\ day)$
where $\Delta \chi^2$ is defined above (criterion C6) and where
$t_E$ characterizes the event time scale. 
\end{itemize}
Figure \ref{qualite} shows the distribution of $\delta_{fit}$
versus $(\Delta \chi^2_B+\Delta \chi^2_R)/(t_E/1\ day)$
for the data satisfying the filtering conditions, for the final candidates and for the
simulated sample (see  Sect. \ref{sec:simulation}).
Within our final sample, three subgroups are apparent:
\begin{itemize}
\item a few (6) selected events have both a large positive
$\delta_{fit}$ and $(\Delta \chi^2_B+\Delta \chi^2_R)/t_E$. 
These are events for which a non-standard microlensing fit provides a better interpretation. 
Each of them is discussed in the remarks of table \ref{candidates}.
\item the bulk of our final sample (20) are events with a large 
$(\Delta \chi^2_B+\Delta \chi^2_R)/t_E$ and $\delta_{fit}$ compatible with 0, as
expected for standard microlensing events (and as is the case for our simulated sample). 
\item 
Event GSA-u1 has a small $(\Delta \chi^2_B+\Delta \chi^2_R)/t_E$.
After visual inspection (see Annex A, last event),
we cannot exclude a microlensing interpretation,
but the long duration and the lack of a reliable base 
make it very uncertain, considering the relatively low value of
$\Delta \chi^2_B+\Delta \chi^2_R$ (only 73.).
The status of this candidate remains pending until
further observations over a longer time range can be made.
%
%
%
One should keep in mind that confirmed events
of this type would give a major contribution to the optical depth
(GSAu1 would contribute for $\sim 0.5\times 10^{-6}$ towards $\tmus$).
%
\end{itemize}
\subsection{Comparison with the EROS 3 year analysis (paper II)}
\label{sec:comp3yr}
We first checked the coherence between the present results and those
of paper II.
Three additional candidates (GSA8, 13 and 25) with a maximum occurring
during the first three years have been found.
GSA13 and GSA25 belong to subfields that were not analyzed in paper II.
GSA8 is located at the border of two subfields, and was missed by
our previous analysis that did not systematically explore the
overlapping regions between subfields.

Four of the 7 candidates, all towards $\gsct$, found in paper II
are now rejected for the following reasons:
\begin{itemize}
\item
GSA4 and GSA7 both showed a second fluctuation
after the first three years.
\item
For GSA5, the $\chi^2$ improvement when replacing a constant fit
by a microlensing fit is no longer significant enough, due to the low
signal to noise ratio that prevails for its
light curve during the 7 years of data taking.
\item
GSA6 was found to have an impact parameter of $0.98\pm 0.04$
in paper II. Taking
into account the full light curve, the new fitted value is $u_0=1.03\pm 0.07$,
now just above our threshold. Incidentally, $\Delta\chi^2$ is also much
smaller than our threshold (60), indicating that the previous selection
of this event could have been due to a fluctuation.
\end{itemize}
One notices that these rejected candidates were the low
signal/noise ones towards $\gsct$.
Clearly, 7 years of observations
allow a much better noise reduction than 3 years.
\subsection{Overlap with other published surveys}
A very small region of $\gnor$ overlaps with the OGLE II microlensing
survey (\cite{webogle}).
No event from this region was reported in the latter survey (\cite{ogle2000a}).

A small region of our survey
overlaps the MACHO fields (\cite{Thomas}).
Amongst the 9 MACHO candidates or alerts found around $\gsct$,
3 are located within one of our monitored fields, but have not been
selected in our analysis for the following reasons:
\begin{itemize}
\item
MACHO alert number 302.44928.3523 is too faint to be measured in $B_{EROS}$
and no measurement was made in $R_{EROS}$ within 40 days of the
magnification maximum. Nevertheless, an object clearly appears in the
$B_{EROS}$ images around the maximum magnification date.
\item
MACHO alert number 301.45445.840 is too faint to be in the EROS catalog.
Furthermore, EROS missed the event as its time of maximum magnification
was 106 days before the first EROS observation of the corresponding field.
\item
MACHO alert number 302.45258.1038 was very close to one of the gaps
located between CCDs. Thus many measurements are missing.
Our standard procedure does not try to recover complete light-curves
in such a case, and the standard light-curve failed our selection process.
Nevertheless, we confirm the presence of the bump at
the right time, with the maximum magnification occurring during
the very first days of the
EROS data taking.
\end{itemize}

%
\subsection{Statistical properties of the candidate parameters}
\subsubsection{The lens configurations}
\begin{figure}
\begin{center}
\includegraphics[width=9.7cm]{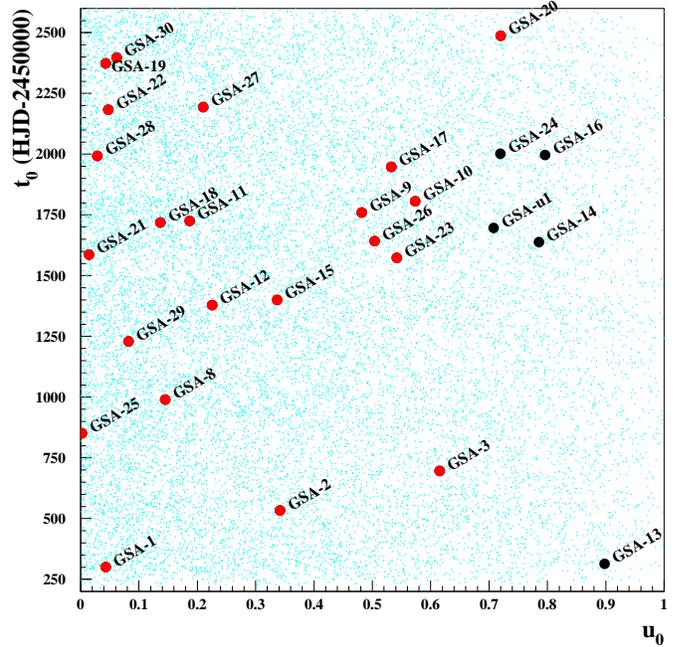}
\caption[]{$t_0$ versus fitted $u_0$ for the simulated events satisfying
the analysis criteria (small dots) and for the detected candidates (big dots).
Red dots correspond to events with fitted $u_0<0.7$ (``standard'' fit).
In the case of complex events, the {\bf best fit} $u_0$ value is plotted.
}
\label{t0vsu0}
\end{center}
\end{figure}

Microlensing events occur
with a flat-distributed impact parameter and minimum approach time.
The sample of observed microlensing event
($t_0$, $u_0$) configurations should be statistically representative of
such a distribution after taking into account
our detection efficiencies.
This is illustrated in Fig. \ref{t0vsu0}, where 
simulated events are generated as described in Sect. \ref{sec:simulation}.
\subsubsection{The microlensed star population}
\label{sec:population}
\begin{figure}
\begin{center}
\includegraphics[width=9.5cm]{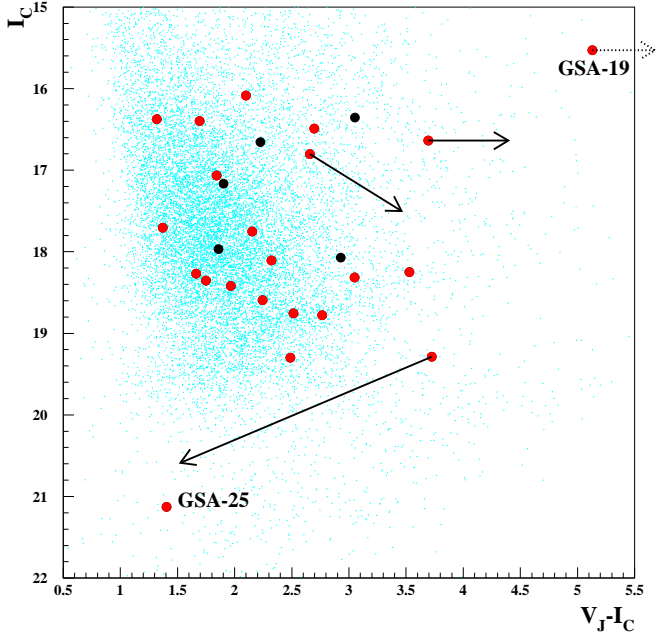}
\includegraphics[width=9.5cm]{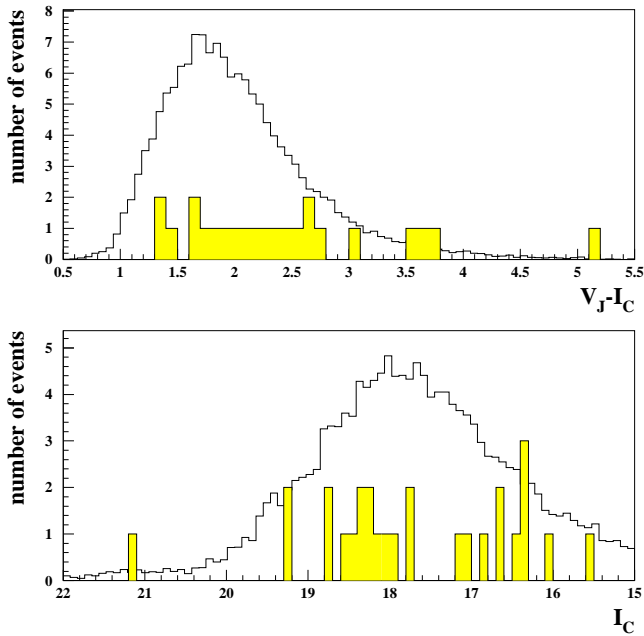}
\caption[]{
Color-magnitude diagram and projections of the simulated events satisfying
the analysis criteria (small dots) and the detected candidates (big dots).
The arrows show ($V_J-I_C,I_C$) of the magnified component in the case of
blending (see notes of table \ref{candidates}).
The red dots represent those events that are used for the optical
depth estimates.
The histograms of these events are superimposed on the projections
(not normalized).
}
\label{diagHRxeff}
\end{center}
\end{figure}

\begin{figure}
\begin{center}
\includegraphics[width=4.4cm]{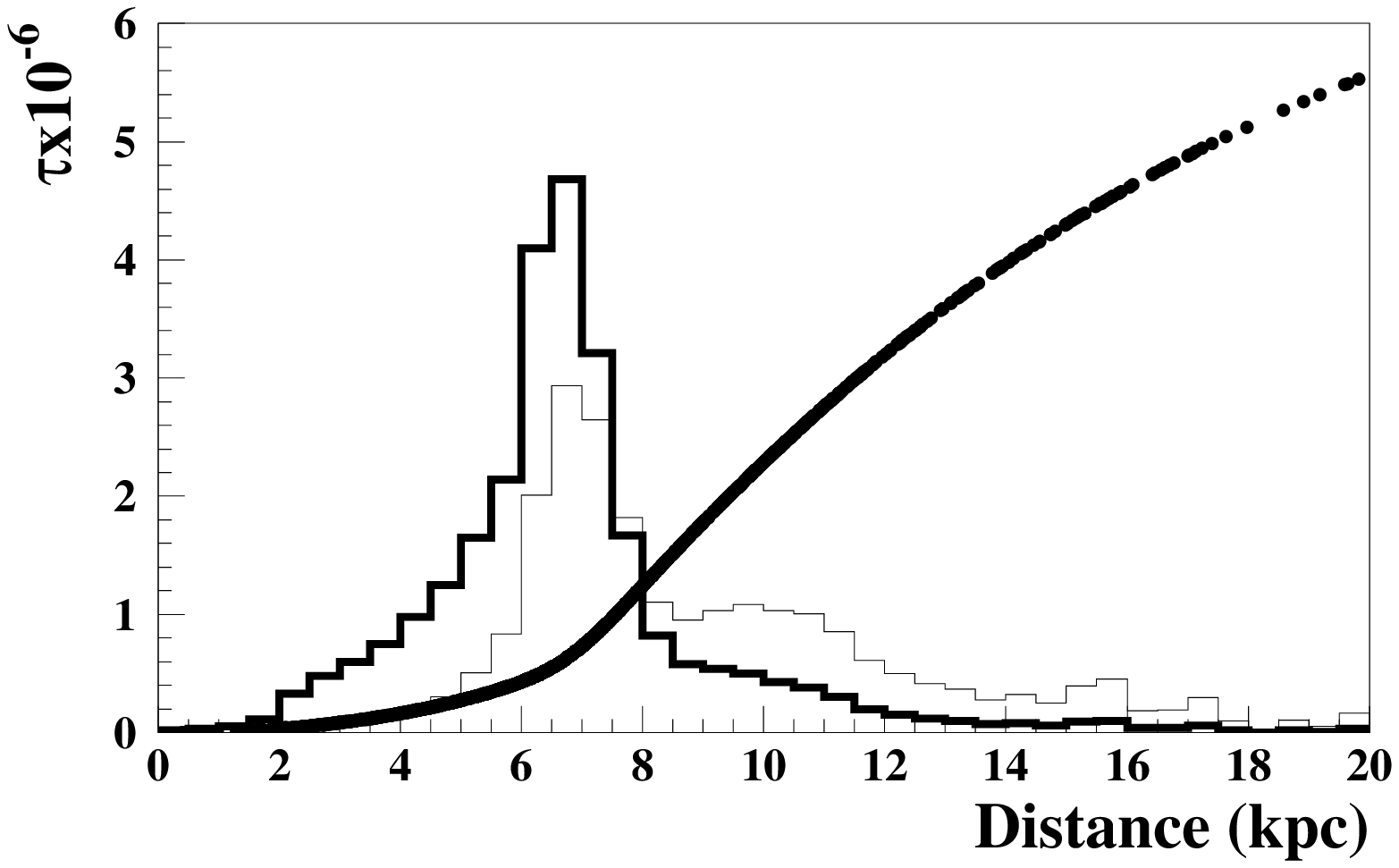}
\includegraphics[width=4.4cm]{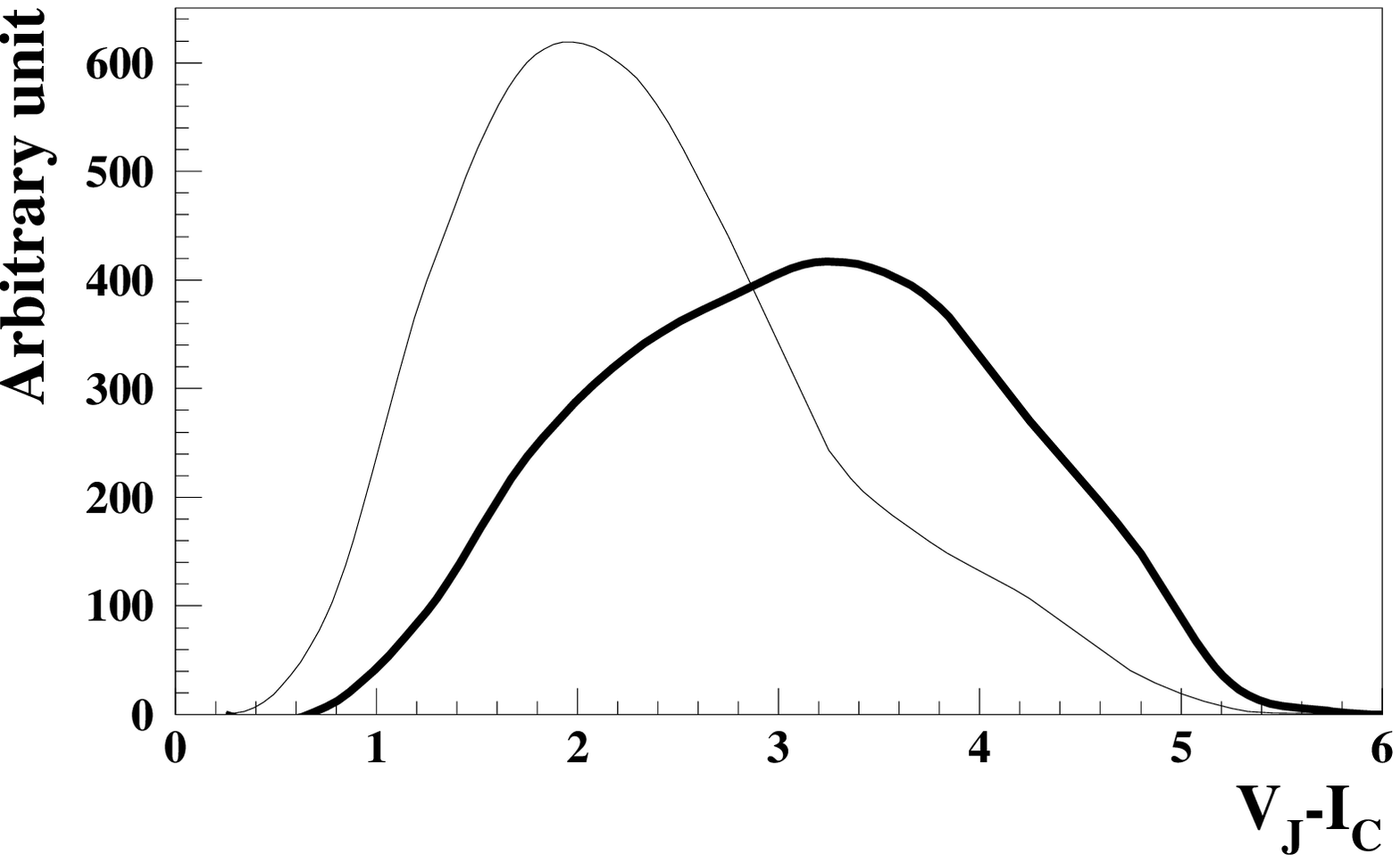}
\caption[]{
Left: Optical depth as a function of the distance (thick line),
source distance distribution of a simulated catalog
(thick histogram), and distance distribution weighted by
the optical depth (thin histogram).
\newline
Right: color distribution of the stars of the simulated catalog (thin line) and
expected color distribution of the {\it lensed} stars (thick line).
}
\label{diagHRdist}
\end{center}
\end{figure}

The microlensed star population should also be representative of
the monitored population weighted by the microlensing detection
efficiencies
and by the optical depth that may vary from source to source.
As the sources are likely to be distributed along the line of sight,
a possible variation of the optical depth with distance
must be considered in the data analysis.
As the light of a remote source is expected to be more reddened
than the light of a close one,
the optical depth $\tau$ should increase on
average with the color index.
Figure \ref{diagHRxeff} shows our color-magnitude diagram,
weighted by the microlensing efficiencies and {\it assuming the same
optical depth for all stars}.
It is directly obtained from the simulated events
that satisfy the analysis requirements.
The distribution of the observed candidates is less
peaked than the simulated one in the low color index region
because the most reddened stars are more likely to be lensed.
We were able to qualitatively confirm this color bias
through the catalog produced with a simple simulation
towards $\gsct$
described in Sect. \ref{sec:synthesizing},
that takes into account the source distance distribution
(Fig. \ref{diagHRdist} left).
%
Fig. \ref{diagHRdist}(right) shows that the color distribution
of the {\it lensed} sources (obtained by weighting with the optical depth)
is significantly biased towards the red color with respect to the simulated
distribution of detected sources.
We conclude that the distance scattering of the sources can explain
the observed bias of the lensed stars towards red color.
A more complete interpretation will be provided in a forthcoming publication
accordingly to the guidelines given in Sect. \ref{sec:guidelines}.
%
%

Two outliers need a specific comment. 
On closer inspection, it appears that GSA25 is a genuine
microlensing candidate of a very faint star.
It was detected because of the very strong magnification. This
is a rare case, but there is no reason to discard it from our list.
GSA19 is a very bright and very red object. It could be a
strongly absorbed nearby star lensed by a closer object.
%
%

\begin{figure*}[htbp]
\begin{center}
\mbox{\includegraphics[width=4.5cm]{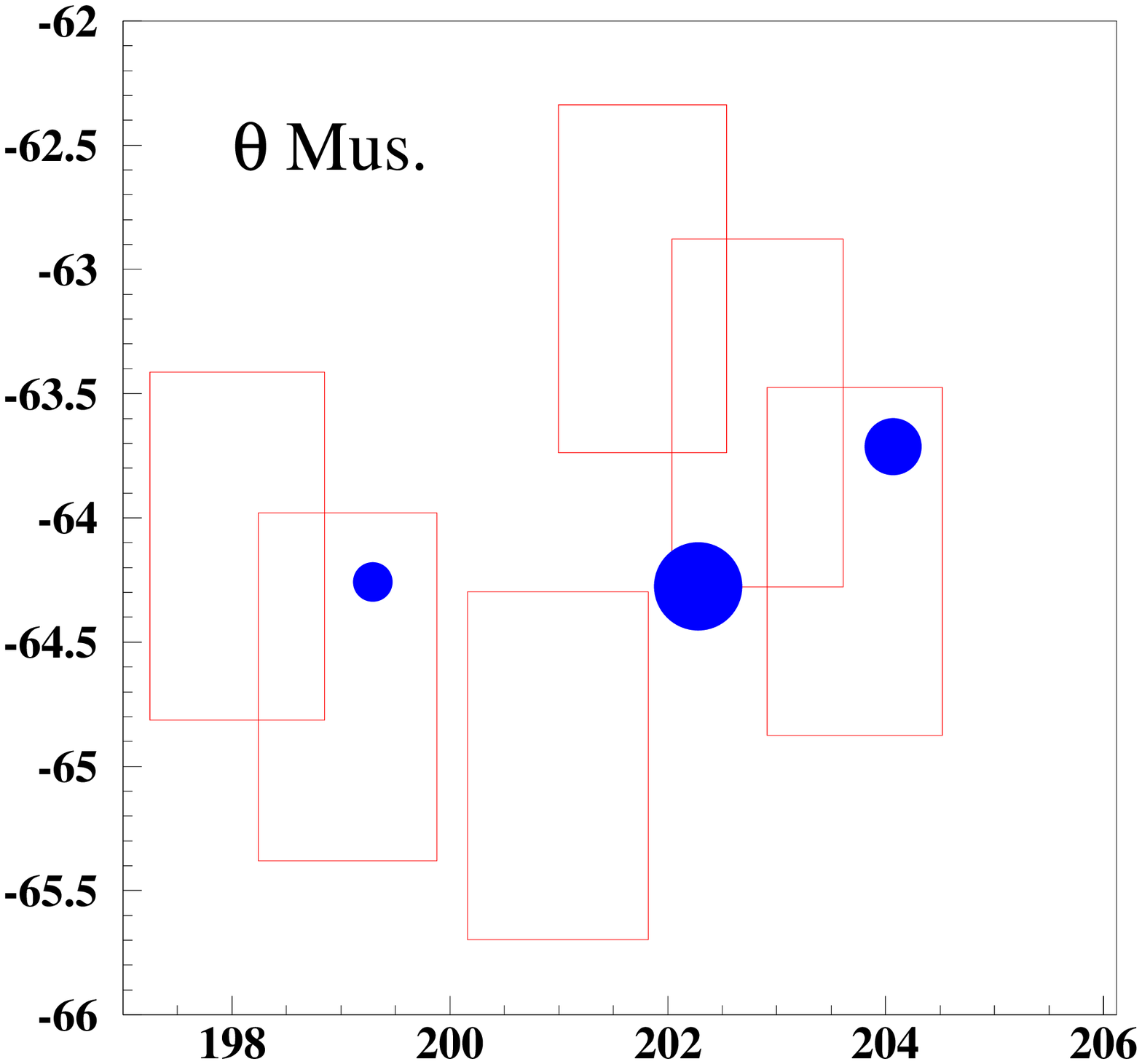}
\includegraphics[width=4.5cm]{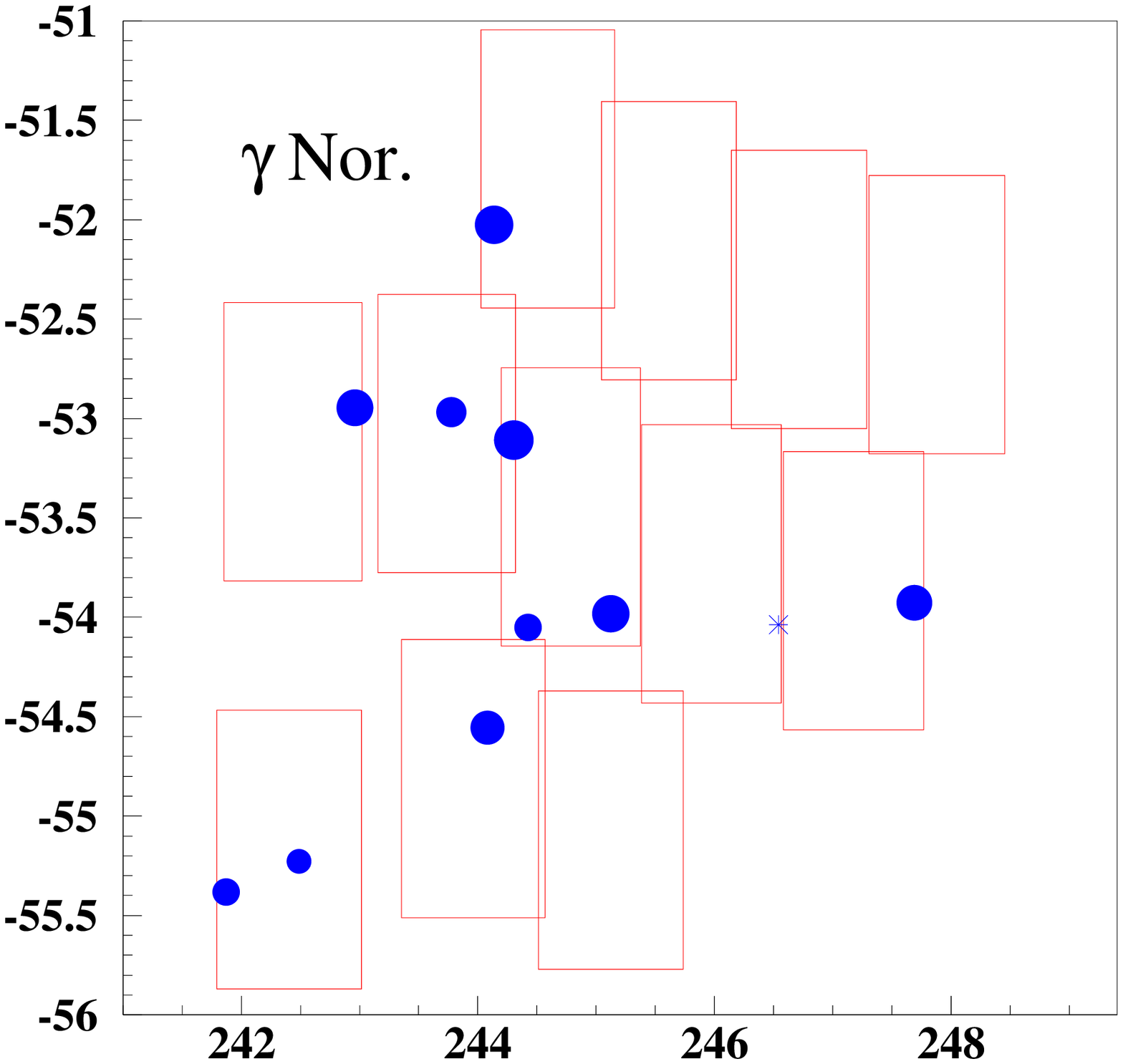}
\includegraphics[width=4.5cm]{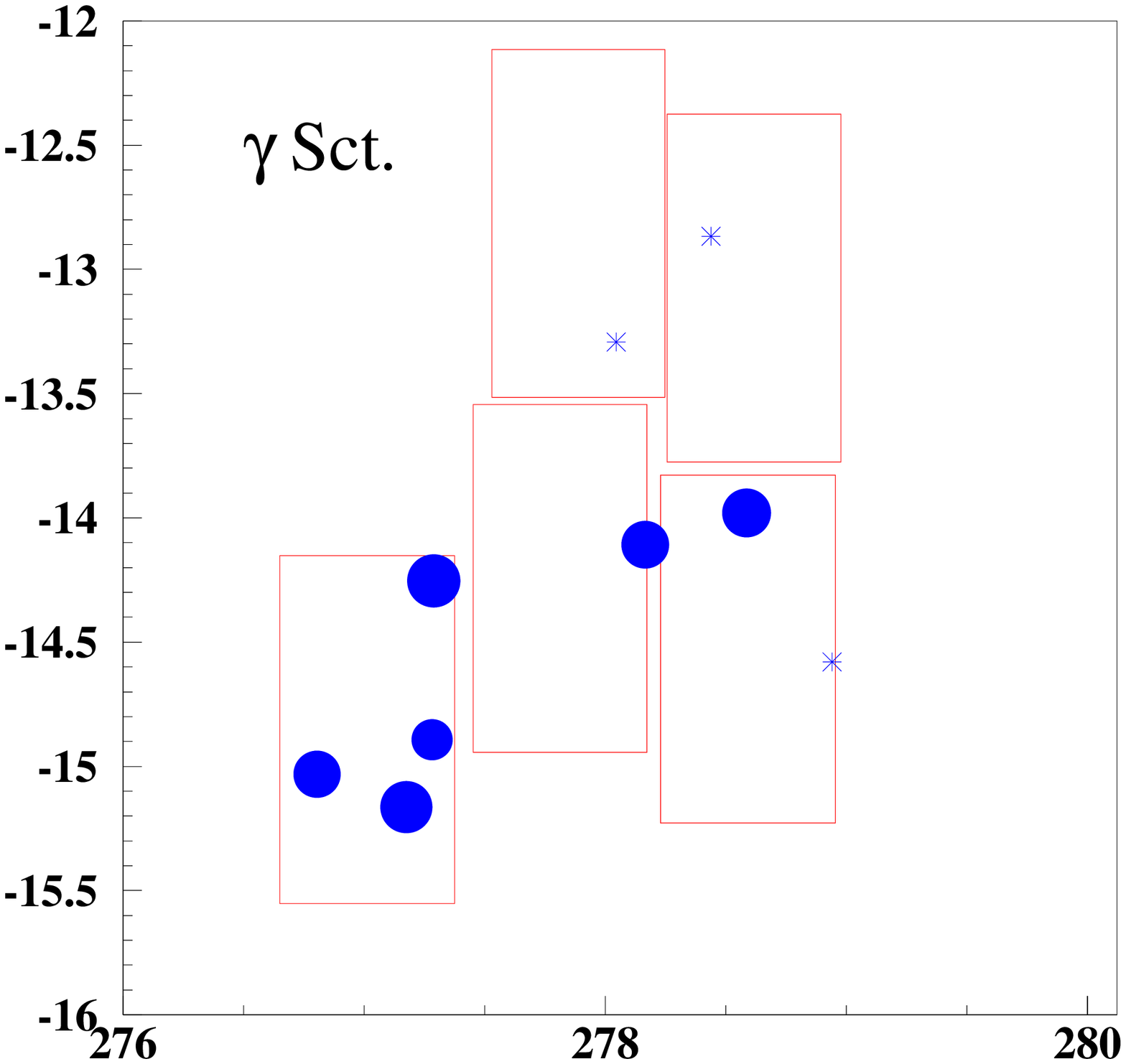}
\includegraphics[width=4.5cm]{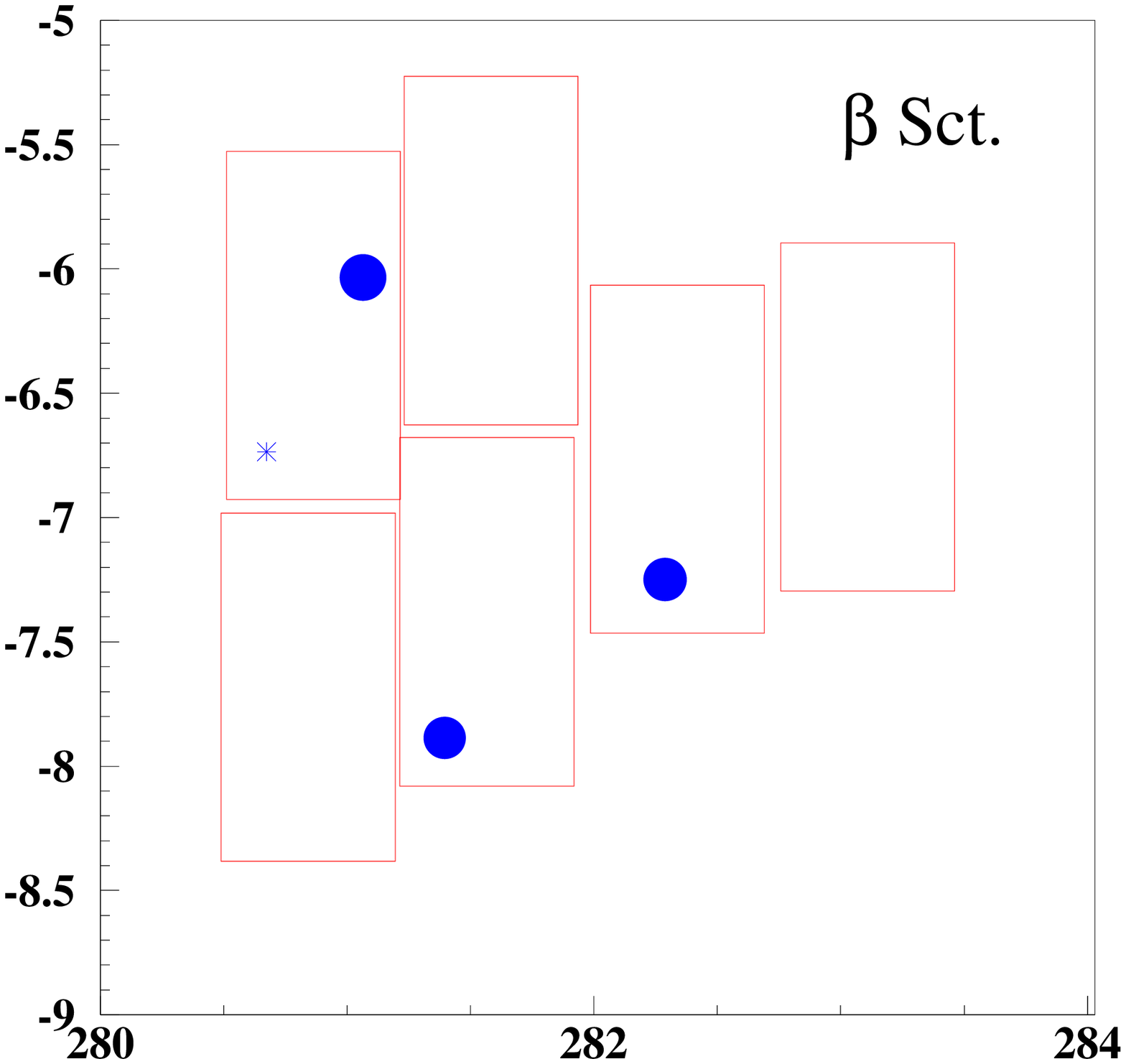}
}
\caption[]{From left to right: spatial distribution $(\alpha,\delta)$ of the 27 candidates
in the monitored fields towards $\tmus$, $\gnor$, $\gsct$ and $\bsct$.
East is right, North is up.
The area of a marker is proportional to the contribution of the
candidate to the optical depth towards the corresponding target.
Events with $u_0>0.7$ are marked with a star.
}
\label{spatial}
\end{center}
\end{figure*}

The spatial distribution of the candidates shown in
Fig. \ref{spatial} does not indicate any remarkable concentration.

\subsection{Domain of sensitivity of the analysis}
The {\bf C1} and {\bf C2} cuts that use the $\Delta T_{u<2}$
duration of the stage with magnification $A>1.06$
mainly affect the light-curves
which show long bumps. Removing these two cuts
adds 6 candidates, all of very long duration,
that have a poor signal to noise ratio.
As is the case for the GSA-u1 event, only a very long
monitoring could change the status of such candidates and
improve or degrade their signal to noise ratio.
Therefore one should keep in mind that the optical depths we
publish in this paper are almost insensitive to events
with $t_E>700\ days$ (cf. the detection efficiency versus
$t_E$ curve in Fig. \ref{efficacite}).

\section{Optical depth}
To obtain reliable optical depth values, we use a sub-sample
of good quality candidates which should be almost free of the microlensing
like variable objects that have been identified towards other EROS
targets (\cite{ErosLMCfinal}; \cite{Hamadache}) and in Sect. \ref{sec:comp3yr}.
For this purpose, we will only keep those candidates that have $u_0<0.7$
in the fit that assumes a point-like source and a point-like deflector
with a constant speed. This is approximately equivalent to requesting
$A_{max}>1.68$.
Events GSA13, 14, 16, 24 and GSAu1 are then discarded for the optical depth
analysis.
\subsection{Microlensing detection efficiency}
\label{sec:simulation}
\begin{figure*}
\begin{center}
\includegraphics[width=18cm,height=22cm]{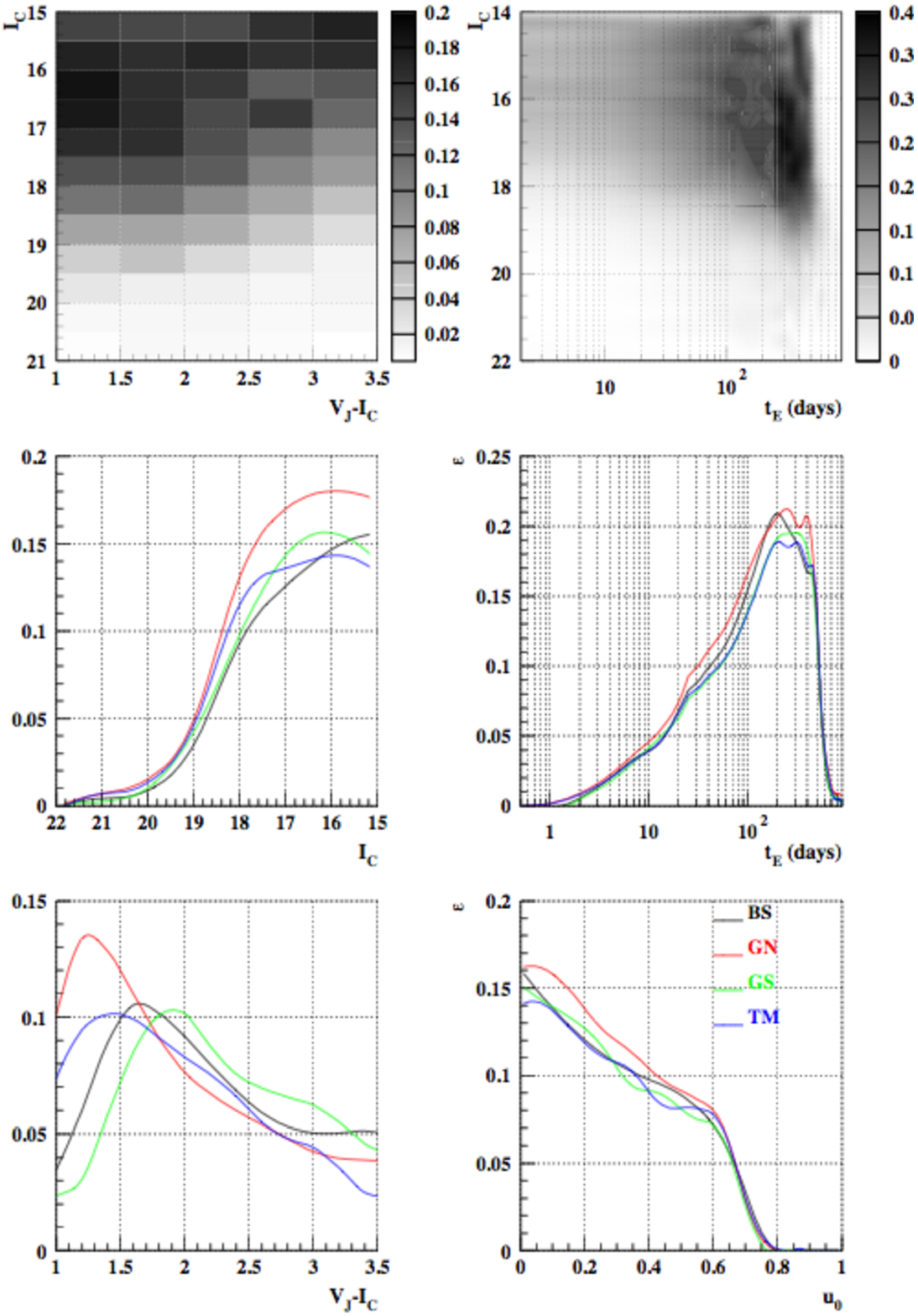}

\caption[]{Microlensing detection efficiency.
\newline
Upper left: the efficiency in the ($V_J-I_C, I_C$) plane
averaged over $u_0$, $t_0$, $t_E$ and over the 4 monitored fields.
\newline
Upper right: the efficiency in the ($t_E,I_C$) plane
averaged over $u_0$, $t_0$, $V_J-I_C$ and over the 4 monitored fields.
\newline
Other panels: efficiency as a function of the
lensed star magnitude $I_C$, as a function of the color index $(V_J-I_C)$
and as a function of the event parameters $t_E$ and $u_0$.
Each curve gives the efficiency averaged
over all the other parameters.}
\label{efficacite}
\end{center}
\end{figure*}

We present here the efficiency calculation for the detection
of events with $u_0<0.7$.
As for our previous papers, we calculate our detection efficiency
by superimposing simulated events on
measured light curves from an unbiased sub-sample of our catalog.
Events are simulated as point-source, point-lens constant velocity
microlensing events, with parameters uniformly spanning a domain
largely exceeding the domain of EROS sensitivity
($u_0$ up to 2, $1\ day<t_E<900\ days$, $t_0$ generated from 150 days
before the first observation to 150 days after the last).
Efficiency is defined as the ratio of events satisfying the selection
cuts to the number of events generated up to $u_0=1$.
Figure \ref{efficacite} (upper left panel) shows the EROS efficiency
as a function of
the source position in the color-magnitude diagram
averaged over all the other parameters and over all directions.
The other frames of Fig. \ref{efficacite} show the
efficiency as a 2D-function of $I_C$ of the lensed star and $t_E$,
and as a 1D-function of $t_E$, $u_0$, $I_C$ and $V_J-I_C$,
averaged over all the other parameters, for each monitored direction.
The efficiency is significantly better towards $\gamma$ Nor, because
of the higher sampling of the light curves in this direction.

\subsection{Optical depth determination}
\label{sec:determination}
The optical depth values obtained from the 22 events that satisfy $u_0<0.7$
are given in the first part of table \ref{table:predictions}.
\begin{table}[h!]
\begin{center}
\begin{tabular}{|c|cccc|c|}
\cline{2-5}
\multicolumn{1}{c|}{} & \bf $\tmus$ &  \bf $\gnor$ & \bf $\gsct$ & \bf $\bsct$ & \multicolumn{1}{|c}{}\\ \cline{1-5}
 $\bar{b\degree}$     & -1.46 & -2.42 & -2.09 & -2.15 & \multicolumn{1}{|c}{}\\
 $\bar{l\degree}$     & 306.56 & 331.09 & 18.51 & 26.60 & \multicolumn{1}{|c}{}\\ \hline
\multicolumn{1}{|c}{} & \multicolumn{4}{c|}{\bf Observations} & All \\ \hline
 $\tau\times 10^6$    &$.67^{+.63}_{-.52}$&$.49^{+.21}_{-.18}$&$.72^{+.41}_{-.28}$&$.30^{+.23}_{-.20}$ & $.51^{+.13}_{-.13}$\\  
 $N_{events}$              &   3   &       10              &        6             & 3  & 22  \\
 $\bar{t_{\rm E}}$ (days) &  $97\pm 47$ &   $57\pm 10$   &   $47\pm 6$     & $59\pm 6$ & $60\pm 9$    \\
 $\sigma_{t_{\rm E}}$ & 80  & 29  & 13   & 10  & 40  \\
%
$\epsilon$ corrected $\bar{t_{\rm E}}$ & $65\pm 45$ & $43\pm 10$ & $45\pm 6$ & $58\pm 6$ & $48\pm 9$ \\
$\epsilon$ corrected $\sigma_{t_{\rm E}}$ & 75 & 31 & 13 & 9 & 38 \\
 $median\ t_E$	& 74.2	& 64.	& 47.2	& 55.	&  56.5 \\
\hline
\multicolumn{6}{c}{} \\
\hline
\multicolumn{1}{|c}{} & \multicolumn{4}{c|}{\bf $\tau\ (\times 10^6)$ from models} & $\chi^2_{model}$ \\ \hline
A      		& 0.32  & 0.48                 & 0.79                 & 0.60 & 2.3 \\
+spiral      	& 0.56  & 0.69                 & 1.07                 & 0.83 & 7.9 \\
B      		& 0.34  & 0.51                 & 0.85                 & 0.64 & 2.9 \\
+spiral      	& 0.61  & 0.72                 & 1.13                 & 0.90 & 10.2 \\
C      		& 0.47  & 0.78                 & 1.11                 & 0.95 & 12.5 \\
+spiral      	& 0.71  & 1.18                 & 1.43                 & 1.23 & 36.1 \\
D		& 0.32	& 0.56		       & 0.41		      & 0.38 & 1.4 \\
model 1		& 0.42  & 0.52                 & 0.71                 & 0.57 & 1.8 \\
model 2         & 0.54  & 0.68                 & 0.90                 & 0.74 & 5.4 \\
\hline
\multicolumn{6}{c}{} \\
\cline{1-5}
\multicolumn{1}{|c}{} & \multicolumn{4}{c|}{\bf Predictions from model 1} & \multicolumn{1}{|c}{} \\ \cline{1-5}
 $N_{events}$                  &  2.8  &  9.9                 & 7.1                  & 6.3 & \multicolumn{1}{|c}{} \\
 $\bar{t_{\rm E}}$   &  73.8   &  67.9                  & 37.9                   & 60.2 & \multicolumn{1}{|c}{} \\
 $\sigma_{t_{\rm E}}$ &  63   &  54                  & 36                   & 48 & \multicolumn{1}{|c}{} \\
 $median\ t_E$	& 54.5	& 52.5	& 28.0	& 46.5	& \multicolumn{1}{|c}{} \\
\cline{1-5}
\multicolumn{1}{|c}{} & \multicolumn{4}{c|}{\bf $\bar{t_{\rm E}}$ from published models} & \multicolumn{1}{|c}{} \\
\cline{1-5}
C (no effic.)		& 45	& 28	& 25	& 27 & \multicolumn{1}{|c}{} \\
C+spiral		& 38	& 38	& 44	& 40 & \multicolumn{1}{|c}{} \\
D 		& 80.9	& 86.5	& 76.9	& 77.4 & \multicolumn{1}{|c}{} \\
\cline{1-5}
\cline{1-5}
\end{tabular}
\caption[]
{
%
%
\newline
- Observed optical depth $\tau$, number of events $N_{events}$, 
average $\bar{t_{\rm E}}$, dispersion $\sigma_{t_{\rm E}}$,
efficiency corrected $\bar{t_{\rm E}}$ and $\sigma_{t_{\rm E}}$
and median time scale
for each monitored direction.
Here, we consider only the events used for the optical depth
estimates ({\it i.e.} with $u_0<0.7$).
\newline
- Optical depth predictions from
models A, B, C,
including or not a spiral structure (see text), 
and from model D,
and estimates up to {\rm 7 kpc} from the two
simple Galactic models described in table \ref{table:tabmodel}.
The $\chi^2$'s quantify the adequation with the observations
(see Sect. \ref{sec:comparisons}).
\newline
- Expected numbers of events and duration distribution parameters
from model 1 and from models C and D.
$\bar{t_{\rm E}}$ from model C do not take into account the
detection efficiency.
}
\label{table:predictions}
\end{center}
\end{table}
The average over all directions is defined as
the proportion of stars covered by an Einstein disk. It is given by
$$\bar{\tau}_{fields}=\frac{\sum_i N_*^i \tau_i}{\sum N_*^i}\ {\rm with}\ 
\tau_i=\frac{\pi}{2}\frac{1}{N_*^i \Delta T_{obs}^i}\sum_{events}\frac{t_E}{\epsilon^i(t_E)},$$
where $N_*^i,\ \Delta T_{obs}^i ,\ \epsilon^i(t_E)$ are
respectively the number of stars monitored, the observation
duration and the microlensing selection efficiency towards direction $i$.
As usual, selection efficiency is relative to events with $u_0<1$
(even though the efficiency is almost zero for $u_0>0.7$).
As explained at the end of Sect. \ref{sec:complications}, we use
$\frac{t_E(best\ fit)}{\epsilon(t_E(standard\ fit))}$
in the expression of $\tau$ for non-standard microlensing events.
The statistical uncertainties are estimated from the definition of the
classical $68\%$ confidence intervals (\cite{feldman}), multiplied by
the factor
\begin{equation}
\frac{\sqrt{<t^2_E/\epsilon^2(t_E)>}}{<t_E/\epsilon(t_E)>}\; ,
\end{equation}
following \cite{Han-1995}.
According to the discussion of Sect. \ref{sec:catalog}, we
assume a $10\%$ maximum systematic uncertainty due to blending effects,
as in \cite{SMC5ans},
and we account for another $5\%$ uncertainty due to
the statistical limitations in the determination of the
efficiencies. These errors are small compared to the smallest
statistical uncertainty ($\sim 30\%$) estimated in this paper.
\subsection{Robustness of the optical depth values}
\begin{figure}
\begin{center}
\includegraphics[width=8.5cm]{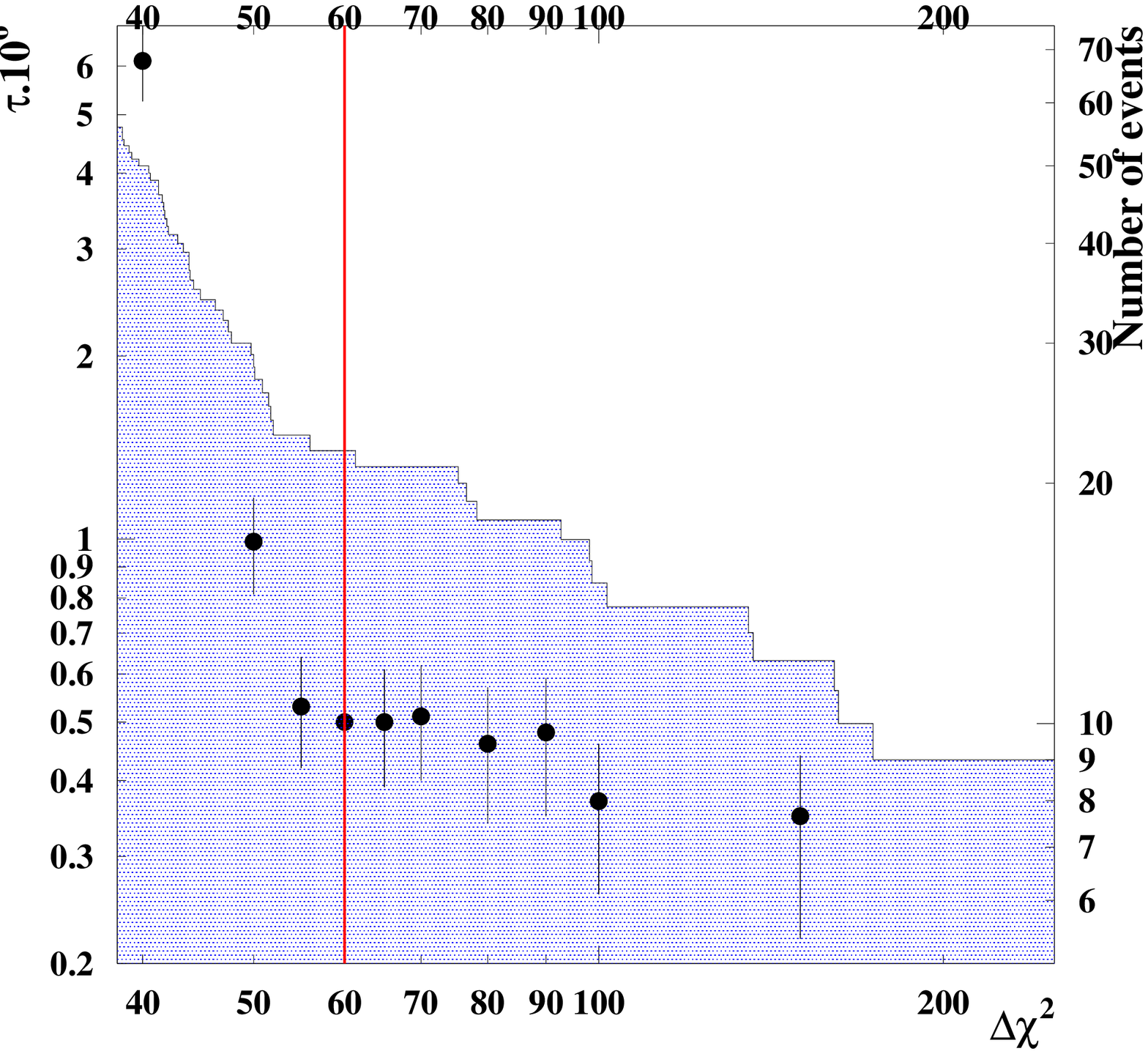}
\caption[]{
Variation of the number of selected candidates (histogram, right scale)
and of $\bar{\tau}_{fields}$ (dots, left scale)
with the $\Delta \chi^2$ threshold. The red vertical line
shows our cut. Only statistical errors are plotted.
}
\label{stabilite}
\end{center}
\end{figure}

We have studied the stability of the optical depth
averaged over all fields by changing the selection cuts.
Fig. \ref{stabilite} gives the variation of $\bar{\tau}_{fields}$
with the $\Delta \chi^2$ threshold.
Relaxing this threshold makes the optical depth increase rapidly,
due to the inclusion of false events,
as a logical result of the analysis optimization.
But using a stricter cut does not significantly change the
optical depth values, as long as the statistics remains significant.

\begin{figure}
\begin{center}
\includegraphics[width=8.5cm]{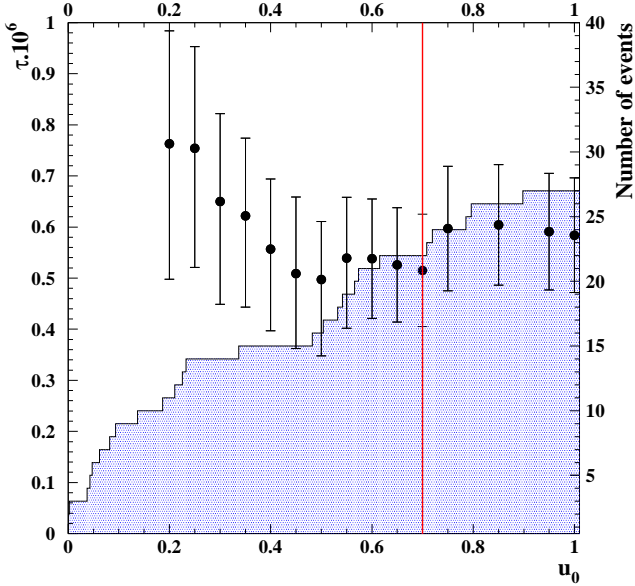}
\caption[]{
Variation of the number of selected candidates (histogram, right scale)
and of $\bar{\tau}_{fields}$ (dots, left scale)
with the $u_0$ threshold. The red vertical line
corresponds to our selection.
Only statistical errors are plotted.
}
\label{stabu0}
\end{center}
\end{figure}

Fig. \ref{stabu0} also gives $\bar{\tau}_{fields}$ as a function
of the $u_0$ threshold. Our result does not depend on this cut
within statistical errors, showing that we probably found an
optimum between the quality of the events and the number of
those kept for our optical depth calculations.

\begin{figure}
\begin{center}
\includegraphics[width=8.5cm]{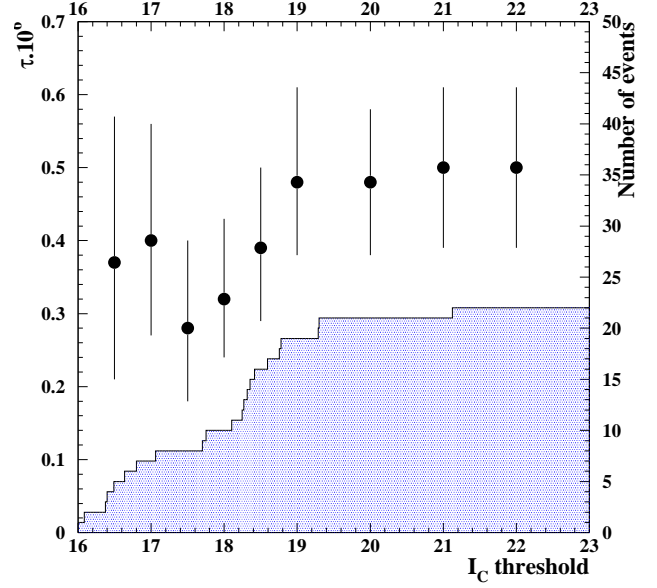}
\caption[]{
Number of selected candidates (histogram, right scale)
and $\bar{\tau}_{fields}$ (dots, left scale)
for the sub-samples of stars brighter than
the $I_C$ threshold. Only statistical errors are plotted.
}
\label{stabIc}
\end{center}
\end{figure}

\begin{figure}
\begin{center}
\includegraphics[width=8.5cm]{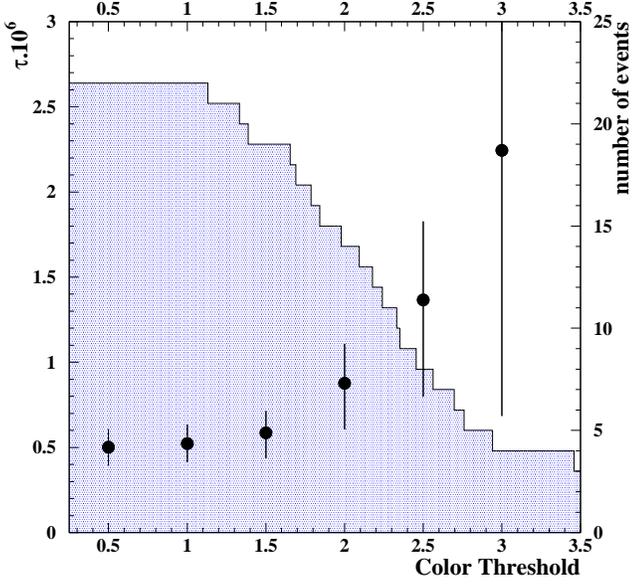}
\caption[]{
Number of selected candidates (histogram, right scale)
and $\bar{\tau}_{fields}$ (dots, left scale)
for the sub-samples of stars redder than the
color index ($V_J-I_C$) threshold. Only statistical errors are plotted.
}
\label{tauvscol}
\end{center}
\end{figure}

Fig. \ref{stabIc} and \ref{tauvscol} show the variation of $\bar{\tau}_{fields}$
with the maximum magnitude $I_c$ of the source population
and with the minimum color index $V_J-I_C$.
There is no evidence for a variation with $I_C$ threshold.
As discussed below in Sect. \ref{sec:further},
this comes from the fact that the variation of
the optical depth with the distance does not result
in a variation with the magnitude as
more distant identified sources
do not appear fainter in our catalog on average.
Interestingly one may use these figures
to extract $\tau$ for specific
stellar populations, in particular the
population of the brightest 
stars with $I_C<18.5$ 
that are better identified and measured, and that
suffer less blending (see Sect. \ref{sec:catalog}).
Moreover, our catalog is almost
complete up to this magnitude, as our star detection efficiency is
large 
(see Fig. \ref{effdet}).
Using such a sub-catalog of bright stars
should make the interpretation easier within a Galactic model framework
as will be discussed hereafter.
\section{Discussion: comparisons with simple models}
\label{sec:comparisons}
\subsection{Optical depth}
\begin{figure}
\begin{center}
\includegraphics[width=8.5cm]{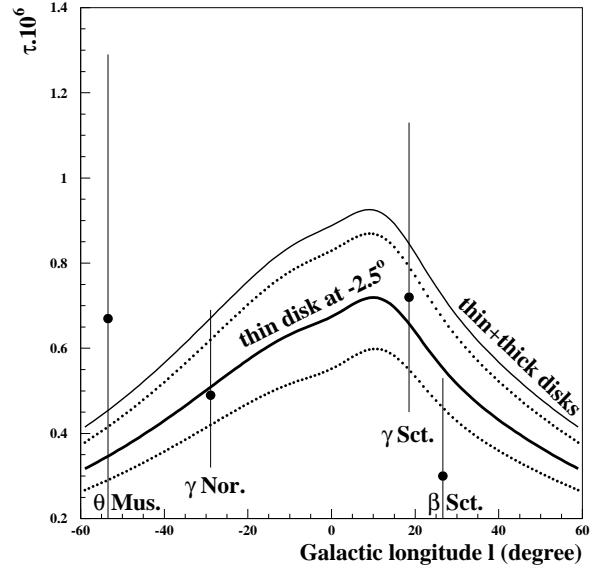}
\caption[]{
Expected optical depth up to $7\ kpc$ at a Galactic latitude
$b=-2.5\degree$
for model 1 (thick curve) and model 2 (thin curve)
which assumes an additional thick disk.
The dotted lines
show the excursion of the model 1 predictions when varying the
latitude by $\pm 1\degree$. The measured optical
depths are given for our 4 targets.
}
\label{optdepth}
\end{center}
\end{figure}

We will consider here 4 published optical depth calculations and
our own calculations based on simple
Galactic models (model 1 without a thick disk, model 2 with a thick disk)
that we already used for discussions in papers I and II.
The main revision to these 2 models since our previous
papers comes from the bulge inclination;
we now take 
$\Phi=45\degree$ instead of $15\degree$ (\cite{Hamadache} and \cite{picaud})
as the angle of the outer bulge with respect
to the line of sight towards the Galactic center.
The only impact of this change is a little variation of our optical
depth value towards $\gsct$.
We also completely
neglect any contribution from the halo to the optical
depth, in the light of the latest EROS results towards the Magellanic Clouds
(\cite{ErosLMCfinal}).
As in our previous papers, we performed simple optical depth calculations
assuming all the sources to be at the same distance. More sophisticated
modelling based on the guidelines discussed in Section \ref{sec:guidelines}
will be considered in a forthcoming paper.
We give in table \ref{table:tabmodel} the list of the
geometrical and kinematical parameters used in these models I and II.
\begin{table}
\begin{center}
\caption[]{Parameters of the Galactic models 1 and 2 used in this article. 
\label{table:tabmodel}
}
\begin{tabular}{|c|l|c|c|}
\hline
        & Parameter                     &model 1&model 2 \\ \hline
	& $R_{\odot}\ ({\rm kpc})$		& \multicolumn{2}{c|}{8.5} \\ \hline
        & $\Sigma\ (M_{\odot} {\rm pc}^{-2})$   & \multicolumn{2}{c|}{50} \\ 
        & $H\ ({\rm kpc})$                      & \multicolumn{2}{c|}{0.325} \\ 
       	& $R\ ({\rm kpc})$                      & \multicolumn{2}{c|}{3.5}  \\
Thin disk        & $M_{thin}(\times 10^{10}M_{\odot})$   & \multicolumn{2}{c|}{4.3} \\
        & $\sigma_r\ (km\ s^{-1})$              & \multicolumn{2}{c|}{34.} \\
        & $\sigma_{\theta}\ (km\ s^{-1})$       & \multicolumn{2}{c|}{28.} \\
        & $\sigma_z\ (km\ s^{-1})$              & \multicolumn{2}{c|}{20.}\\ \hline
        & $\Sigma\ (M_{\odot} {\rm pc}^{-2})$ & -    & 35      \\
        & $H\ ({\rm kpc})$                      & -     & 1.0     \\ 
        & $R\ ({\rm kpc})$                      & -    & 3.5   \\
Thick disk      & $M_{thick}(\times 10^{10}M_{\odot})$  & -     & 3.1 \\
        & $\sigma_r\ (km\ s^{-1})$              & -     & 51. \\
        & $\sigma_{\theta}\ (km\ s^{-1})$       & -     & 38. \\
        & $\sigma_z\ (km\ s^{-1})$              & -     & 35.\\ \hline
        & $a\ ({\rm kpc)}$      & \multicolumn{2}{c|}{1.49}     \\ 
        & $b\ ({\rm kpc})$      & \multicolumn{2}{c|}{0.58}     \\ 
Bulge   & $c\ ({\rm kpc})$	& \multicolumn{2}{c|}{0.40}       \\
	& Inclination $\Phi$	& \multicolumn{2}{c|}{$45 \degree$}       \\
        & $M_{B} (\times 10^{10}M_{\odot})$     & \multicolumn{2}{c|}{1.7} \\
        & $\sigma_{bulge}\ (km\ s^{-1})$        & \multicolumn{2}{c|}{110.} \\ \hline
\noalign{\smallskip}
\end{tabular}
\end{center}
\end{table}
The disk densities are modeled by a double exponential expressed
in cylindrical coordinates:
\begin{eqnarray*}
\rho_{D}(R,z) = \frac{\Sigma}{2H} \exp 
\left(\frac{-(R-R_{\odot})}{h} \right) \exp 
\left( \frac{-|z|}{H} \right) \ ,
\end{eqnarray*}
where $\Sigma$ is the column density of the disk at the Sun position $R_{\odot}$,
$H$ the height scale and $h$ the length scale of the disc.
The density distribution for the bulge -~a bar-like triaxial model~-
is taken from \cite{Dwek} model G2, given in Cartesian 
coordinates:
\begin{eqnarray*}
\rho_{B} &=&  \frac{M_{B}}{6.57 \pi abc} e^{-r^{2}/2} \ , \ 
r^{4} = \left[ \left( \frac{x}{a} \right)^{2} + 
	       \left( \frac{y}{b} \right)^{2} \right]^{2} + 
	\frac{z^{4}}{c^{4}} \ ,
\end{eqnarray*}
where $M_{B}$ is the bulge mass, and $a$, $b$, $c$ the length
scale factors.
Fig. \ref{optdepth} shows
the measured optical depth as a function of the
Galactic longitude, with the expectations from models 1 and 2
at $7\ kpc$ and at a Galactic latitude $b=-2.5\degree$.
We also show in this figure the effect of a
Galactic latitude change by $\pm 1\degree$.

The optical depth predictions of model 1 and 2 and of the 4 following models  
are reported in table \ref{table:predictions}:
\begin{itemize}
\item
{\bf Model A}, from \cite{Binneyetal}, revised by \cite{Bissantz}, has a
cuspy and flat Galatic bar, inclined by $\Phi\sim 20\degree$.
\item
{\bf Model B}, described in \cite{Dwek}, has a wider cuspy bar, inclined by
$\Phi\sim 24\degree$.
\item
{\bf Model C}, described in \cite{freudenreich}, has a more
extended and diffuse bar, inclined by $\Phi\sim 14\degree$.
\end{itemize}
The optical depths towards the directions monitored by EROS have been
calculated for these 3 models and discussed by \cite{evans}, who have
also included the contribution from the disk and
considered separately the impact of a spiral structure.
These 3 models have been normalized to have the same total mass
($1.5\times 10^{10}\Msol$) within 2.5 kpc. Therefore, they mainly
differ by the shape details and the orientation of the bar.
\begin{itemize}
\item
{\bf Model D}, described in \cite{Grenacher-1999}, has a bar that is similar
to the one of our models 1 or 2, but inclined by $\Phi\sim 20\degree$,
with a combination of a light thin disk plus a thick disk and a halo
contribution.
\end{itemize}
The adequation of all these models with the observed optical depths $\tau$
can be compared through the value of
\begin{equation}
\chi^2_{model}=\sum_{targets\ i}\frac{(\tau_i(model)-\tau_i(observed))^2}{\sigma^2_i},
\end{equation}
where $\sigma_i$ is the error interval of the $\tau_i$ determination
(as errors are asymmetrical, we consider the largest one for each measurement).
The numbers reported in table \ref{table:predictions} show that
our model 1 is clearly favored by the data, and also models A and D without
spiral structure, which are the ones that predict the smallest optical depths.

We cannot draw more conclusions about our model 1,
as we know that it is not a realistic description,
since all targets are supposed to be at the same distance.
One may also notice that the extrapolation of any
{\it bulge} model to the relatively distant region
that we monitored is very uncertain.
Nevertheless, it seems that ``heavy'' models trying to include
a thick disk or any spiral structure are not favored.
These results confirm the conclusions of \cite{Hamadache},
in particular for model C, that is also disfavored by the present
data.
\subsection{Event duration distribution}
\begin{figure}
\begin{center}
\includegraphics[width=10cm]{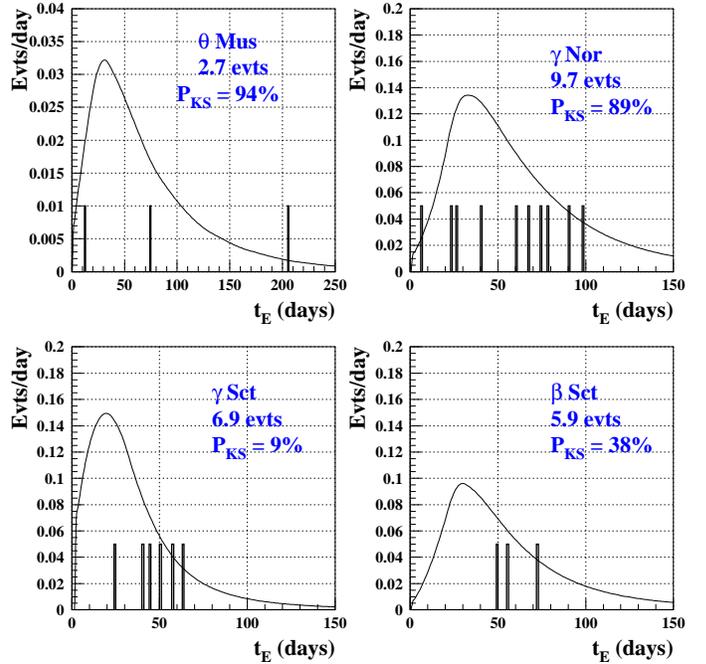}
\caption[]{
measured $t_E$ distributions compared to the one calculated with
our model 1 for one day intervals.
$P_{KS}$ gives the Kolmogorov-Smirnov test probability of compatibility
of the shapes. Notice that the scales are different for $\tmus$.
The $t_E$ values of the observed events are given by the vertical bars.
}
\label{durees}
\end{center}
\end{figure}

Fig. \ref{durees} gives the $t_E$ distributions for the observed
events, compared to the expected distributions from model 1.
The procedure to build these distributions is the same as
in paper II:
\begin{itemize}
\item
The mass function for the lenses is taken
from \cite{Gould-1997} for both the disk and the bulge.
\item
The solar motion with respect to the disk is taken from (\cite{delhaye}):
\begin{equation}
v_{\odot R}=-10.4,\ v_{\odot \theta}=14.8,\ v_{\odot z}=7.3\ \ 
({\rm km}/{\rm s}).
\end{equation}
\item
The global rotation of the disk is given by
\begin{equation}
V_{rot}(r) = V_{rot,\odot} \times \left[ 
1.00762 \left( \frac{r}{R_{\odot}} \right)^{0.0394} + 0.00712 \right] \ ,
\end{equation}
where $V_{rot,\odot} = 220$ {\rm km/s} (\cite{rotdisc}).
\item
The peculiar velocity of disk stars is described by
an anisotropic Gaussian distribution and a velocity dispersion
given in table \ref{table:tabmodel}.
\item
The velocity distribution of the bulge stars is given by
\begin{equation}
f_{T}(v_{T}) = 
\frac{1}{\sigma_{bulge}^{2}} v_{T} \exp \left( -\frac{v_{T}^{2}}{2\sigma_{bulge}^{2}} 
\right),
\end{equation}
with $\sigma_{bulge} \sim 110 \ {\rm km/s}$.
\end{itemize}


The Kolmogorov-Smirnov tests
indicate that model 1 produces
$t_E$ distributions that are fully compatible with the observations,
except for $\gsct$ where the approximation of a single distance for all
the sources is the most questionable (due to the contribution of the sources
belonging to the elongated bulge).
Nevertheless, the supplement of short events towards this direction
that is visible in the $t_E$ distribution corresponds to the
expectations from bulge lenses.
We found that our simulation results are not dramatically changed if we
change the contribution of the lightest lens objects (around $0.1\Msol$) by
$50\%$. This is mainly due to the low detection efficiency for
the short events expected from low mass lenses
(this loss of efficiency is specially noticiable for low mass bar lenses
because their velocity is larger).

Table \ref{table:predictions}
gives the observed averages, dispersions and medians of ${t_E}$,
with the predictions from model 1 and models C and D.
Only mean duration predictions have been estimated for model C
(with no efficiency corrections)
and D (taking into account EROS microlensing detection efficiency).
The general tendency that emerges from the observations and that
is confirmed by the predictions
is that $\bar{t_{\rm E}}$ increases with the Galactic longitude. This
is due to the fact that the direction of global motion of the lenses
tends to align with the line of sight when the longitude increases, and
consequently the transverse speed becomes smaller.

As the variation of the microlensing detection
efficiency with $t_E$ is not taken into account in the predictions of
model C, we should compare here
the $\bar{t_{\rm E}}$ predicted values with the efficiency corrected means given by
\begin{equation}
\frac{\sum_{events} t_E/\epsilon (t_E)}
{\sum_{events} 1/\epsilon(t_E)}.
\end{equation}
The observed corrected means
are significantly larger than the predictions of model C
with no spiral structure.
This discrepancy could be explained (at least partially) by the absence
of very short time scale events that have a negligible detection efficiency and
can be totally missed in our statistically limited sample;
a more precise comparison could be done between the observed $t_E$ distribution
and a modified predicted distribution taking into account the
microlensing detection efficiency as a function of $t_E$ (Fig. \ref{efficacite}).
Keeping in mind the possibility of undetectable events,
it seems that model C without spiral structure is disfavoured.
\cite{evans} have emphasized the impact of streaming in the bar on the
$t_E$ distribution around the Galactic center and the impact of the spiral
structure on $\bar{t_{\rm E}}$ in every direction.
Such a streaming and the spiral structure have a serious impact
on the $\bar{t_{\rm E}}$ values, and could
reconcile the model expectations with the $t_E$ data, but lighter disk and
spiral structure should be considered to fit the optical
depth data as well.
There is probably some margin of freedom for doing this, as
model C is obtained from the extrapolation of the light
distribution at latitudes larger than $5\degree$.

The mean durations predicted by \cite{Grenacher-1999} (model D)
take into account the EROS efficiencies.
Therefore, we can directly compare the predictions
with our (uncorrected) $\bar{t_{\rm E}}$ values.
This model predicts longer durations than observed, but
here again it seems that the predictions
could be adjusted, as they are sensitive to the value of the minimal lens mass.
\section{Guidelines for further interpretation}
\label{sec:guidelines}
\begin{figure}
\begin{center}
\includegraphics[width=8.5cm]{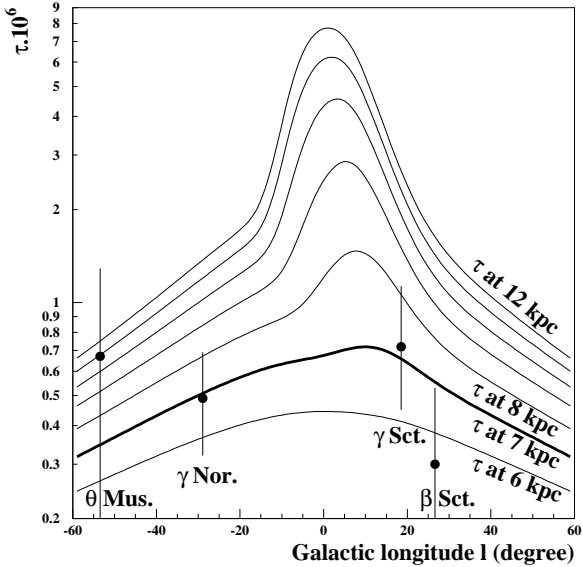}
\caption[]{
Model 1 predictions for the optical depth at
6, 7, 8, 9, 10, 11 and 12 kpc (from lowest to highest curve)
at a Galactic latitude
$b=-2.5\degree$ as a function of the Galactic longitude.
The measured optical depths are given for our 4 targets.
}
\label{tauvsldist}
\end{center}
\end{figure}

As we always emphasized when presenting previous results towards the
spiral arms, the fact that the distance distribution of the target
sources is poorly known complicates the optical depth interpretation.
Fig. \ref{tauvsldist} shows the expected optical depth as a function of the
Galactic longitude $l$, for different target distances, using model 1.
This figure shows first that the impact of the bulge
on the optical depth is significant only towards $\gsct$.
One also sees that the optical depth to distances smaller than $12\ kpc$
is larger for $l>0$, the near side of the bulge,
because the number density of {\it lenses} is larger on this side;
but on the contrary, the OGLE II collaboration (\cite{ogle2000a}) reported
a larger rate for $l<0$ than for $l>0$
in their catalog of microlensing events in the Galactic bulge.
This is due to the fact that
the optical depth {\it averaged}
over all distances can be larger for $l<0$, the far side of the bulge,
because there are more distant {\it sources} on this side, with a
larger optical depth. The observed asymmetry should then depend on the
selection of the monitored sources.
This illustrates the difference between the optical depth up to a
given distance, and the average optical depth over a set of
stars located at various distances.
Therefore, we provide
here guidelines to interpret our optical depth values
within a Galactic model framework.
\subsection{The concept of ``catalog optical depth''}
%
%
The EROS measured optical depth is 
the average over the distance distribution
of the monitored sources.
Establishing this distance distribution
through individual spectrophotometric
measurements would require an enormous amount of complementary observations.
This leads us to define the concept of ``catalog optical depth''
$\tau_{cat}$, which is relative to a particular catalog of monitored stars:
$\tau_{cat}$ is
defined as the fraction of stars {\it of a given catalog} that undergo a
magnification $A>1.34$.
Our {\it measured} optical depth can be compared with
the depth derived from a lens and source distribution model
as follows:
first, one has to generate a synthetic source catalog
that matches our own catalog, taking into account
the EROS star detection efficiency (Fig. \ref{effdet});
then one can use the generated source distance distribution
to estimate the average optical depth and compare it with the
measurements. This procedure is described in more detail below.
\subsection{Synthesizing a catalog that matches the EROS one}
\label{sec:synthesizing}
Let $S_{model}(D,M_I,M_V-M_I)$ be the source number density
predicted by the model
as a function of the distance $D$ and of the {\it absolute} magnitude $M_I$
and color $M_V-M_I$;
let $A_I(D)$ and $A_V(D)$ be the predicted absorptions in I and V.
The pre-requisite for an optical depth interpretation is that
$n_{cat}(I,V-I)$,
the density per solid angle per {\it apparent} magnitude and color index
of the synthesized source catalog, fits the observed one
within the visible color-magnitude diagram.
This density is related to $S_{model}(D,M_I,M_V-M_I)$ as follows:
\begin{equation}
n_{cat}(I,V-I)\!=\!
\!\!\int_{0}^{\infty}\!\!\!\!S_{model}(D,M_I,M_V-M_I)
\epsilon_{star}(B_{EROS})D^2 \ud \mathit D
\end{equation}
where $\epsilon_{star}(B_{EROS})$ is our star detection efficiency
given in Fig. \ref{effdet} as a function of
$B_{EROS}=I+0.6(V-I)$ {\it apparent} $B$ magnitude in the EROS system.
For given $I$ and $V-I$ apparent magnitude and color,
$M_I$ and $M_V-M_I$ depend on the distance $D$ as follows:
\begin{eqnarray}
M_I&=&I-2.5 \log (D/10kpc)-A_I(D), \label{magabs} \\
M_V-M_I&=&V-I-A_V(D)+A_I(D). \nonumber
\end{eqnarray}
For the comparison with the observations,
the color-magnitude diagrams $n_{eros}(I,V-I)$ of our
catalogs (Fig. \ref{HRdiagrams})
should be used.
They provide the observed stellar density per unit solid angle
in the effectively monitored field, that is corrected for
the dead regions of our CCDs, the overlap between fields and
the blind regions around the brightest objects.

\begin{figure}[htbp]
\begin{center}
\includegraphics[width=8.5cm]{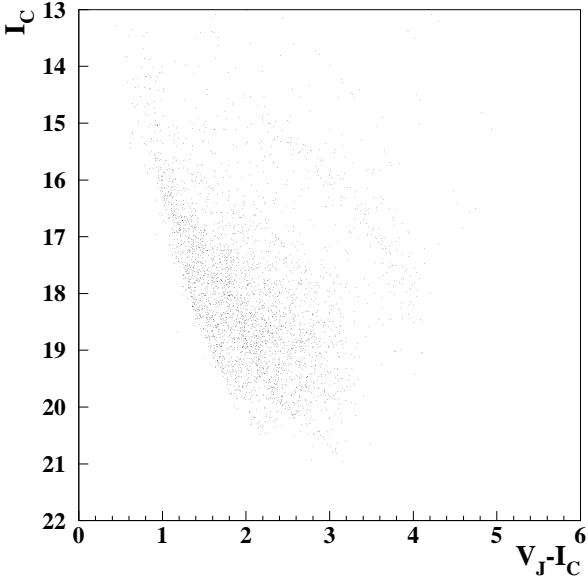}
\caption[]{
Simulated color-magnitude diagram towards $\gsct$ (see text).
}
\label{diaghrsimu}
\end{center}
\end{figure}

We show here the color-magnitude diagram of a preliminary catalog
obtained with this procedure (Fig. \ref{diaghrsimu}).
This catalog has been produced for
the qualitative understanding of the observed diagrams
(Sect. \ref{sec:colmag}, Fig. \ref{HRdiagrams})
and also of the color bias of the lensed population
(Sect. \ref{sec:population}, Fig. \ref{diagHRdist}).
For this simulation, we considered a stellar population
identical to the one of the solar neighborhood (\cite{hipp}, \cite{Turon}),
distributed according to the disk and bulge mass distributions described
in Sect. \ref{sec:comparisons};
the reddening was calculated assuming the following absorption law
(\cite{Weingartner}):
\begin{equation}
A(V)/N_H=5.3\times 10^{-22} cm^2\ \ \ \ A(I)/N_H=2.6\times 10^{-22} cm^2,
\end{equation}
where $N_H$ is the column-density of neutral hydrogen (atomic plus molecular),
assumed to represent $4\%$ of the mass density.
The catalog was simulated using our star detection efficiency of Fig. \ref{effdet}.
Fig. \ref{diagHRdist}(left) shows the source distance distribution
of the catalog thus obtained. 
This catalog will be finalized in a forthcoming paper.
\subsection{Estimating the average optical depth}
\label{sec:further}
Once the synthesis of a catalog is performed,
one can compute the average optical depth over the synthesized catalog:
\begin{equation}
<\tau_{model}>
=\frac{\int_0^{\infty} n_{model}(D)\tau_{model}(D)D^2\, \ud D}
{\int_0^{\infty} n_{model}(D) D^2\, \ud D},
\end{equation}
where $\tau_{model}(D)$ is the predicted optical depth at distance $D$ and
\begin{eqnarray}
&&n_{model}(D)=  \\
&&\int\!\!\!\int\!S_{model}(D,M_I,M_V-M_I)\cdot
\epsilon_{star}(B_{EROS})\, \ud M_I\, \ud (M_V-M_I) \nonumber
\end{eqnarray}
is the number density of stars (per solid angle and distance unit)
located at distance $D$ that enters the catalog.
$B_{EROS}=I+0.6(V-I)$ is obtained from $M_I$, $M_V-M_I$ and $D$
through equations (\ref{magabs}).

This procedure assumes that the average microlensing detection efficiency
does not vary significantly with the distance of the sources.
This is indeed the case because both the efficiency of our star detector
and the average microlensing detection efficiency depend on the
{\it apparent} magnitudes.
The microlensing detection efficiency averaged on
distant {\it detected} stellar populations
is then similar to the one averaged on nearby detected populations
because their apparent magnitude distributions are both peaked
around the limiting magnitude.
We have checked these features with simulations
using our star detection efficiencies.

The observed optical depths in Table \ref{table:predictions} are relative
to our entire catalogs. We recall here the average value for the 4 targets
($12.9\times 10^6$ stars):
$$\bar{\tau}_{full\ catalog}=0.51\pm 0.13\times 10^{-6}.$$
For easier comparisons, we extract from Fig. \ref{stabIc}
the following optical depth
$$\bar{\tau}(I_c<18.5)=0.39\pm .11\times 10^{-6},$$
relative to
the sub-set of 6.52 million stars brighter than $I_c=18.5$,
for which our catalog is close to being complete.

\section{Conclusions}
The microlensing event search of EROS2 towards transparent windows of
the spiral arms leads to
optical depths that are consistent with a very simple Galactic model.
The possibility of a long bar that was proposed
in our paper II as a possible explanation of the observed optical depth
and duration asymetries is not confirmed; indeed, the inclination of the bar
was revised since the time of this publication, and the final optical
depth measured towards $\gsct$ is also smaller (but statistically compatible).
A more complete interpretation that would take into account the distance distribution
of the monitored sources needs a model that allows one to synthesize
the EROS catalogs of these sources.
With such a model, one would also be able to make use of the
event duration distributions.

The VISTA project with its wide field infrared camera appears to offer
an excellent opportunity to improve
our knowledge of the microlensing towards the Galactic plane.
The infrared light will allow the observers to monitor stars through dust, making
them free of the transparent windows that were limiting the EROS fields.
\begin{acknowledgements}
We are grateful to the technical staff at ESO, La Silla for the support
given to the EROS-2 project.
We thank J-F. Lecointe and A. Gomes for the assistance with the online computing
and the staff of the CC-IN2P3, especially the team in charge of the HPSS storage
system, for their help with the data management.
\end{acknowledgements}

\section{Annex A}
\begin{figure*}
\begin{center}
\begin{tabular}{ccc}
\begin{minipage}[m]{3cm}
\includegraphics[width=3cm,bb=104 21 509 425]{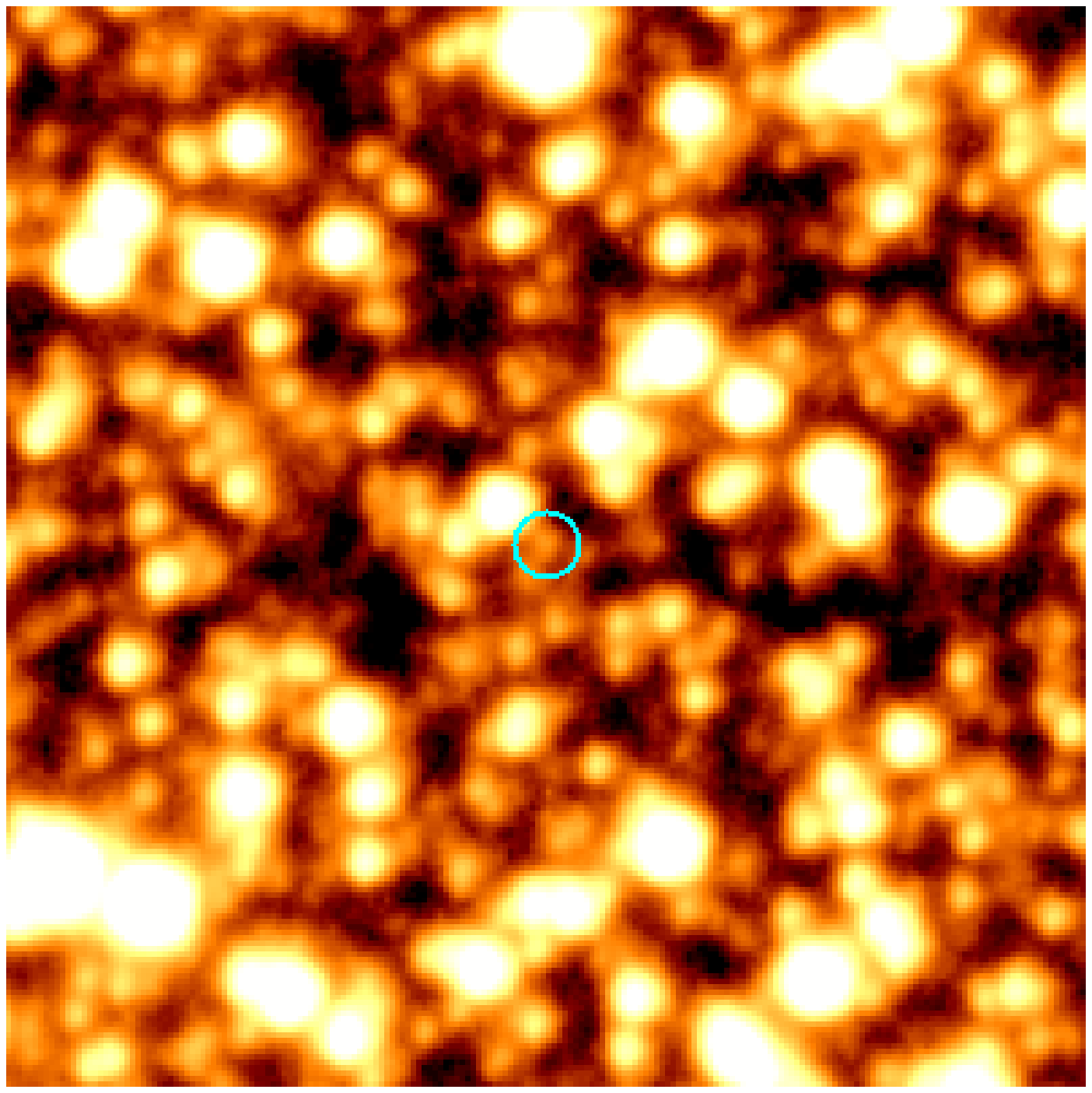}
\end{minipage}
&\begin{minipage}[m]{6.5cm}\includegraphics[width=6cm,bb=0 15 800 400,clip=true]{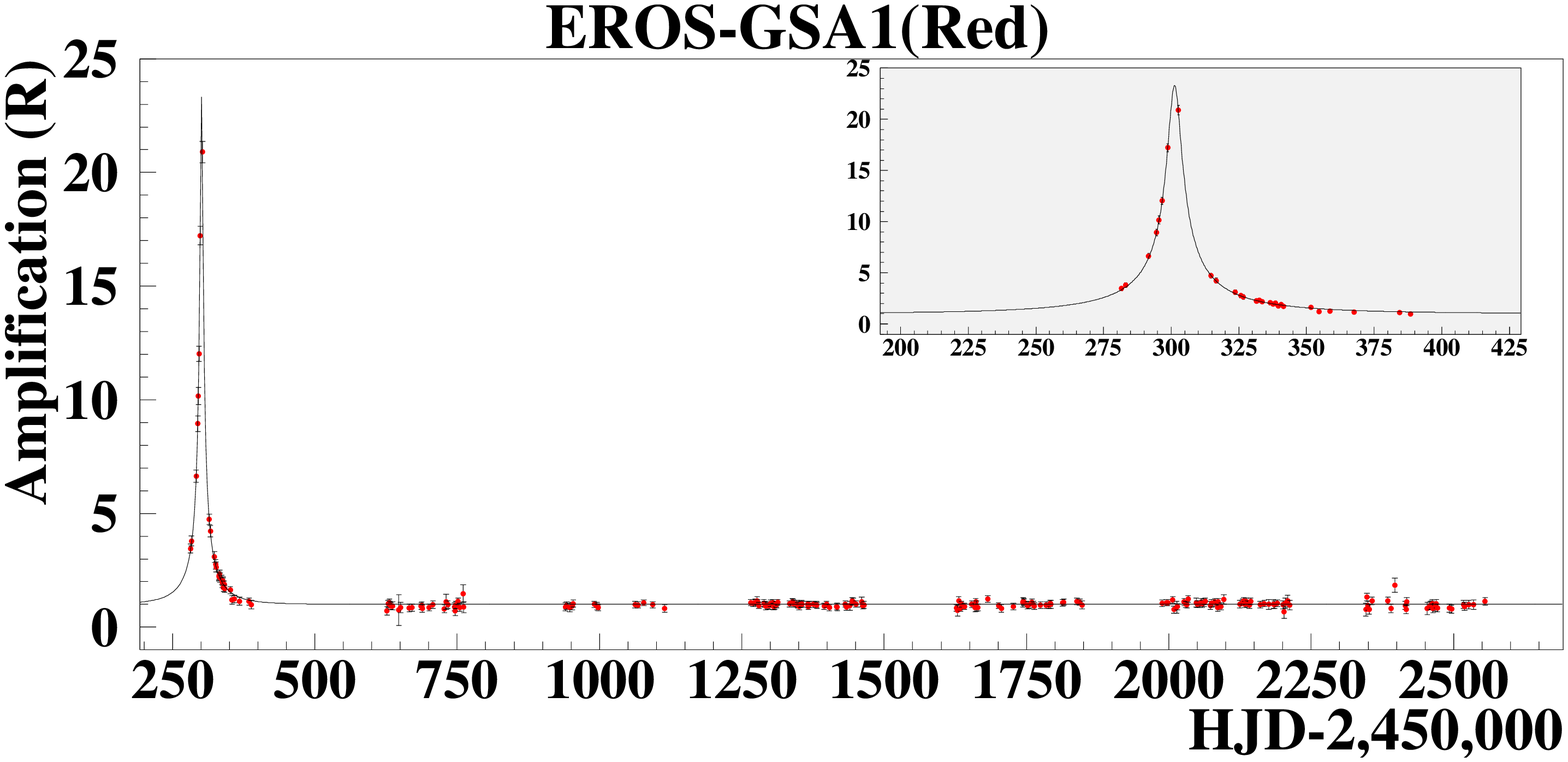}\end{minipage}
&\begin{minipage}[m]{6.5cm}\includegraphics[width=6cm,bb=0 15 800 400,clip=true]{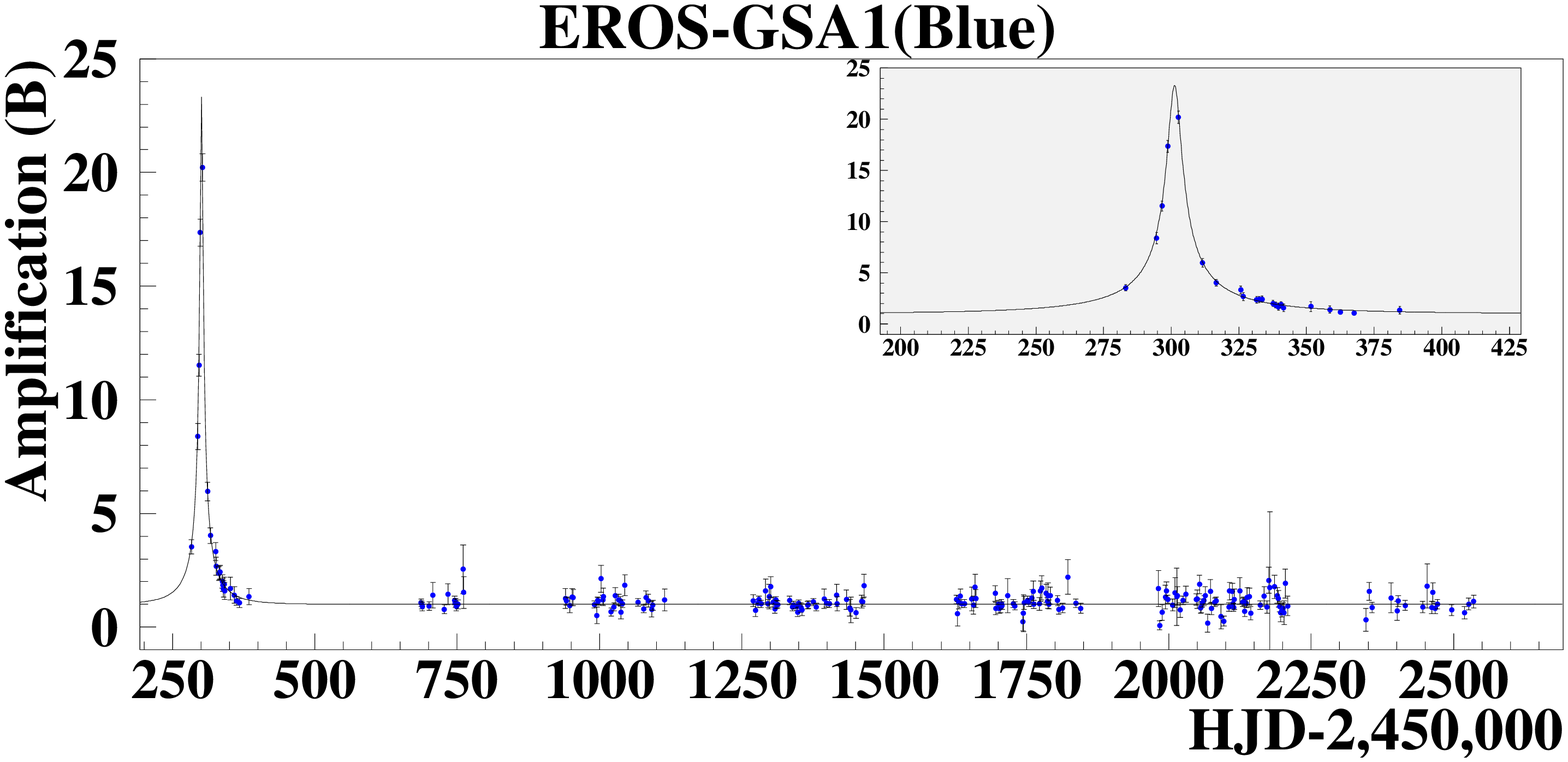}\end{minipage}
\\
\begin{minipage}[m]{3.cm}
\includegraphics[width=3.cm,bb=104 21 509 425]{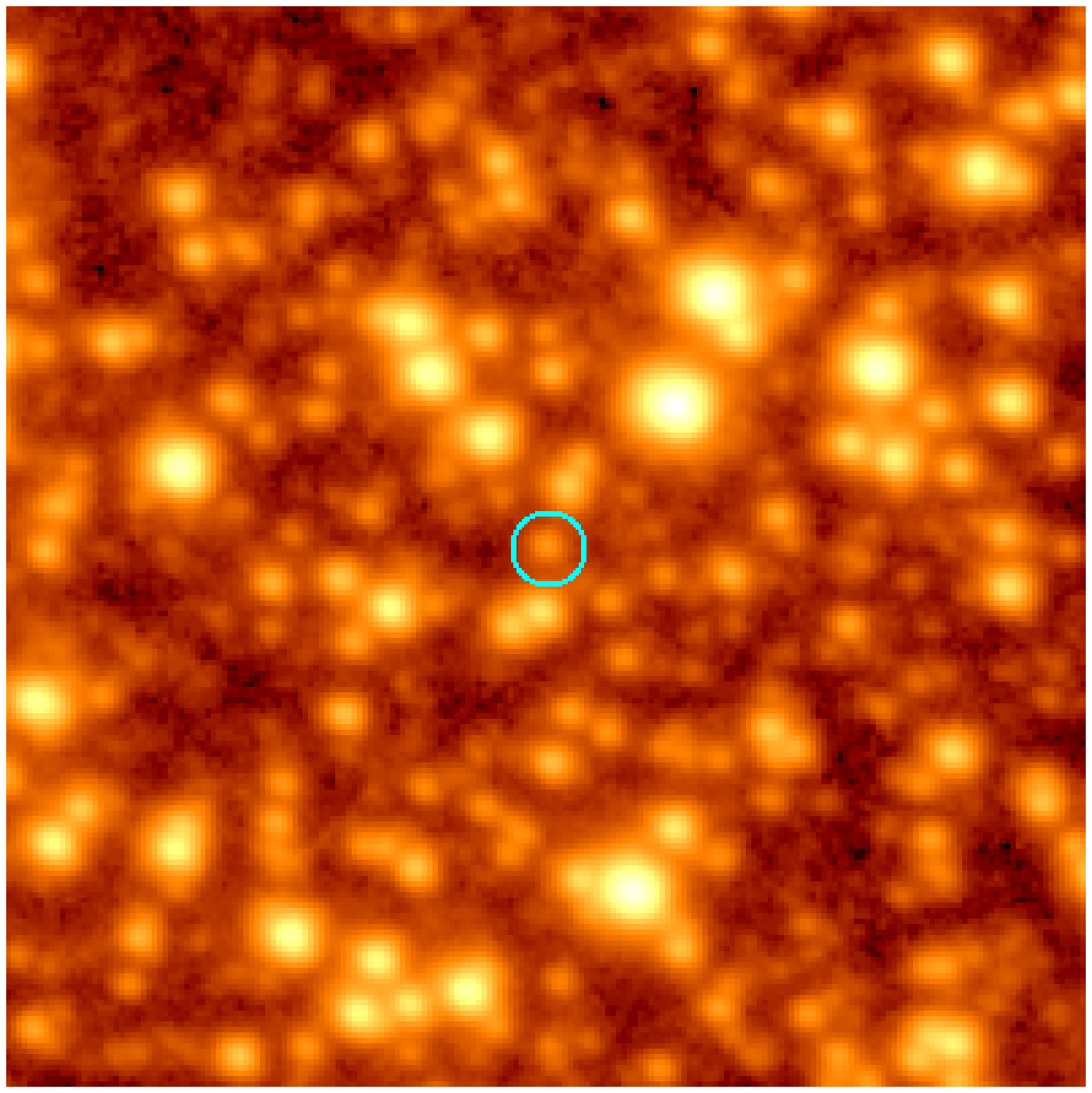}
\end{minipage}
&\begin{minipage}[m]{6.5cm}\includegraphics[width=6cm,bb=0 15 800 400,clip=true]{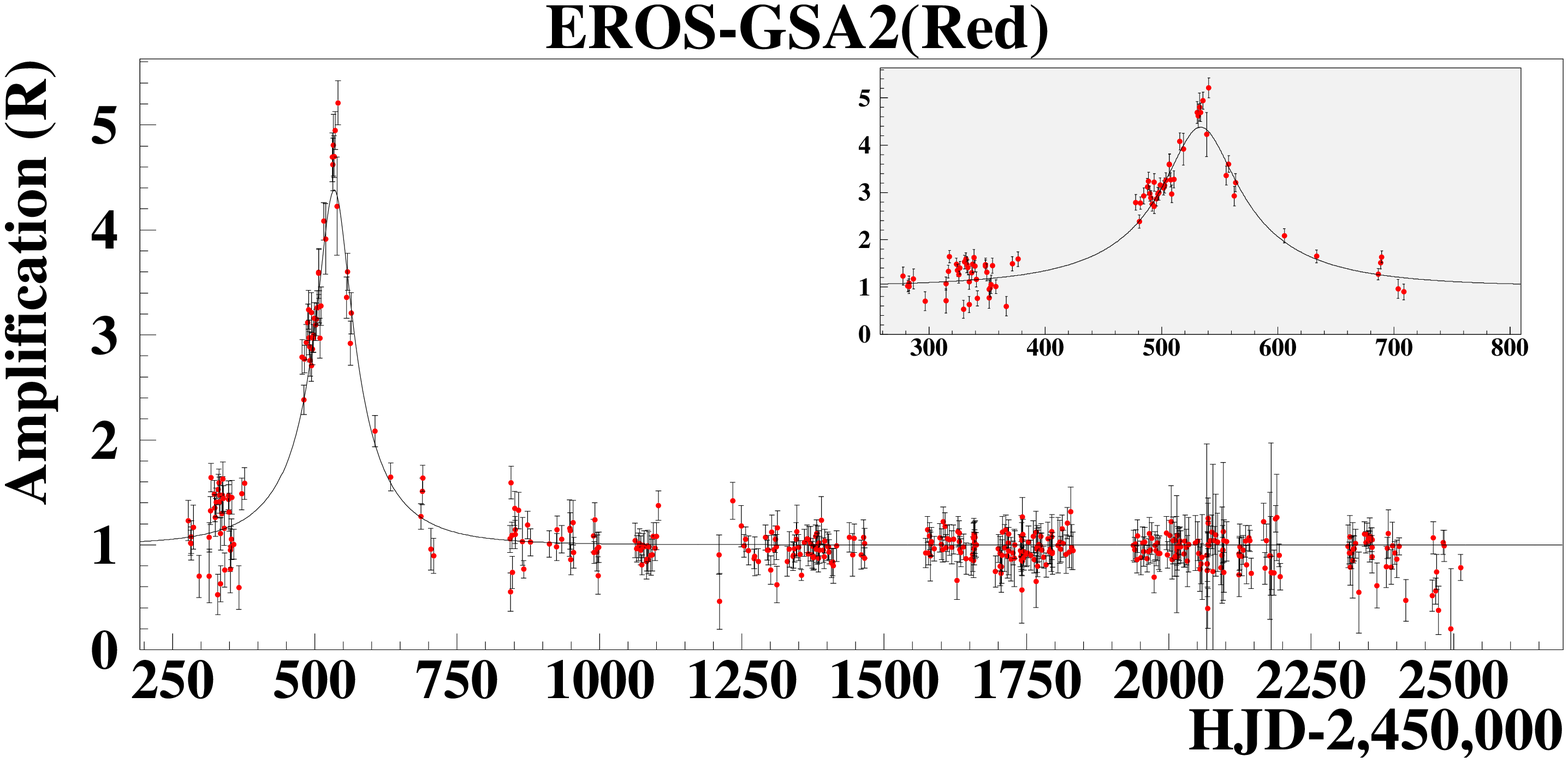}\end{minipage}
&\begin{minipage}[m]{6.5cm}\includegraphics[width=6cm,bb=0 15 800 400,clip=true]{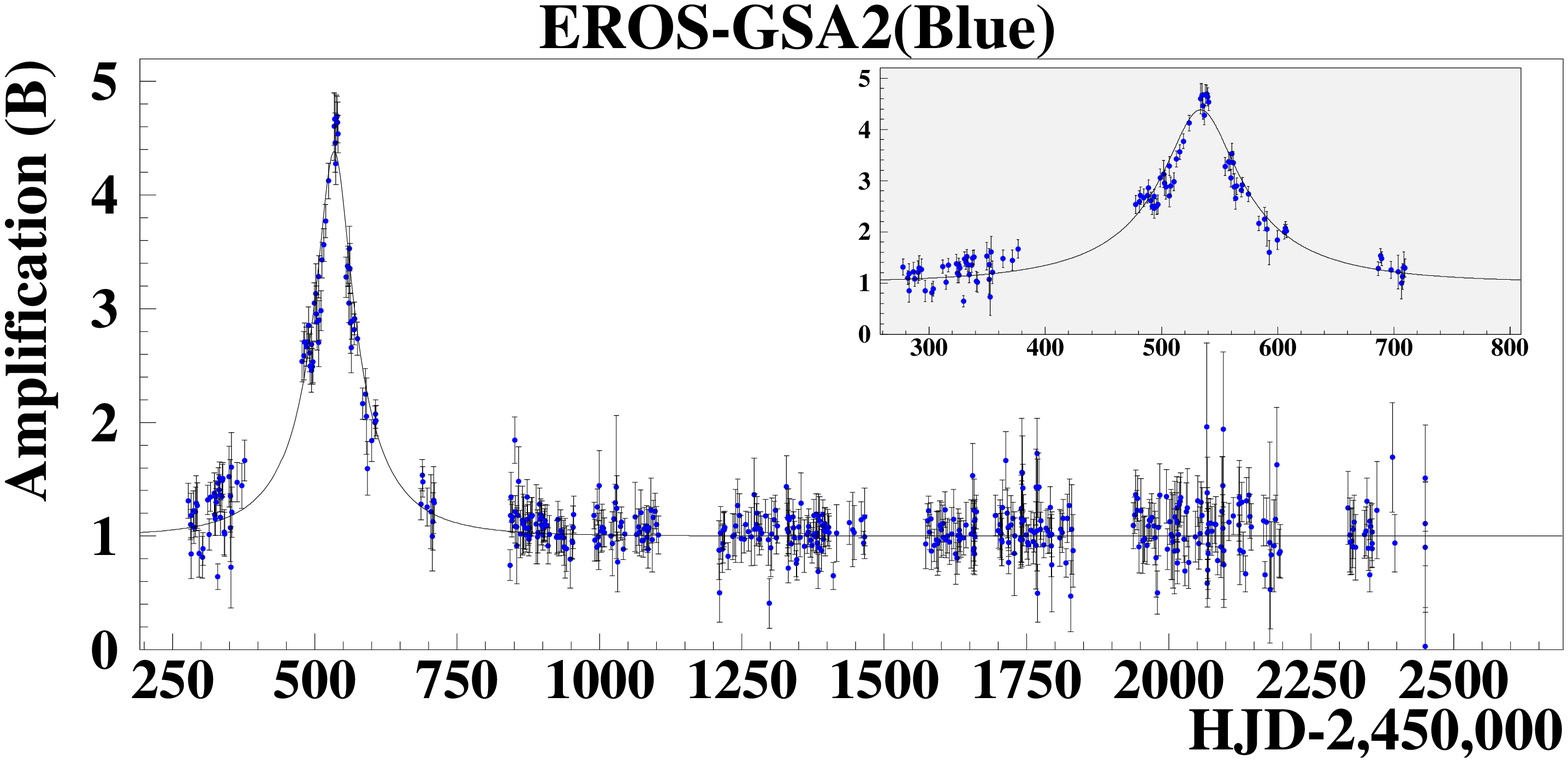}\end{minipage}
\\
\begin{minipage}[m]{3.cm}
\includegraphics[width=3.cm,bb=104 21 509 425]{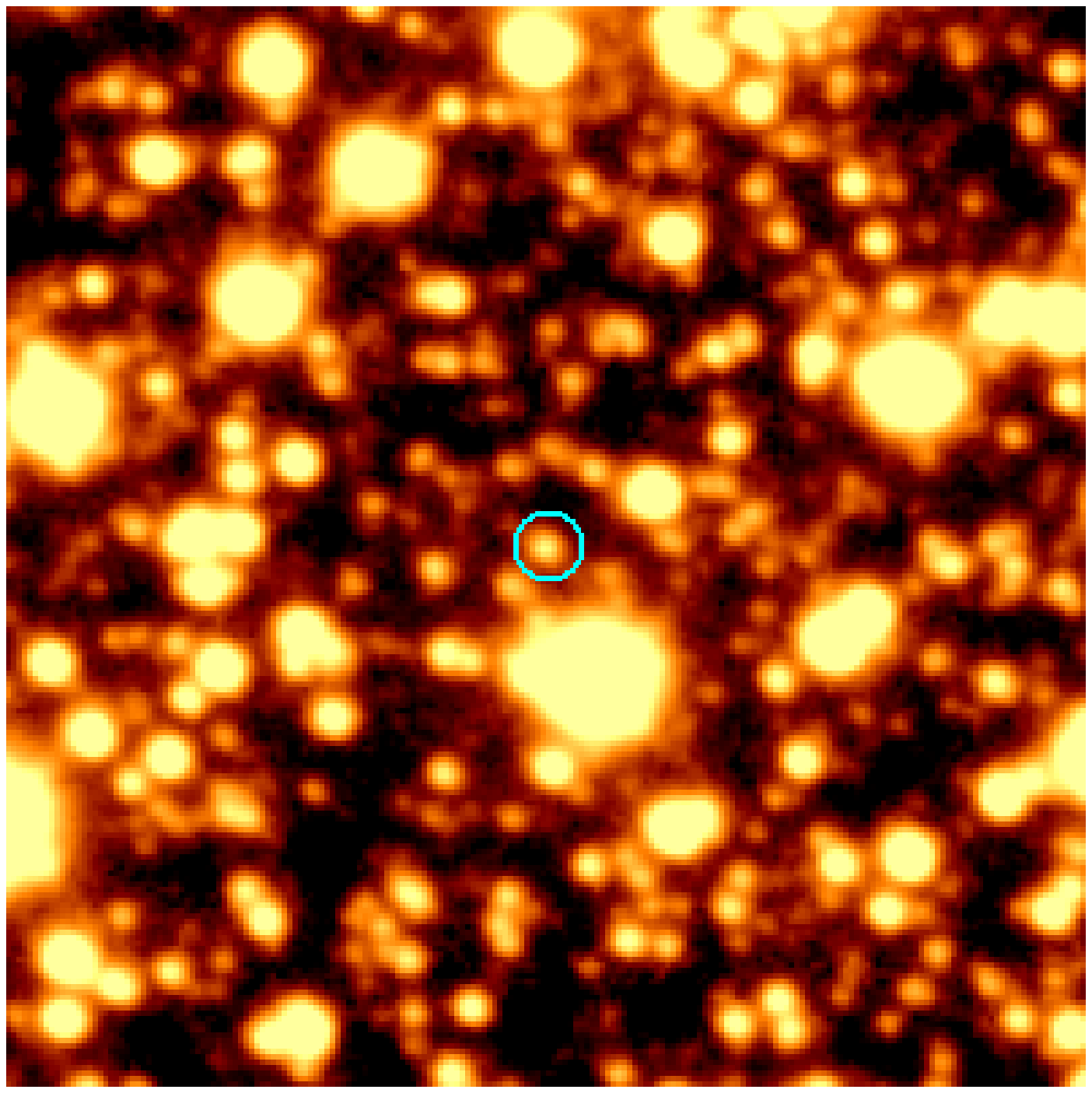}
\end{minipage}
&\begin{minipage}[m]{6.5cm}\includegraphics[width=6cm,bb=0 15 800 400,clip=true]{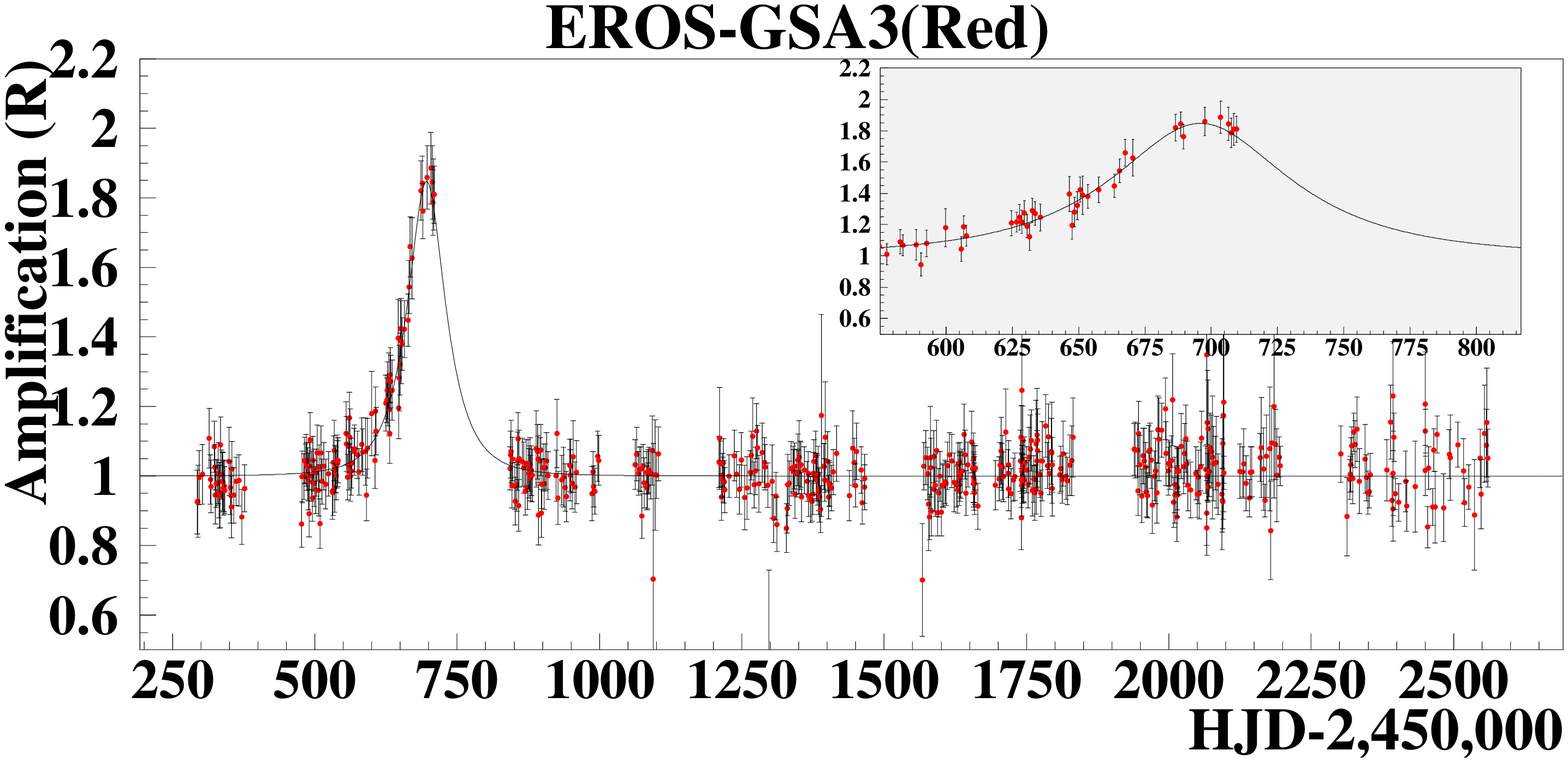}\end{minipage}
&\begin{minipage}[m]{6.5cm}\includegraphics[width=6cm,bb=0 15 800 400,clip=true]{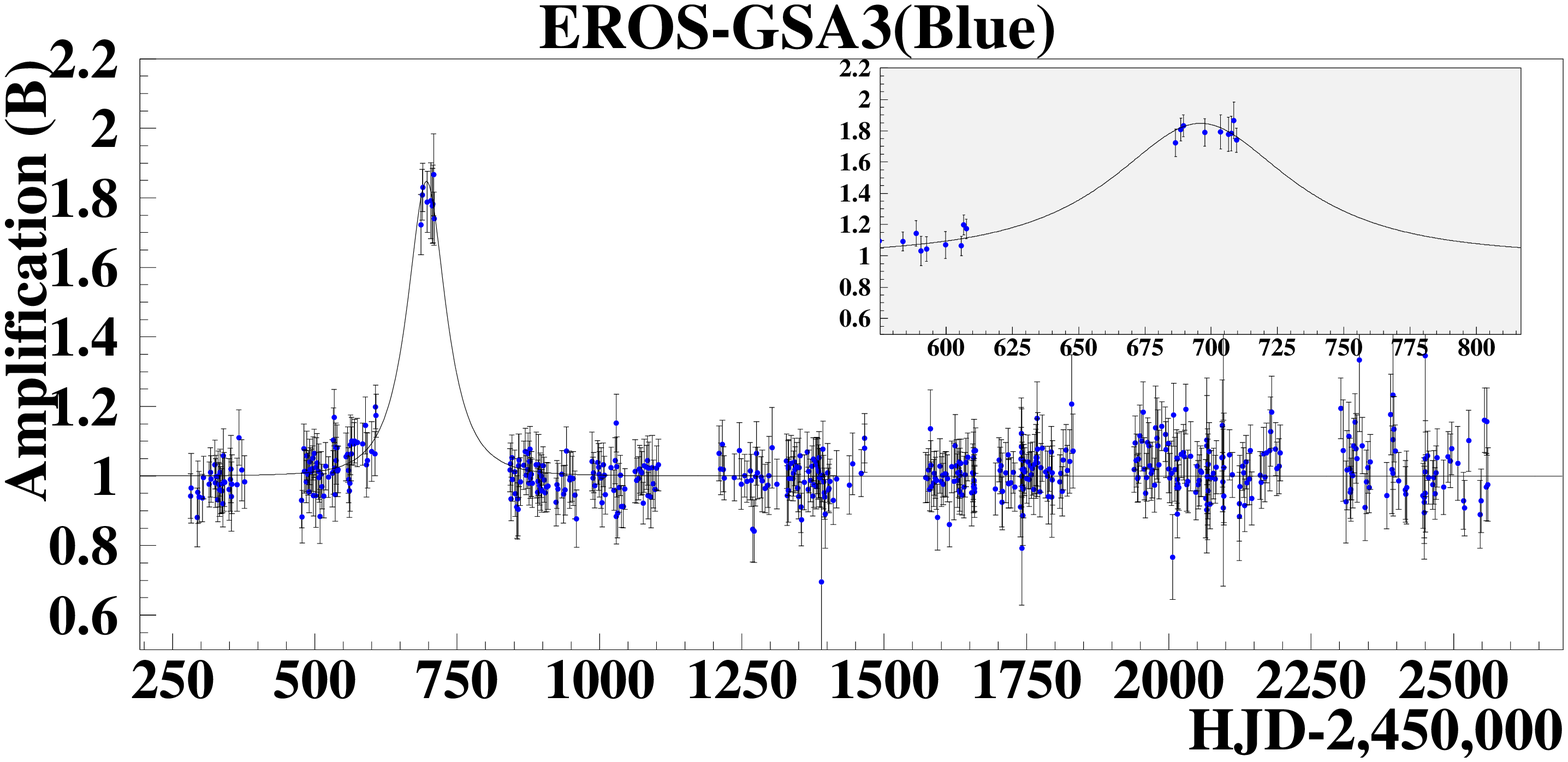}\end{minipage}
\\
\begin{minipage}[m]{3.cm}
\includegraphics[width=3.cm,bb=104 21 509 425]{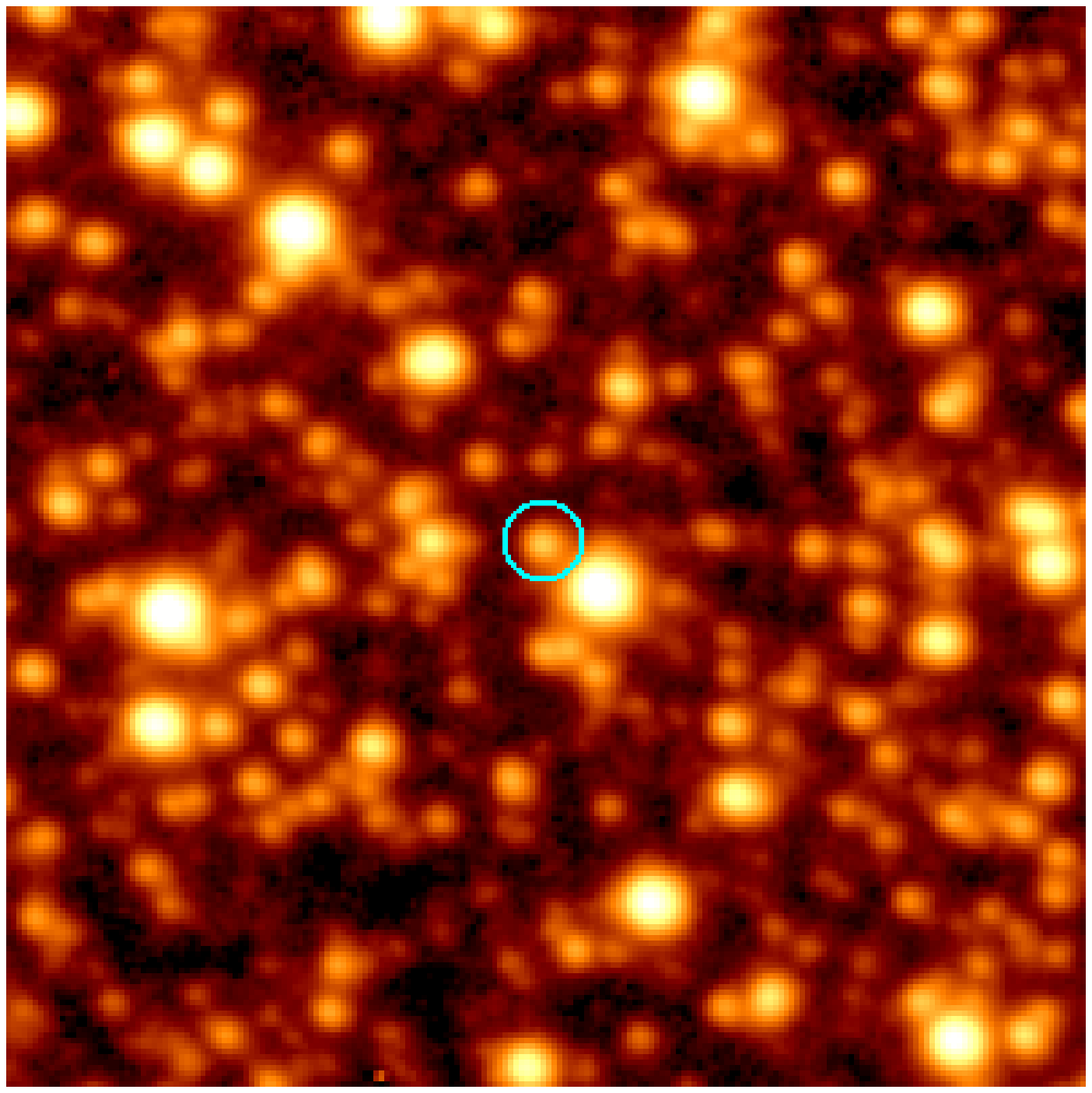}
\end{minipage}
&\begin{minipage}[m]{6.5cm}\includegraphics[width=6cm,bb=0 15 800 400,clip=true]{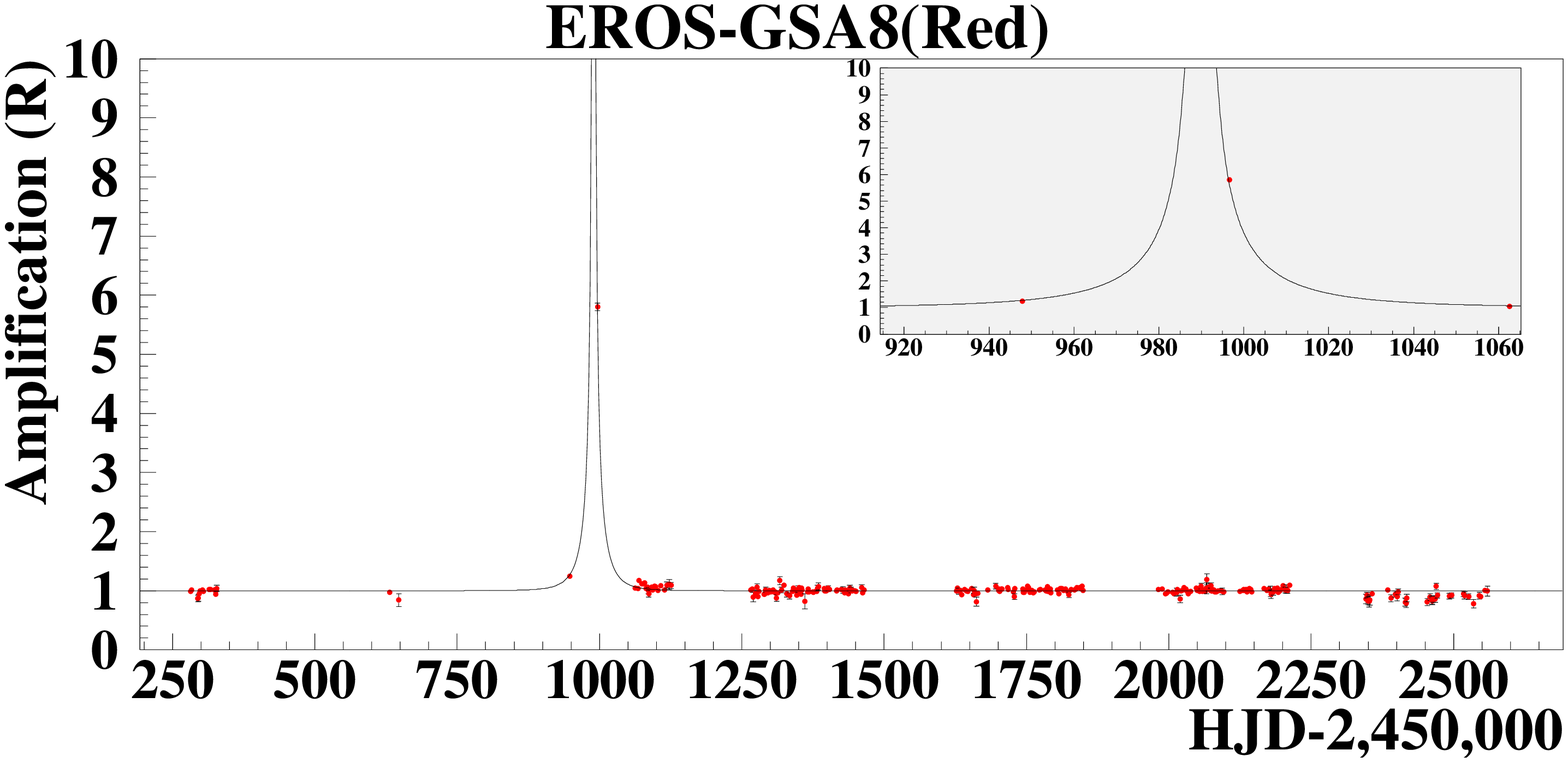}\end{minipage}
&\begin{minipage}[m]{6.5cm}\includegraphics[width=6cm,bb=0 15 800 400,clip=true]{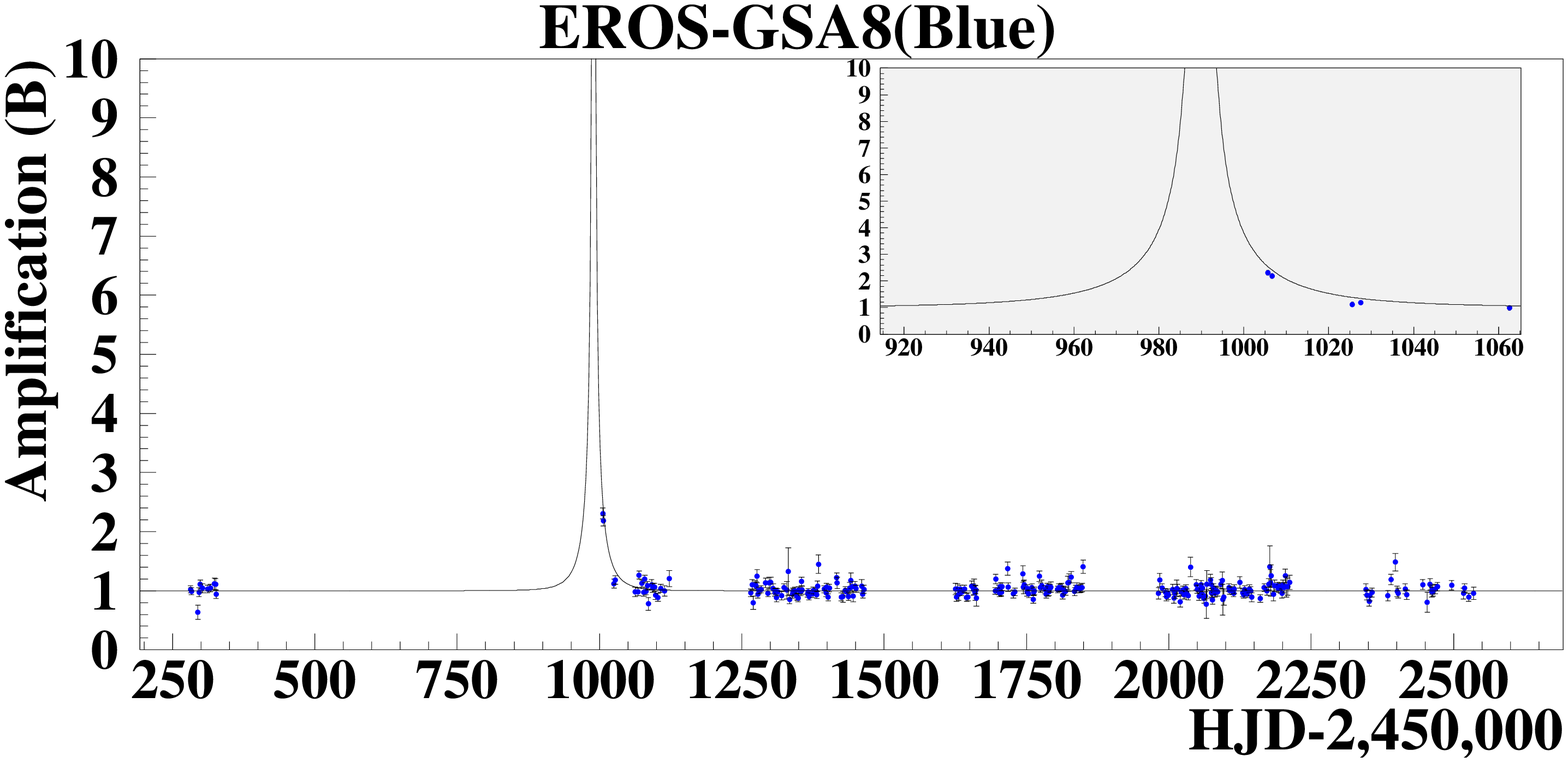}\end{minipage}
\\
\begin{minipage}[m]{3.cm}
\includegraphics[width=3.cm,bb=104 21 509 425]{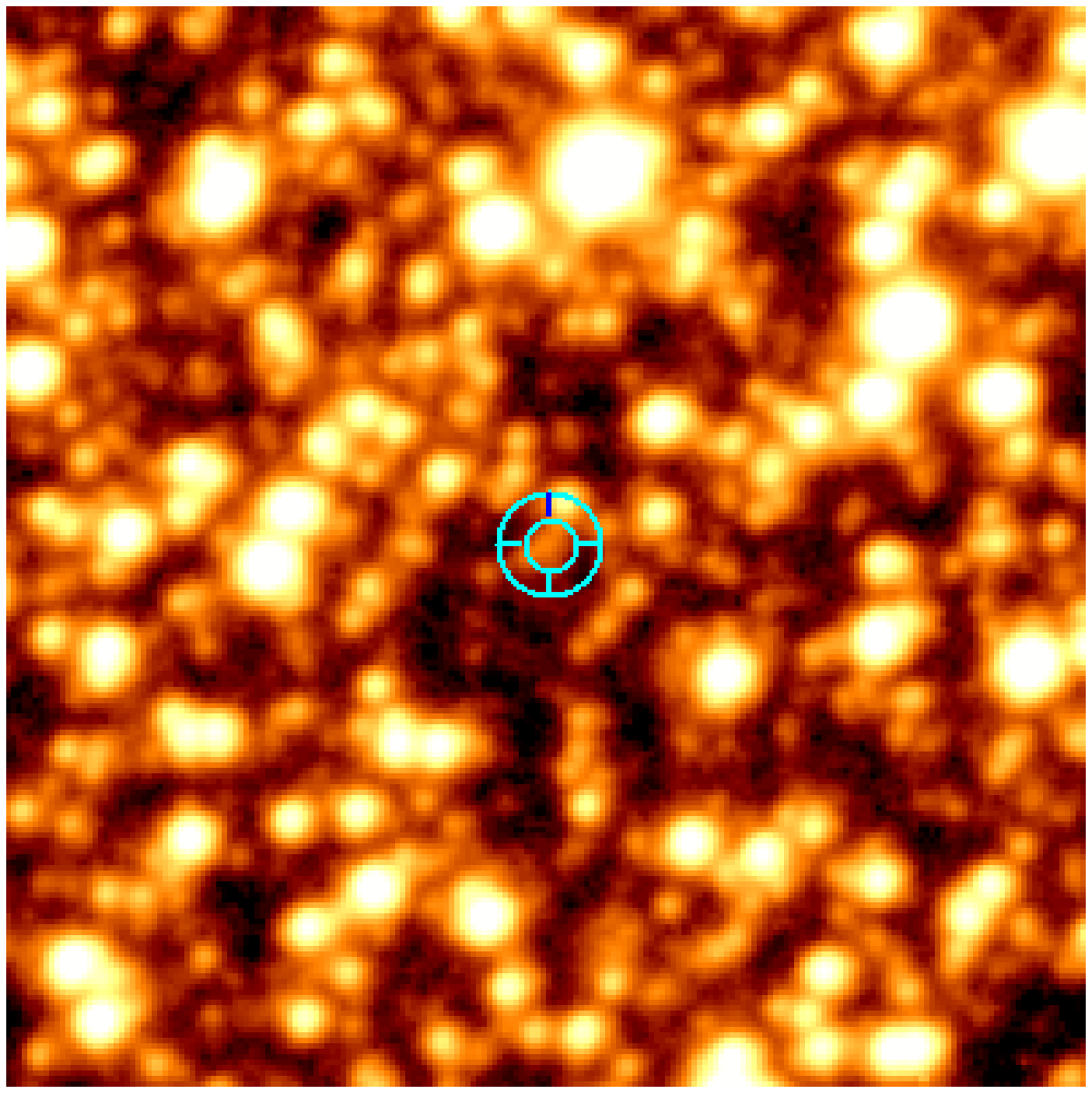}
\end{minipage}
&\begin{minipage}[m]{6.5cm}\includegraphics[width=6cm,bb=0 15 800 400,clip=true]{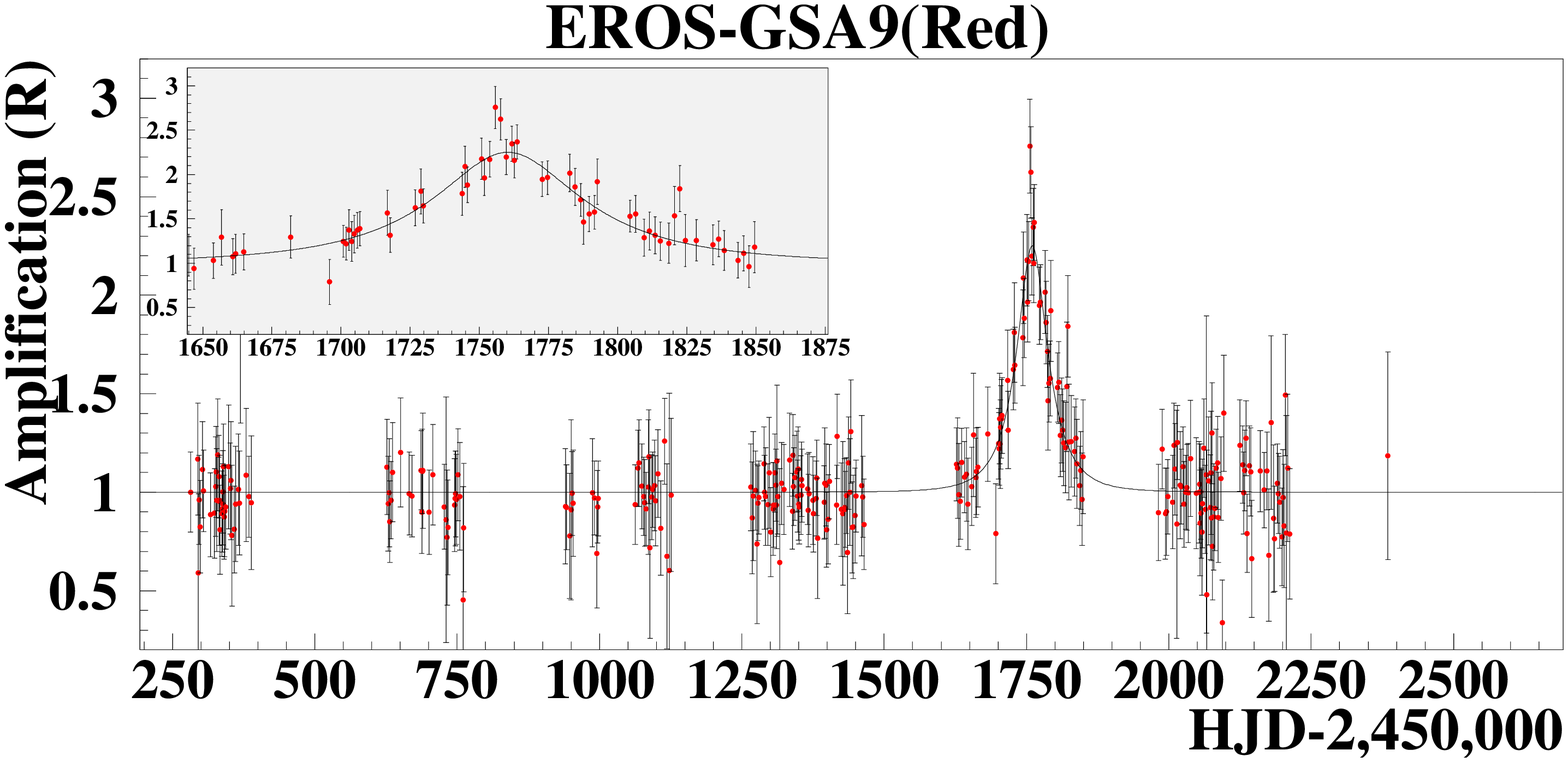}\end{minipage}
&\begin{minipage}[m]{6.5cm}\includegraphics[width=6cm,bb=0 15 800 400,clip=true]{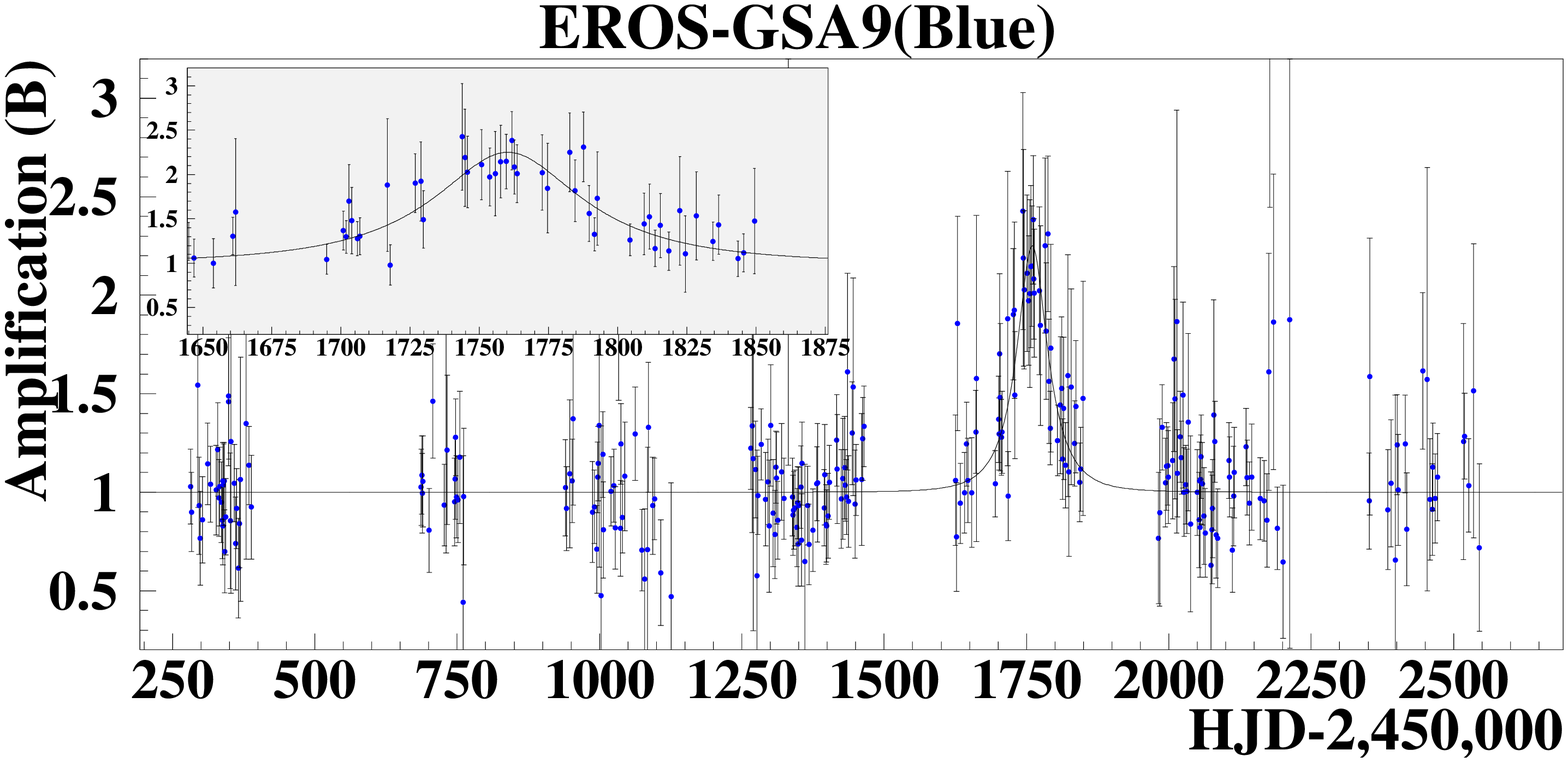}\end{minipage}
\\
\begin{minipage}[m]{3.cm}
\includegraphics[width=3.cm,bb=104 21 509 425]{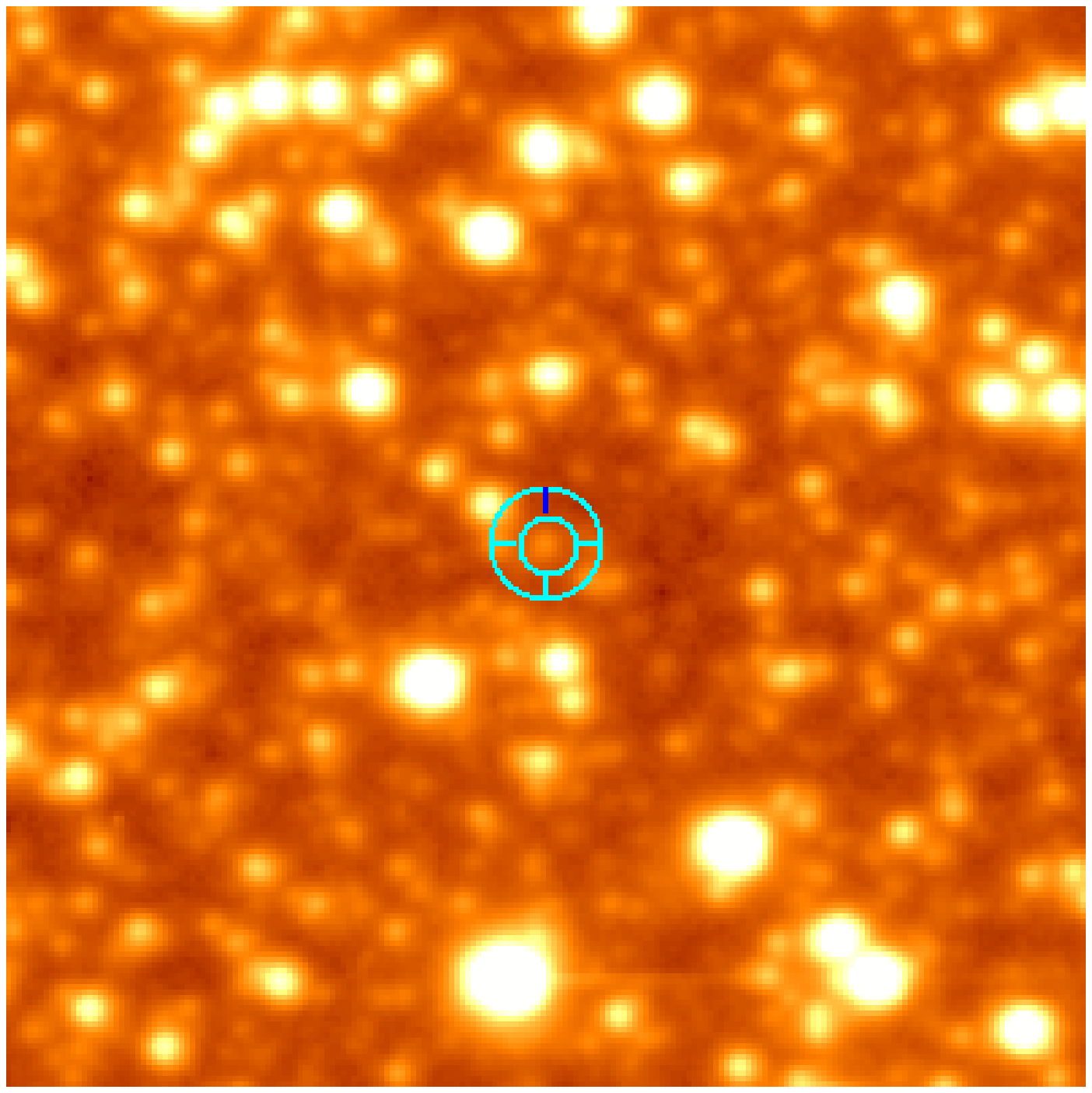}
\end{minipage}
&\begin{minipage}[m]{6.5cm}\includegraphics[width=6cm,bb=0 15 800 400,clip=true]{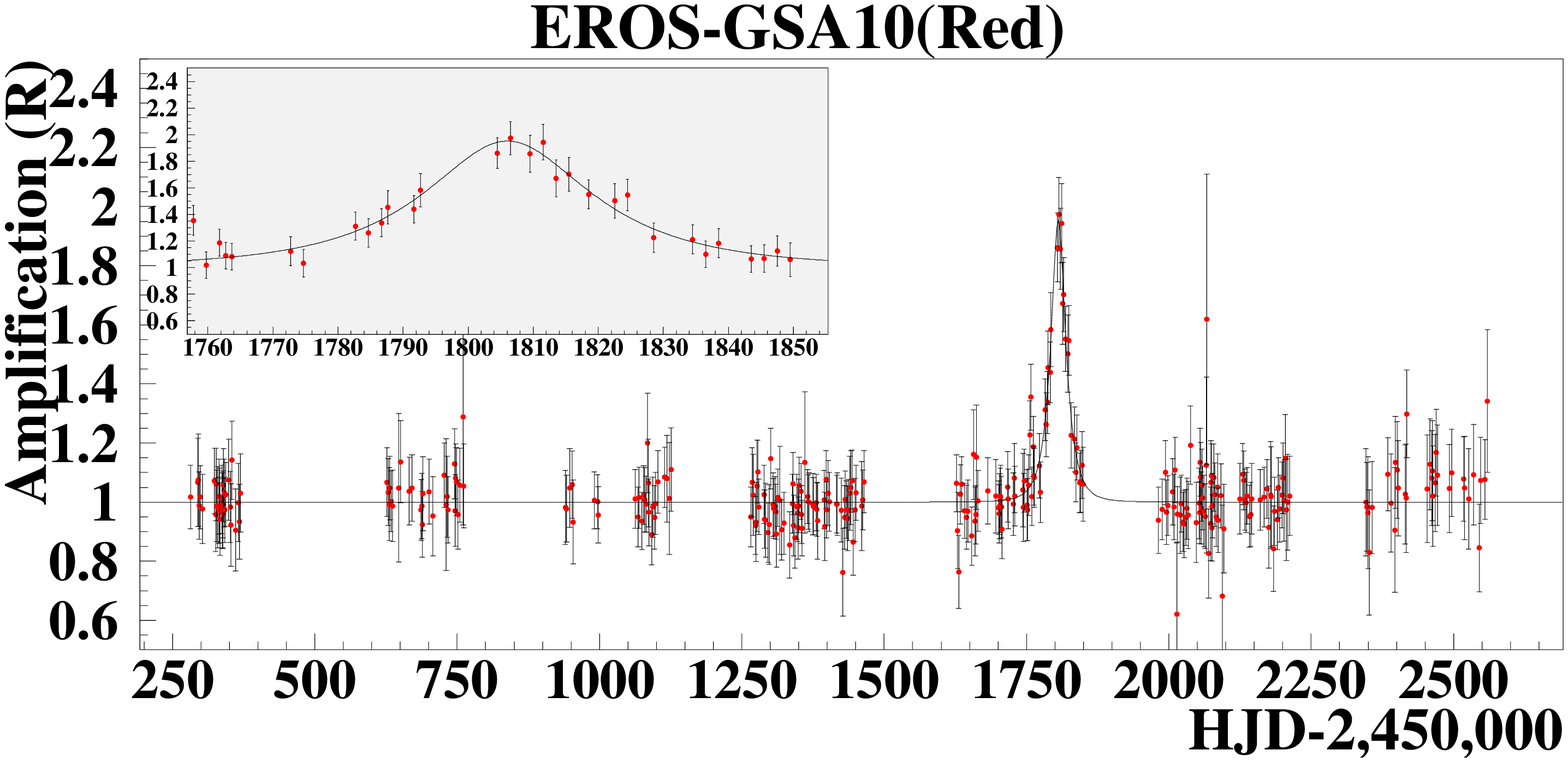}\end{minipage}
&\begin{minipage}[m]{6.5cm}\includegraphics[width=6cm,bb=0 15 800 400,clip=true]{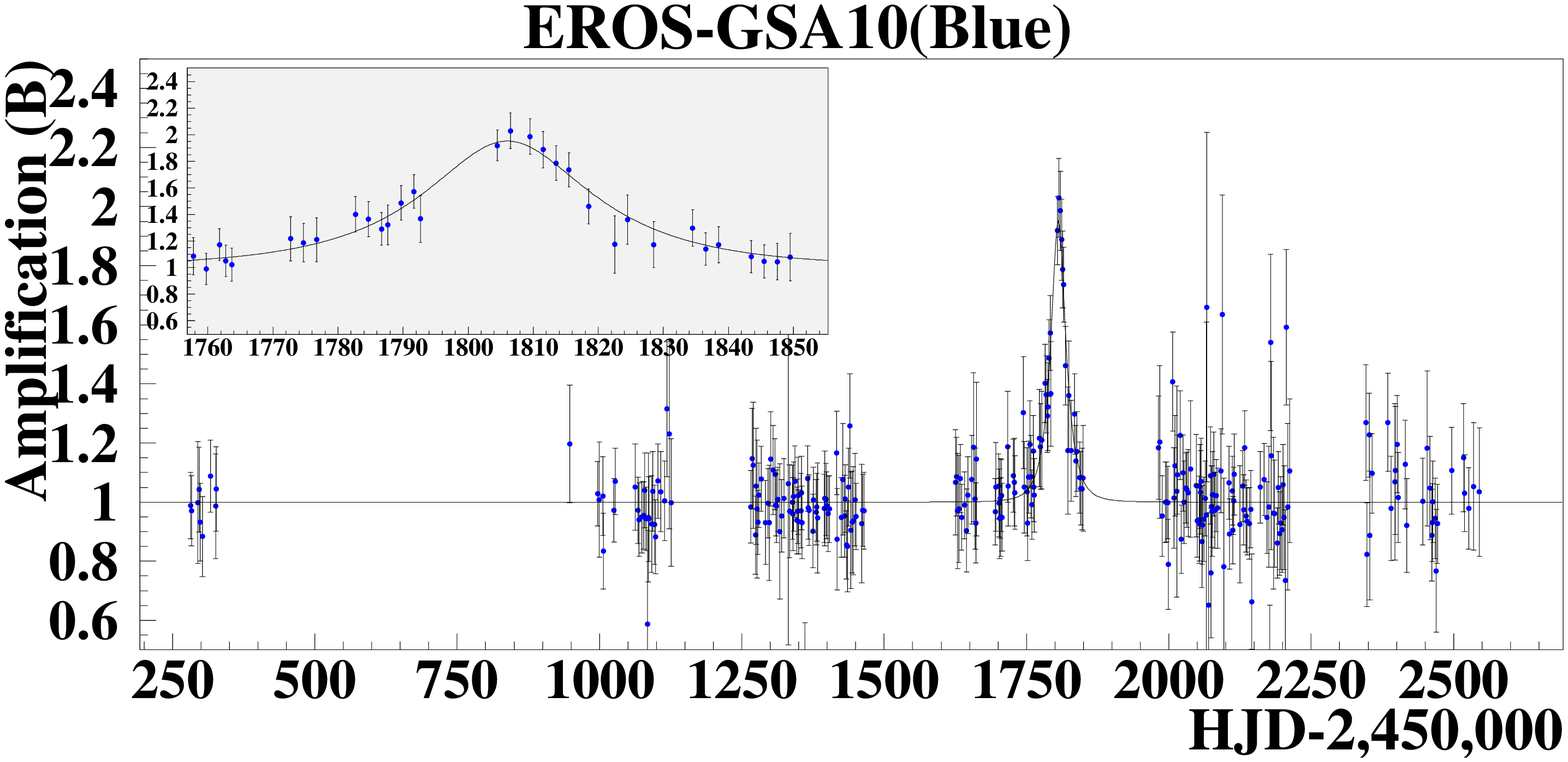}\end{minipage}
\\
\begin{minipage}[m]{3.cm}
\includegraphics[width=3.cm,bb=104 21 509 425]{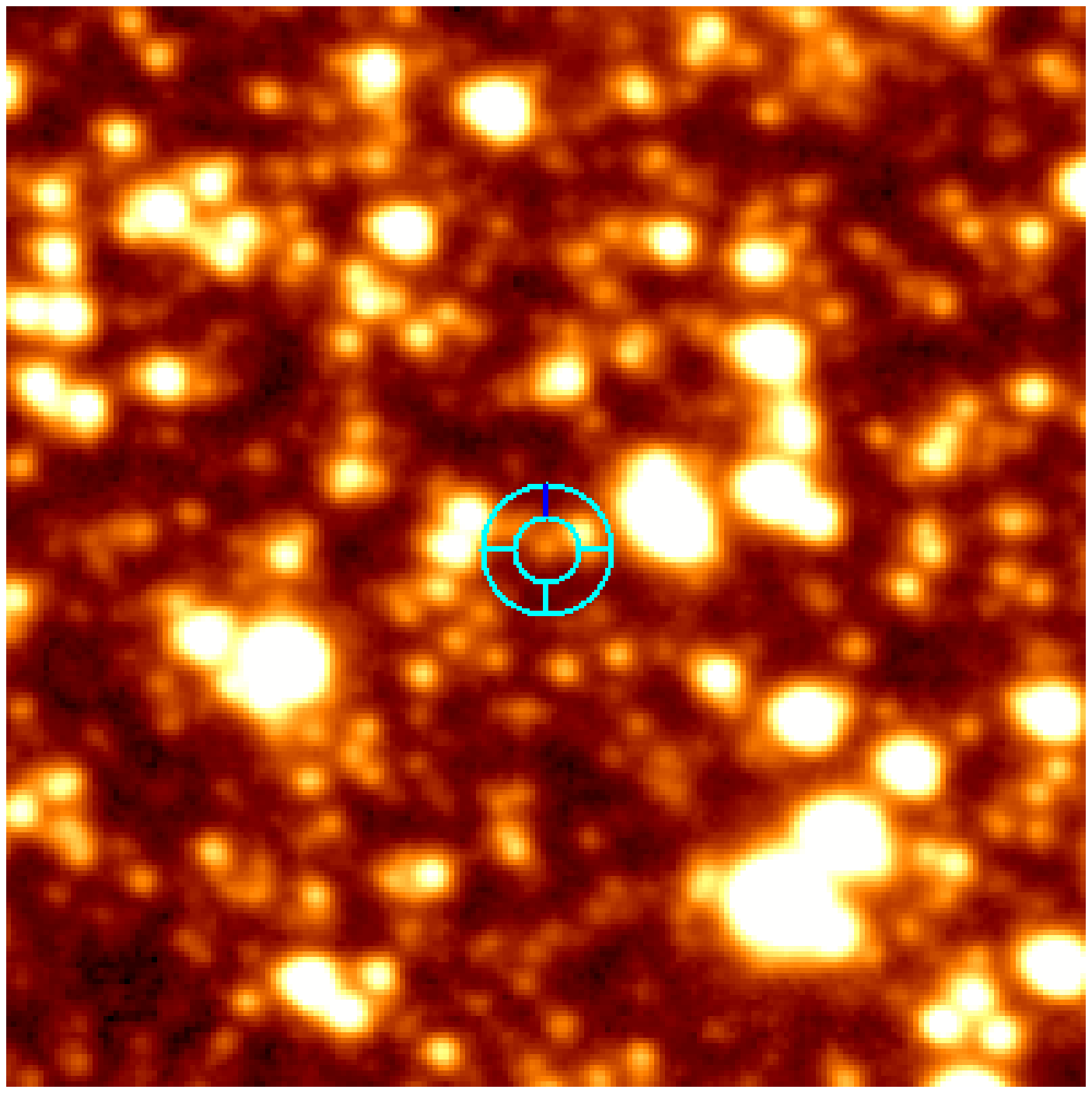}
\end{minipage}
&\begin{minipage}[m]{6.5cm}\includegraphics[width=6cm,bb=0 15 800 400,clip=true]{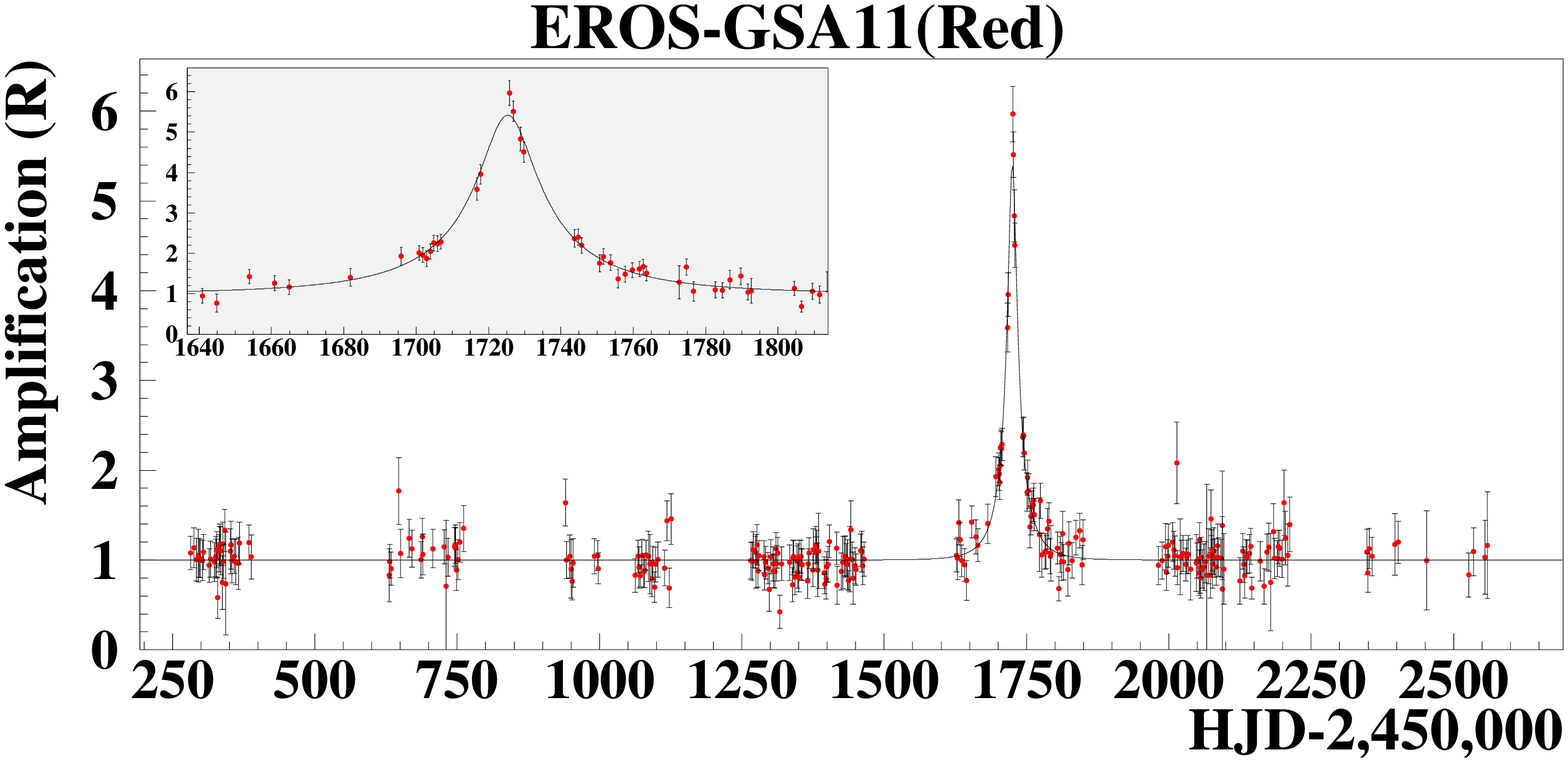}\end{minipage}
&\begin{minipage}[m]{6.5cm}\includegraphics[width=6cm,bb=0 15 800 400,clip=true]{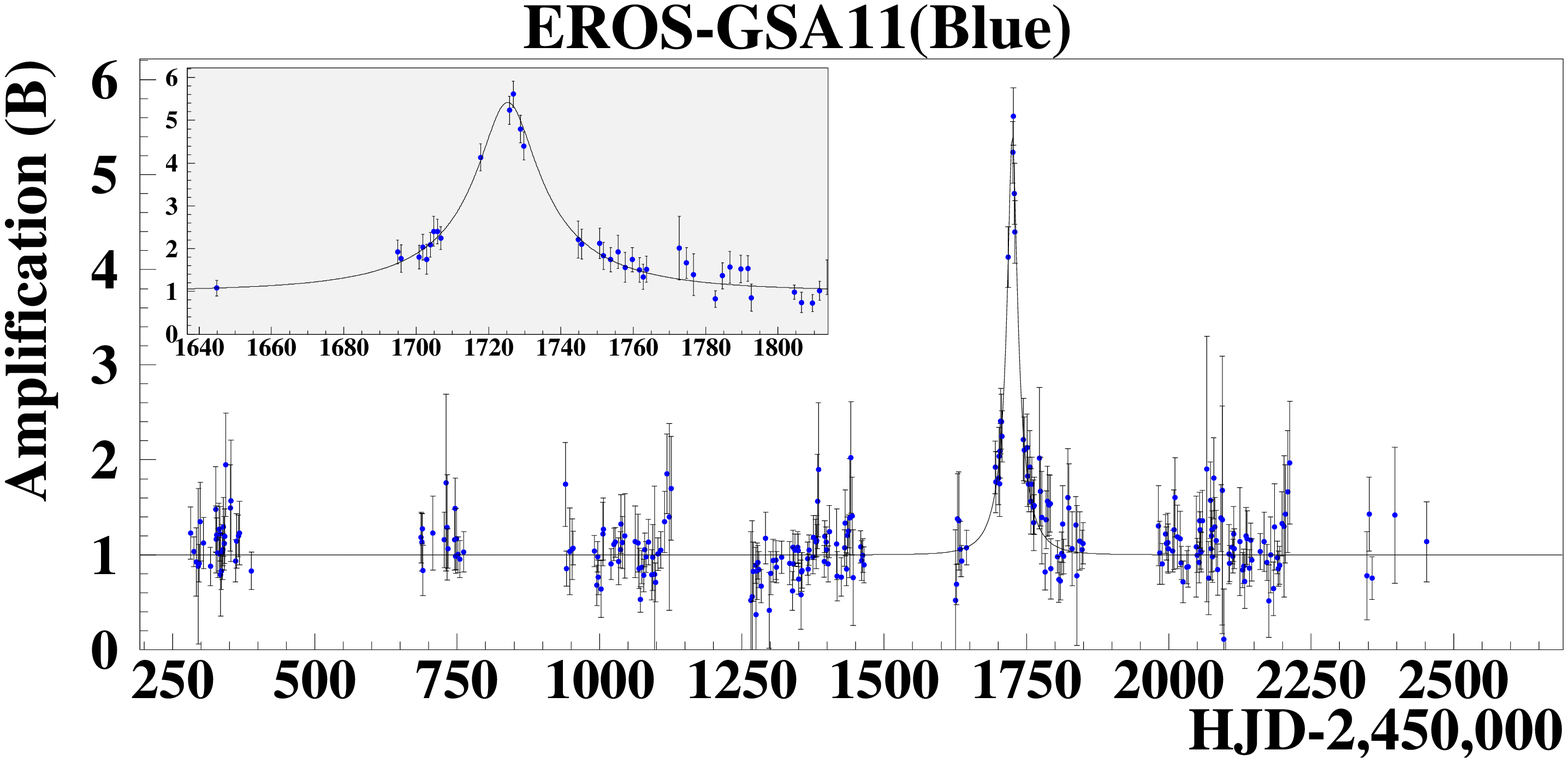}\end{minipage}
\\
\begin{minipage}[m]{3.cm}
\includegraphics[width=3.cm,bb=104 21 509 425]{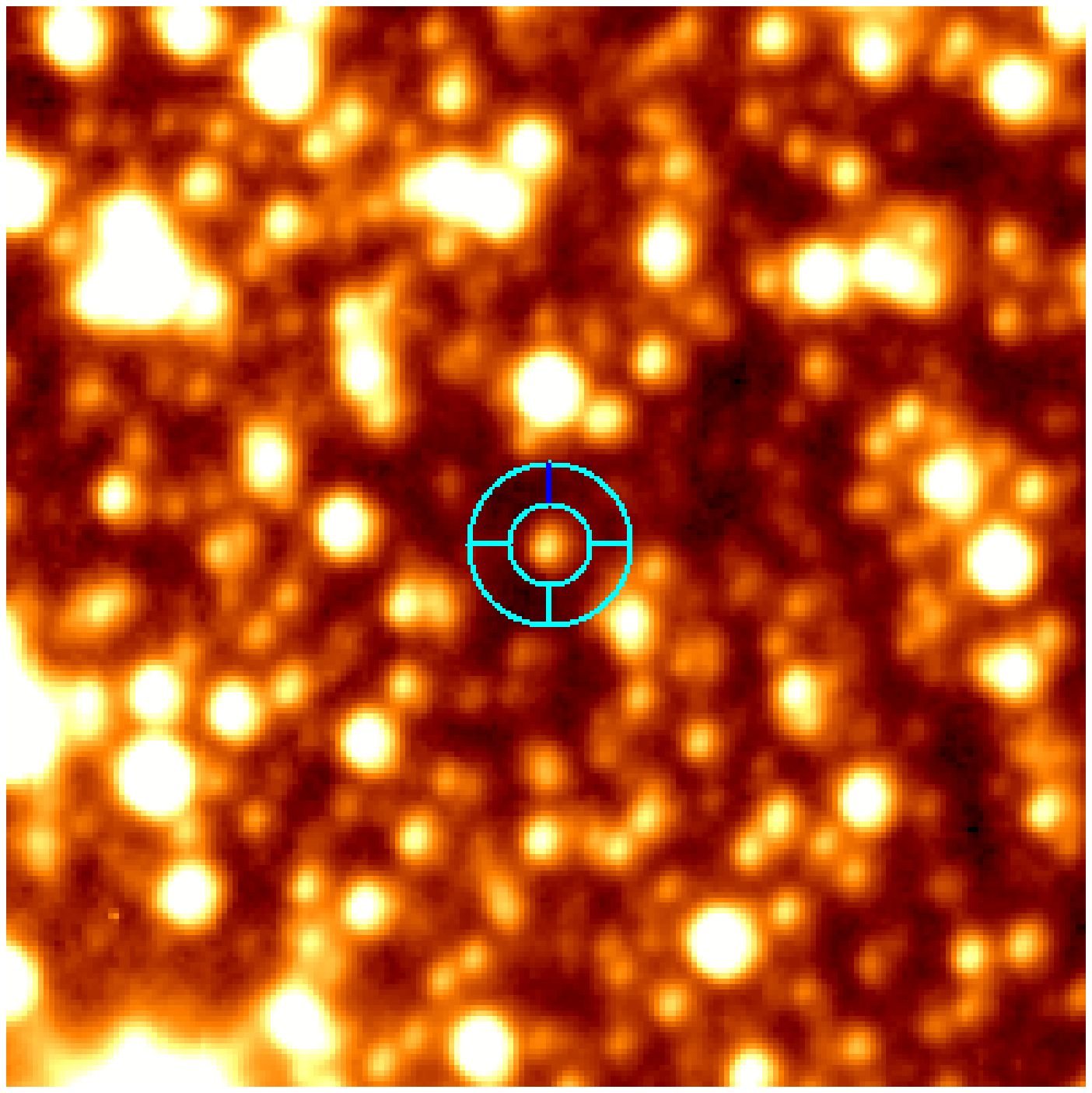}
\end{minipage}
&\begin{minipage}[m]{6.5cm}\includegraphics[width=6cm,bb=0 15 800 400,clip=true]{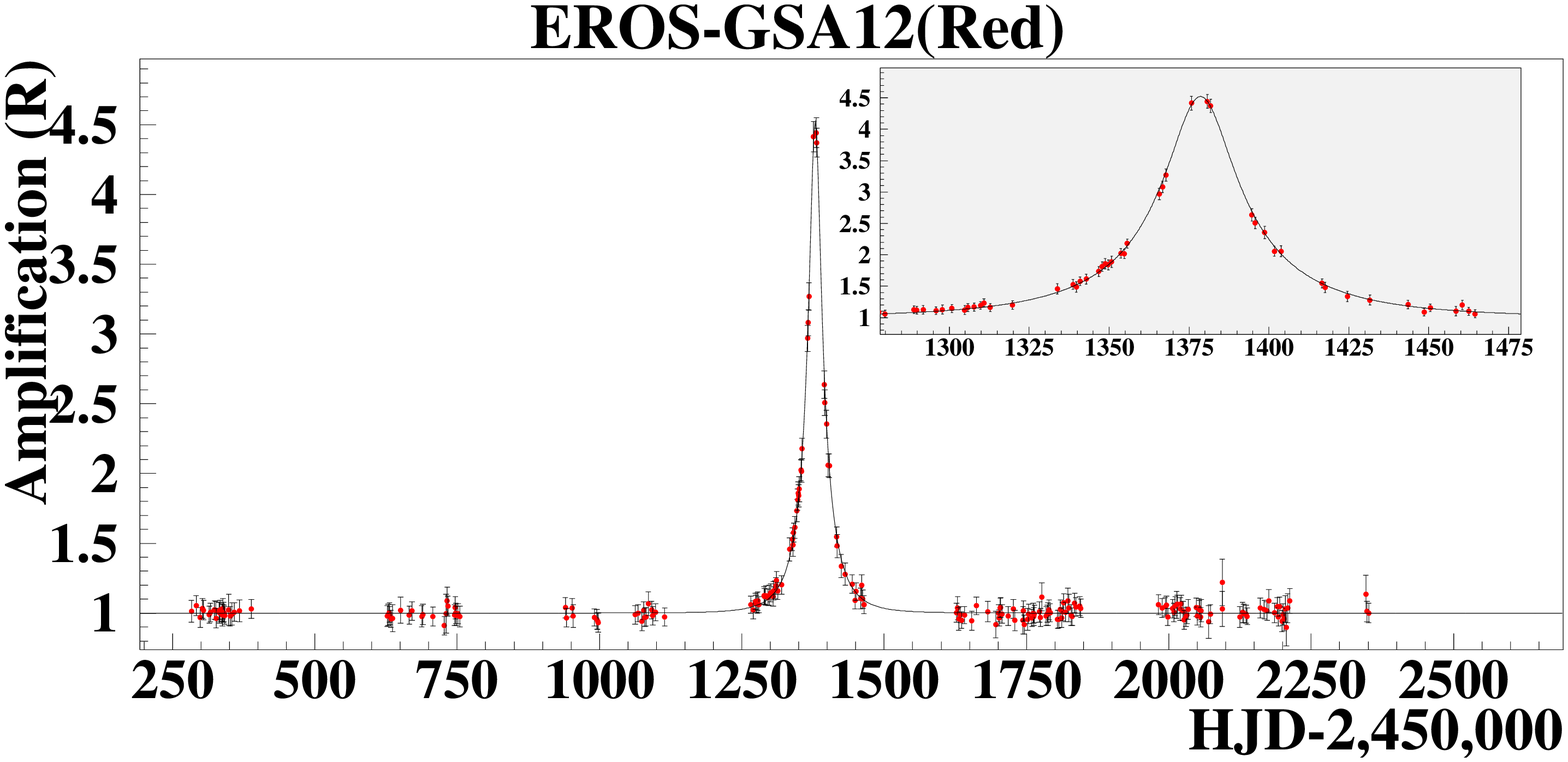}\end{minipage}
&\begin{minipage}[m]{6.5cm}\includegraphics[width=6cm,bb=0 15 800 400,clip=true]{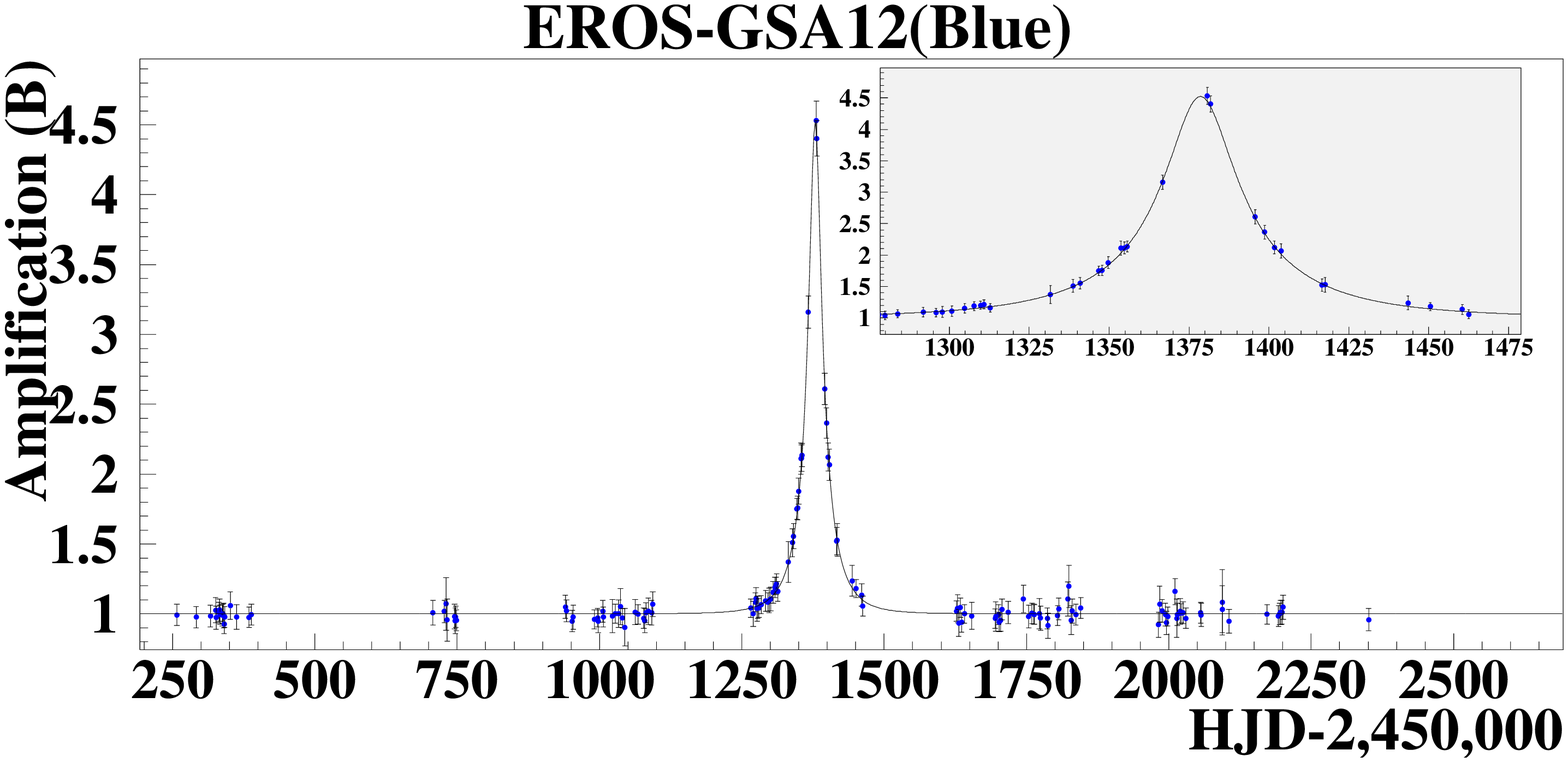}\end{minipage}
\end{tabular}
\caption[]{Finding charts ($84''\times84''$, N-up, E-left) and magnification
of the candidates as a function of the Heliocentric Julian Day
($HJD-2,450,000$).
In the case of multi-detected events, we display only the best sampled light curve.}
\label{figclum}
\end{center}
\end{figure*}

\begin{figure*}
\begin{center}
\begin{tabular}{ccc}
\begin{minipage}[m]{3.cm}
\includegraphics[width=3.cm,bb=104 21 509 425]{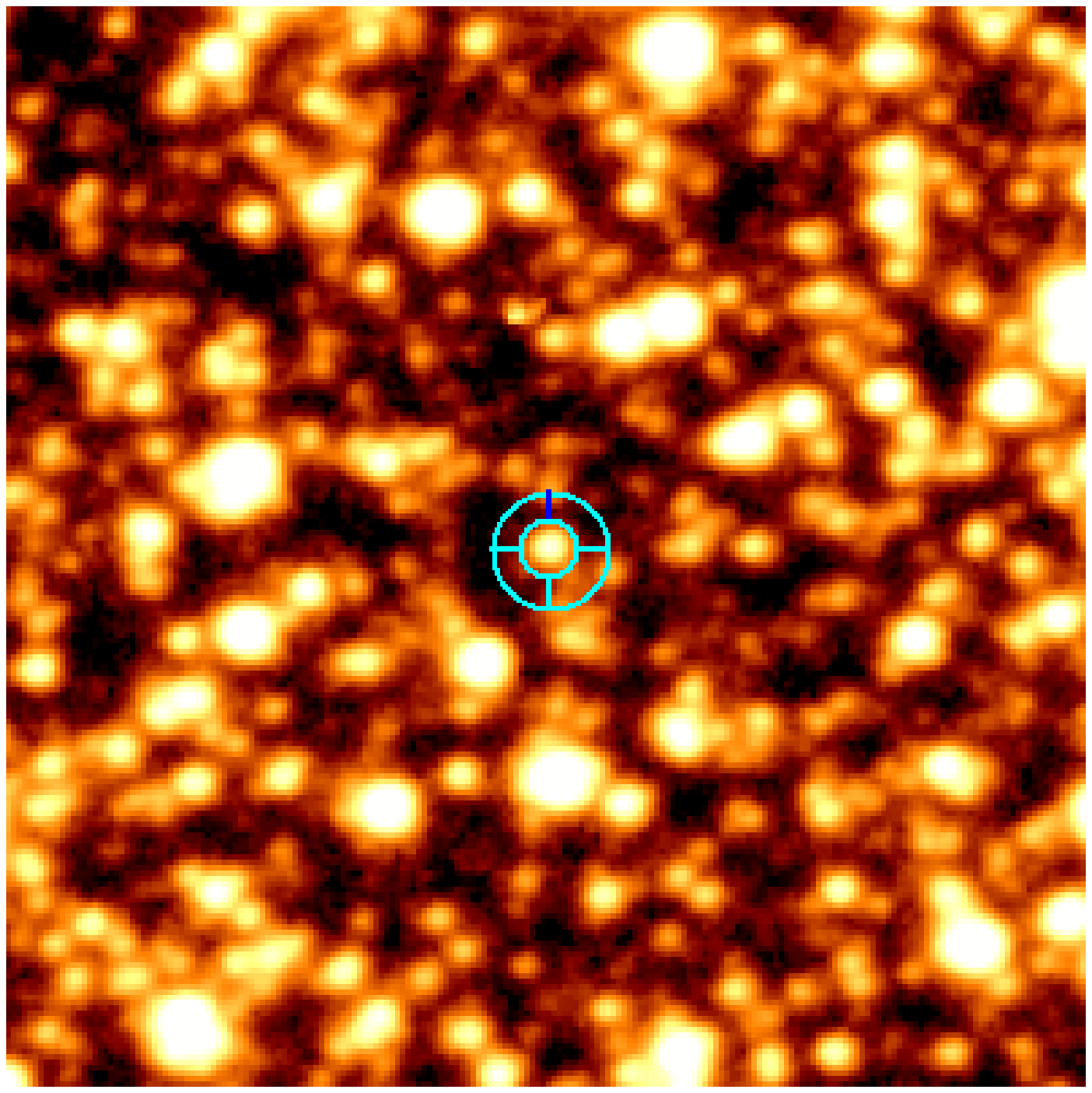}
\end{minipage}
&\begin{minipage}[m]{6.5cm}\includegraphics[width=6cm,bb=0 15 800 400,clip=true]{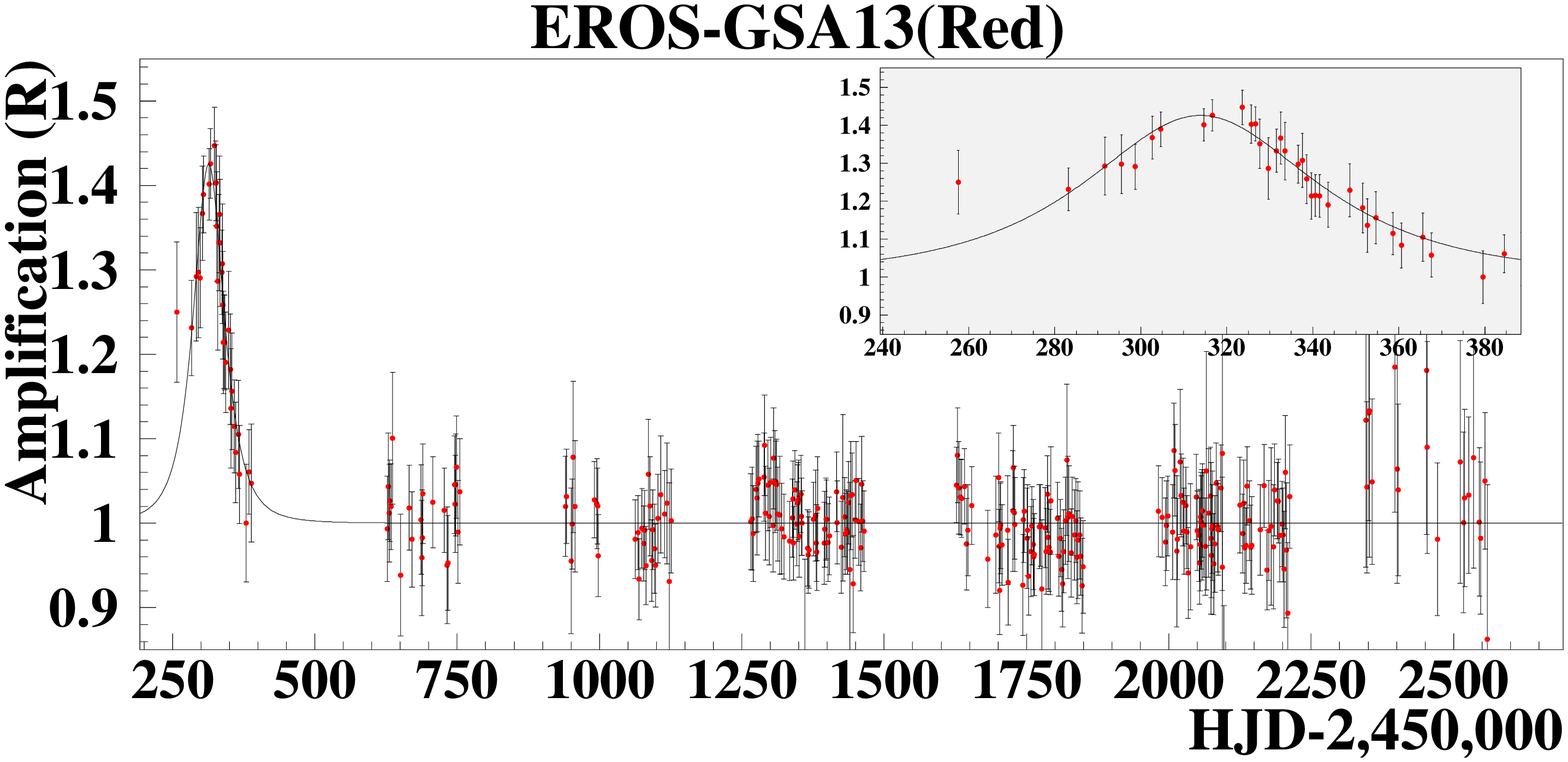}\end{minipage}
&\begin{minipage}[m]{6.5cm}\includegraphics[width=6cm,bb=0 15 800 400,clip=true]{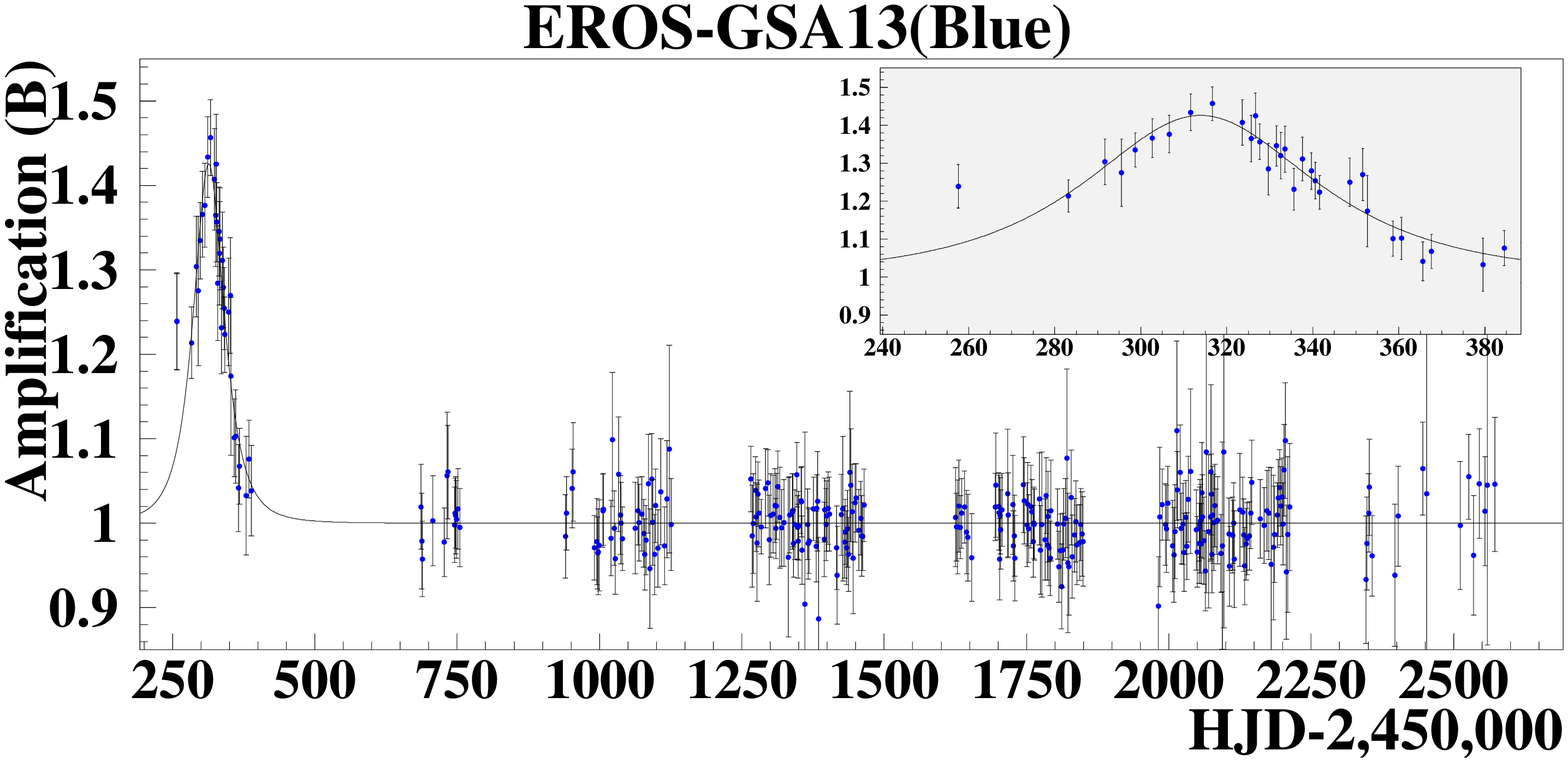}\end{minipage}
\\
\begin{minipage}[m]{3.cm}
\includegraphics[width=3.cm,bb=104 21 509 425]{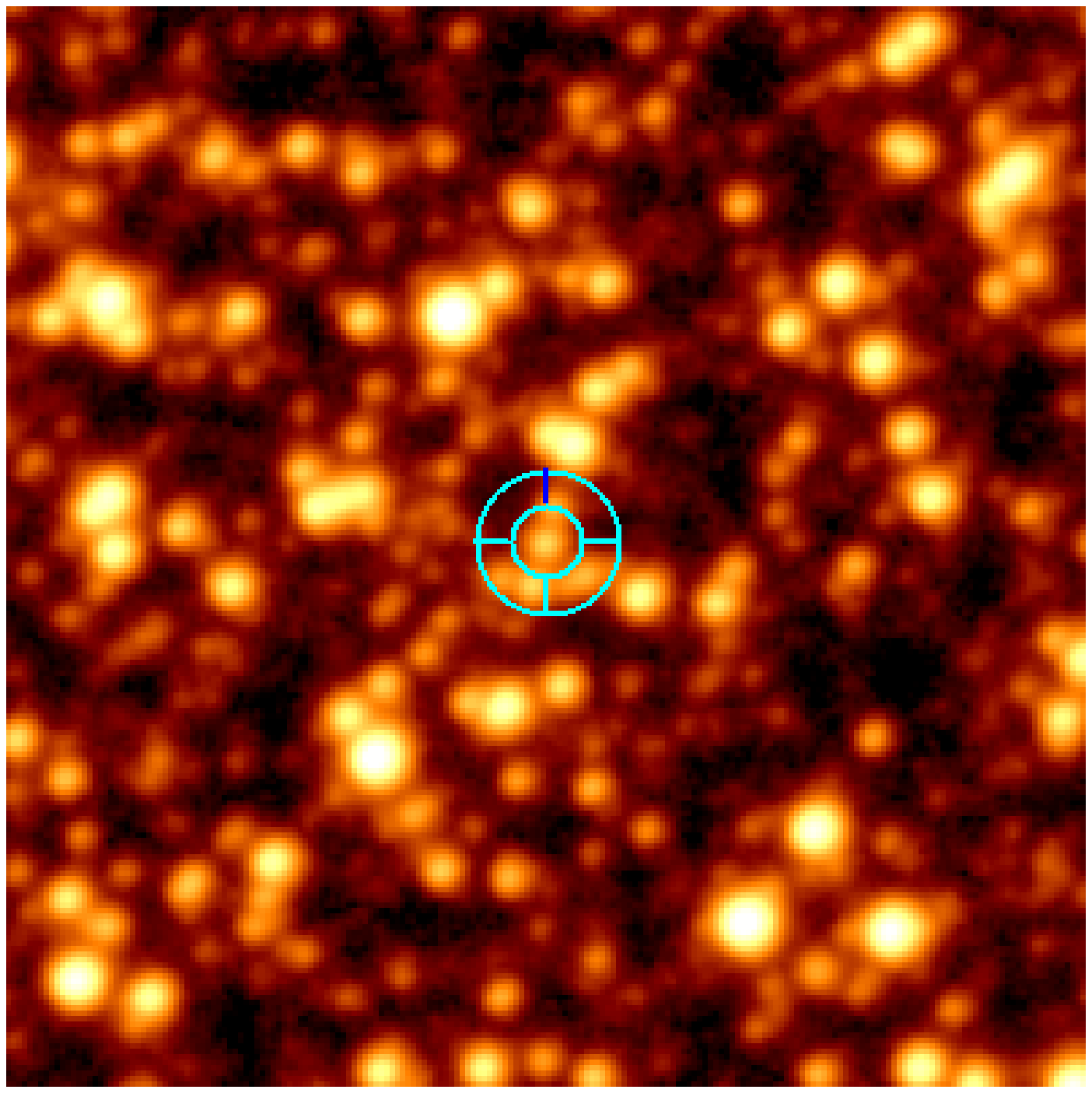}
\end{minipage}
&\begin{minipage}[m]{6.5cm}\includegraphics[width=6cm,bb=0 15 800 400,clip=true]{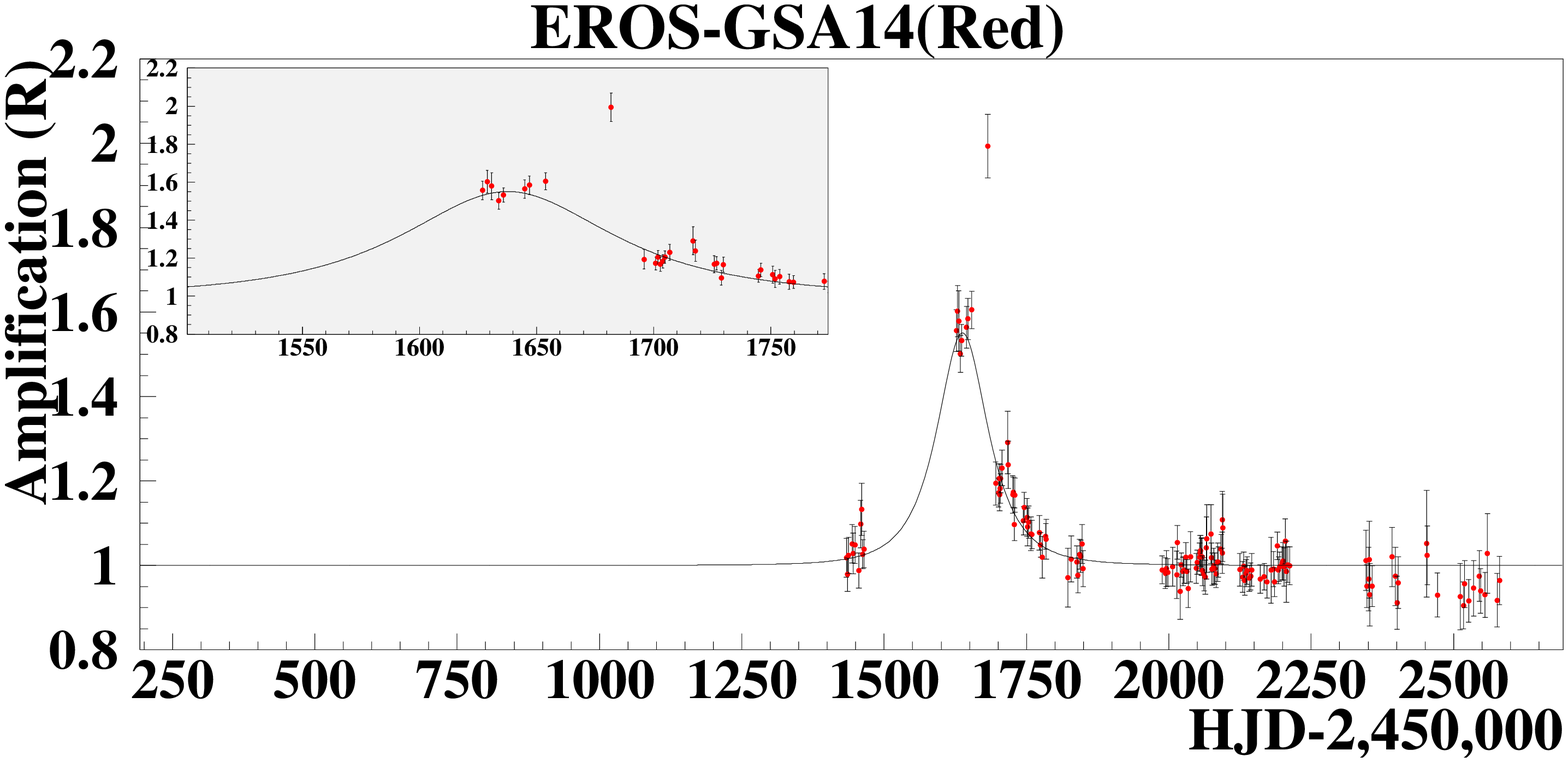}\end{minipage}
&\begin{minipage}[m]{6.5cm}\includegraphics[width=6cm,bb=0 15 800 400,clip=true]{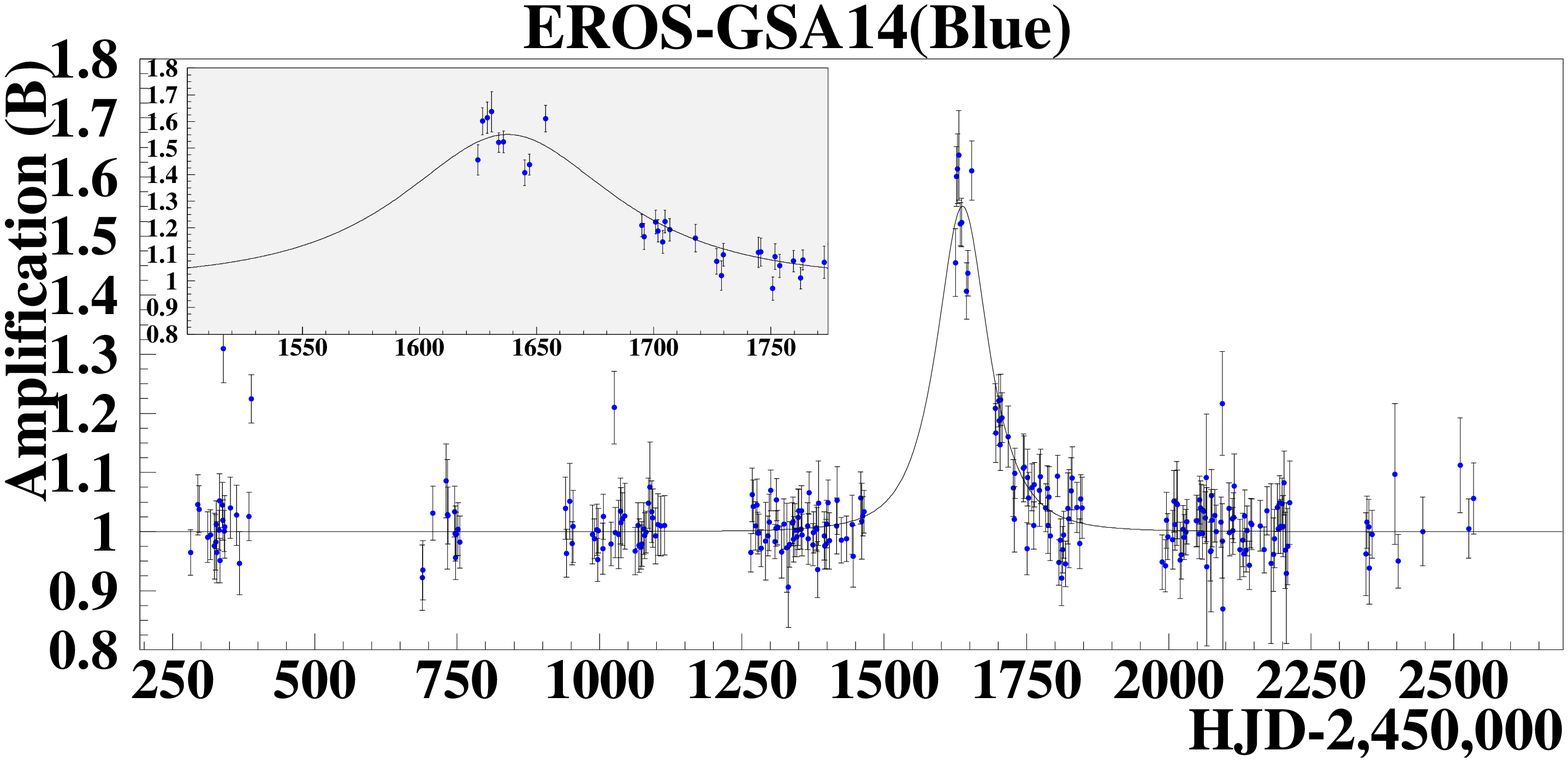}\end{minipage}
\\
\begin{minipage}[m]{3.cm}
\includegraphics[width=3.cm,bb=104 21 509 425]{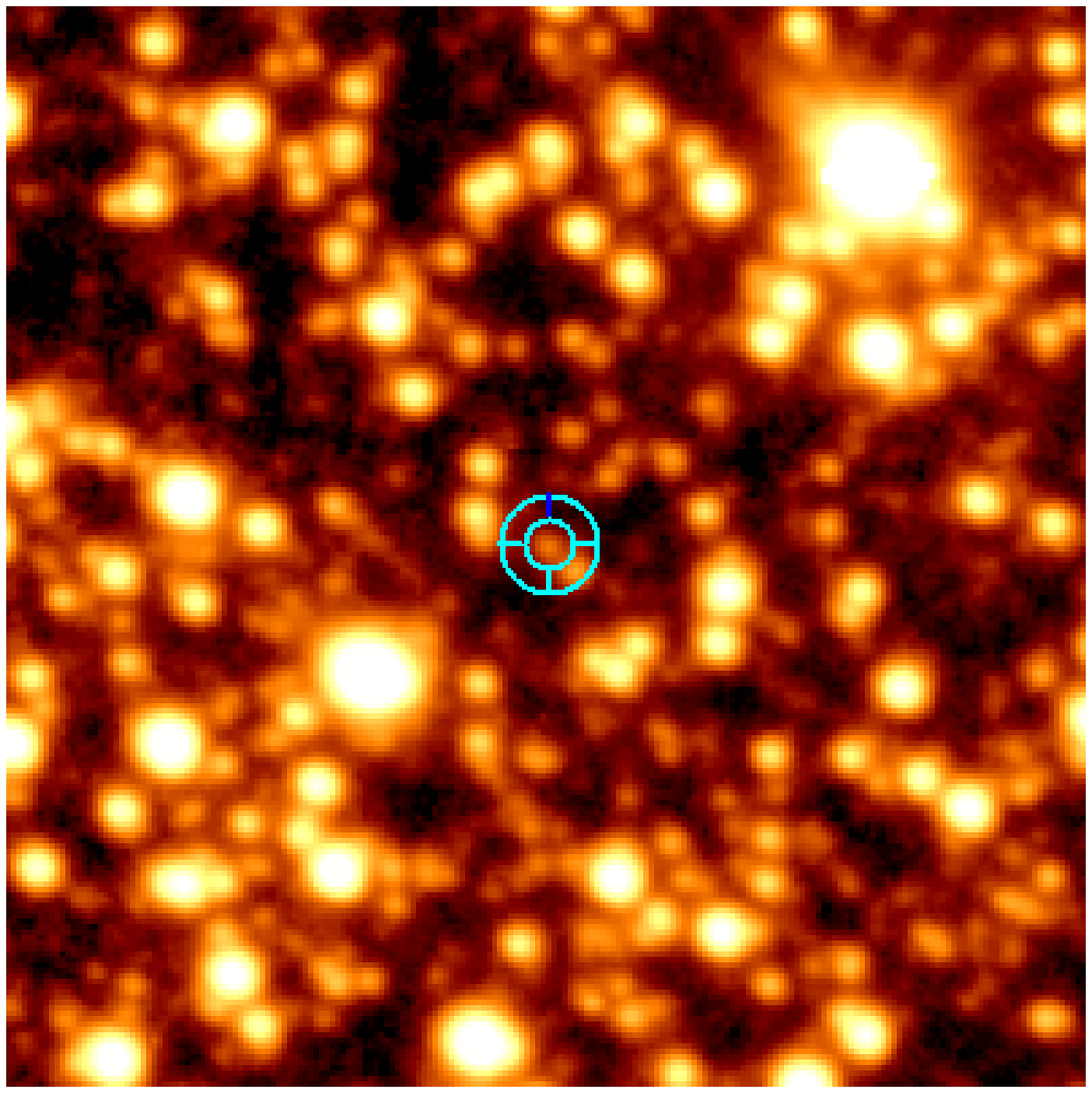}
\end{minipage}
&\begin{minipage}[m]{6.5cm}\includegraphics[width=6cm,bb=0 15 800 400,clip=true]{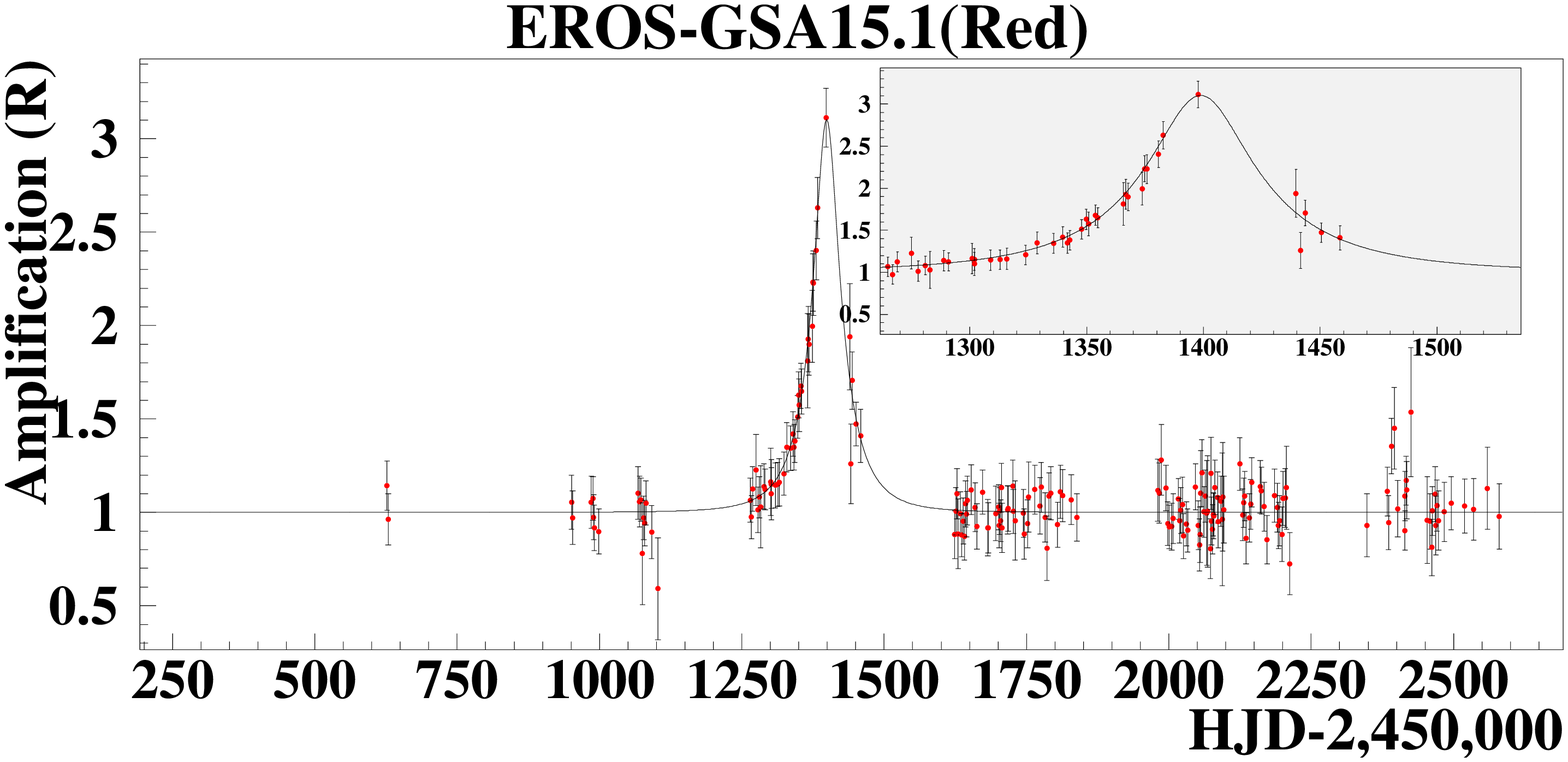}\end{minipage}
&\begin{minipage}[m]{6.5cm}\includegraphics[width=6cm,bb=0 15 800 400,clip=true]{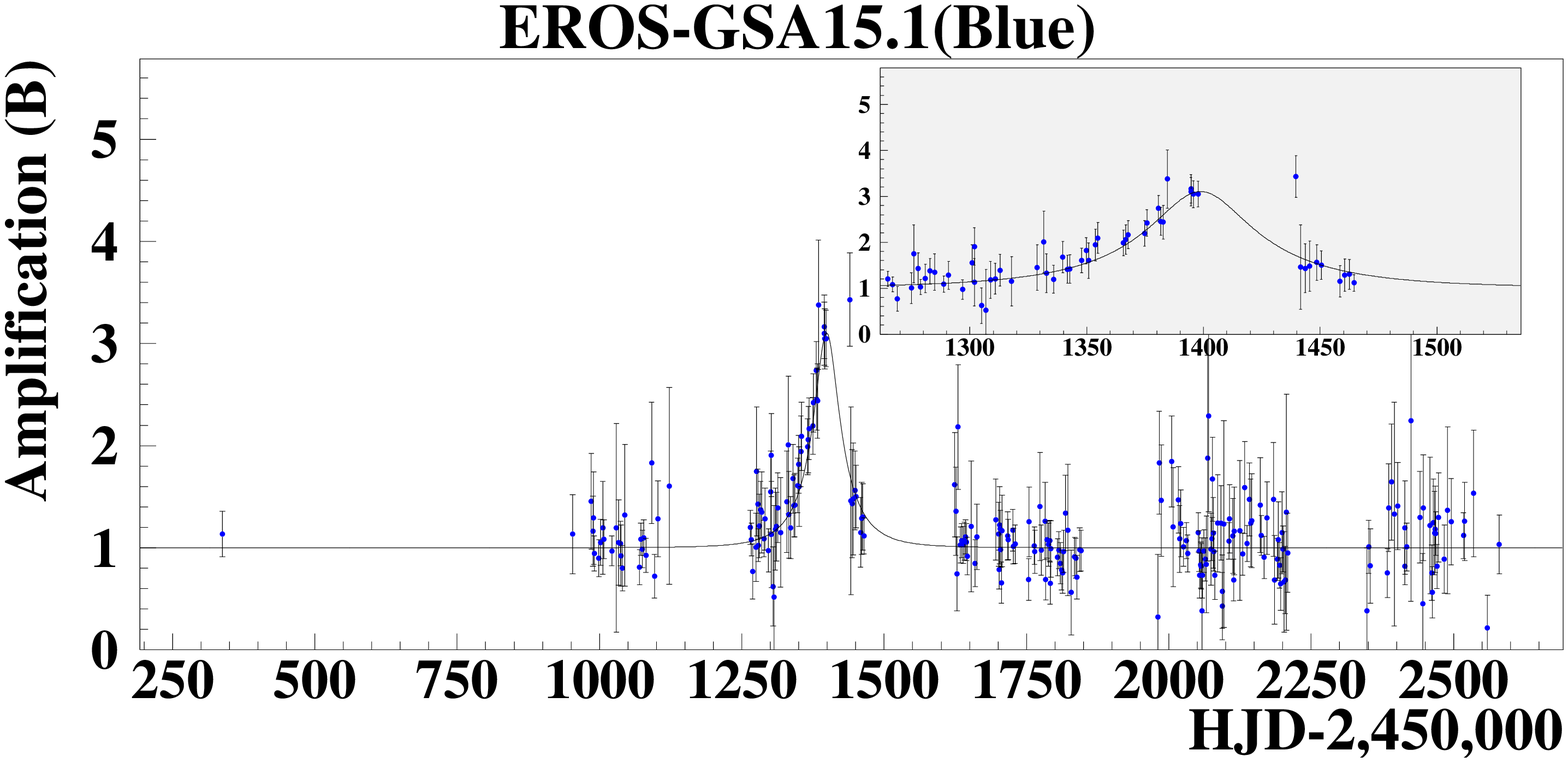}\end{minipage}
\\
\begin{minipage}[m]{3.cm}
\includegraphics[width=3.cm,bb=104 21 509 425]{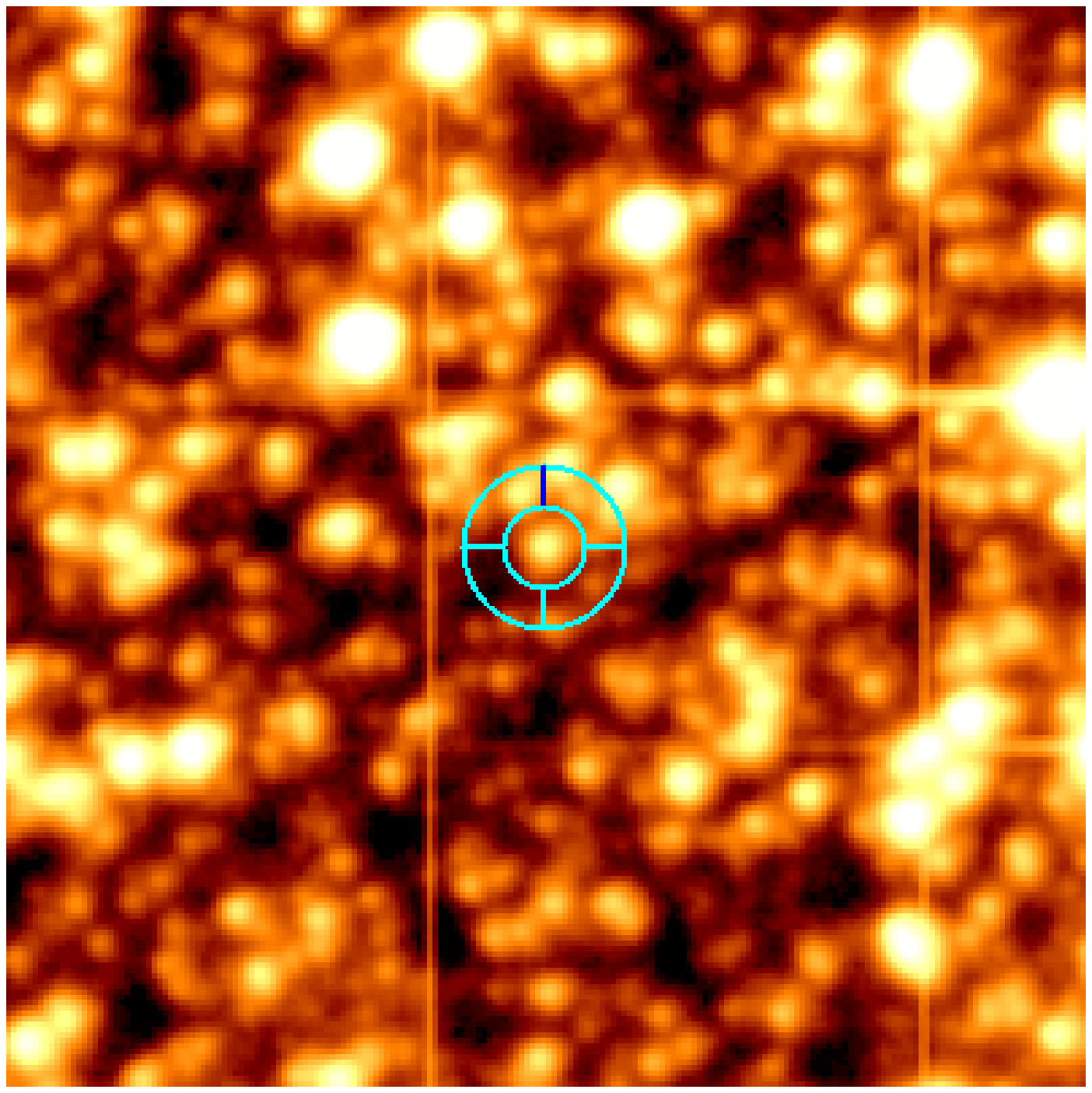}
\end{minipage}
&\begin{minipage}[m]{6.5cm}\includegraphics[width=6cm,bb=0 15 800 400,clip=true]{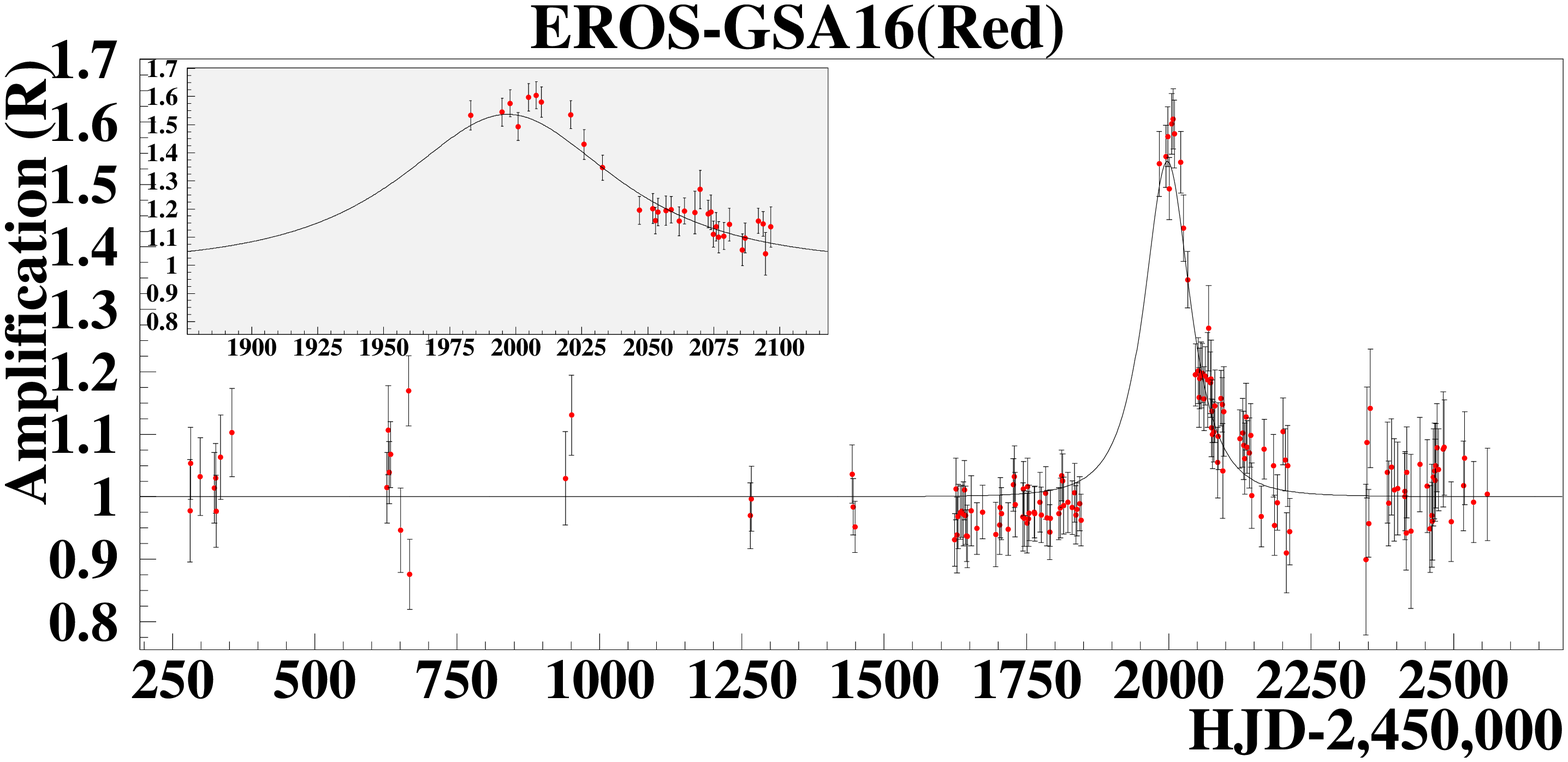}\end{minipage}
&\begin{minipage}[m]{6.5cm}\includegraphics[width=6cm,bb=0 15 800 400,clip=true]{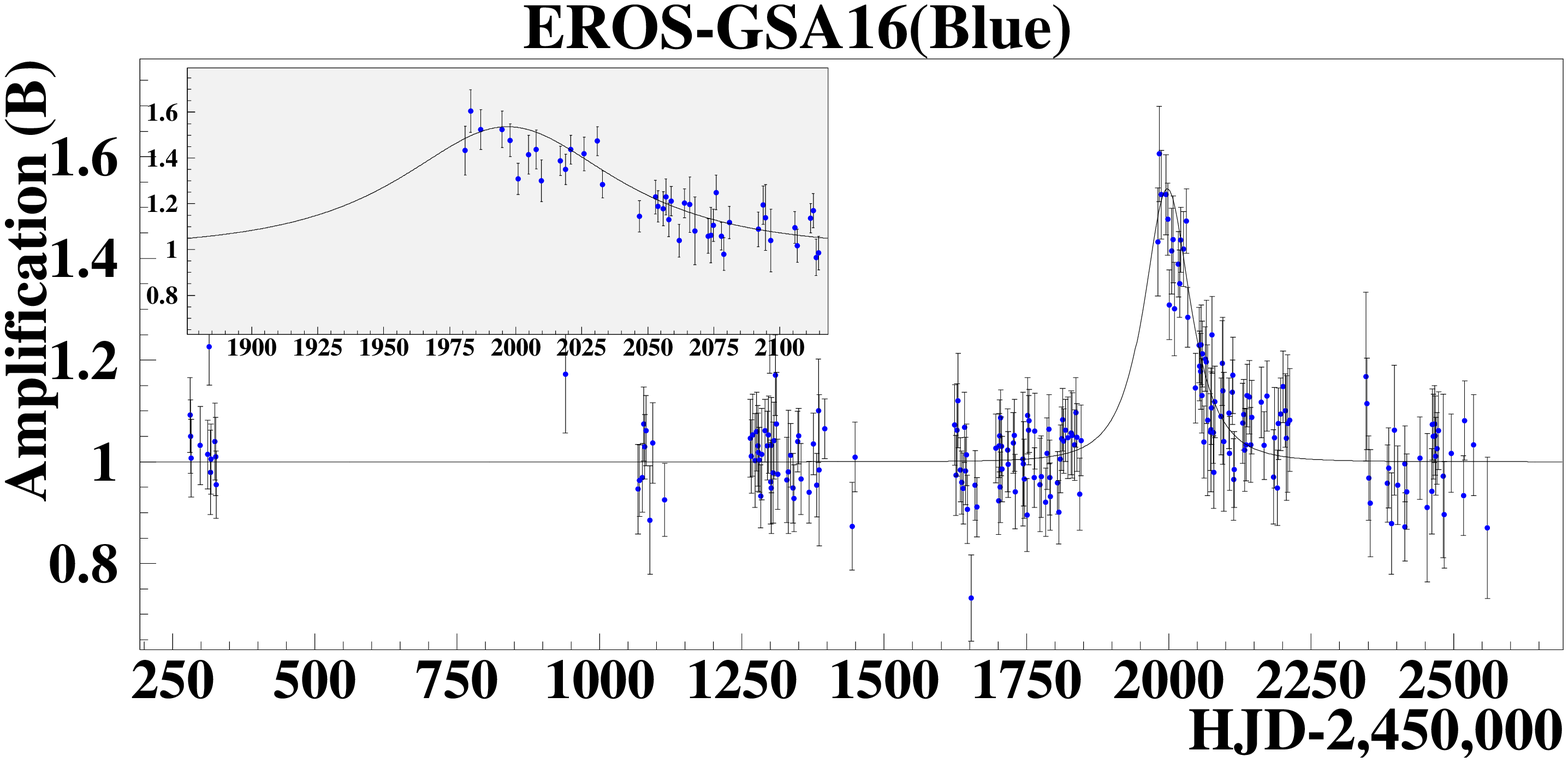}\end{minipage}
\\
\begin{minipage}[m]{3cm}
\includegraphics[width=3cm,bb=104 21 509 425]{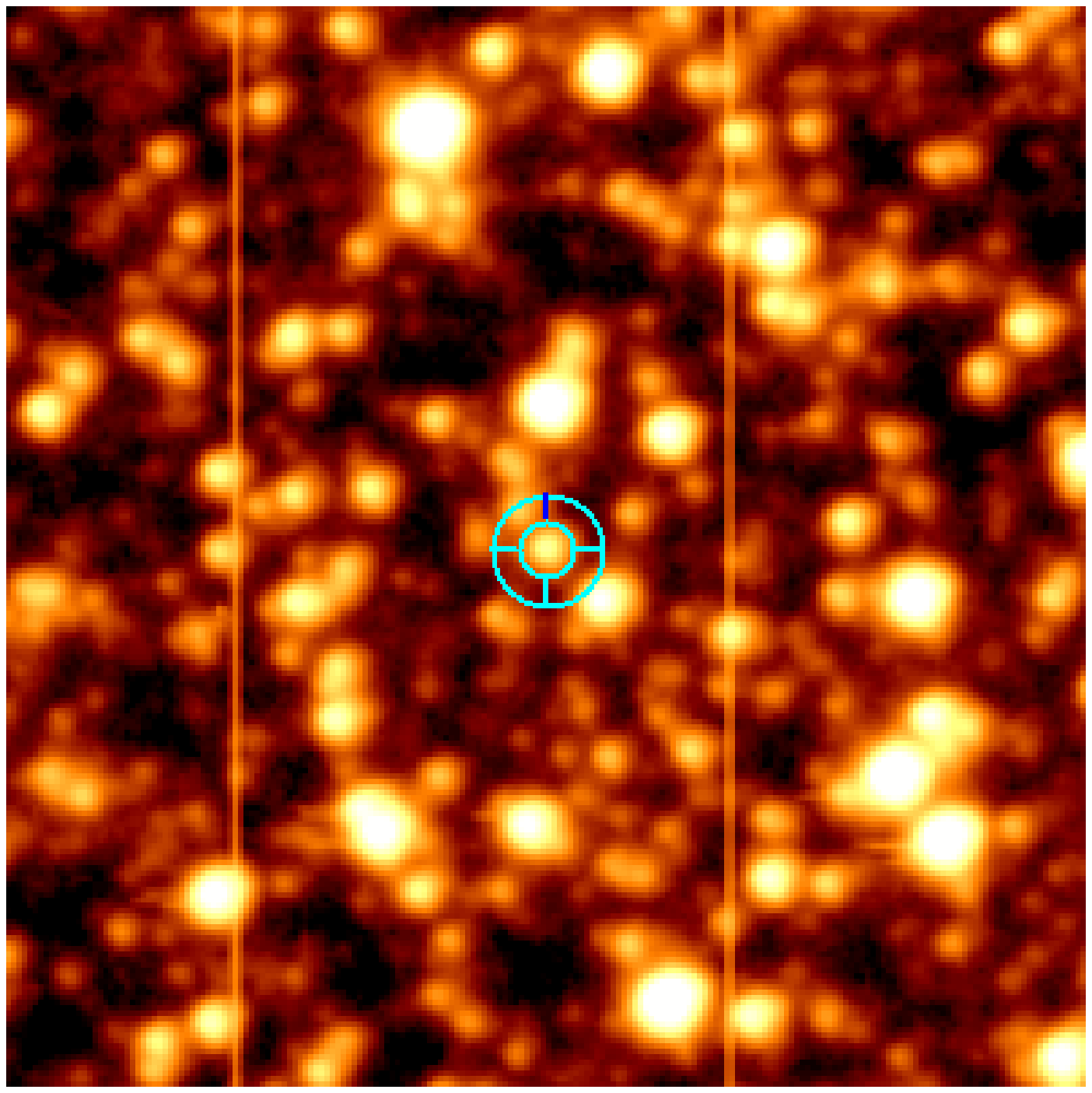}
\end{minipage}
&\begin{minipage}[m]{6.5cm}\includegraphics[width=6cm,bb=0 15 800 400,clip=true]{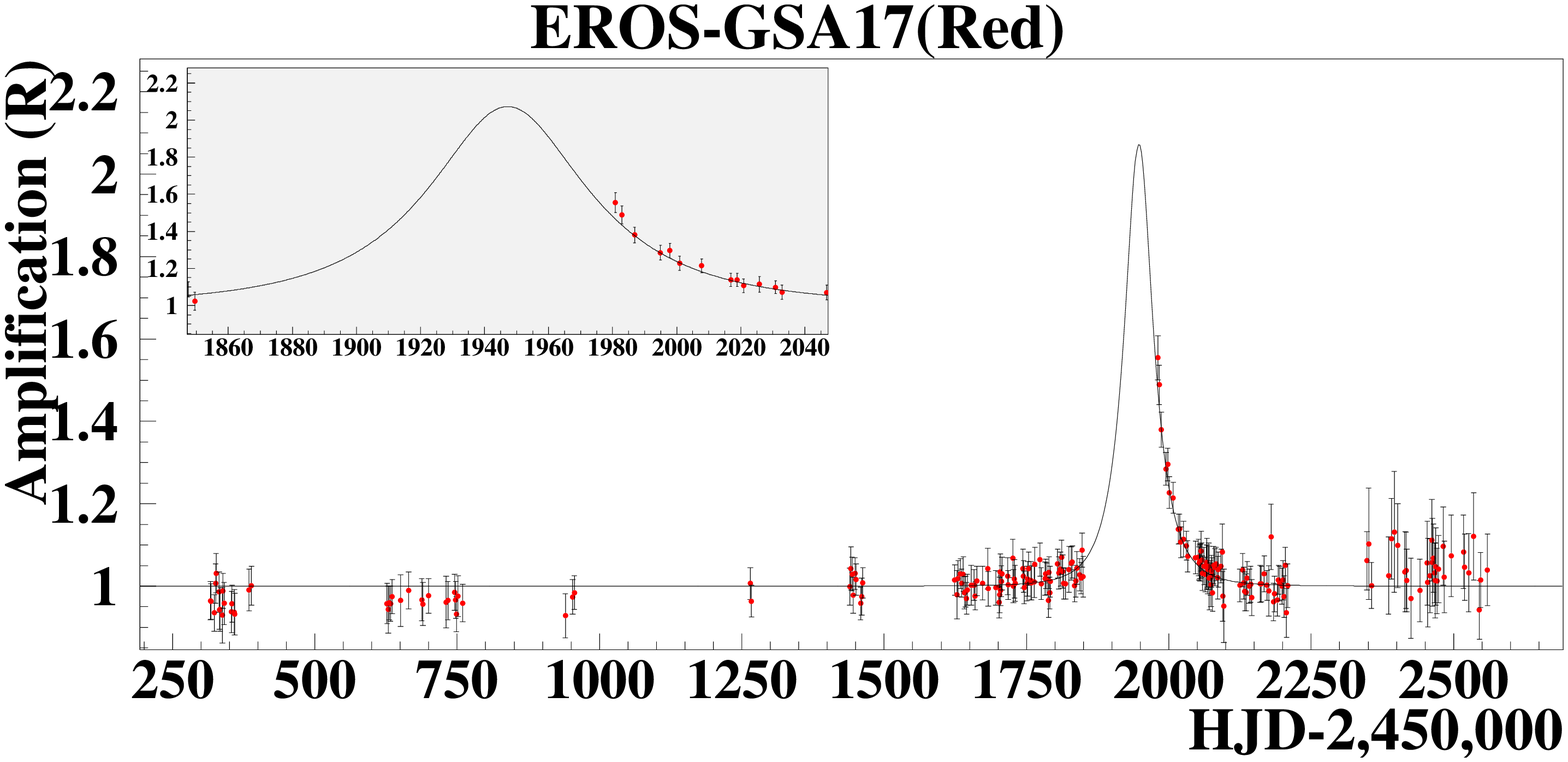}\end{minipage}
&\begin{minipage}[m]{6.5cm}\includegraphics[width=6cm,bb=0 15 800 400,clip=true]{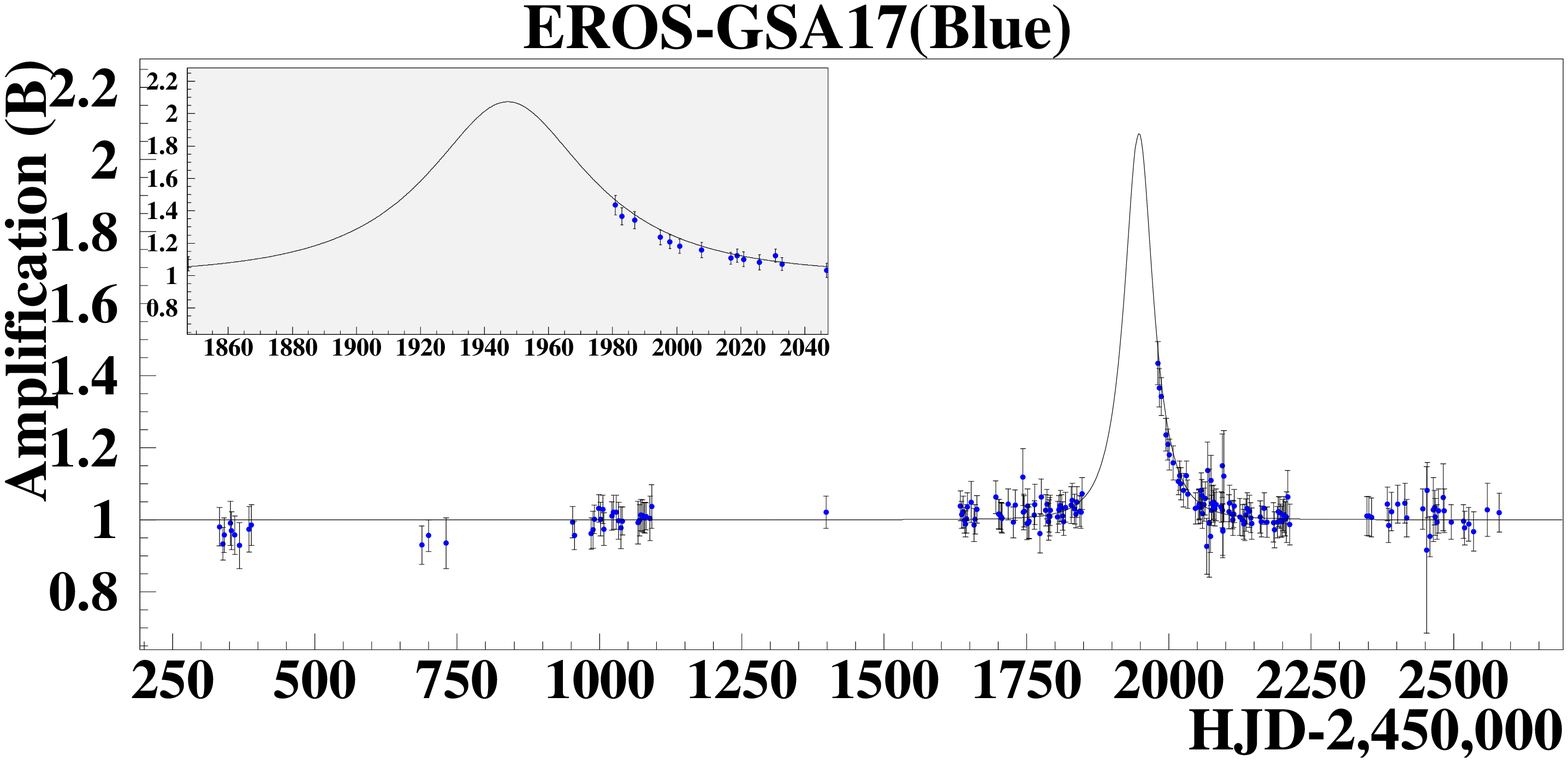}\end{minipage}
\\
\begin{minipage}[m]{3.cm}
\includegraphics[width=3.cm,bb=104 21 509 425]{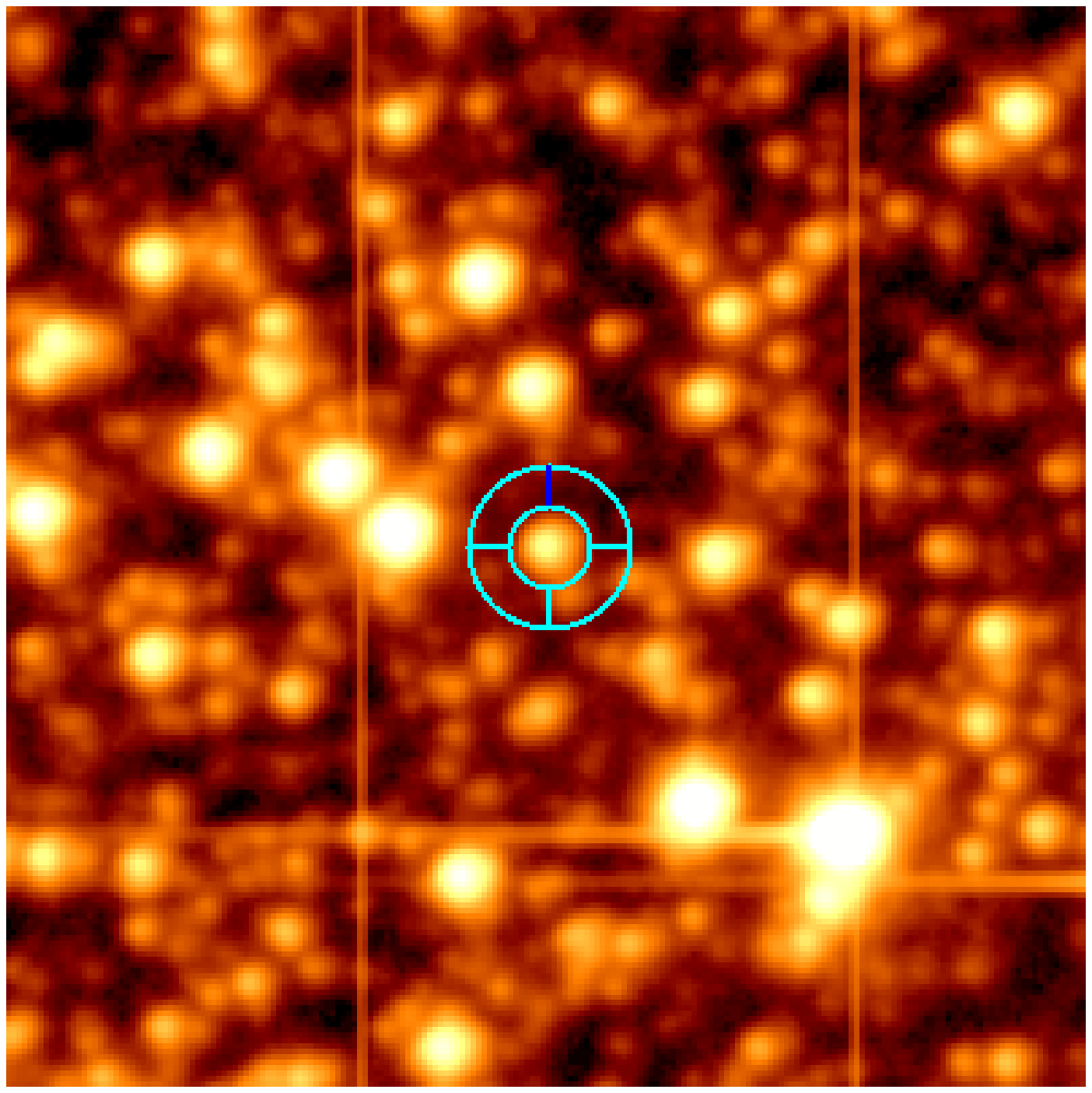}
\end{minipage}
&\begin{minipage}[m]{6.5cm}\includegraphics[width=6cm,bb=0 15 800 400,clip=true]{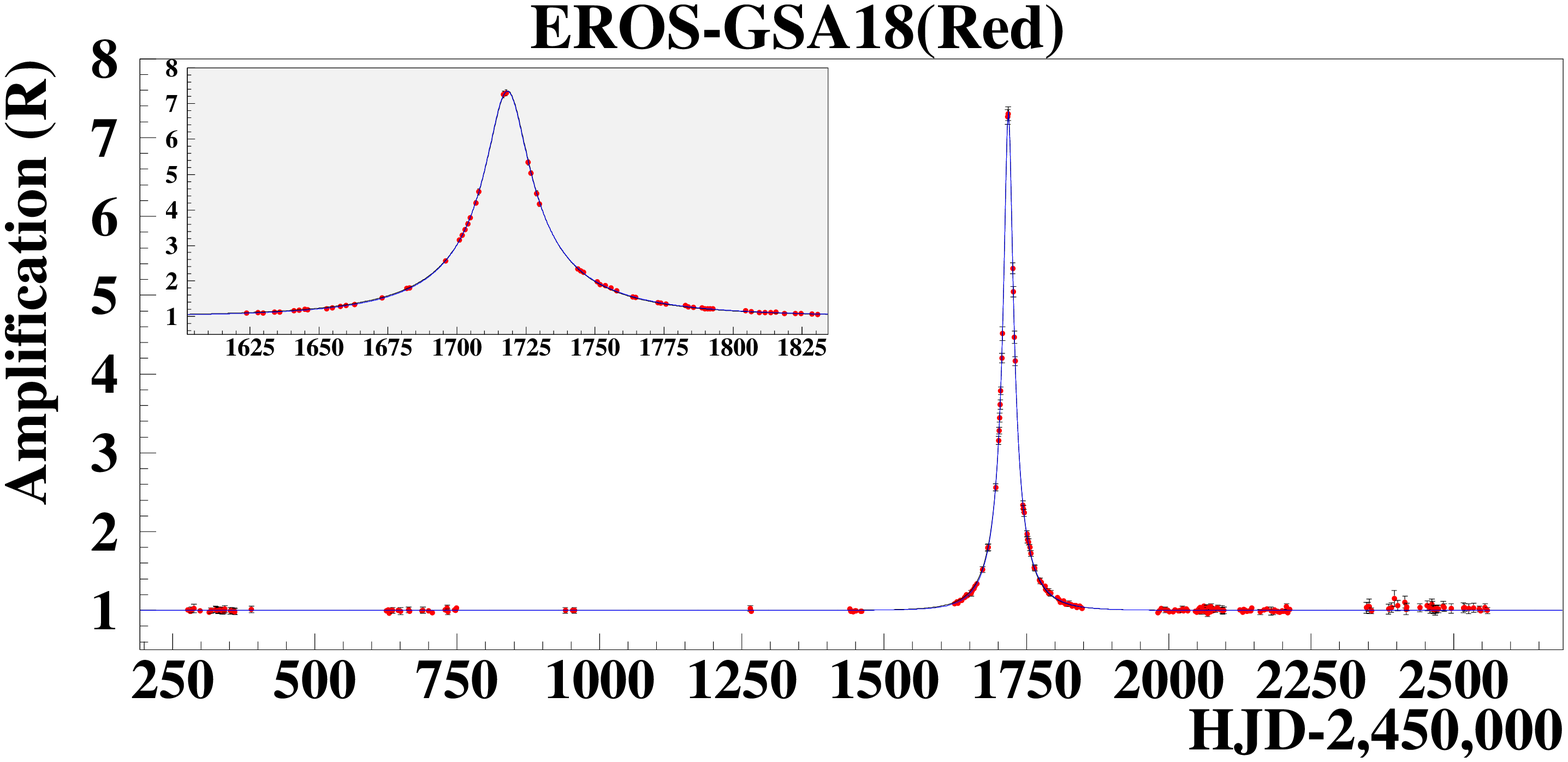}\end{minipage}
&\begin{minipage}[m]{6.5cm}\includegraphics[width=6cm,bb=0 15 800 400,clip=true]{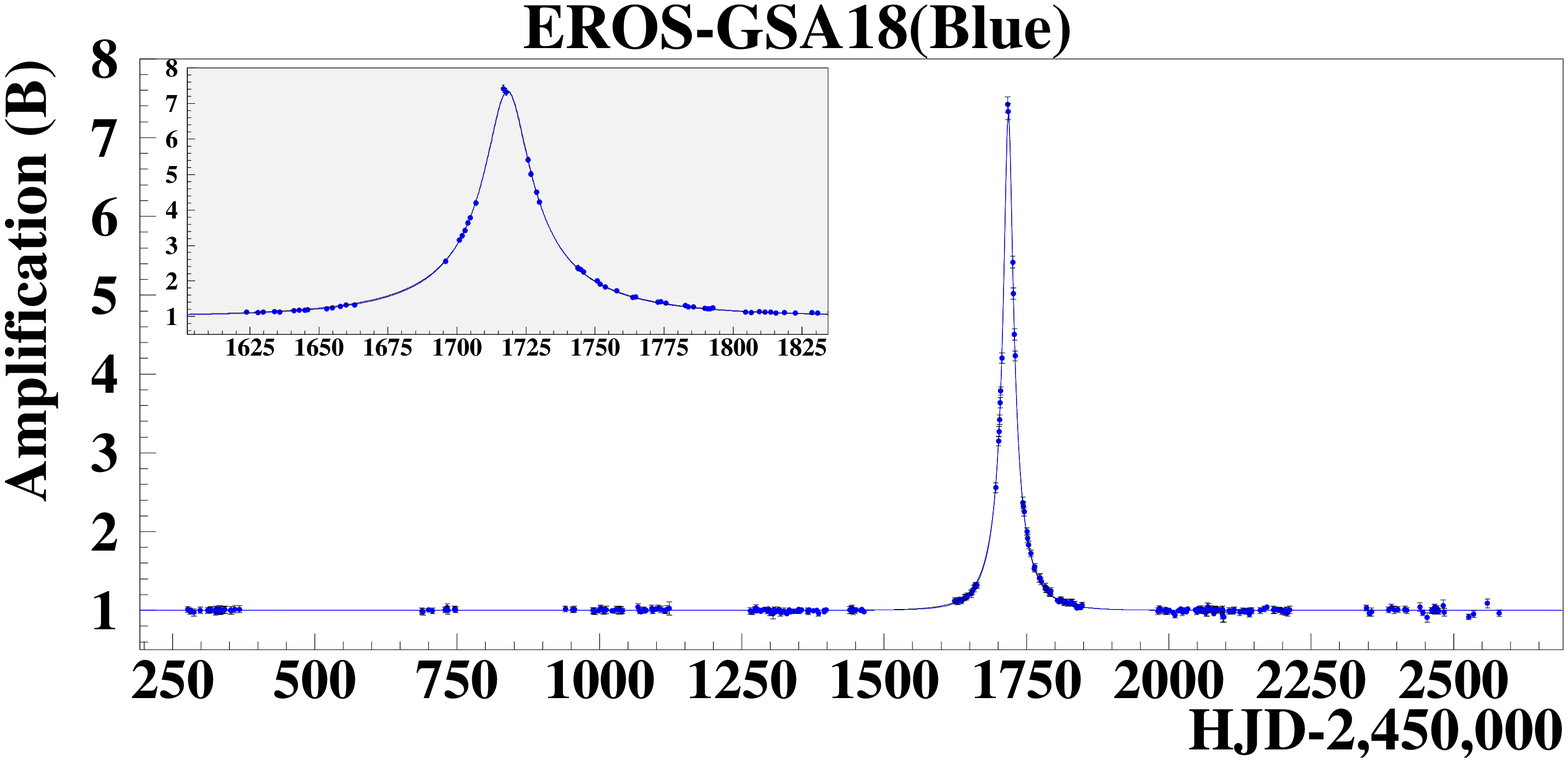}\end{minipage}
\\
\begin{minipage}[m]{3.cm}
\includegraphics[width=3.cm,bb=104 21 509 425]{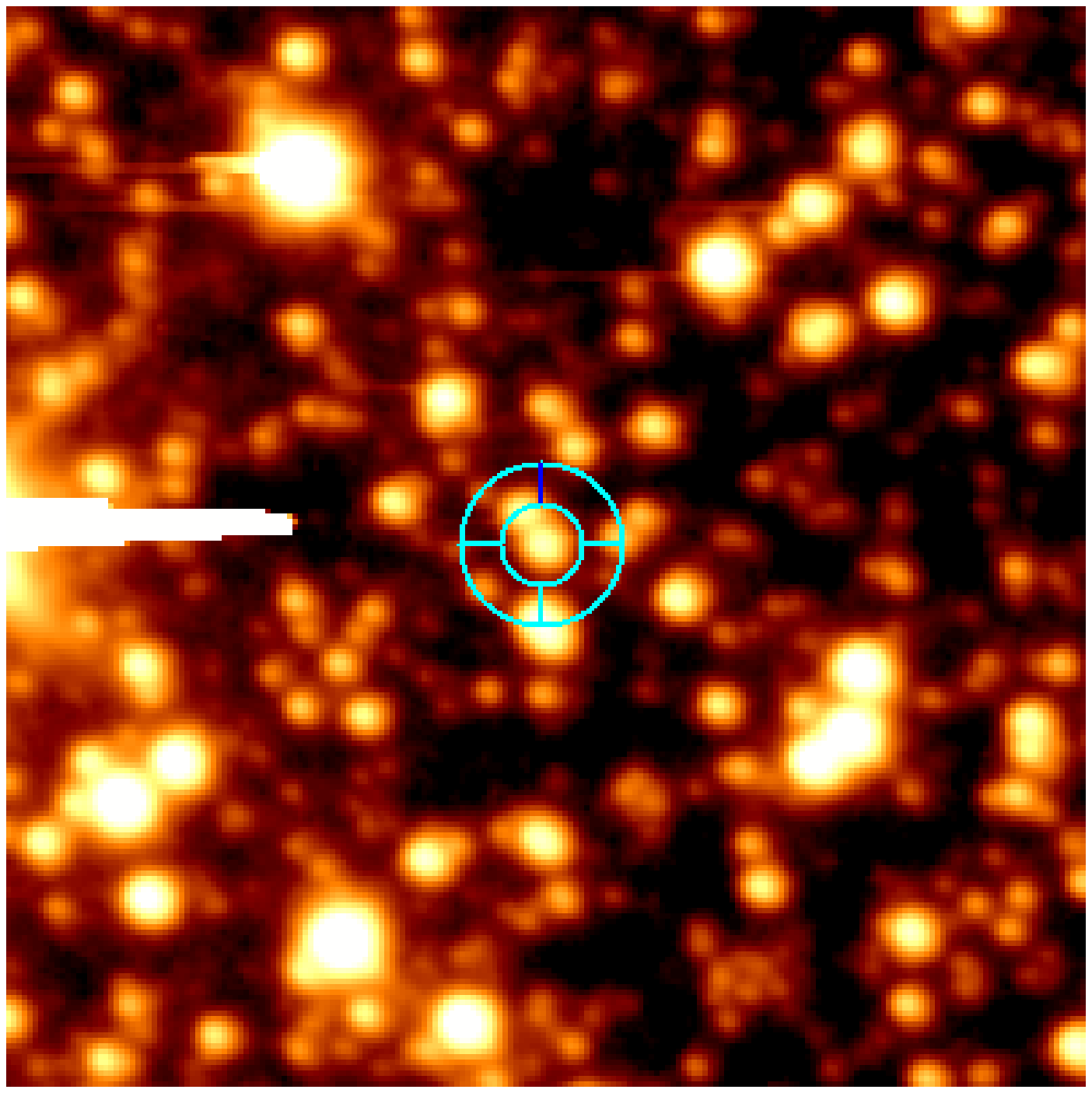}
\end{minipage}
&\begin{minipage}[m]{6.5cm}\includegraphics[width=6cm,bb=0 15 800 400,clip=true]{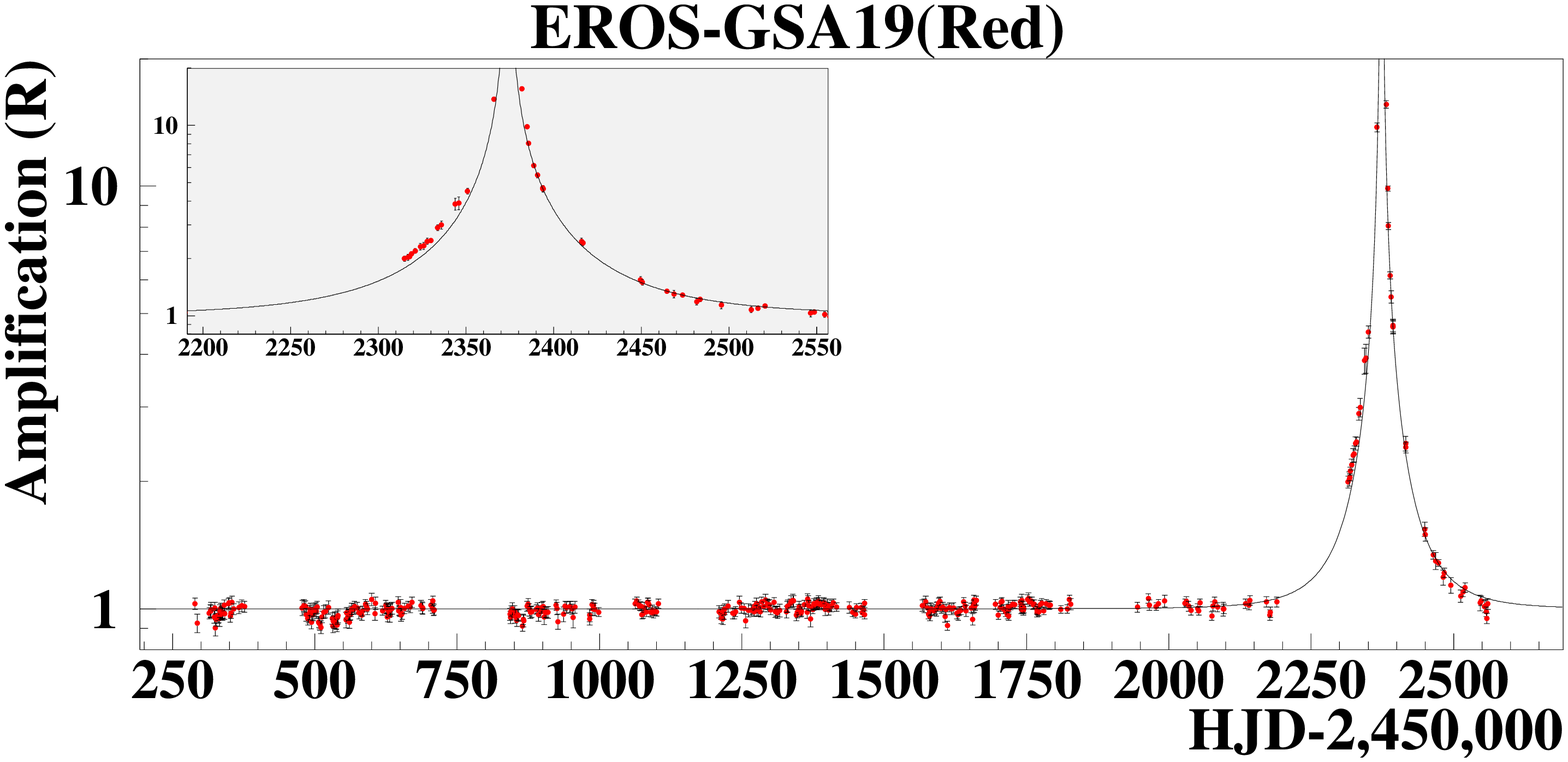}\end{minipage}
&\begin{minipage}[m]{6.5cm}\includegraphics[width=6cm,bb=0 15 800 400,clip=true]{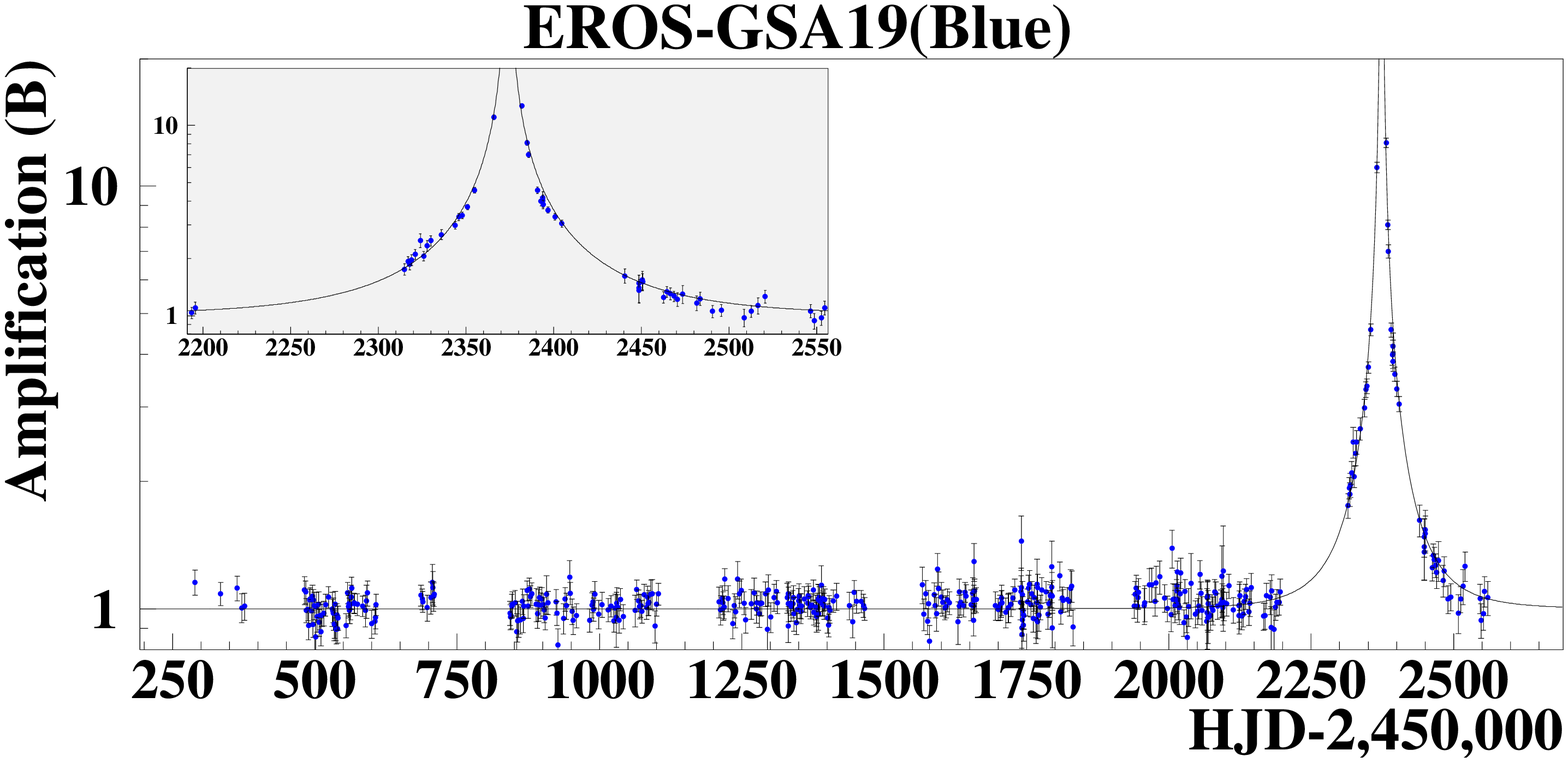}\end{minipage}
\\
\begin{minipage}[m]{3.cm}
\includegraphics[width=3.cm,bb=104 21 509 425]{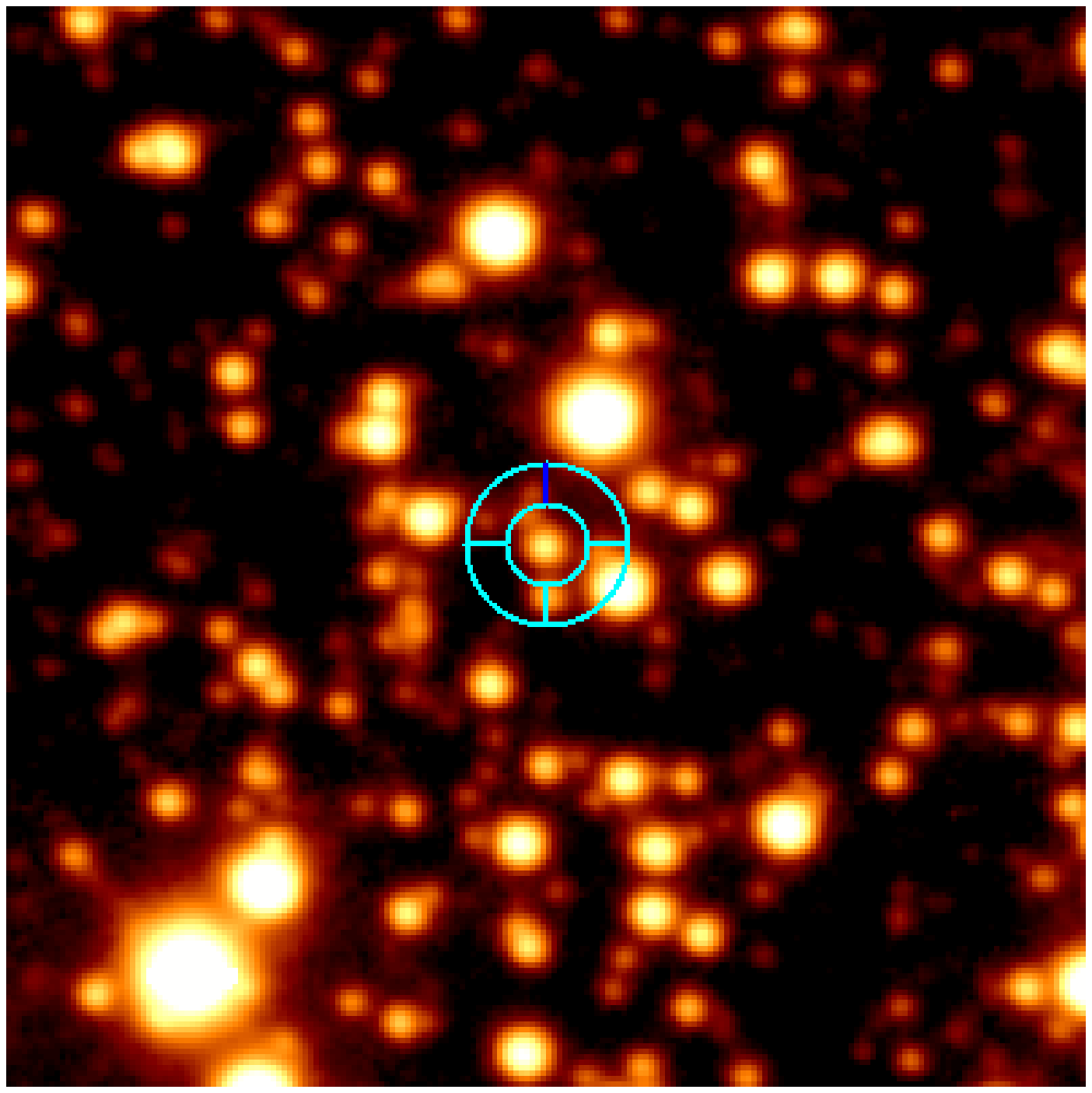}
\end{minipage}
&\begin{minipage}[m]{6.5cm}\includegraphics[width=6cm,bb=0 15 800 400,clip=true]{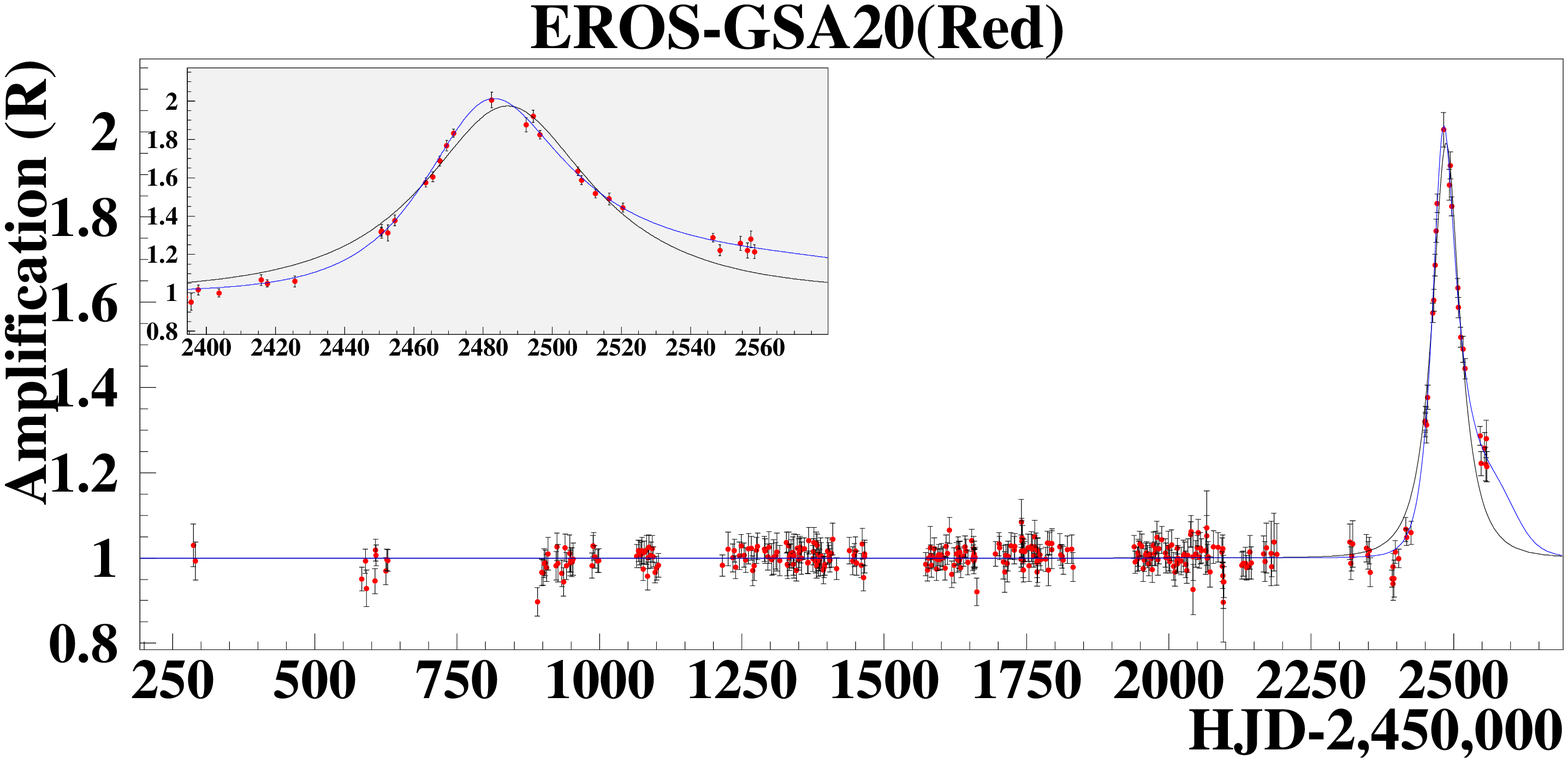}\end{minipage}
&\begin{minipage}[m]{6.5cm}\includegraphics[width=6cm,bb=0 15 800 400,clip=true]{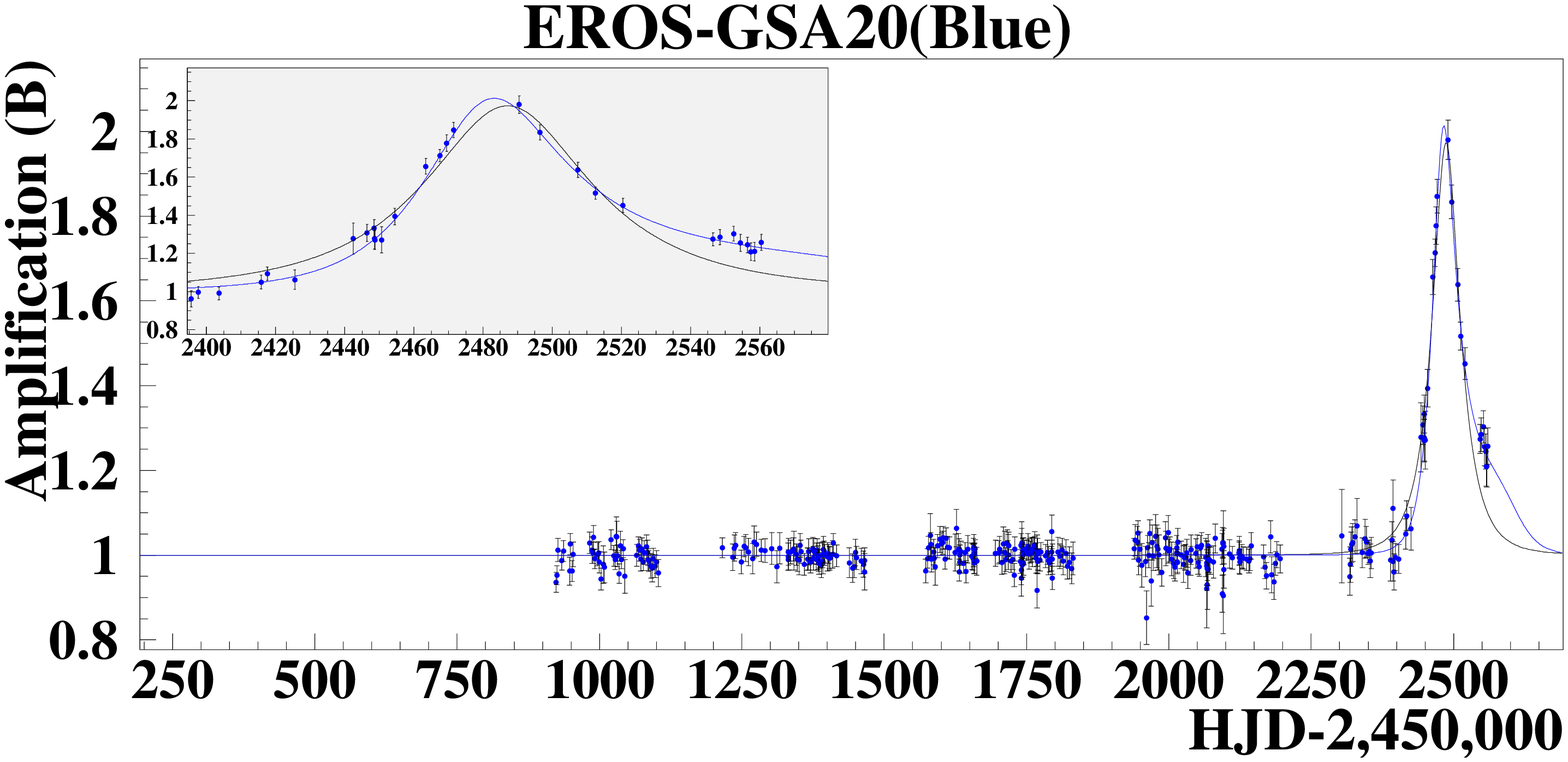}\end{minipage}
\end{tabular}
\caption[]{Light curves and finding charts (continued)}
\label{figclum2}
\end{center}
\end{figure*}

\begin{figure*}
\begin{center}
\begin{tabular}{ccc}
\begin{minipage}[m]{3.cm}
\includegraphics[width=3.cm,bb=104 21 509 425]{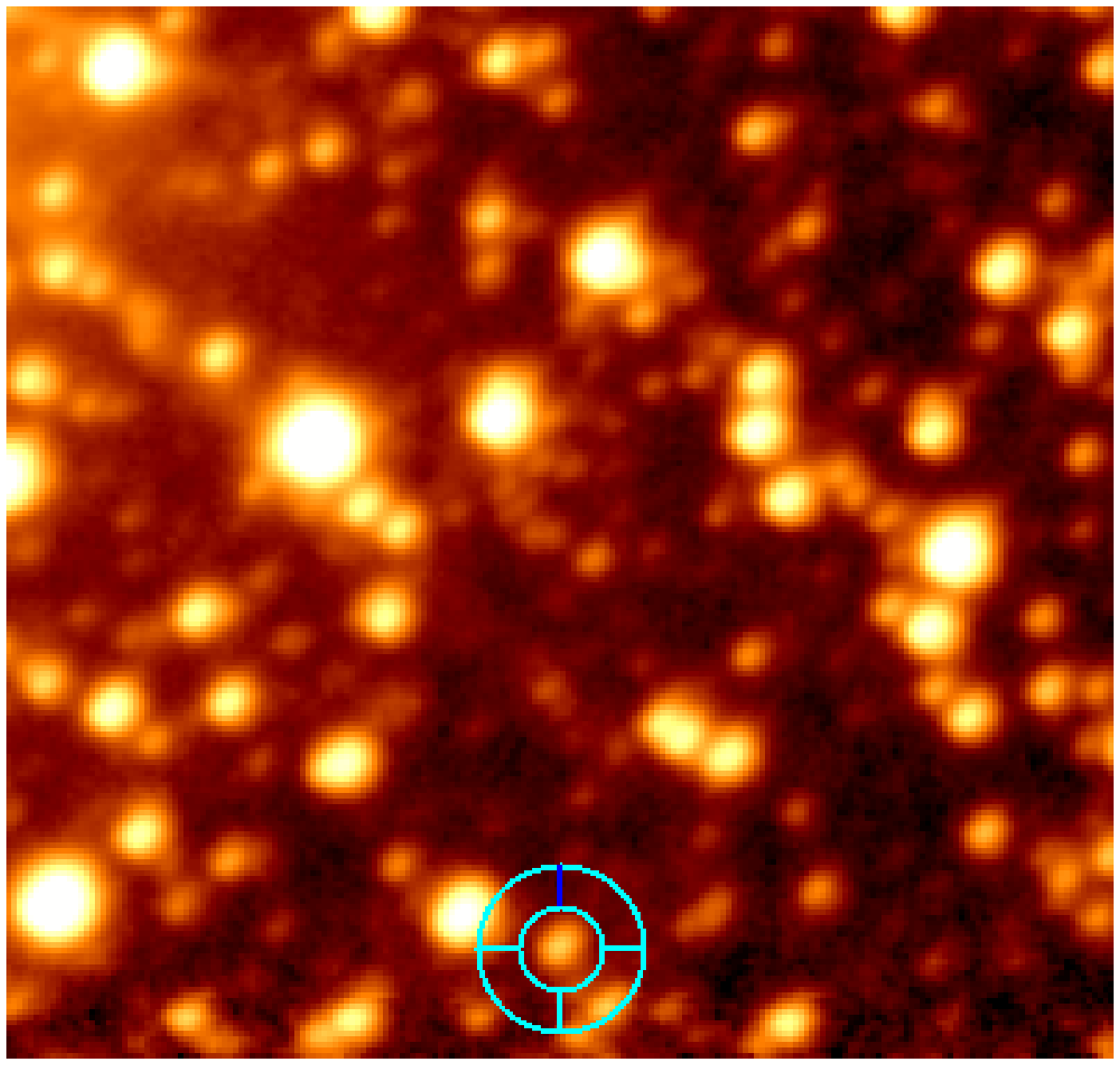}
\end{minipage}
&\begin{minipage}[m]{6.5cm}\includegraphics[width=6cm,bb=0 15 800 400,clip=true]{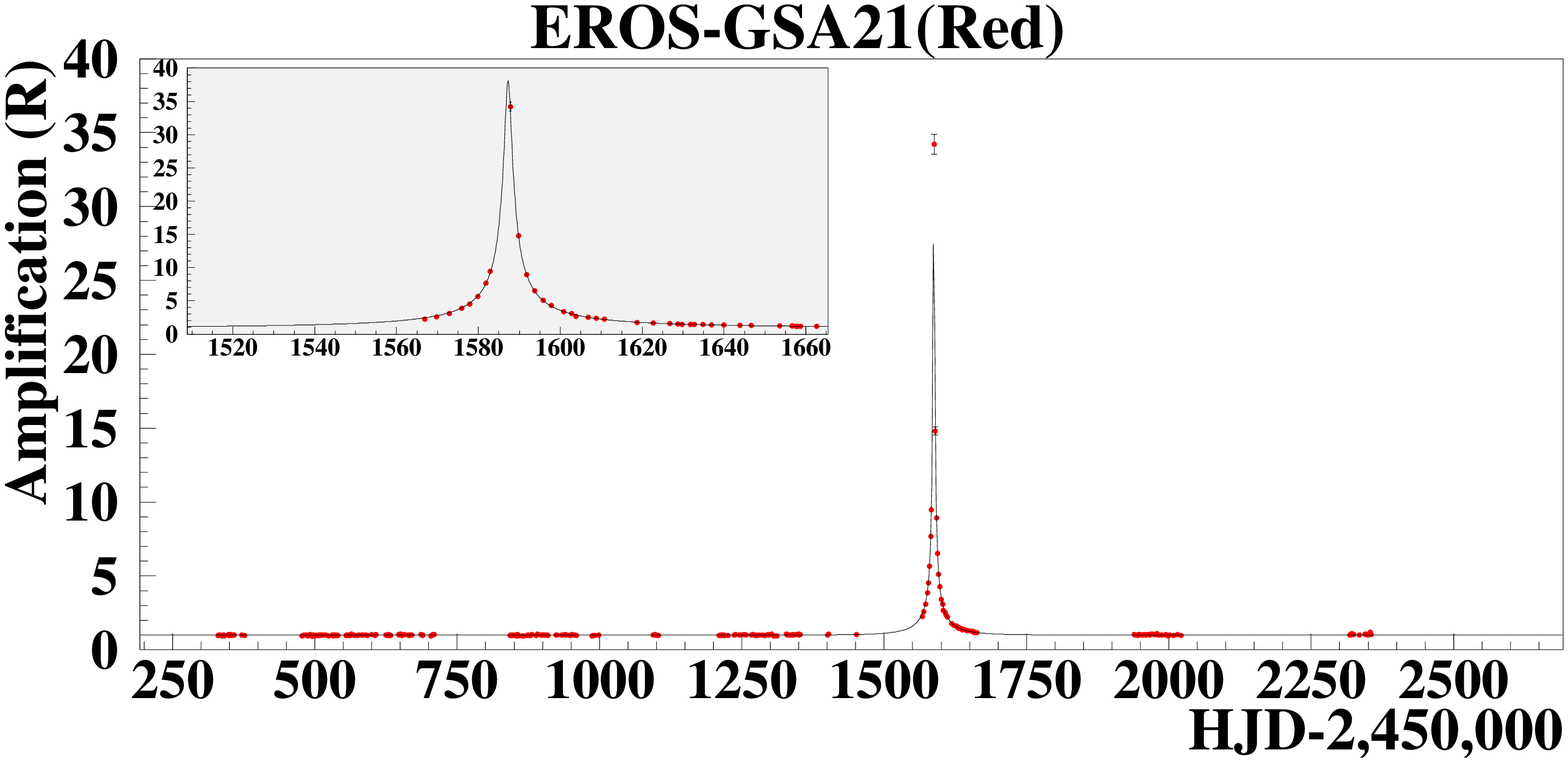}\end{minipage}
&\begin{minipage}[m]{6.5cm}\includegraphics[width=6cm,bb=0 15 800 400,clip=true]{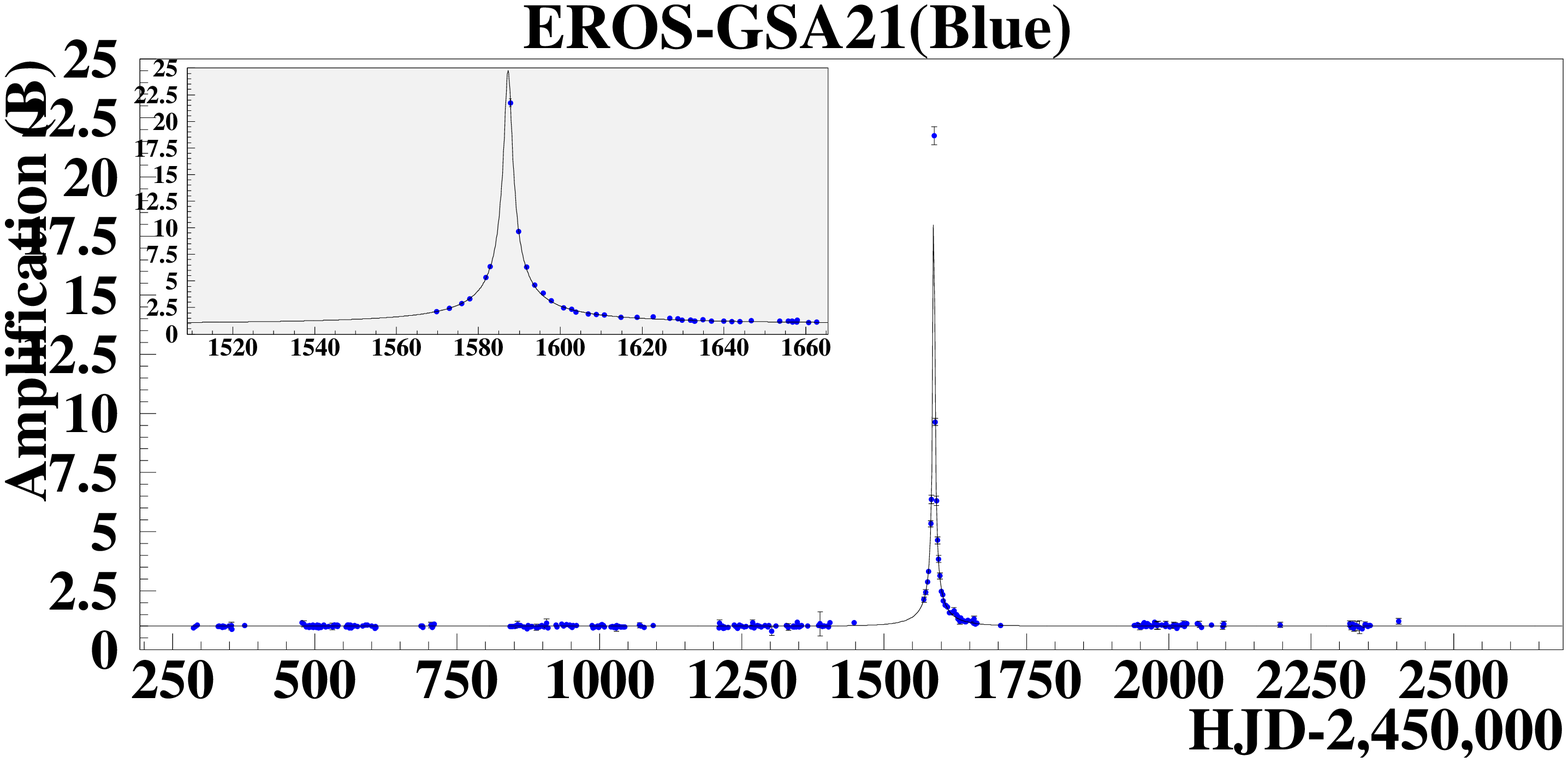}\end{minipage}
\\
\begin{minipage}[m]{3.cm}
\includegraphics[width=3.cm,bb=104 21 509 425]{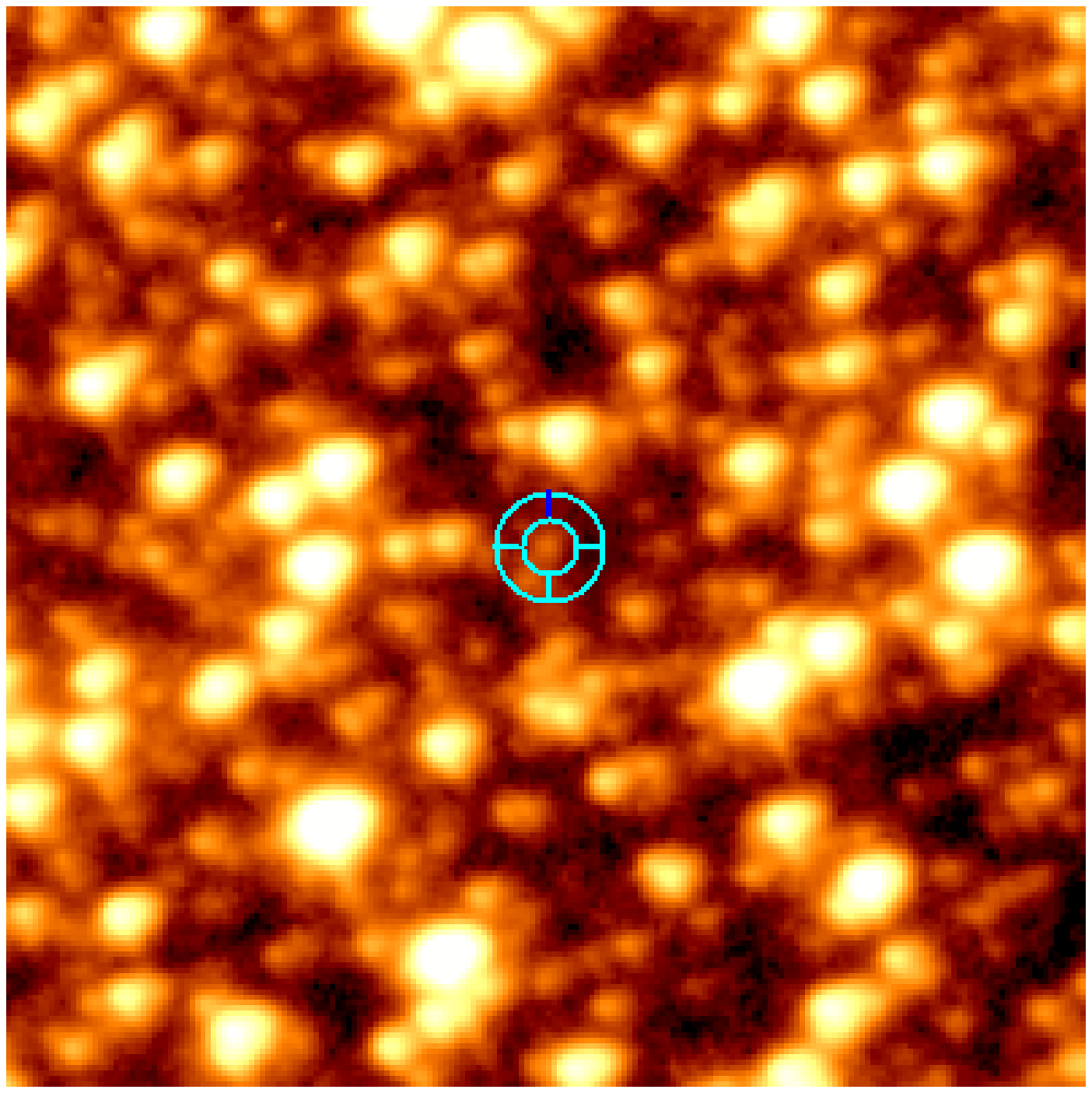}
\end{minipage}
&\begin{minipage}[m]{6.5cm}\includegraphics[width=6cm,bb=0 15 800 400,clip=true]{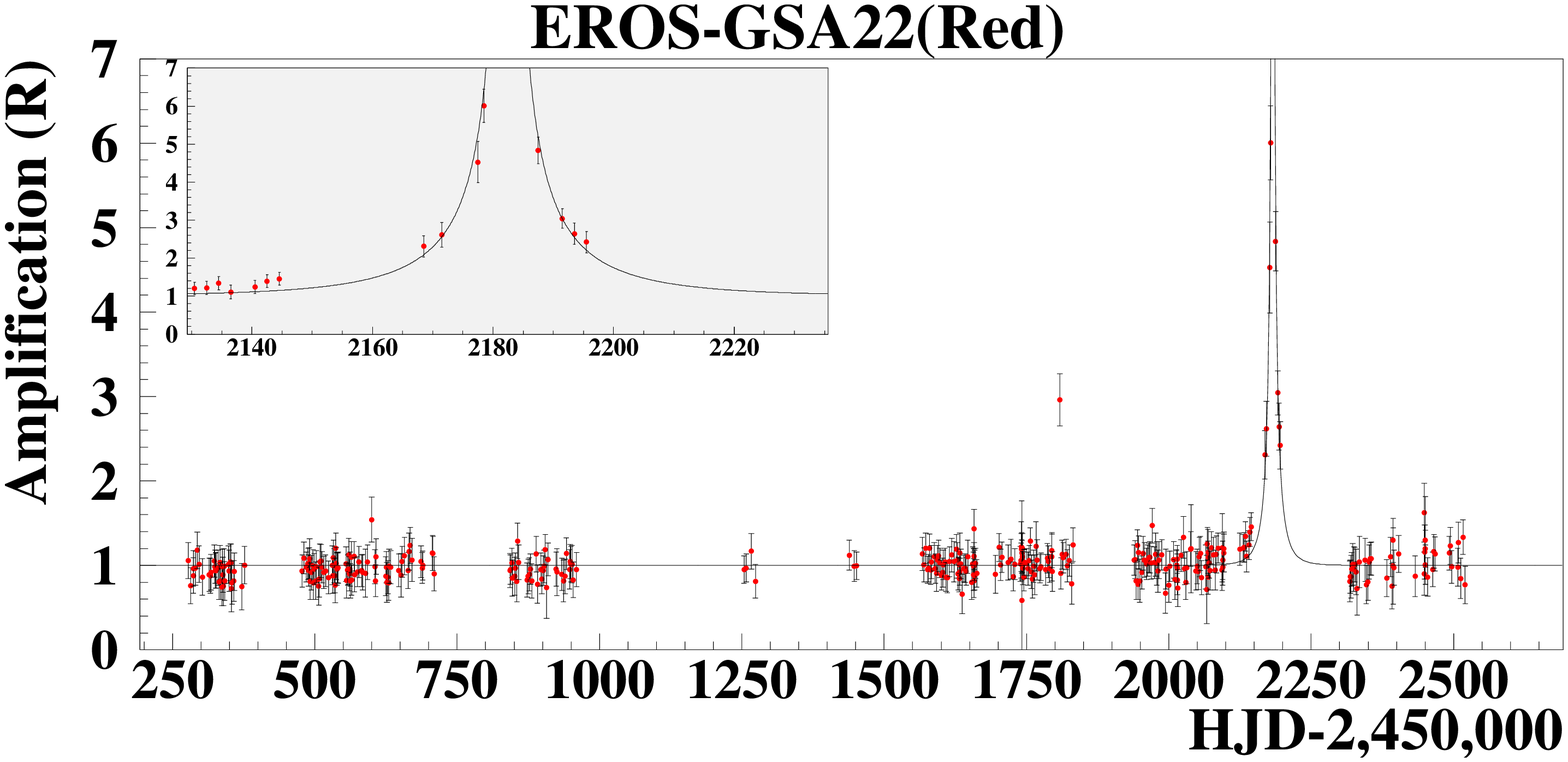}\end{minipage}
&\begin{minipage}[m]{6.5cm}\includegraphics[width=6cm,bb=0 15 800 400,clip=true]{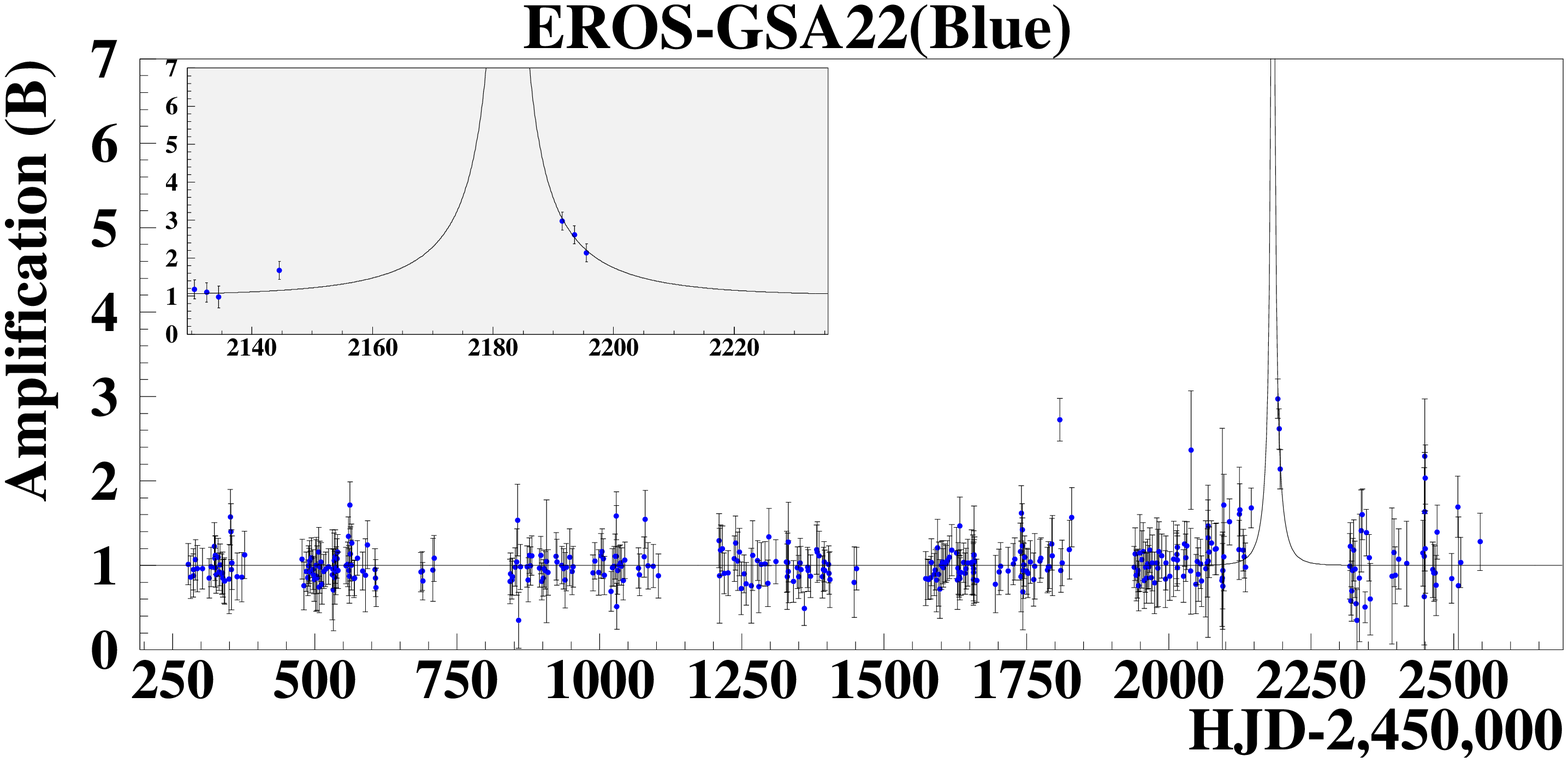}\end{minipage}
\\
\begin{minipage}[m]{3.cm}
\includegraphics[width=3.cm,bb=104 21 509 425]{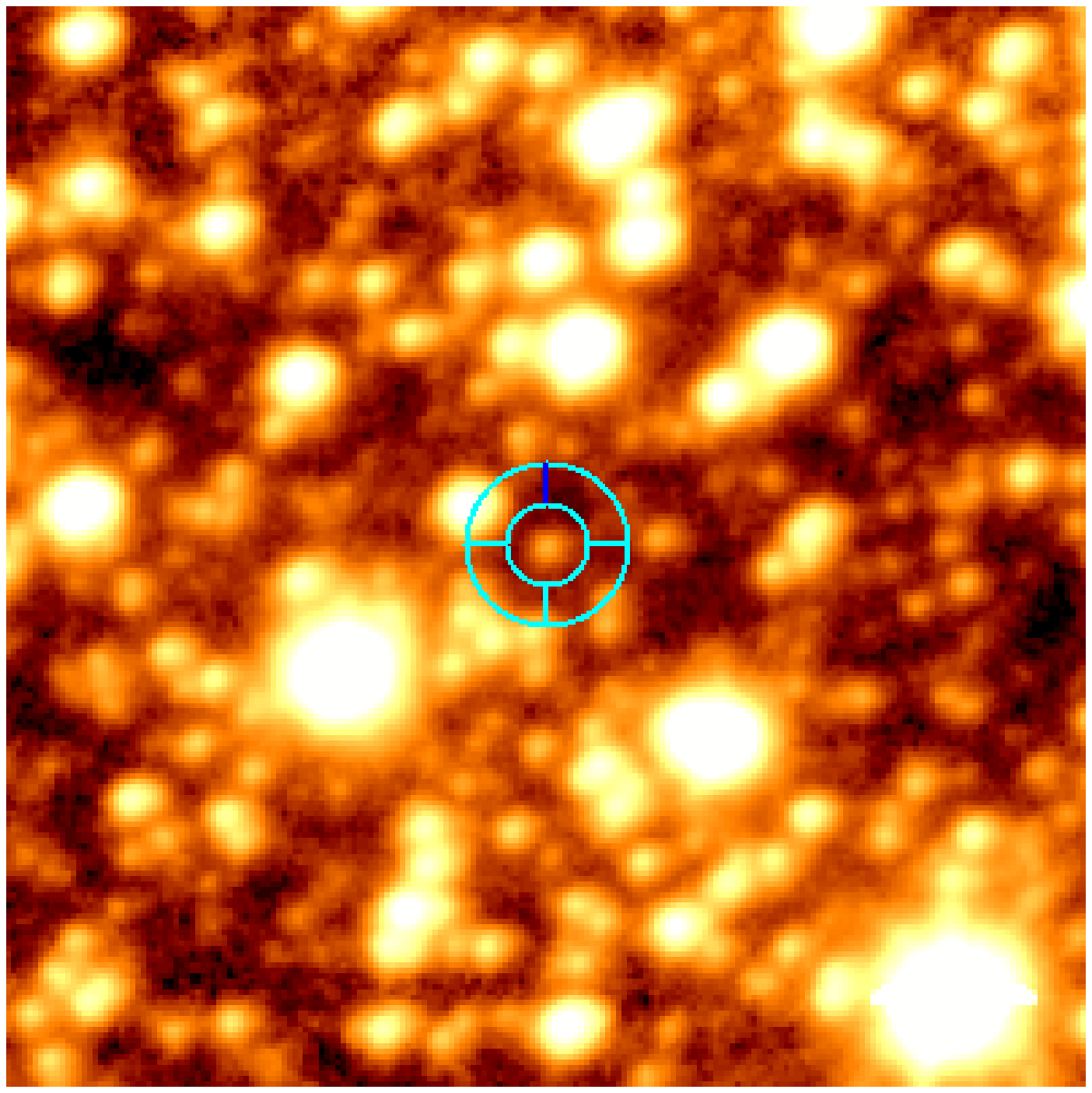}
\end{minipage}
&\begin{minipage}[m]{6.5cm}\includegraphics[width=6cm,bb=0 15 800 400,clip=true]{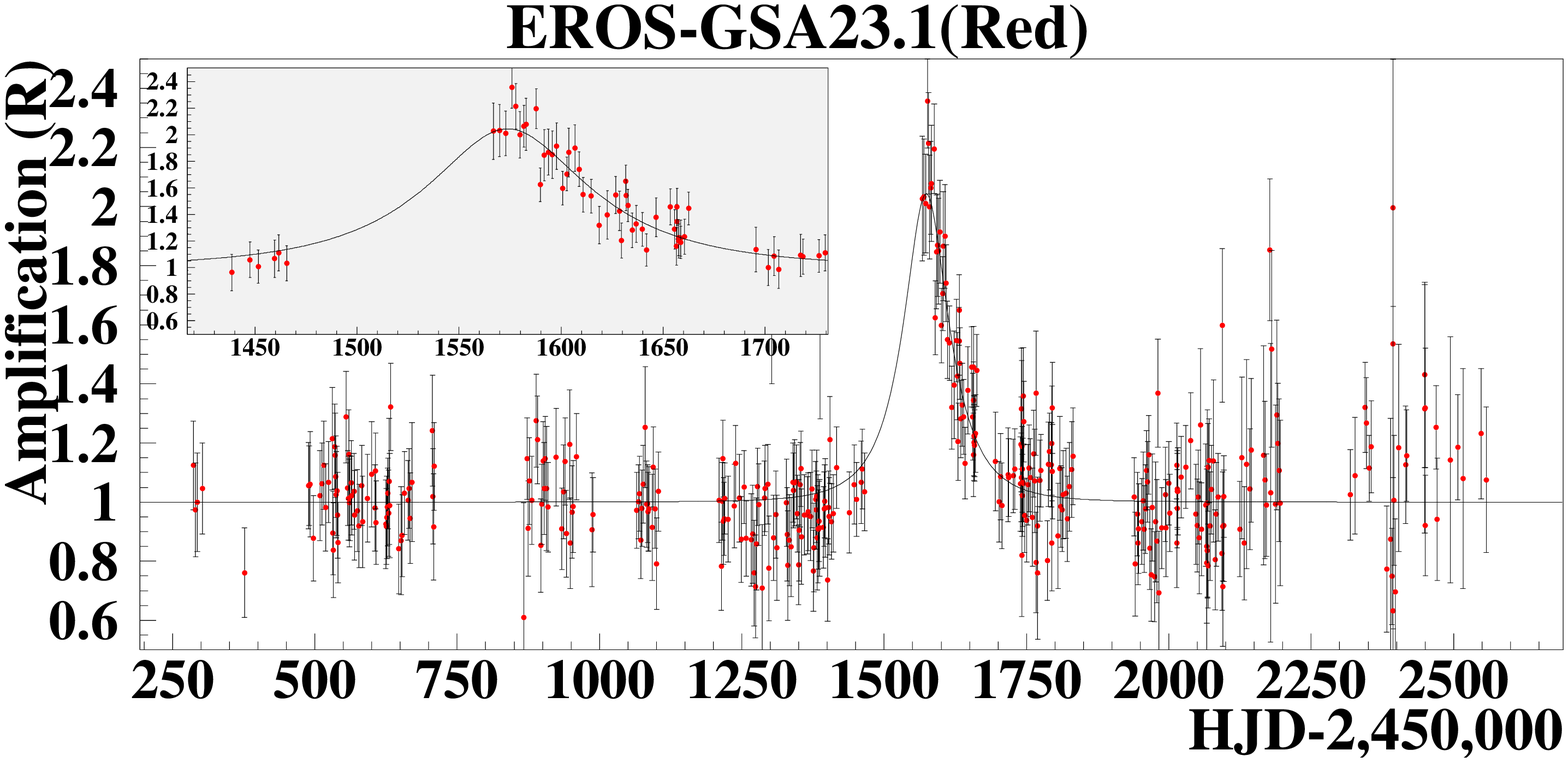}\end{minipage}
&\begin{minipage}[m]{6.5cm}\includegraphics[width=6cm,bb=0 15 800 400,clip=true]{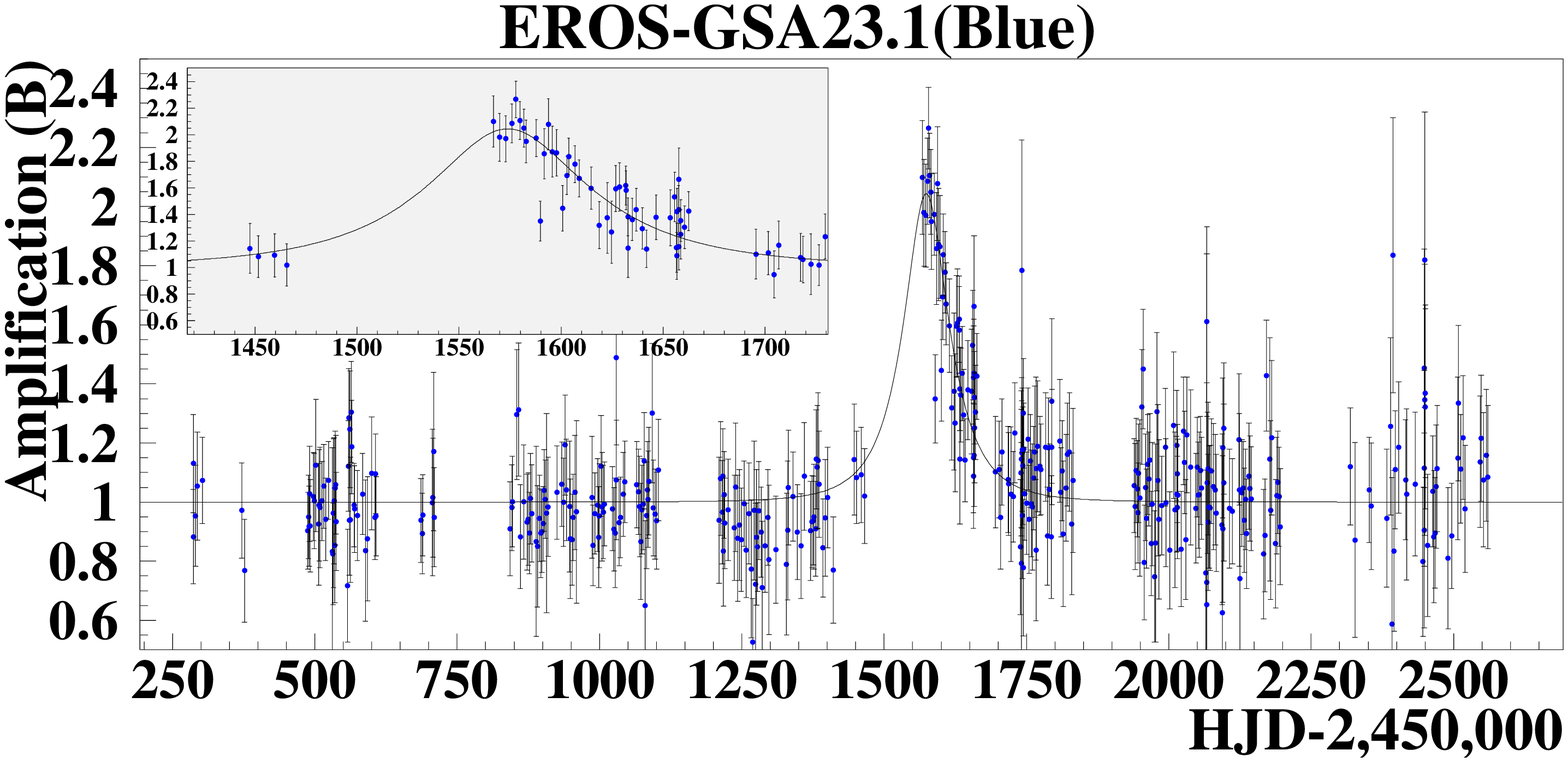}\end{minipage}
\\
\begin{minipage}[m]{3.cm}
\includegraphics[width=3.cm,bb=104 21 509 425]{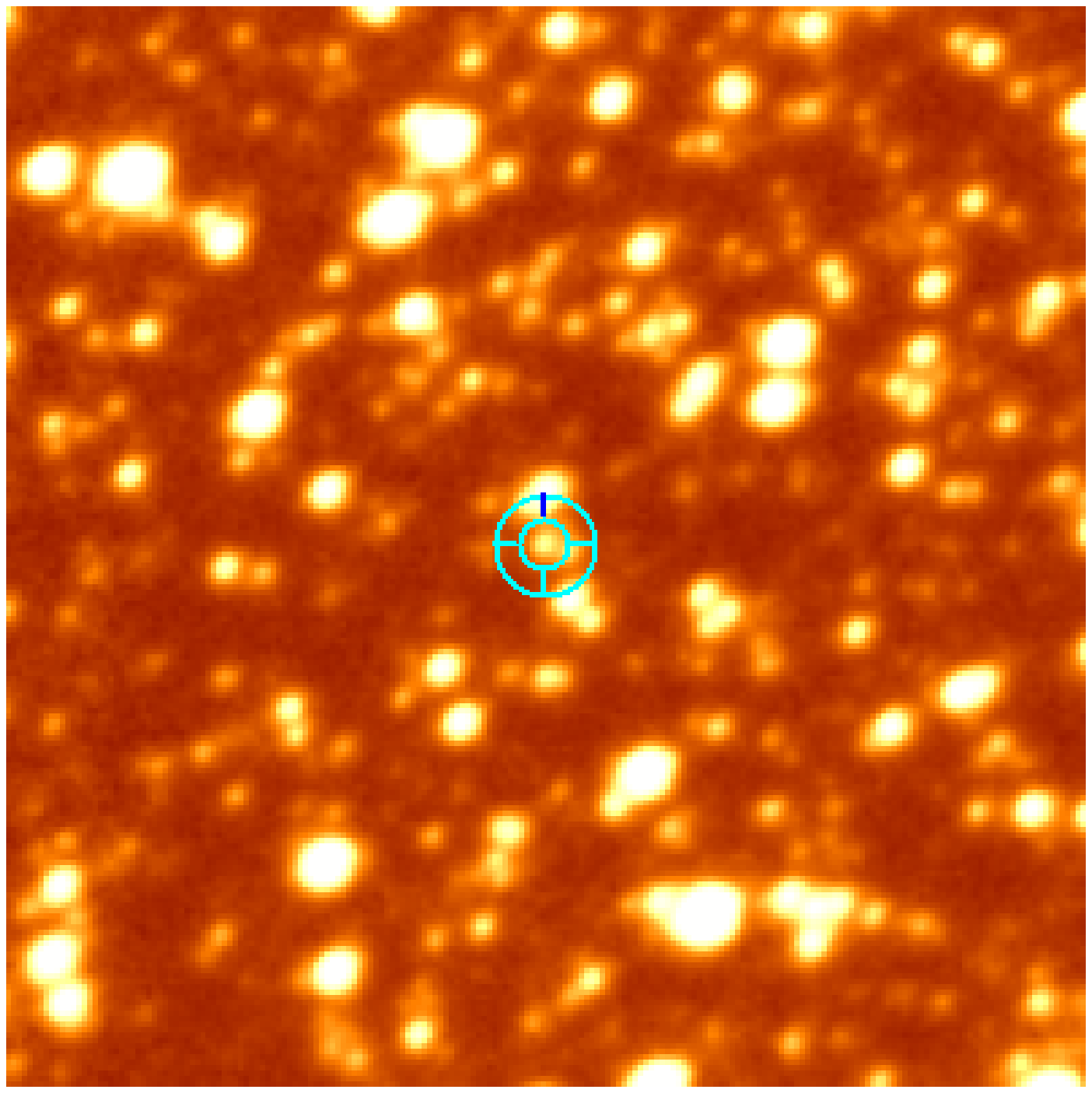}
\end{minipage}
&\begin{minipage}[m]{6.5cm}\includegraphics[width=6cm,bb=0 15 800 400,clip=true]{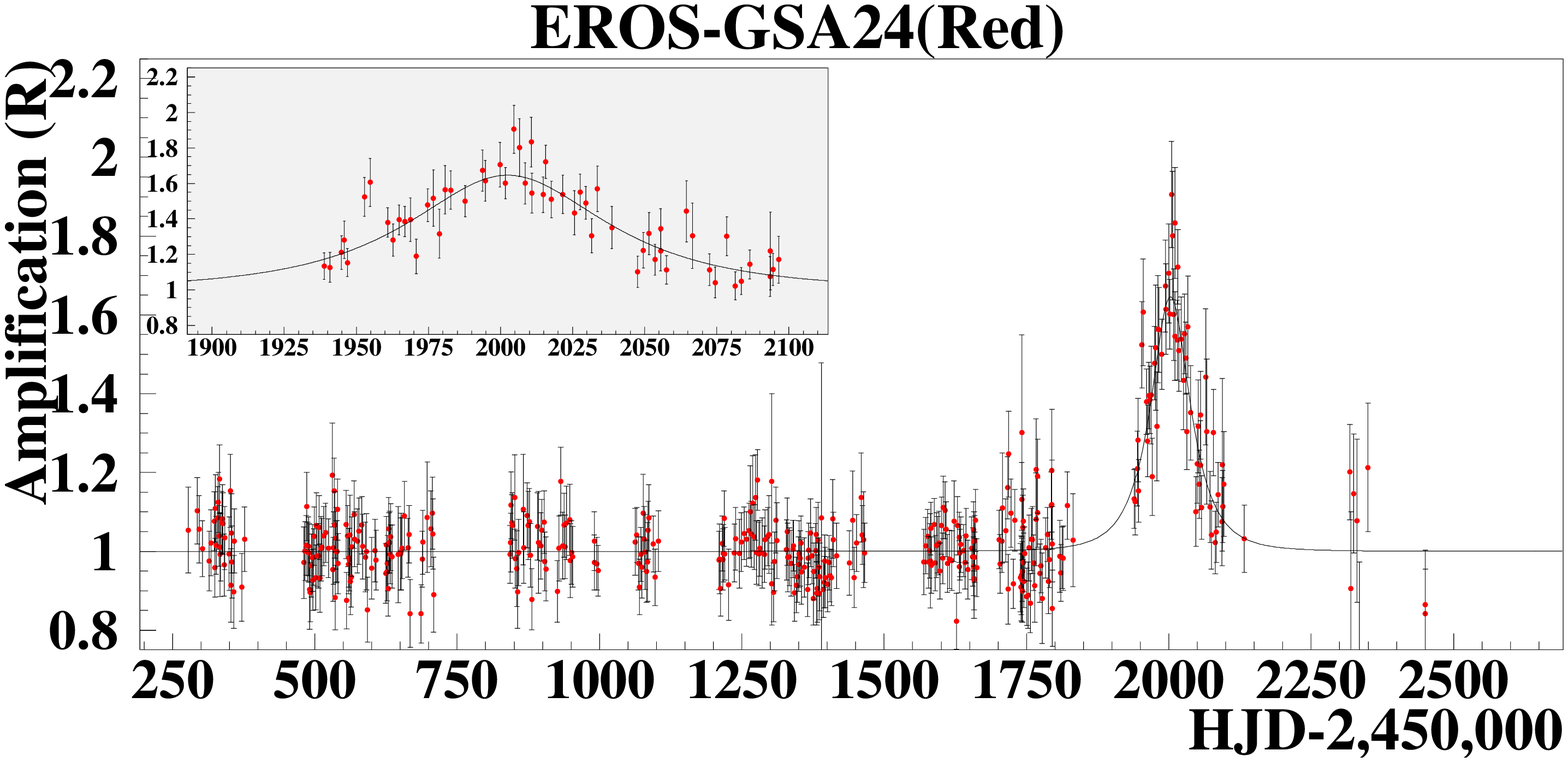}\end{minipage}
&\begin{minipage}[m]{6.5cm}\includegraphics[width=6cm,bb=0 15 800 400,clip=true]{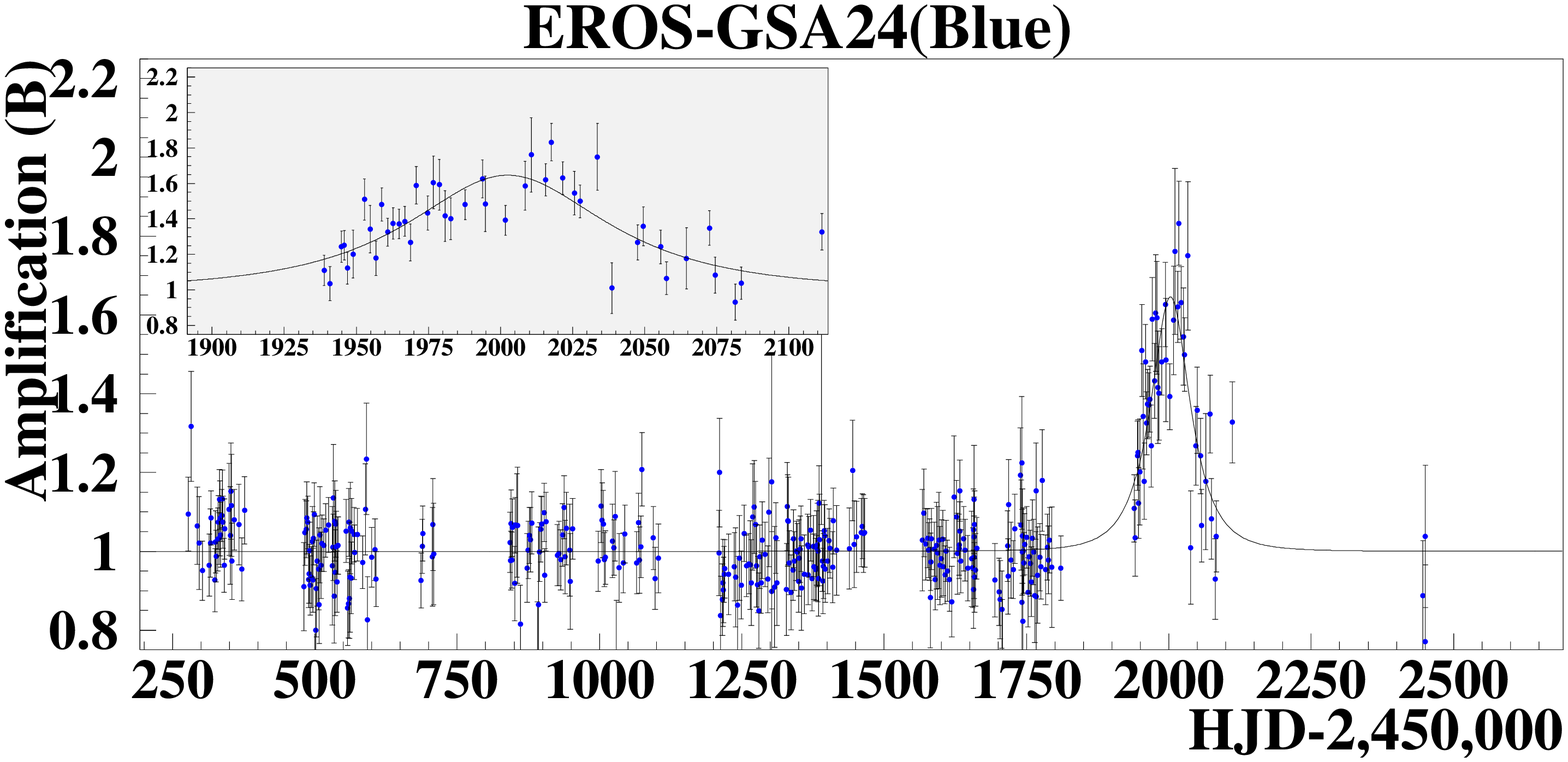}\end{minipage}
\\
\begin{minipage}[m]{3.cm}
\includegraphics[width=3.cm,bb=104 21 509 425]{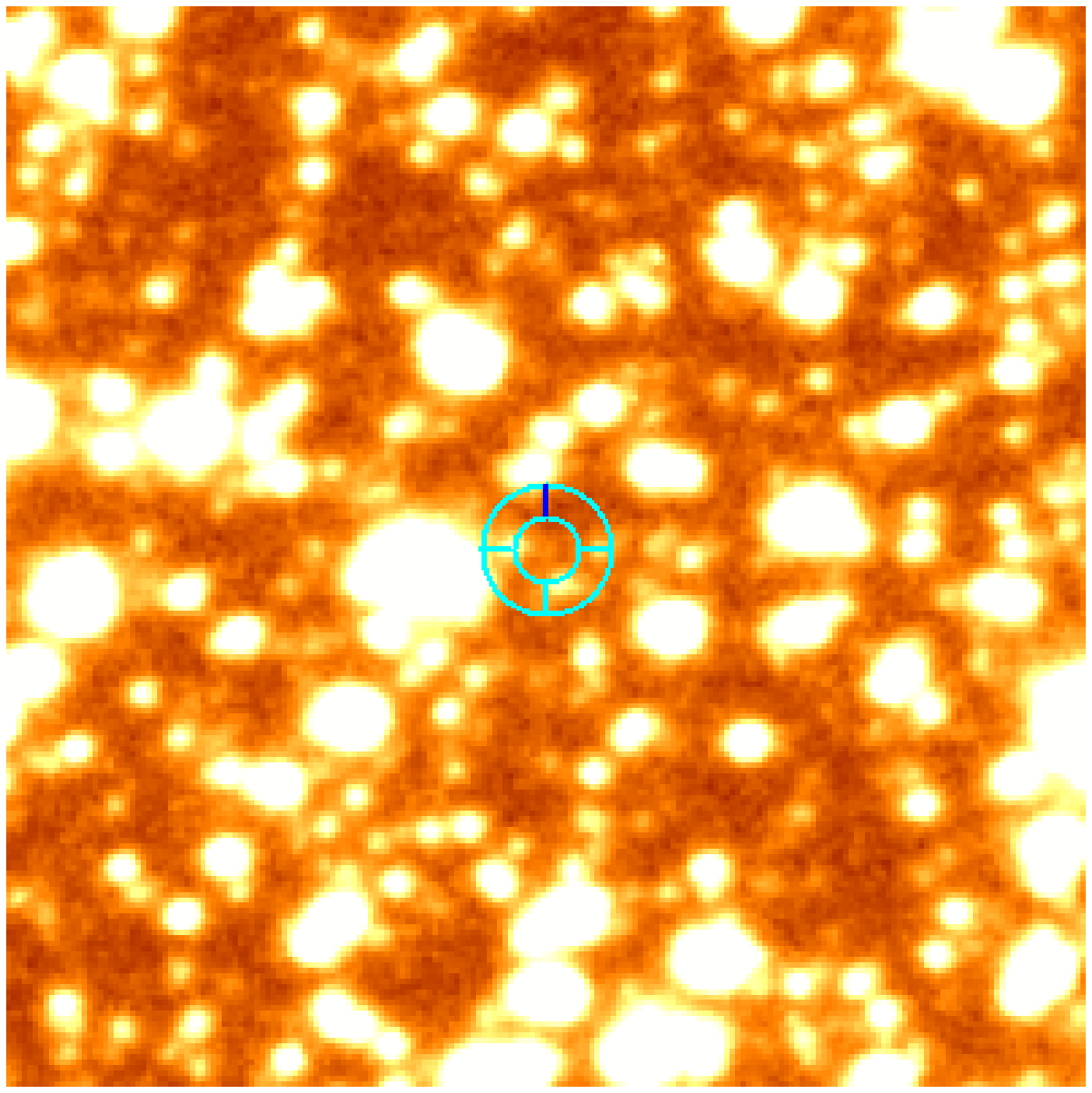}
\end{minipage}
&\begin{minipage}[m]{6.5cm}\includegraphics[width=6cm,bb=0 15 800 400,clip=true]{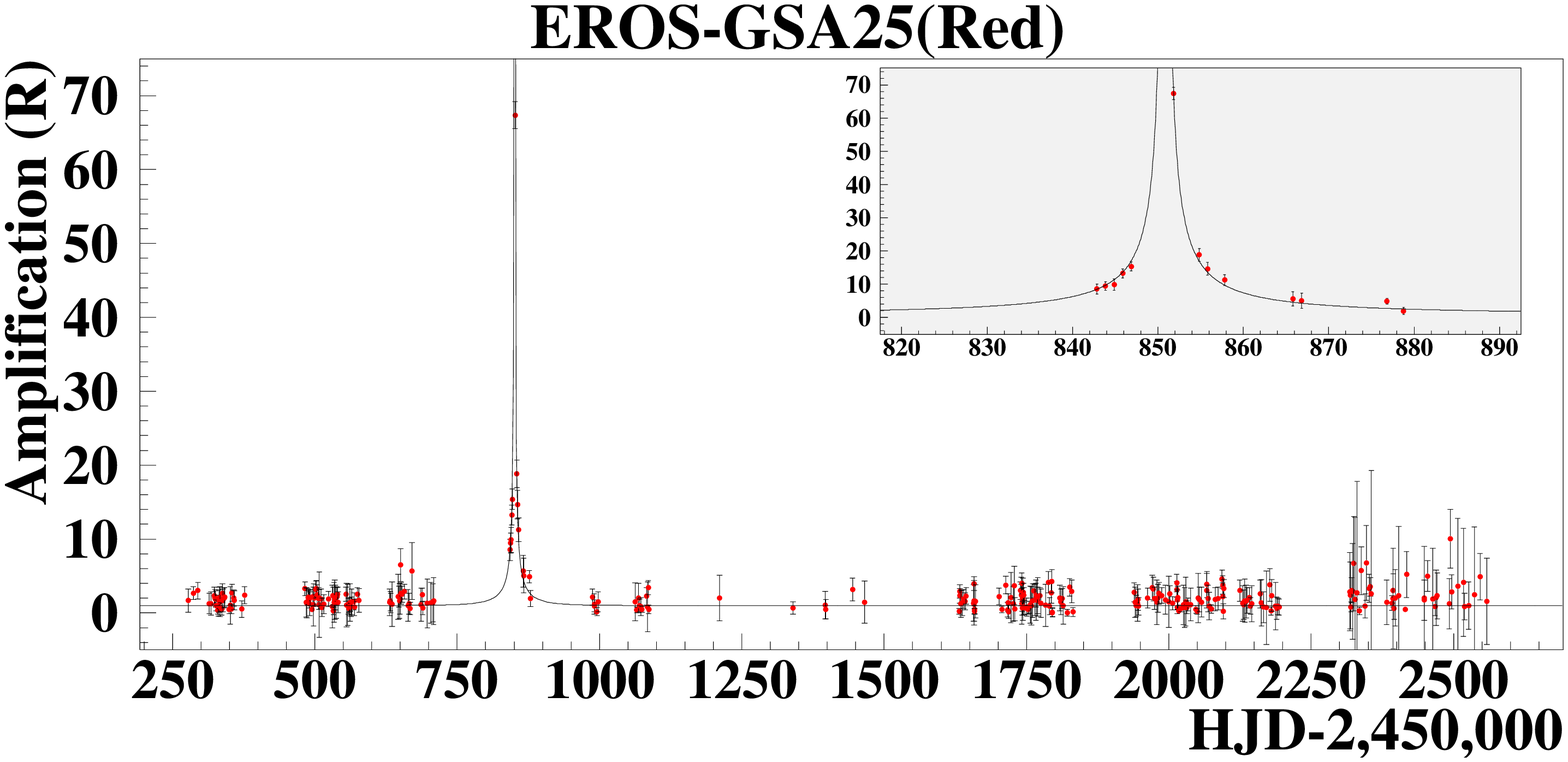}\end{minipage}
&\begin{minipage}[m]{6.5cm}\includegraphics[width=6cm,bb=0 15 800 400,clip=true]{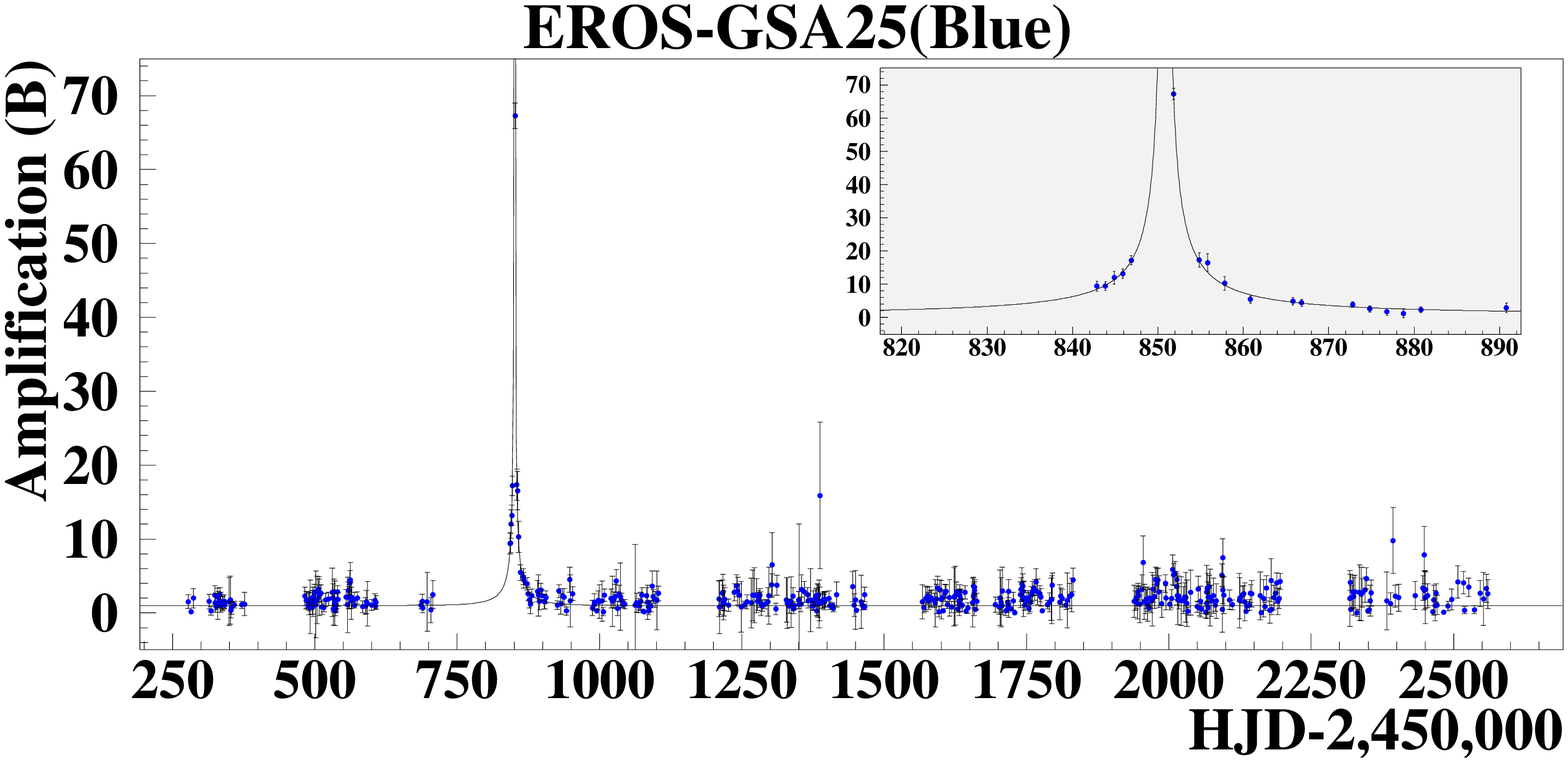}\end{minipage}
\\
\begin{minipage}[m]{3.cm}
\includegraphics[width=3.cm,bb=104 21 509 425]{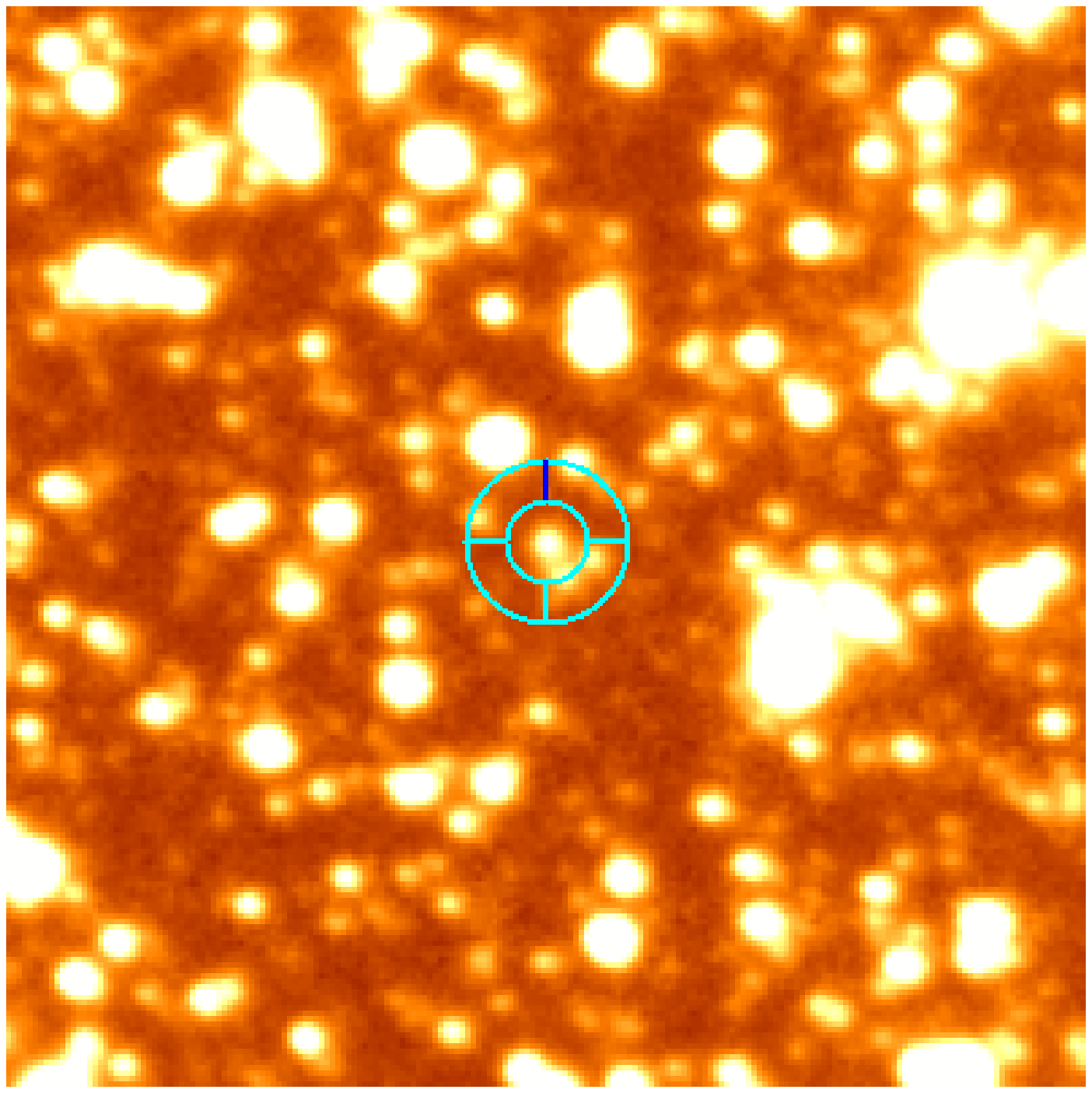}
\end{minipage}
&\begin{minipage}[m]{6.5cm}\includegraphics[width=6cm,bb=0 15 800 400,clip=true]{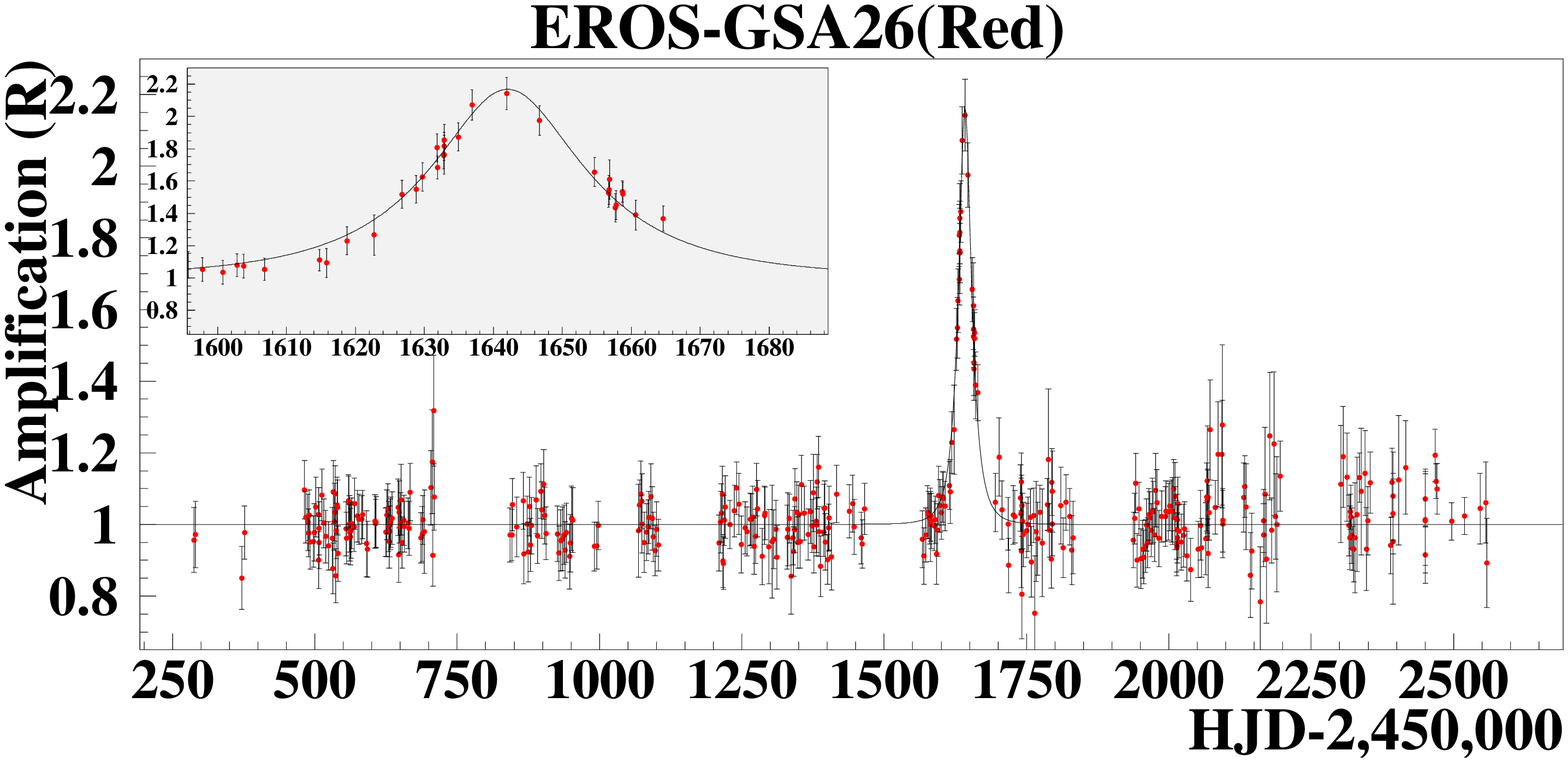}\end{minipage}
&\begin{minipage}[m]{6.5cm}\includegraphics[width=6cm,bb=0 15 800 400,clip=true]{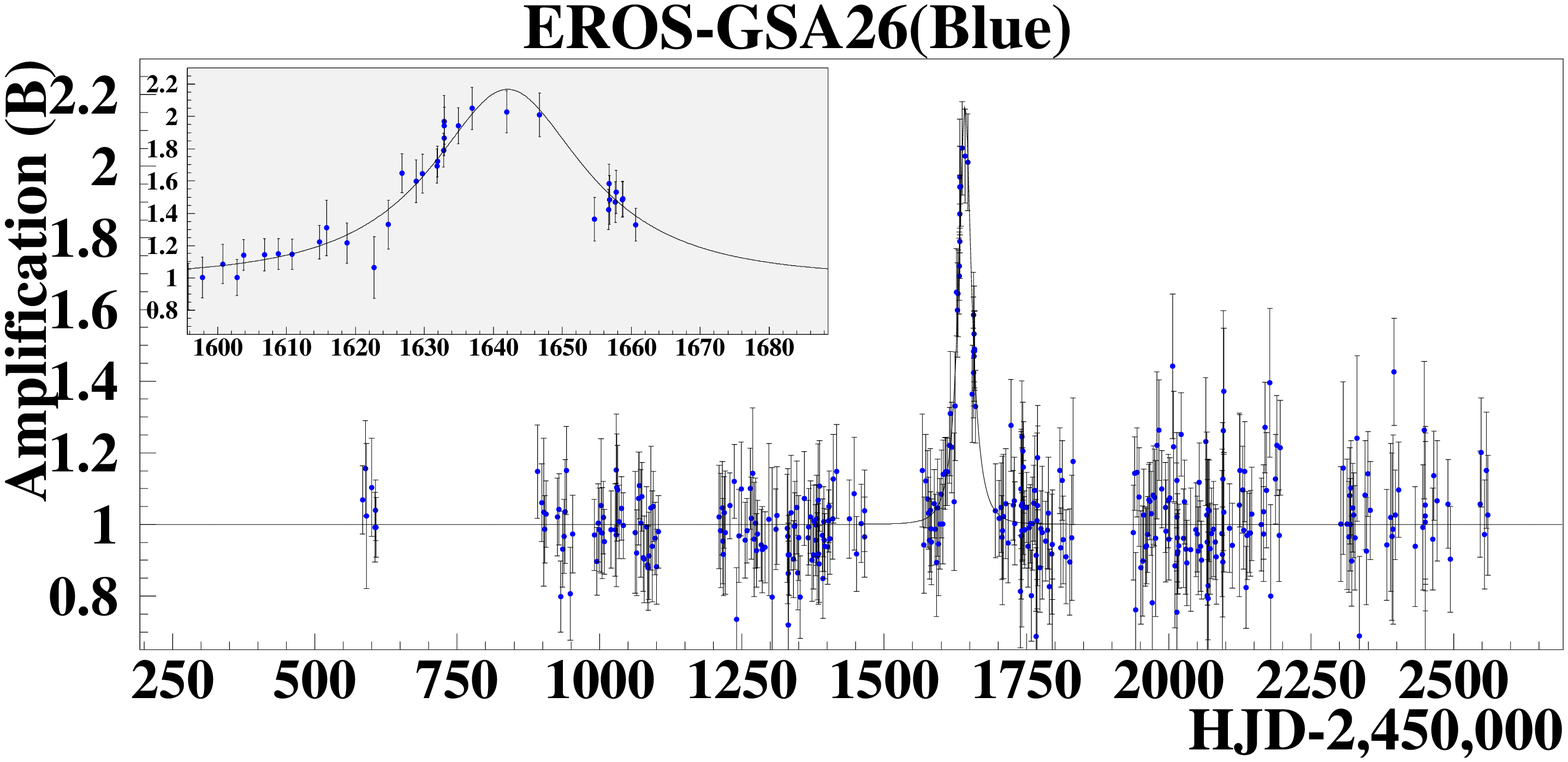}\end{minipage}
\\
\begin{minipage}[m]{3.cm}
\includegraphics[width=3.cm,bb=104 21 509 425]{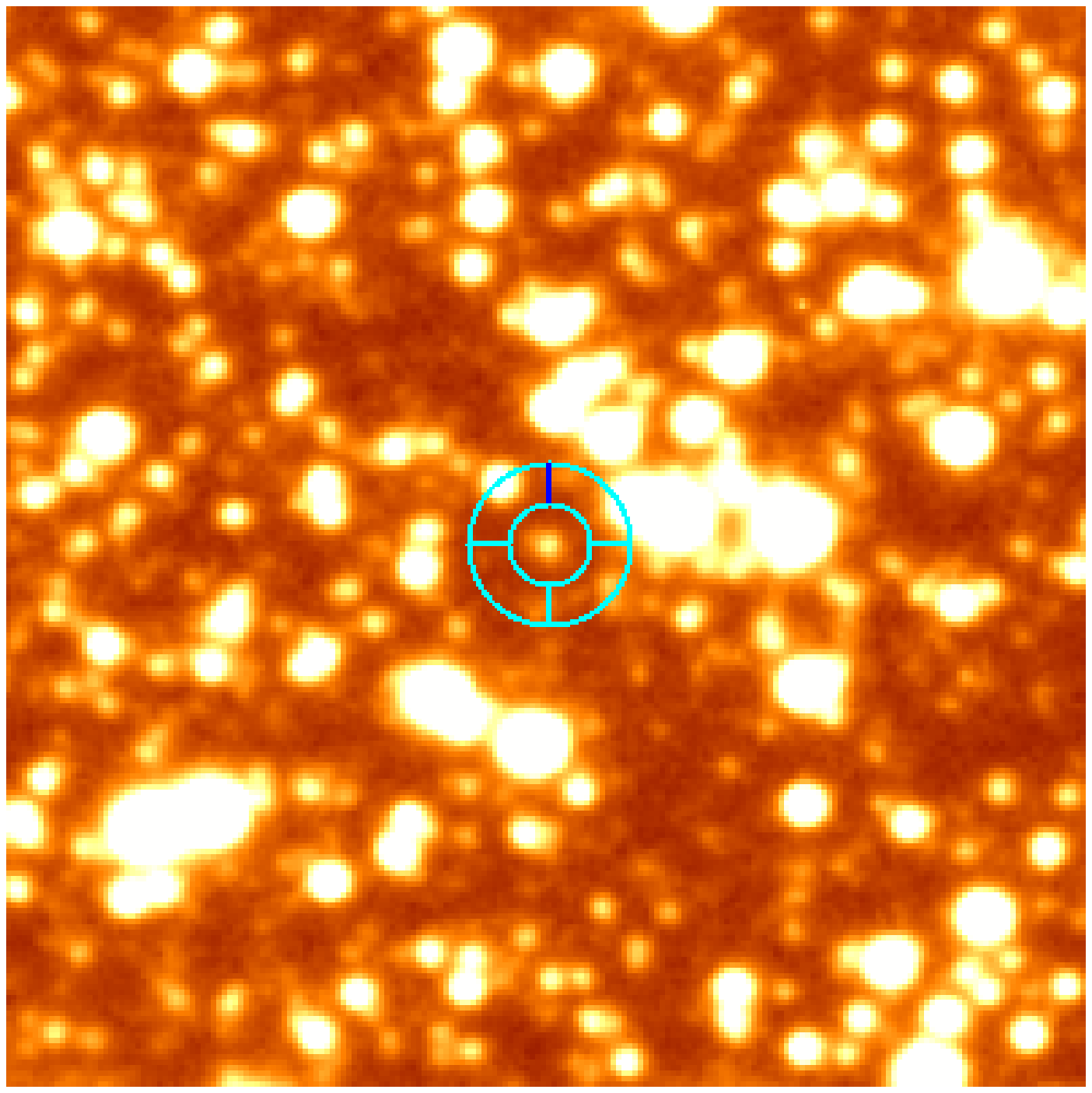}
\end{minipage}
&\begin{minipage}[m]{6.5cm}\includegraphics[width=6cm,bb=0 15 800 400,clip=true]{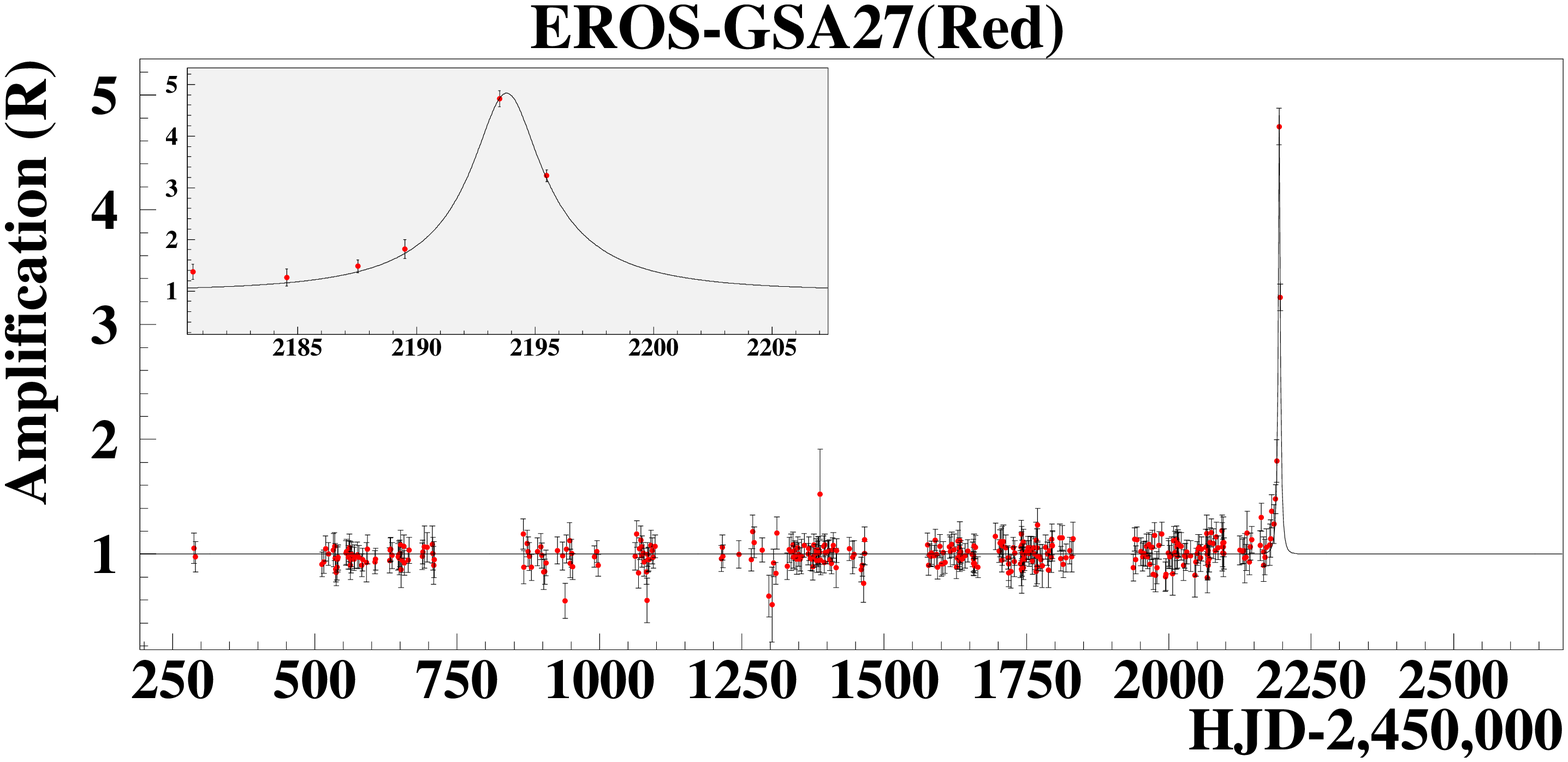}\end{minipage}
&\begin{minipage}[m]{6.5cm}\includegraphics[width=6cm,bb=0 15 800 400,clip=true]{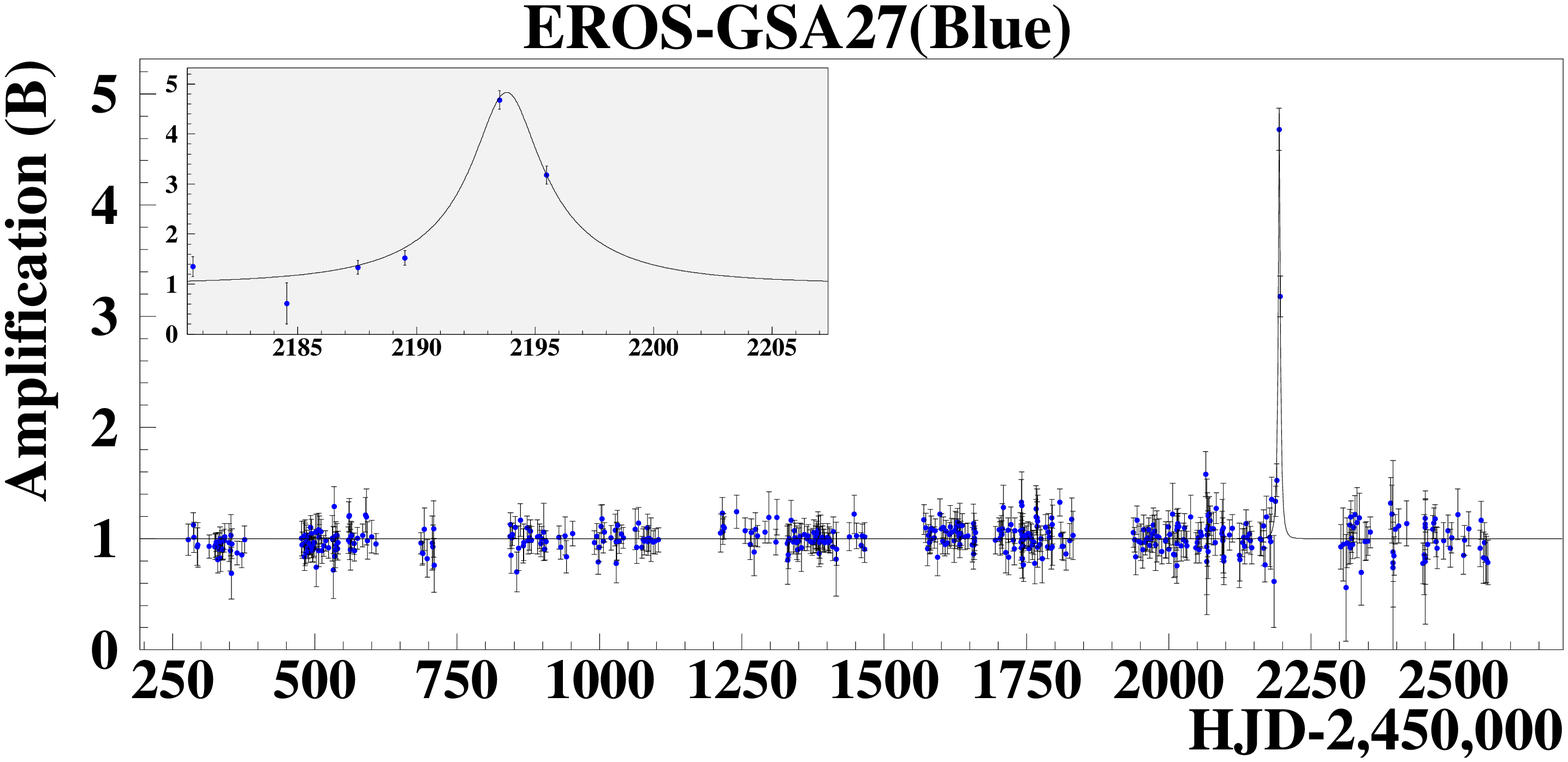}\end{minipage}
\\
\begin{minipage}[m]{3.cm}
\includegraphics[width=3.cm,bb=104 21 509 425]{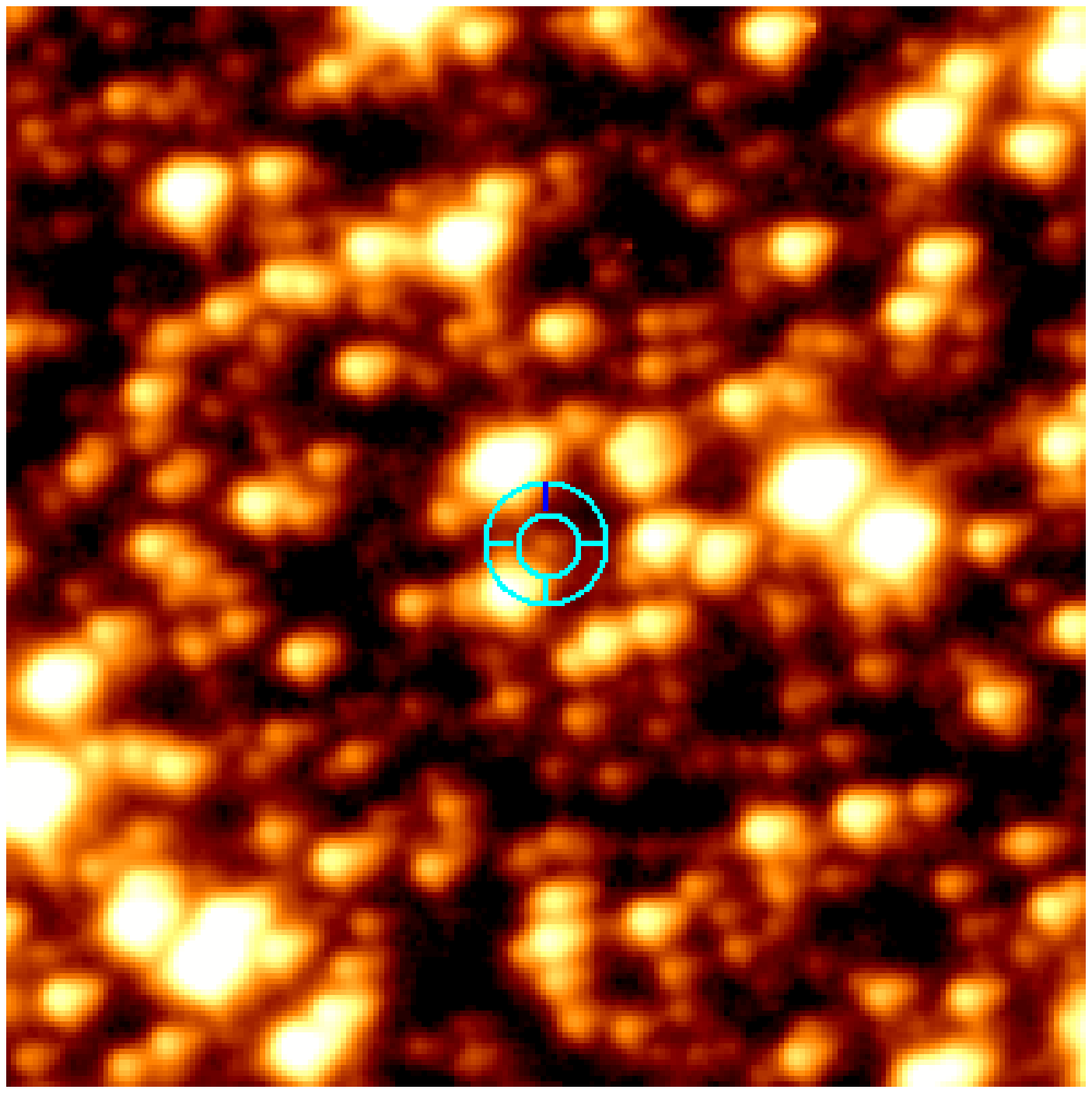}
\end{minipage}
&\begin{minipage}[m]{6.5cm}\includegraphics[width=6cm,bb=0 15 800 400,clip=true]{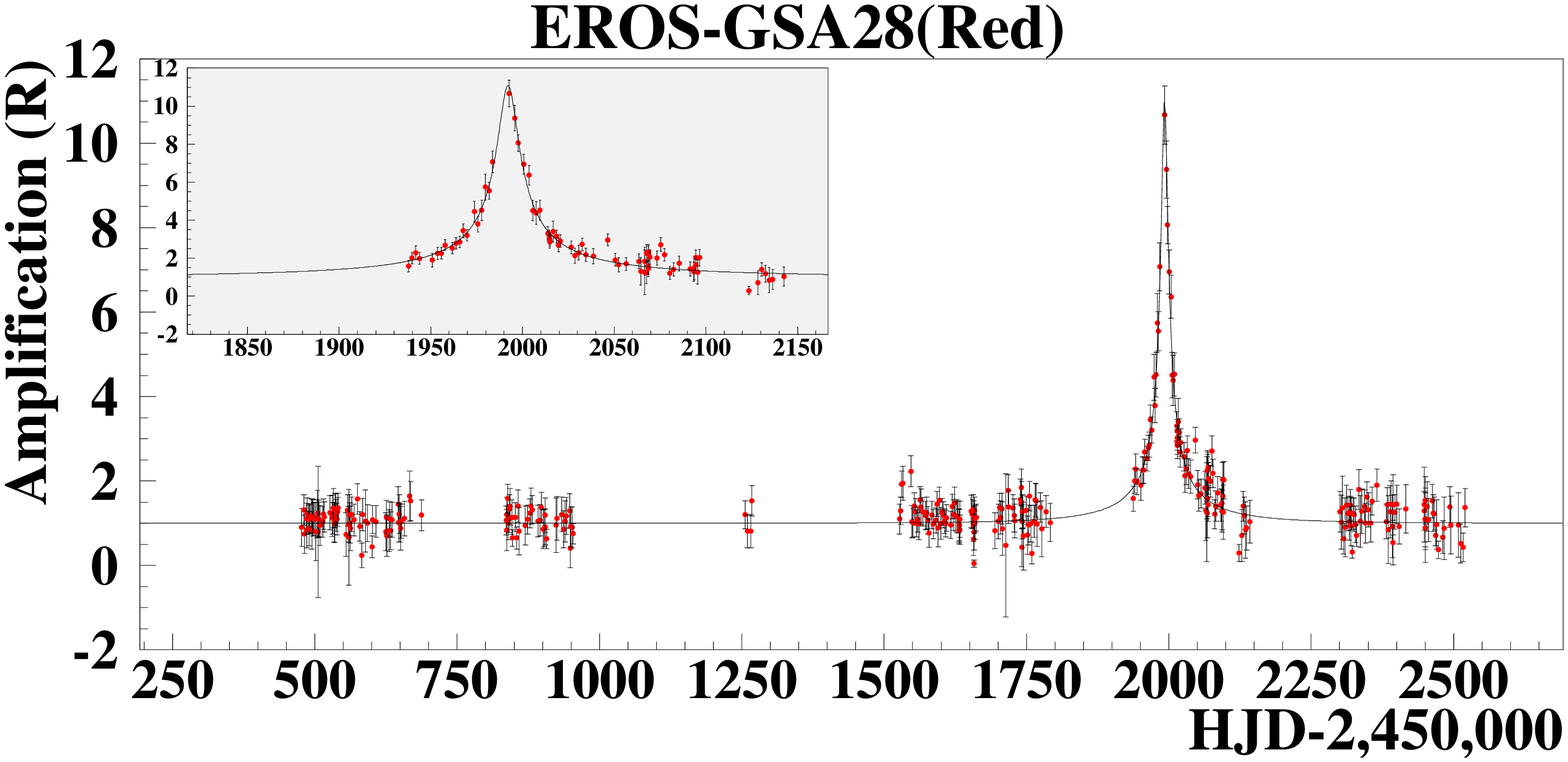}\end{minipage}
&\begin{minipage}[m]{6.5cm}\includegraphics[width=6cm,bb=0 15 800 400,clip=true]{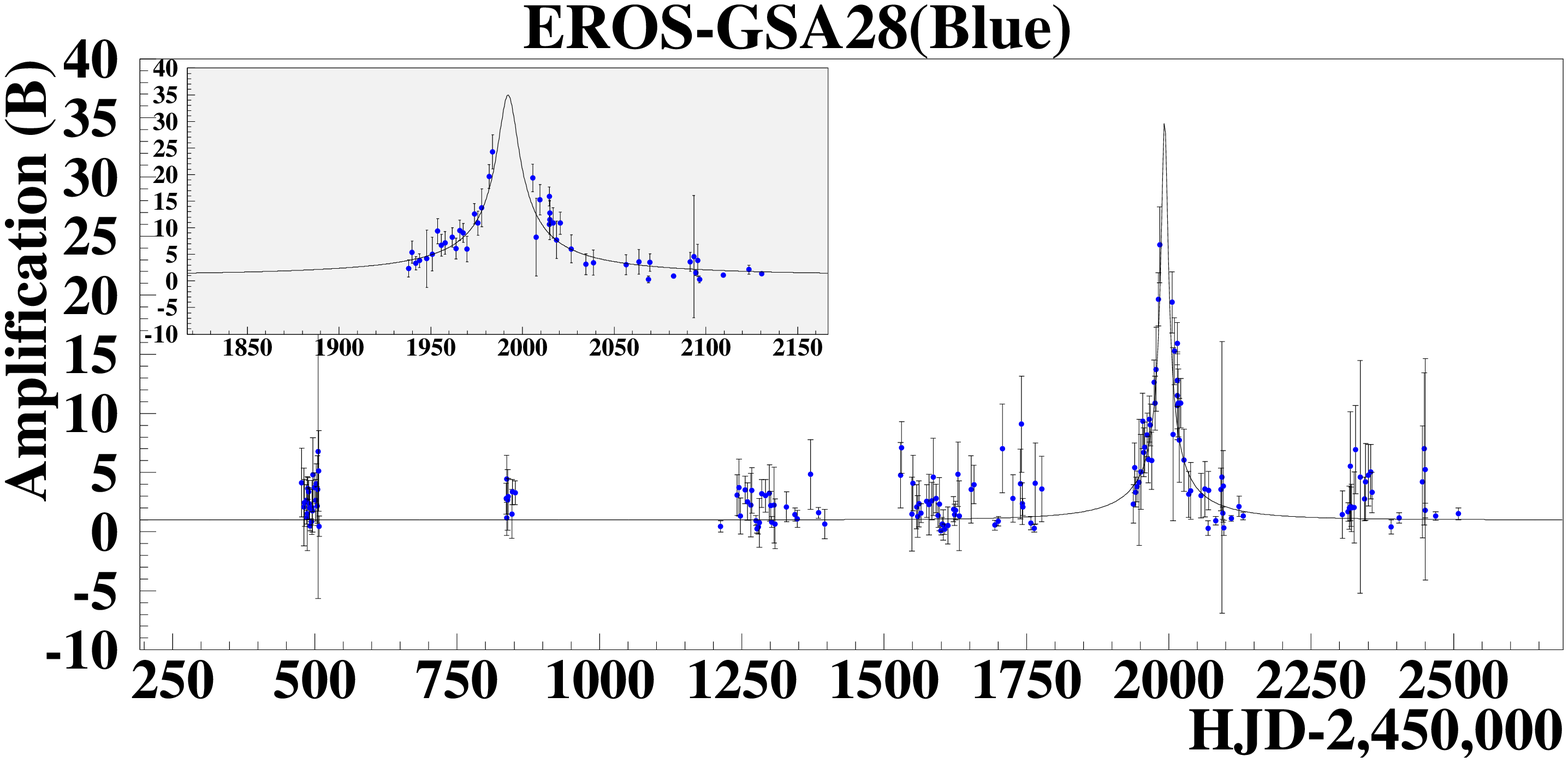}\end{minipage}
\end{tabular}
\caption[]{Light curves and finding charts (continued)}
\label{figclum3}
\end{center}
\end{figure*}

\begin{figure*}
\begin{center}
\begin{tabular}{ccc}
\begin{minipage}[m]{3.cm}
\includegraphics[width=3.cm,bb=104 21 509 425]{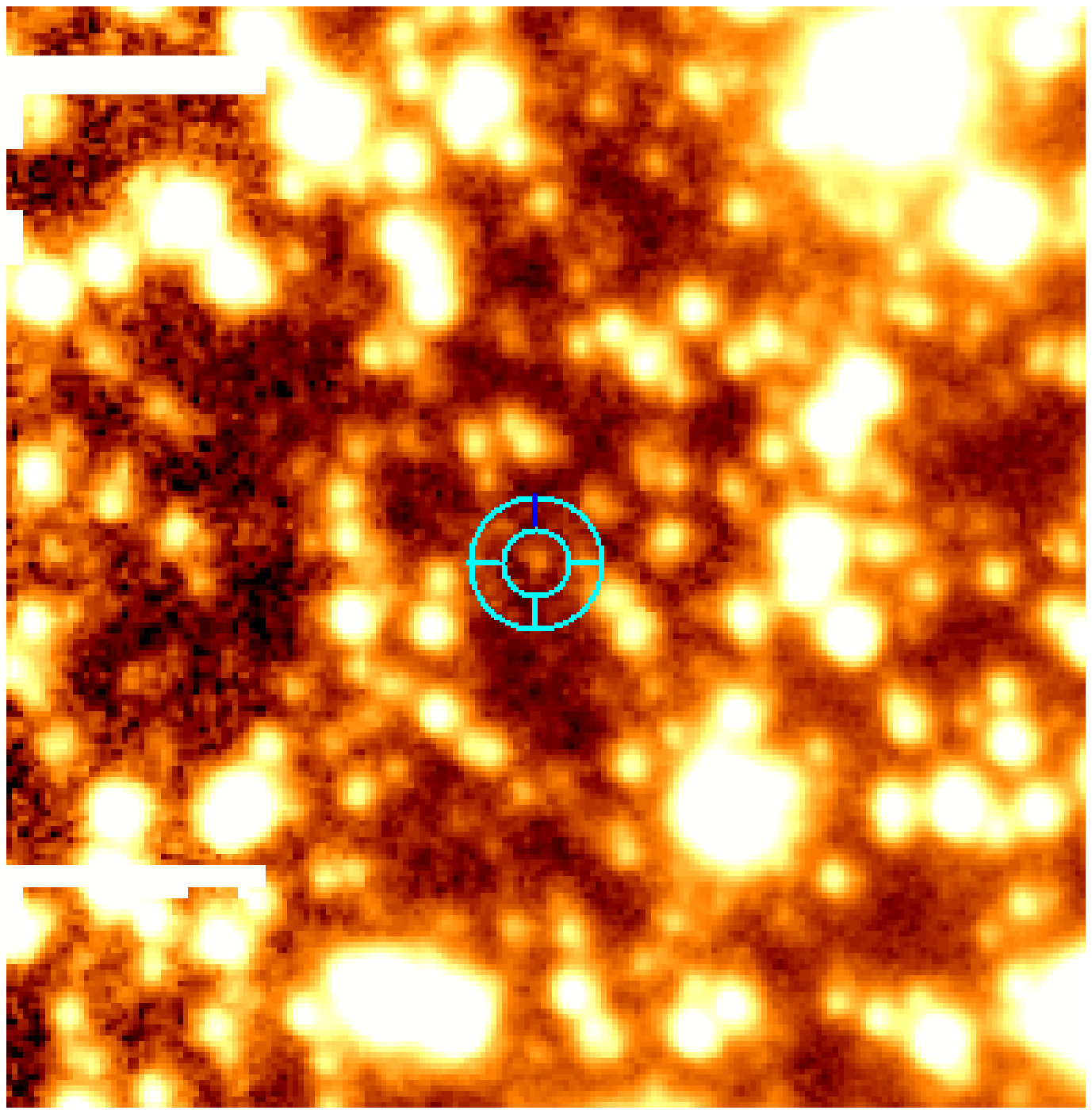}
\end{minipage}
&\begin{minipage}[m]{6.5cm}\includegraphics[width=6cm,bb=0 15 800 400,clip=true]{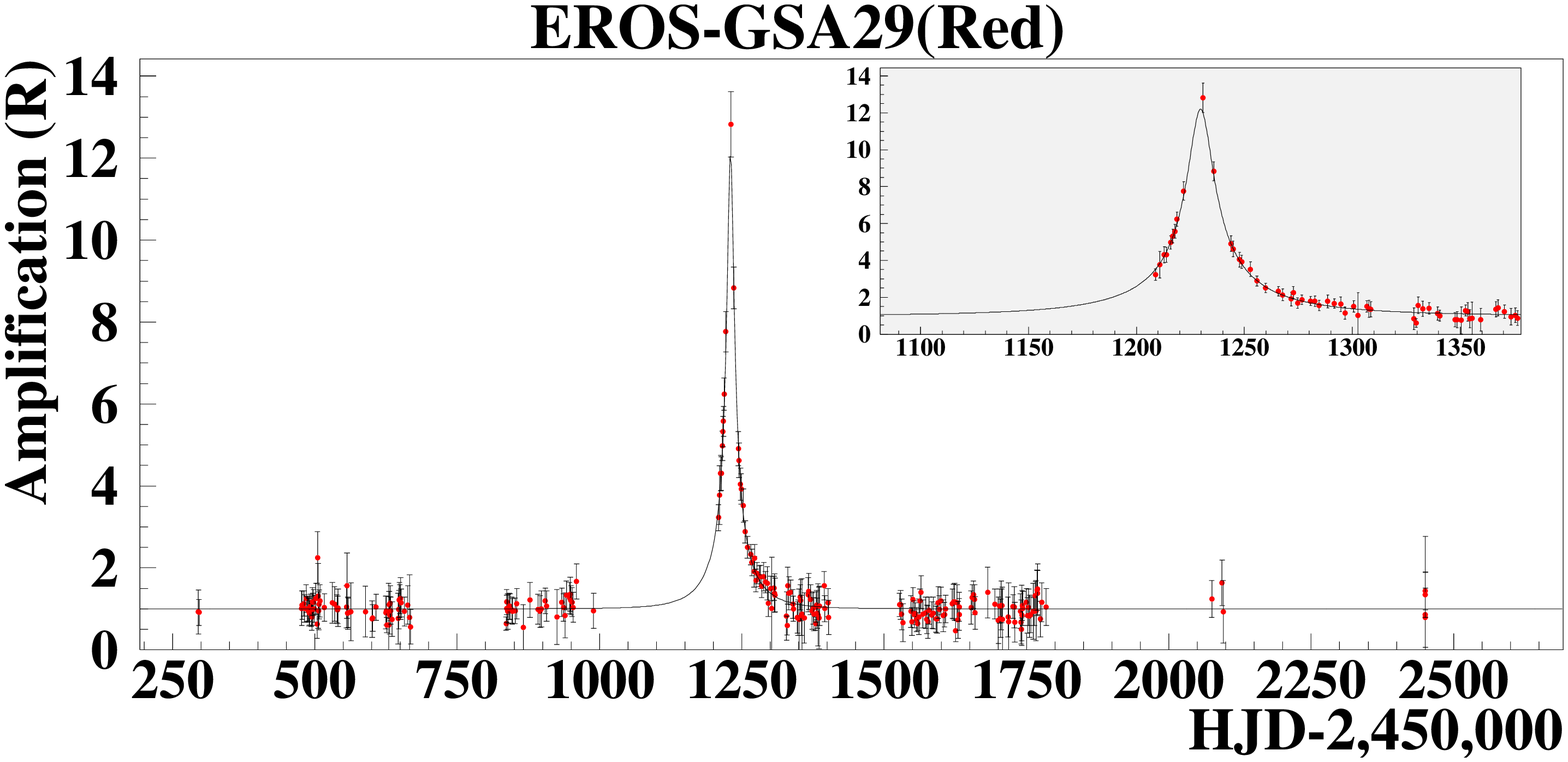}\end{minipage}
&\begin{minipage}[m]{6.5cm}\includegraphics[width=6cm,bb=0 15 800 400,clip=true]{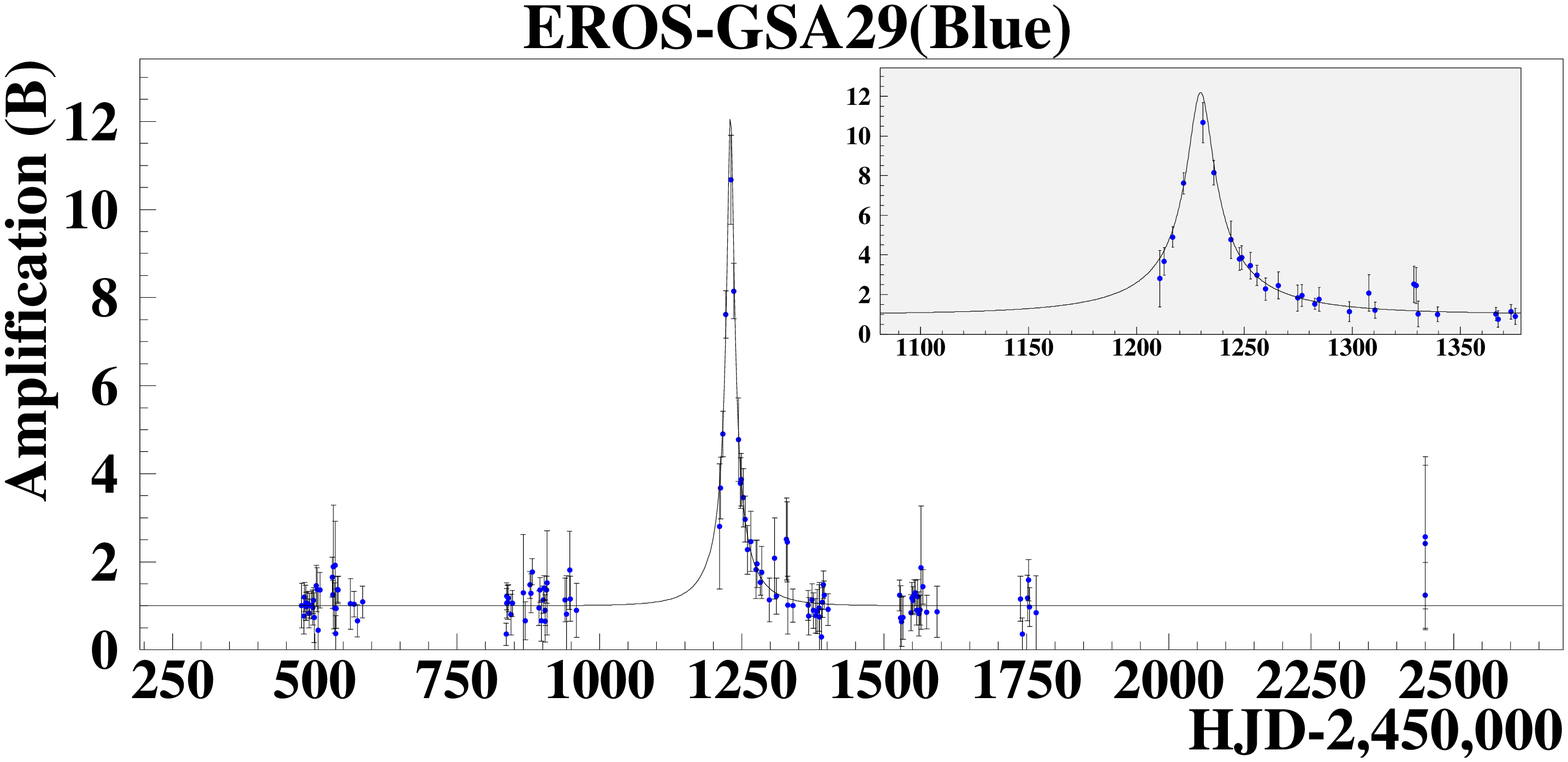}\end{minipage}
\\
\begin{minipage}[m]{3.cm}
\includegraphics[width=3.cm,bb=104 21 509 425]{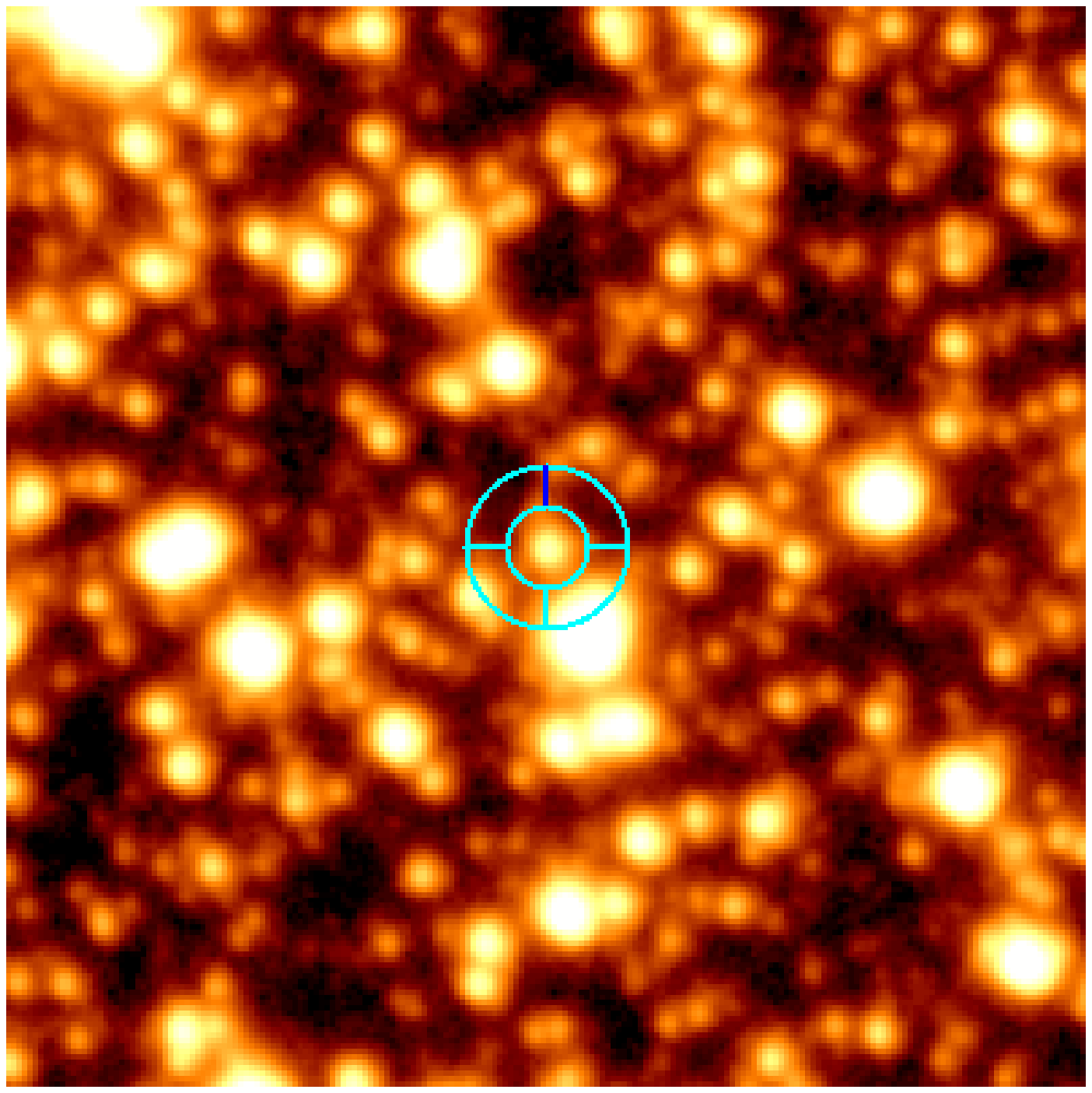}
\end{minipage}
&\begin{minipage}[m]{6.5cm}\includegraphics[width=6cm,bb=0 15 800 400,clip=true]{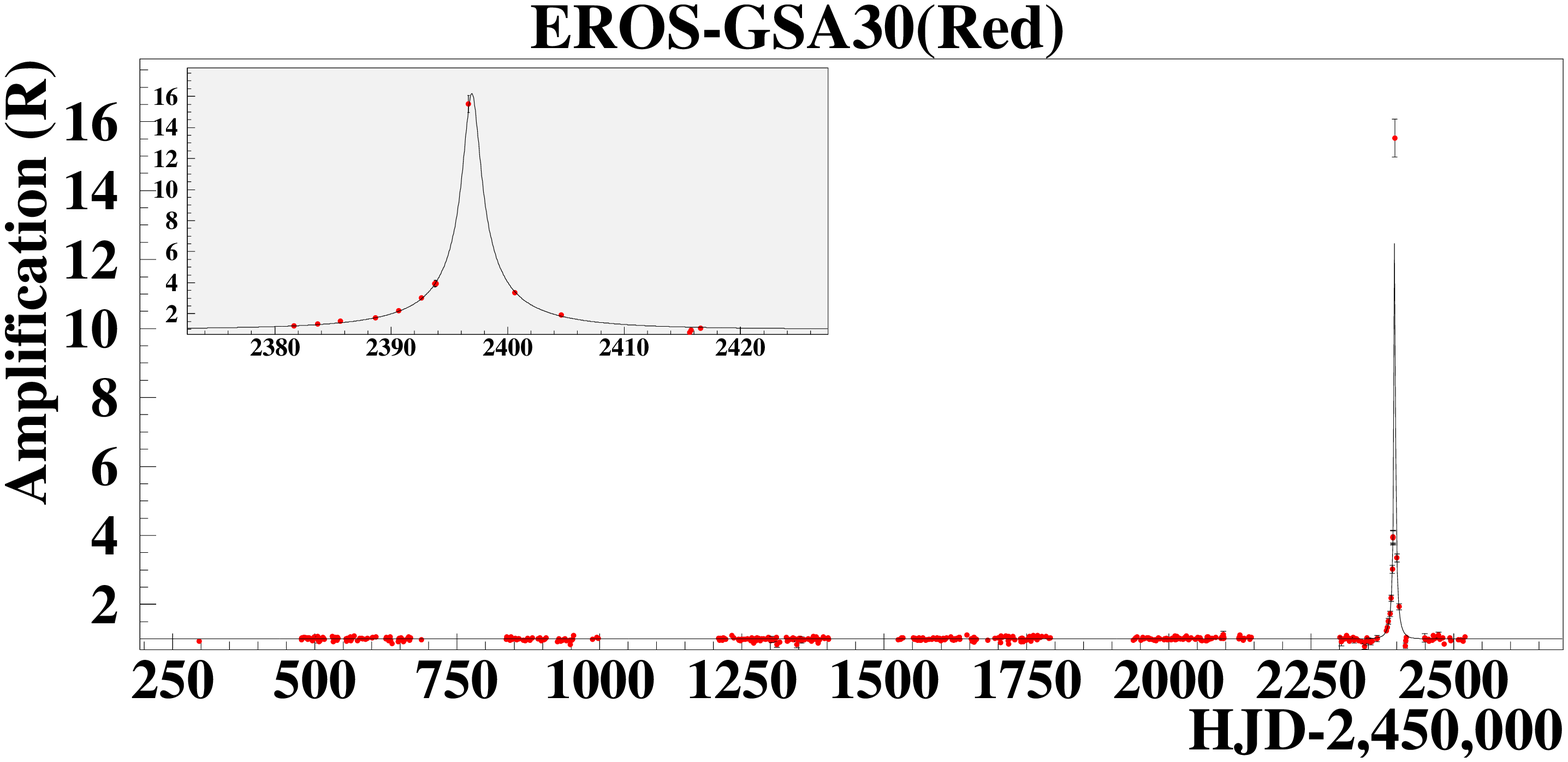}\end{minipage}
&\begin{minipage}[m]{6.5cm}\includegraphics[width=6cm,bb=0 15 800 400,clip=true]{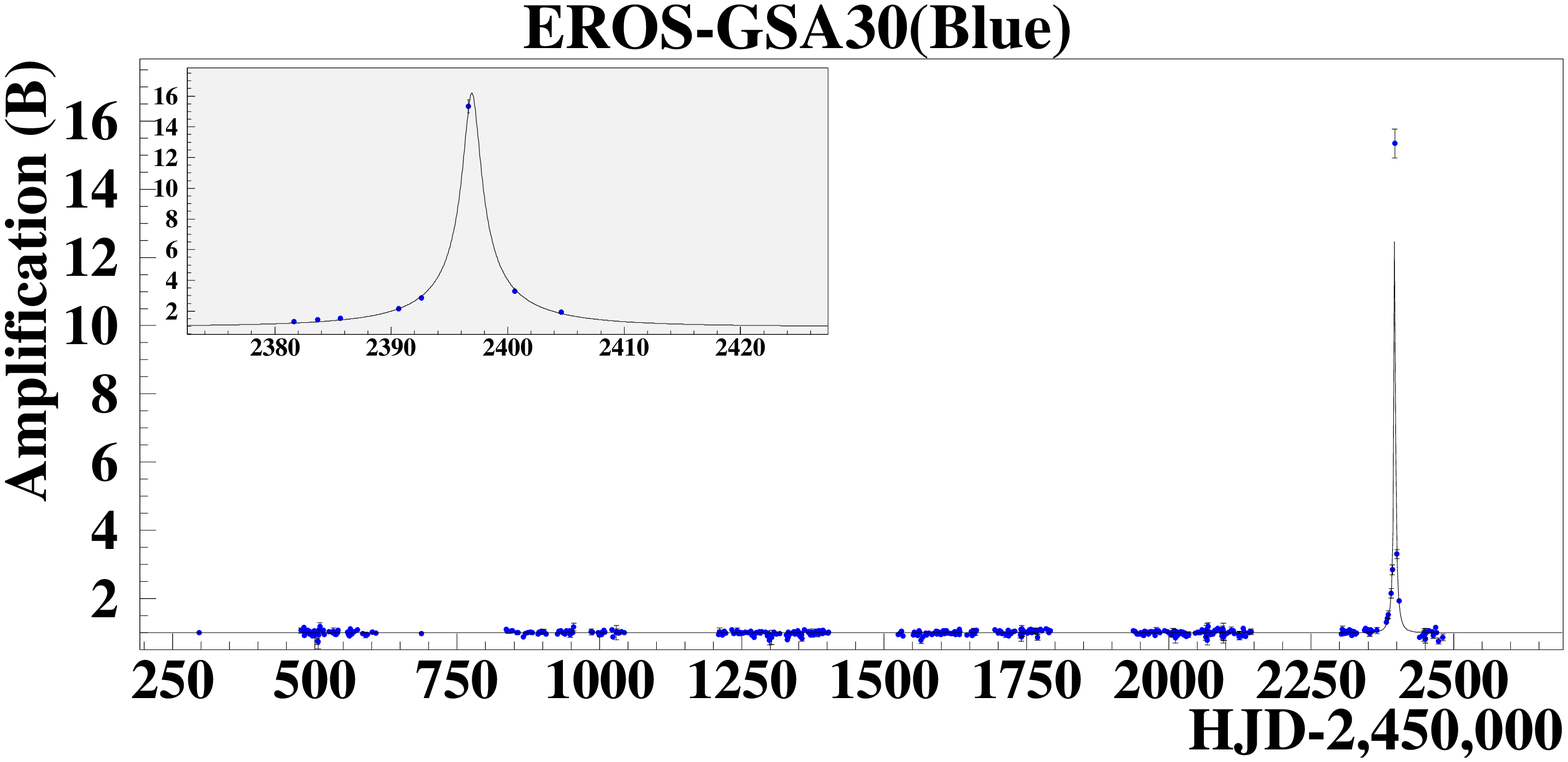}\end{minipage}
\\
\begin{minipage}[m]{3.cm}
\includegraphics[width=3.cm,bb=6.2 1 30.4 26]{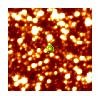}
\end{minipage}
&\begin{minipage}[m]{6.5cm}\includegraphics[width=6cm,bb=0 15 800 400,clip=true]{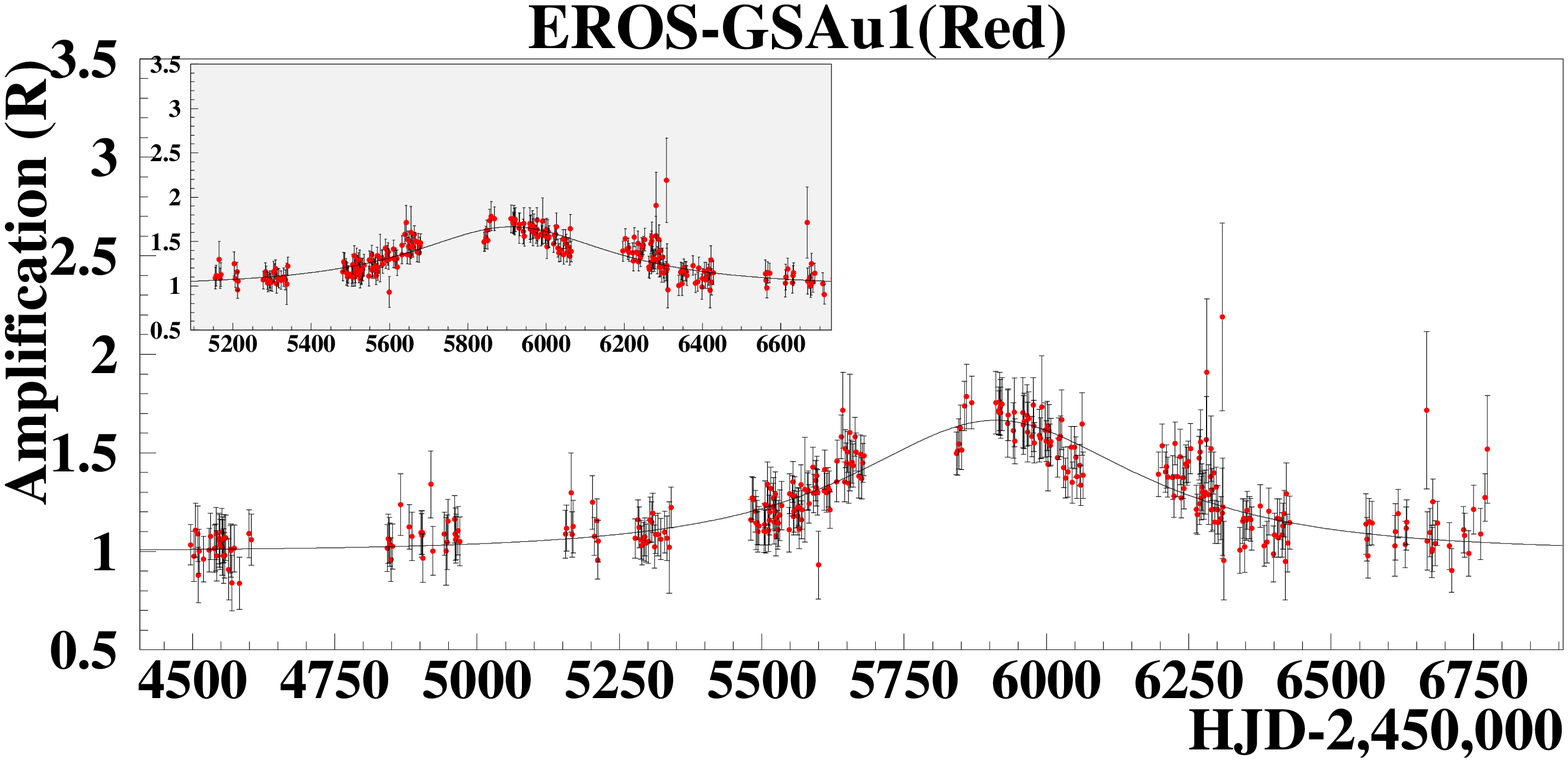}\end{minipage}
&\begin{minipage}[m]{6.5cm}\includegraphics[width=6cm,bb=0 15 800 400,clip=true]{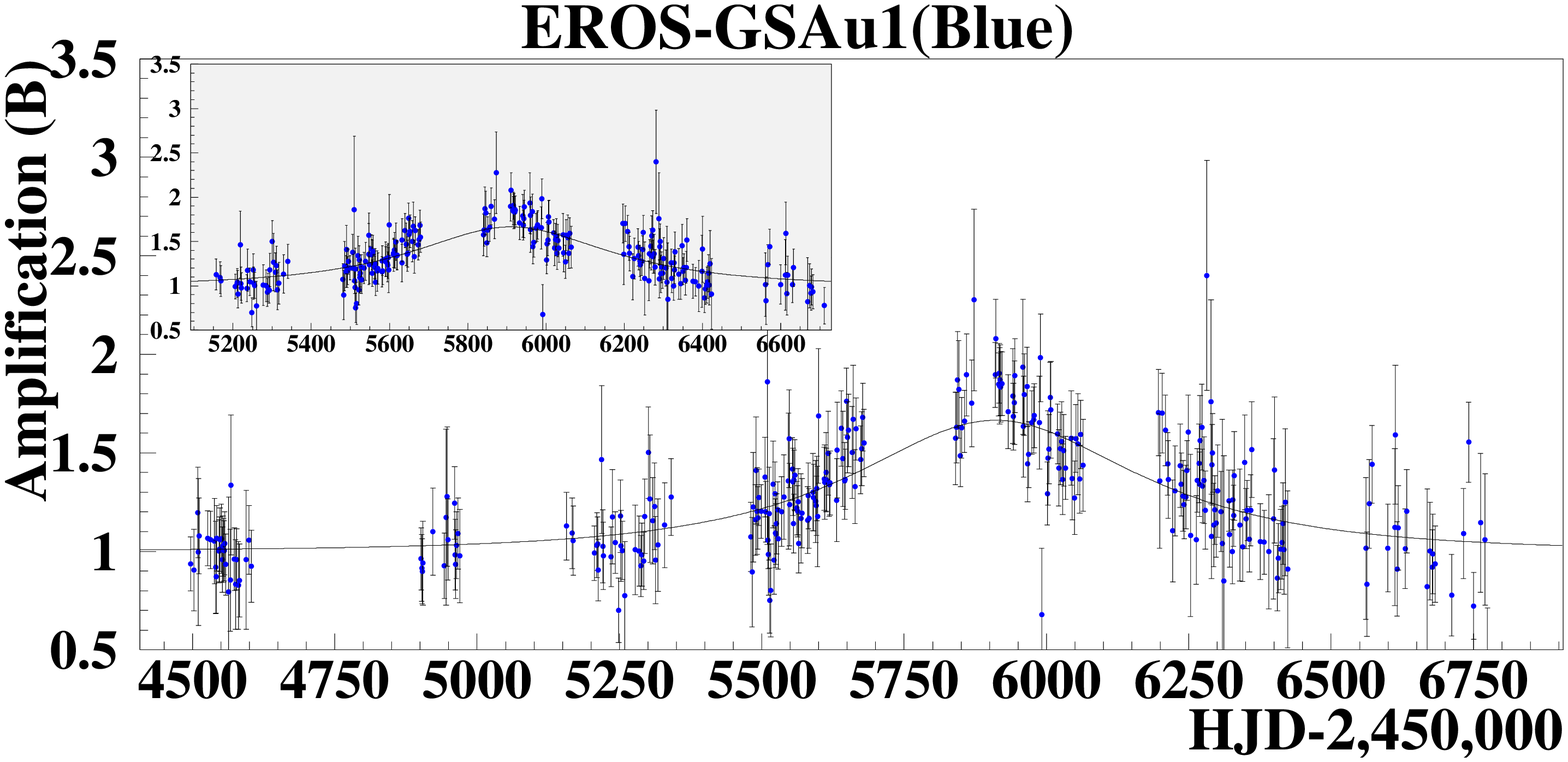}\end{minipage}
\end{tabular}
\caption[]{Light curves and finding charts (continued)}
\label{figclum4}
\end{center}
\end{figure*}


\begin{thebibliography}{}
\bibitem[Afonso {\it et al.} 2003]{SMC5ans}Afonso C., Albert J.-N., Andersen J. {\it et al.} ({\sc EROS} Coll.), 2003, A\&A, 400, 951
\bibitem[Alcock {\it et al.} 1993]{machlmc}Alcock C., Akerlof C.W., Allsman R.A. {\it et al.} ({\sc MACHO} Coll.), 1993, Nat, 365, 621
\bibitem[Alcock {\it et al.} 2000]{macho2000LMC}Alcock C., Allsman R.A., Alves D. {\it et al.}, 2000, ApJ, 542, 281
\bibitem[Ansari 1996]{PEIDA} Ansari R., 1996, Vistas in Astronomy, 40, No 4
\bibitem[Aubourg {\it et al.} 1993]{eroslmc} Aubourg \'E., Bareyre P., Br\'ehin S. {\it et al.} ({\sc EROS} Coll.), 1993, Nat, 365, 623
\bibitem[Bennett 2005]{Bennett} Bennett D. P., 2005, ApJ, 633, 906
\bibitem[Binney {\it et al.} 1997]{Binneyetal} Binney J., Gerhard 0., \& Spergel D., 1997, MNRAS, 288, 365
\bibitem[Bissantz {\it et al.} 1997]{Bissantz} Bissantz N., Englmaier P., Binney J., \& Gerhard 0.E., 1997, MNRAS, 289, 651
\bibitem[Bissantz \& Gerhard 2002]{Bissantz02} Bissantz, N. and Gerhard, O. 2002, MNRAS, 330, 591
\bibitem[Brand \& Blitz 1993]{rotdisc} Brand J., Blitz L., 1993, A\&A, 275, 67
\bibitem[Calchi Novati {\it et al.} 2008]{Calchi} Calchi Novati S., De Luca F., Jetzer Ph. {\it et al.}, 2008, A\&A, 480, 723
\bibitem[Derue {\it et al.} 1999]{GSA2y} Derue F., Afonso C., Alard C. {\it et al.} ({\sc EROS} Coll.), 1999, A\&A, 351, 87
\bibitem[Derue {\it et al.} 2001]{GSA3y} Derue F., Afonso C., Alard C. {\it et al.} ({\sc EROS} Coll.), 2001, A\&A, 373, 126
\bibitem[Derue 1999b]{derueb} Derue F., 1999b, Ph.D. thesis, {\sc CNRS/IN2P3}, {\sc LAL} report 99-14, Universit\'e Paris 11.
\bibitem[Delhaye 1965]{delhaye} Delhaye J. 1965, in Galactic Structure, The University of Chicago Press
\bibitem[Dwek {\it et al.} 1995]{Dwek} Dwek E., Arendt R.G., Hauser M.G. {\it et al.}, 1995, ApJ, 445, 716
\bibitem[Epchtein {\it et al.} 1999]{DENIS} Epchtein N., Deul E., Derriere S. {\it et al.}, 1999, A\&A, 349, 236
\bibitem[Evans \& Belokurov 2002]{evans} Evans N.W. \& Belokurov V., 2002, ApJ, 567, L119
\bibitem[Feldman \& Cousins 1998]{feldman} Feldman G.J. \& Cousins R.D., 1998, Phys.Rev. D, 57, 3873
\bibitem[Freudenreich 1998]{freudenreich} Freudenreich H.T., 1998, ApJ, 492, 495
\bibitem[Gould 1992]{Gould92} Gould, A. 1992, ApJ , 392, 442
\bibitem[Gould, Bahcall \& Flynn 1997]{Gould-1997} Gould A., Bahcall J.N., Flynn C., 1997, ApJ, 482, 913
\bibitem[Grenacher {\it et al.} 1999]{Grenacher-1999} Grenacher L. {\it et al.}, 1999, A\&A, 351, 775
\bibitem[Hamadache {\it et al.} 2006]{Hamadache} Hamadache C., Le Guillou L., Tisserand P. et al., 2006, A\&A, 454, 185
\bibitem[Han \& Gould 1995]{Han-1995} Han C., Gould A., 1995, ApJ, 449, 521
\bibitem[Hardy \& Walker 1995]{Hardy95} Hardy S.J. \& Walker M.A., 1995, MNRAS, 276, L79
\bibitem[ESA 1997]{hipp}The Hipparcos and Tycho Catalogs, ed. M.A.C. Perryman (SP-1200; Noordwijk: ESA)
\bibitem[HST 2002]{HSTarchive}HST archive, https://archive.stsci.edu/
\bibitem[Mao \& Stefano 1995]{mao1995} Mao S., Stefano R.D., 1995, ApJ, 440, 22
\bibitem[M\"ollerach \& Roulet 2002]{Mollerach} M\"ollerach S., Roulet E., 2002, Gravitational lensing and microlensing, World Scientific
\bibitem[OGLE webpage]{webogle} OGLE web page http://www.astrouw.edu.pl/~ogle/ogle2/fields.html
\bibitem[Paczy\'nski 1986]{pacz1986} Paczy\'nski B., 1986, ApJ, 304, 1
\bibitem[Picaud \& Robin 2004]{picaud} Picaud S., Robin A.C., 2004, A\&A, 428, 891
\bibitem[Popowski {\it et al.} 2005]{machobul2005} Popowski, P., Griest, K., Thomas, C. L., {\it et al.}, 2005, ApJ, 631, 879
\bibitem[Rahal 2003]{TheseRahal} Rahal Y. R., 2003, Ph.D. thesis, Universit\'e Paris 6 {\sc LAL-CNRS/IN2P3} report 03-85
\bibitem[Rahvar 2004]{Rahvar} Rahvar, S., 2004, MNRAS, 347, 213
\bibitem[Sumi {\it et al.} 2006]{Sumi2006} Sumi, T., Wozniak, P. R., Udalski, A., {\it et al.} (OGLE coll.), 2006, ApJ, 636, 240
\bibitem[Tisserand 2004]{thesetisserand} Tisserand P., 2004, Ph.D. thesis, Universit\'e de Nice-Sophya Antipolis {\sc CEA} DAPNIA-04-09-T
\bibitem[Tisserand {\it et al.} 2007]{ErosLMCfinal} Tisserand P. {\it et al.},2007, A\&A, 469, 387
\bibitem[Thomas {\it et al.} 2005]{Thomas} Thomas {\it et al}, 2005, ApJ, 631, 906
\bibitem[Turon {\it et al.} 1995]{Turon} Turon C., R\'equi\`eme Y., Grenon M. {\it et al.}, 1995, A\&A 304, 82
\bibitem[Udalski {\it et al.} 1993]{oglpr} Udalski A., Szyma\'nski M., Kaluzny J. {\it et al.} ({\sc OGLE} Coll.), 1993, Act. Astr., 43, 289
\bibitem[Udalski {\it et al.} 2000a]{ogle2000a} Udalski A., Zebrun K., Szyma\'nski M.,{\it et al.} 2000a, Acta. Astr. 50, 1 
\bibitem[Udalski {\it et al.} 2000b]{ogle2000b} Udalski A., Szyma\'nski M., Kubiak M.,{\it et al.} 2000b, Acta. Astr. 50, 307 
\bibitem[Weingartner \& Draine, 2001]{Weingartner}Weingartner J.C., Draine B.T., 2001, ApJ, 548, 296
\end{thebibliography}
\end{document}